%% file: main.tex
\documentclass[a4paper, UKenglish, cleveref, autoref, thm-restate]{lipics-v2021}
\usepackage{multicol,lipsum}
\usepackage{algorithm}
\usepackage{footmisc}
\usepackage{algpseudocode}
\floatname{algorithm}{Algorithm}

\pdfoutput=1 
\hideLIPIcs  

\title{Greedy Matroid Algorithm And Computational Persistent Homology} 

\author{Tianyi Sun and Bradley Nelson}{Department of Statistics, Committee on Computational and Applied Mathematics\\The University of Chicago}{}{}{}

\authorrunning{T. Sun and B. Nelson} 

\Copyright{Tianyi Sun and Bradley Nelson} 



\nolinenumbers 

\EventEditors{John Q. Open and Joan R. Access}
\EventNoEds{2}
\EventLongTitle{42nd Conference on Very Important Topics (CVIT 2016)}
\EventShortTitle{CVIT 2016}
\EventAcronym{CVIT}
\EventYear{2016}
\EventDate{December 24--27, 2016}
\EventLocation{Little Whinging, United Kingdom}
\EventLogo{}
\SeriesVolume{42}
\ArticleNo{23}

\begin{document}

\maketitle

\begin{abstract}
An important problem in computational topology is to calculate the homology of a space from samples. In this work, we develop a statistical approach to this problem by calculating the expected rank of an induced map on homology from a sub-sample to the full space. We develop a greedy matroid algorithm for finding an optimal basis for the image of the induced map, and investigate the relationship between this algorithm and the probability of sampling vectors in the image of the induced map.
\end{abstract}

\input{01intro}
\input{02relatedwork}
\input{03background}
\input{04gma}

\input{05_1experiment}

\input{05_2discussion}
\input{06conclusion}
\newpage
\newpage
\bibliographystyle{plainurl}
\bibliography{lipics-v2021-sample-article}
\input{07appendix} 
\input{08expectedrank}
\end{document}

%% file: 01intro.tex
\section{Introduction}\label{chap:introduction}
A fundamental problem in topological data analysis(TDA) is to estimate the homology of a space from sub-samples. This problem has been studied in the past few decades\cite{edmonds1971matroids, lawler2001combinatorial, borodin2003incremental, henselman2016matroid}.  
This problem could be separated into two steps. 
One is to determine the number of sub-samples required to see the same homology as the full point cloud. 
The other is to determine the size of each sub-sample where the most important homology of full-space can be obtained. 
Estimating homology of a space from samples could be described as determining topological features of a point cloud from sub-samples. 
Computing the homology of a space is useful in various applications such as clustering and cluster analysis\cite{brown2023clustering, 0f7b0bbf, cuevas2001cluster, muller1992excess, muller1991excess, polonik1995measuring}, pattern recognition\cite{devroye1980detection, cuevas1990pattern}, econometrics\cite{tulkens2006public, deprins1983farrell, farrell1957measurement}, text\cite{khasawneh2014stability, rieck2014sam}, and image\cite{fu2022unsupervised, clough2020topological}. Some method has been developed to improve efficiency of topological tools in the setting of cellular\cite{henselman2016matroid}.

A variety of solutions have been proposed to homology estimation. Classical solutions include topological bootstrapping\cite{GS}, matroid filtrations\cite{henselman2016matroid}, Morse theory\cite{mischaikow2013morse}, and estimation of random closed sets\cite{molchanov1990empirical}. Hausdorff distance is the distance measure used to evaluate the performance of these estimators. 
In this paper we wish to study an estimation approach from a matroid perspective. Rather than achieving an accurate estimation for the actual shape of a point cloud, we wish to recover a comprehensive topological properties, i.e., loops and holes. Our observations, Figure \ref{fig:annulus_100_im} and \ref{fig:annulus_300_im},
have shown that minimizing the Hausdorff distance does not guarantee the quality of the homology estimator. 

The motivation for studying the topology of a space comes from the clustering problem. 
From a topological perspective, clustering can be viewed as a question about the homology of the space. The homology of a topological space $X$ is a set of Abelian groups, denoted by ${H_0(X), H_1(X),...}$, where the elements of $H_0(X)$ contain information about the connected components of $X$, and for $k > 0$, the group elements of $H_k(X)$ contain information about ``cycles'' or ``holes'' of different dimensions. 
From the perspective of algebraic topology, the clustering problem is equivalent to recovering $H_0(X)$ where $X$ is a set of point. 
A statistical perspective of the recent efforts in TDA has been to extract topological invariants and homology from random data. 
The idea is that these topological summaries are useful for statistical inference and robust under various transformations. Our goal is to examine homology estimation when the objective is to recover $H_1(X)$.

A classical implication for estimate the homology of a space is recovering the homology of a manifold from a noisy sample and inference the persistent homology of a function.  This problem was previously studied in \cite{pmlr-v22-balakrishnan12a, Niyogi2011ATV}.  Real world point clouds can be high-dimensional and irregularly structured, for example, GPT-$2$ output texts point cloud in Figure \ref{fig:wordcloud}.  A standard approach is computing Vietoris–Rips persistence diagram and persistence barcode on full space.  However, for point cloud with large size this approach is too expensive to compute.

The main objective of this paper is to provide a consistent method for estimating the homology of a space from samples. A standard approach is topological bootstrapping\cite{GS}. The problem with this approach is that due to the discrete nature of homology even a tiny error in the set estimate can introduce an ambiguity in homology. For example, an infinitesimally small region included by mistake can increase the number of components, while a small region excluded by mistake might introduce a hole. Such errors in homology estimation may occur no matter how small the extraneous components and holes are\cite{Bobrowski_2017}. This problem is illustrated in Figure \ref{fig:annulus_100_im}.

The main result in this work is a robust homology estimation approach of noisy point clouds. We consider the induced maps between nested pairs of samples, where the thresholds of Hausdorff distance are fixed $\epsilon>0$. The key object of interest is the induced maps from sub-samples to full space: 
$$\mathbf{IM_n}: \mathbf{C}(X_n) \xrightarrow{\epsilon} \mathbf{C}(X_N),$$
where $n$ is the size of sub-samples and $N$ is the size of full space, $\mathbf{C}$ is simplicial complex, i.e., Rips complex or chain complex, and $\mathbf{IM}$ is induced map. 
Inference of the homology at a single sub-sample is noisy, however the maps $\mathbf{IM_n}$ serves as a filter for the homological noise. \cite{Bobrowski_2017} showed that the image of this map is isomorphic to the homology of the sub-sample.

\begin{figure}
\centering
    \includegraphics[width=0.245\textwidth]{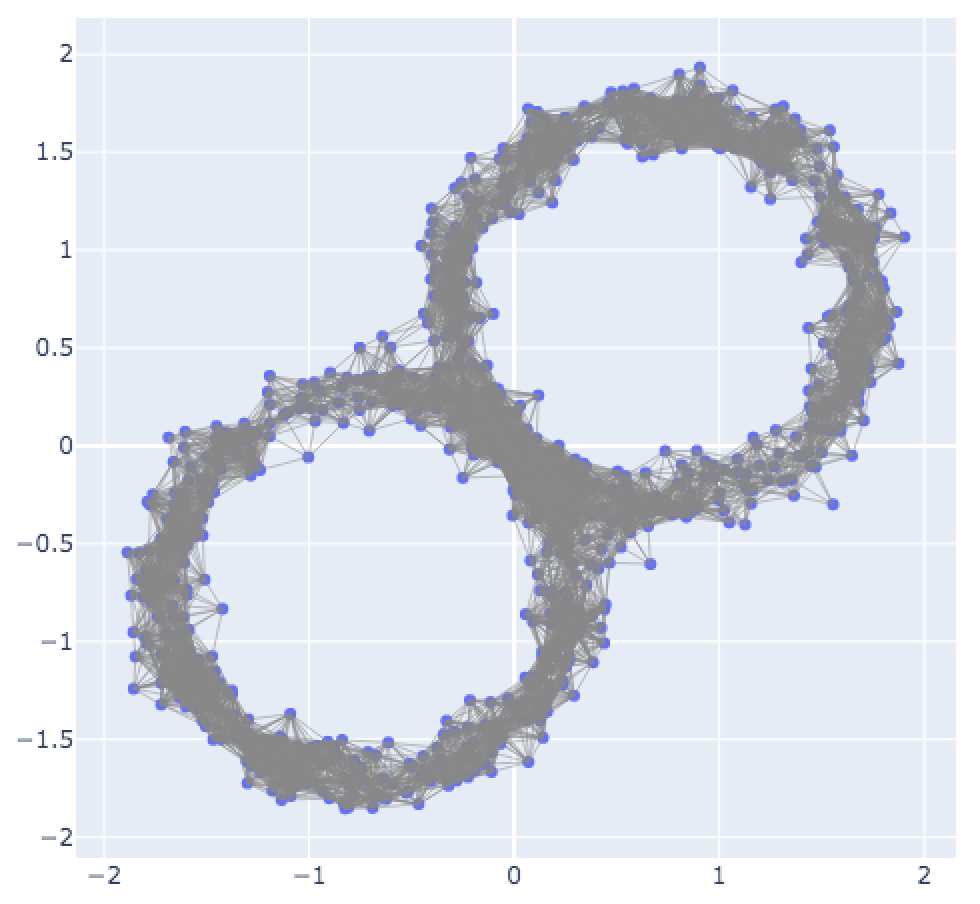}
    \includegraphics[width=0.24\textwidth]{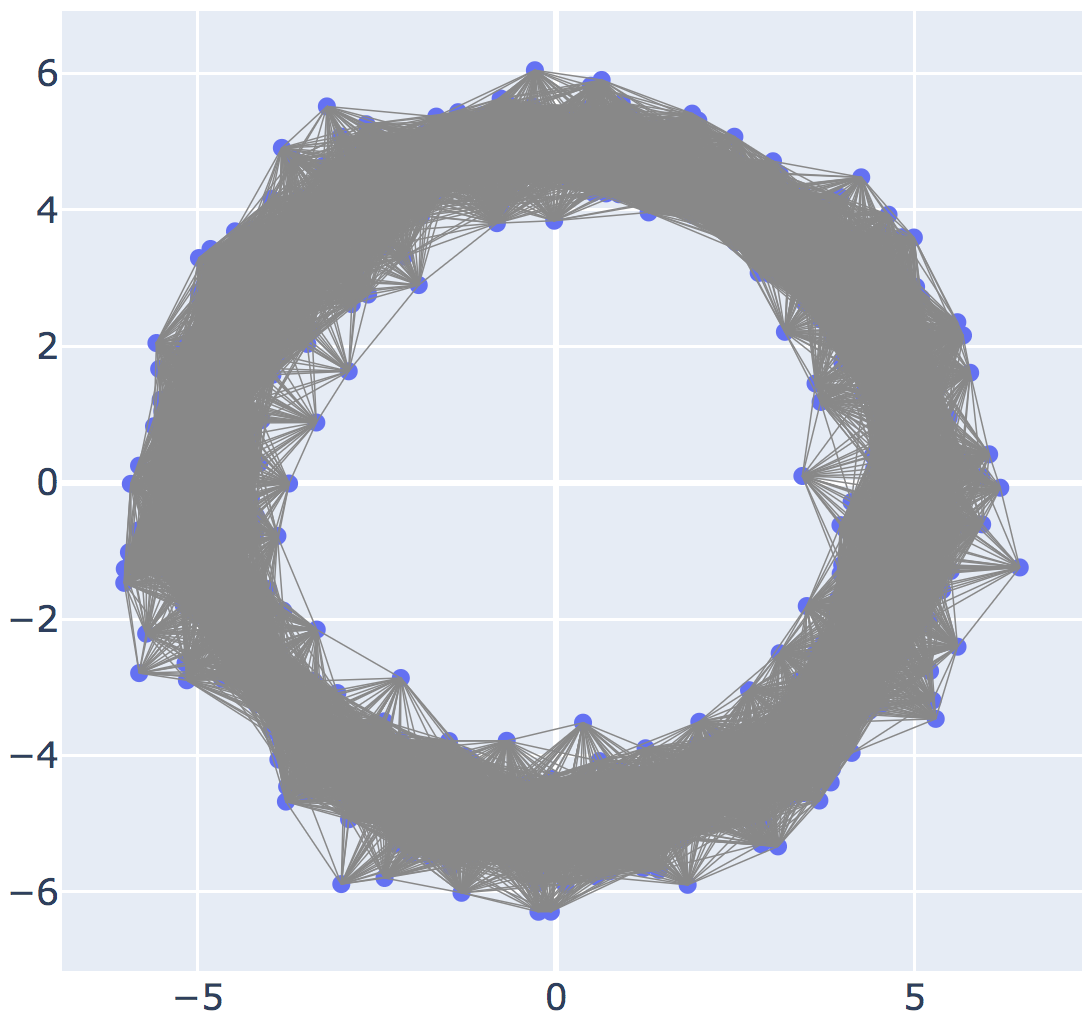}
    \includegraphics[width=0.25\textwidth]{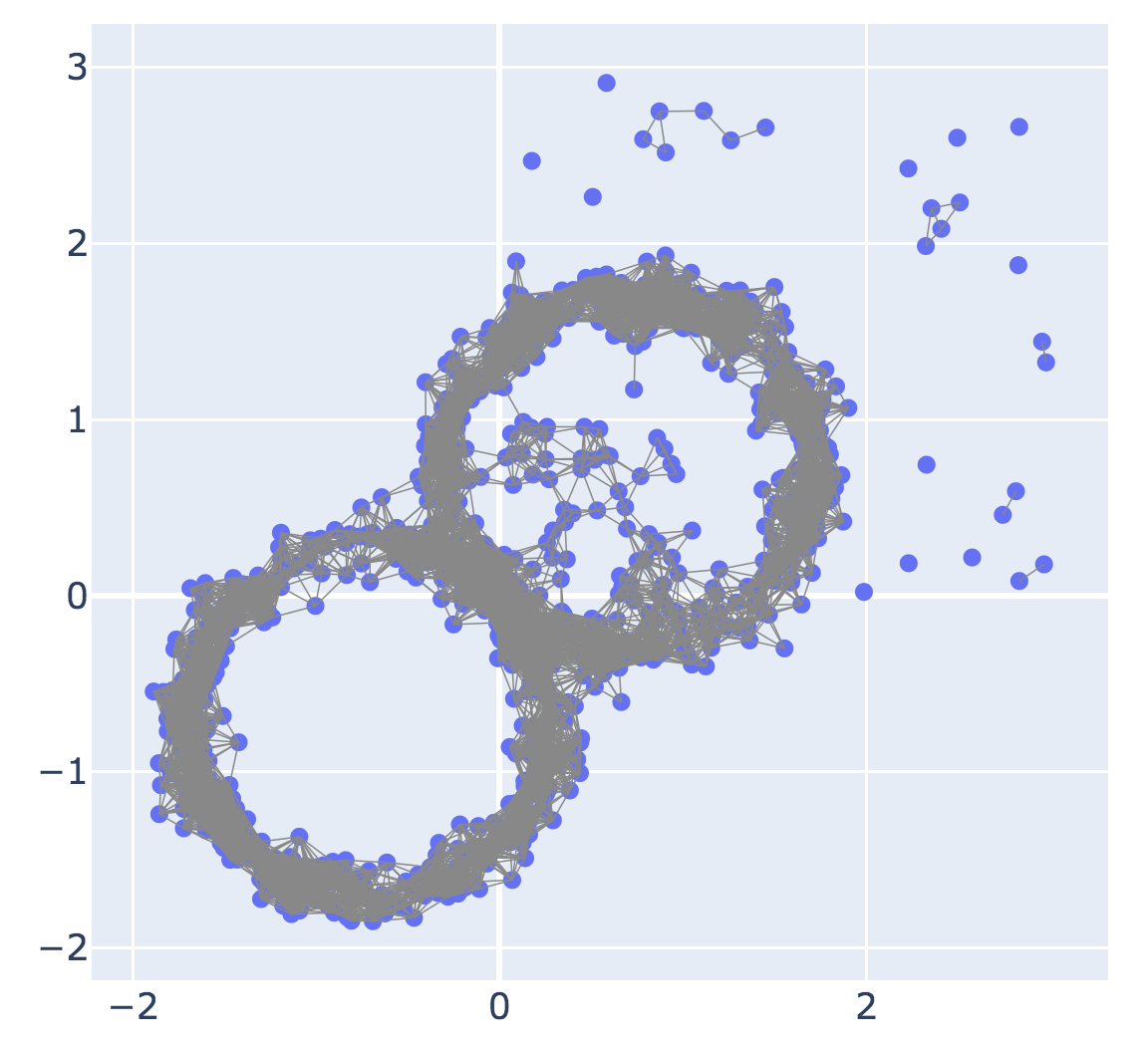}
    \includegraphics[width=0.24\textwidth]{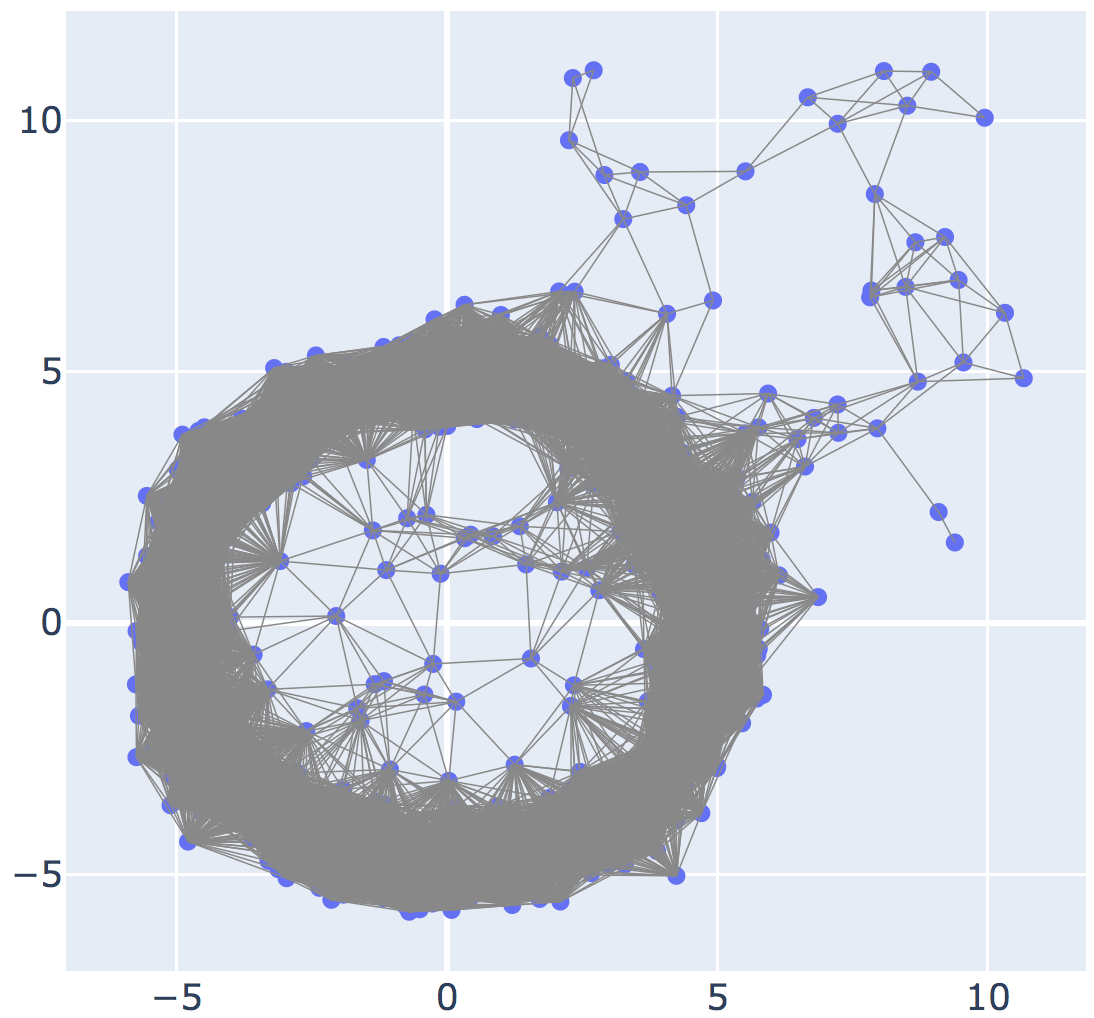}
    \caption{Noisy constructions of Figure-$8$ and Annulus. We compute the Vietoris-Rips complex, where the distance threshold for rips complex is $0.25 + 2*$Hausdorff distance. Point cloud size is 1000 for each construction. We observe that the homology bases of sub-samples of nearly prefect constructions are similar, however, the homology bases of sub-samples of noisy constructions are significantly varied.}
    \label{fig:constructions}
\end{figure}

In this paper, we develop a consistent method to (1) discover the many of sub-samples needed and (2) the size of these sub-samples to address this problem. 
The main idea is a new version of topological bootstrapping\cite{GS}. 
The greedy matroid algorithm is described in Algorithm \ref{alg:gma}. 
In this work, we use two noisy classical constructions in TDA (Figure \ref{fig:constructions}), Figure-8, and Annulus, to illustrate the idea.
The result stated in Section \ref{chap:experiments} shows that this method is more robust than classical solutions in the following aspects:
\begin{itemize}
    \item Greedy matroid algorithm is computationally more efficient than algorithms of computing persistence homology, i.e., Vietoris-Rips persistence diagram and persistence barcode. 
    \item The homology basis computed by our greedy matroid algorithm \ref{alg:gma} gives a comprehensive indication of the basic construction of a given point cloud, which effectively addresses the ambiguity and the problem of missing information in the original topological bootstrapping\cite{GS}.
\end{itemize}

The paper is structured as follows.
In Section \ref{chap:relatedworks}, we state relevant works to this project. 
In Section \ref{chap:background}, we state the topological concepts and necessary definitions used in this paper.
Experiments and results are illustrated in Section \ref{chap:experiments}. 
In Section \ref{chap:alg}, we introduce our method and greedy matroid algorithm. 
We discuss our results in Section \ref{chap:discussion}.
We finally close with a conclusion in Section \ref{chap:conclusion}.
Code and additional experiments are available in Appendix \ref{appendix}.

%% file: 02relatedwork.tex
\section{Related Works}\label{chap:relatedworks}
In this section, we review relevant works for solving the fundamental problem, namely computations of persistence homology and topological bootstrapping.
\subsection{Persistence Homology}
A classical approach is directly computing the persistence homology of the full point cloud using Vietoris-Rips persistence diagram (Figure \ref{fig:rips300}) and persistence barcode.
However, for a point cloud with more than 1000 points, computing persistence homology using Vietoris-Rips persistence diagram and persistence barcode is expensive. Other simplicial complexes, i.e., Cech complex and Delaunay complex, could be an alternative, however, they are even more expensive to be used to compute. 
\begin{definition}[Vietoris-Rips complex\cite{10.1145/160985.161139}]\label{v-rips} Let $X = \{x_1, x_2, ..., x_n\}$ be a collection of points in $\mathbb{R}^d$, and let $r\geq 0.$ The Vietoris-Rips complex $R(X, r)$ is constructed as follows: 
\begin{itemize}
    \item The $0$-simplices (vectors) are the points in $X$.
    \item A $k$-simplex $[x_{i_0}, ..., x_{i_k}]$ is in $R(X, r)$ if $\|x_{i_j} - x_{i_l}\| \leq r$ for all $0\leq j,$ and $l\leq k$. 
\end{itemize}
\end{definition}
\subsection{Topological Bootstrapping}
\begin{figure}
    \centering
    \begin{subfigure}{0.38\linewidth}
    \includegraphics[width=\linewidth]{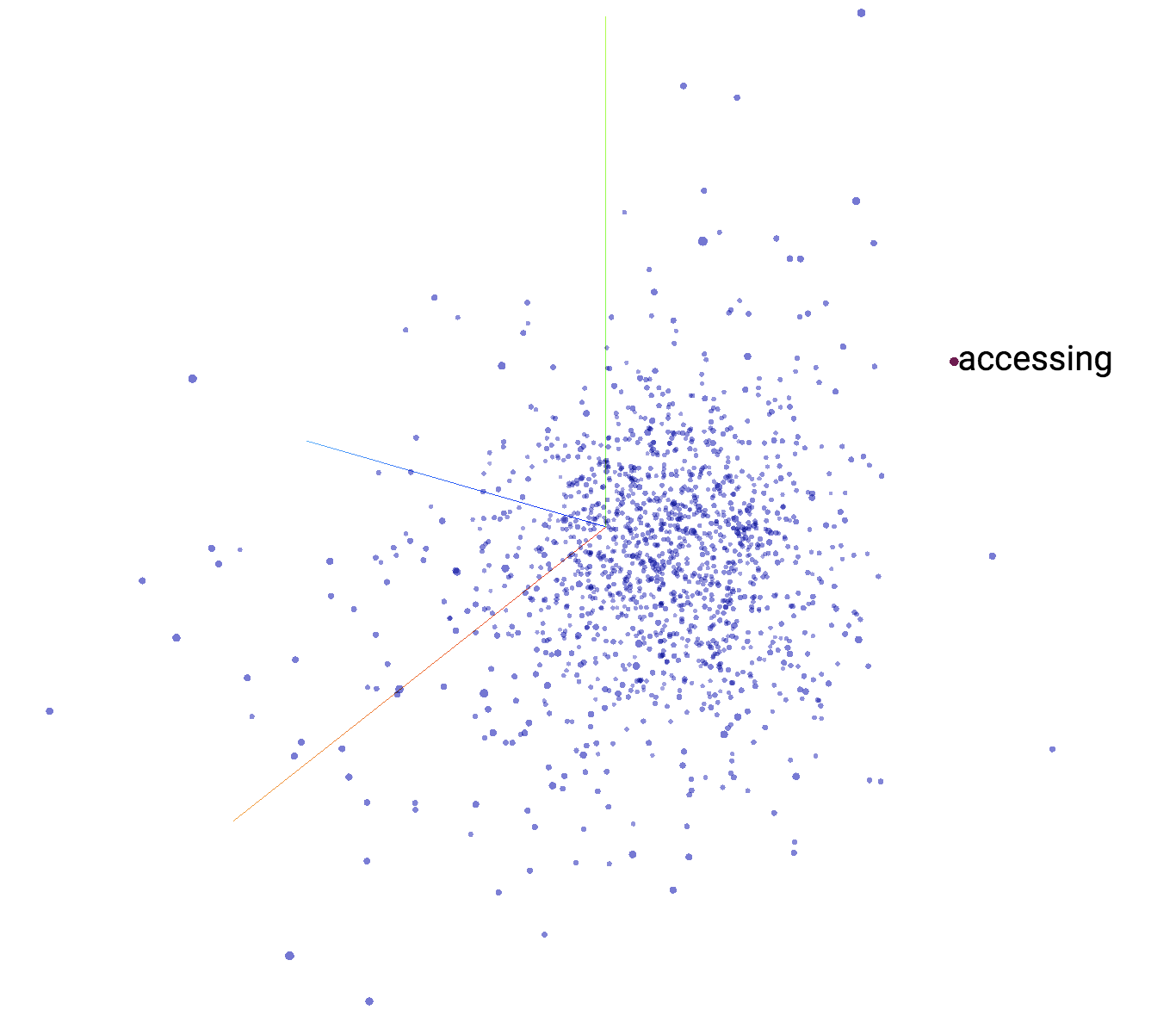}
    \caption{A point cloud of words of the first $100$ texts in GPT-$2$ outputs dataset\cite{gpt2dataset} using Word2Vec\cite{Mikolov2013EfficientEO}. The dimension of a word vector is $128$, we reduced the dimension to $3$ by principle component analysis.}
    \label{fig:wordcloud}
    \end{subfigure}
    \begin{subfigure}{0.3\linewidth}
    \includegraphics[width=\linewidth]{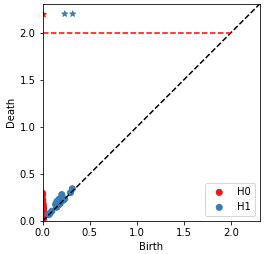}
    \caption{Vietoris - Rips persistence diagram of Figure-8.}
    \label{fig:rips300}
    \end{subfigure}
    \begin{subfigure}{0.3\linewidth}
    \includegraphics[width=\linewidth]{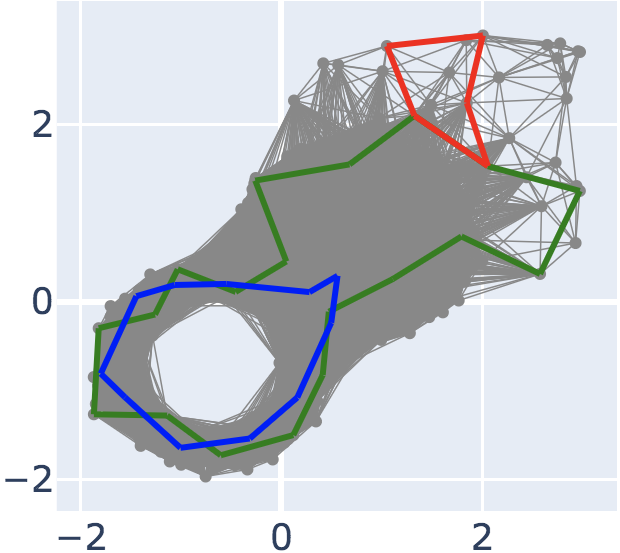}
    \caption{Noisy Figure-8. Red, blue, and green lines are detected representatives.}
    \label{fig:prepresentatives}
    \end{subfigure}
\end{figure}

Another classical approach is topological bootstrapping\cite{GS, tausz2012extensions}. However, topological bootstrapping may detect features that are not important. For example, the red representative in the noisy construction of Figure-8 in Figure \ref{fig:prepresentatives}, which is not the important representative we are expecting to detect.

Given a point cloud $\mathcal{X}$, which is too large to be processed completely. A way of measure the point cloud is through the set of sub-samples. We can take a sequence of sub-samples $\mathcal{X}_1, \cdots, \mathcal{X}_n$, where the sizes are small enough to be efficiently computed on homology through Rips persistence diagram or persistence barcode to get the persistence homology of each sub-sample. However, the problem is that are these homology computed from sub-samples the real homology of the full point cloud? If yes, then $\mathcal{X}$ would have the same homology as these homology. If no, then there would be two cases need to be distinguished\cite{GS}. One is a single feature is detected repeatedly. The other is multiple features are detected randomly, but one at a time. If $n$ features are detected in $\mathcal{X}_i$ on average, then are we detecting $n$ features of $\mathcal{X}$ with detection probability equals to $1$, or $kn$ features with detection probability equals to $\frac{1}{k}$? Therefore, correlating features across different sub-samples of the full construction is required.  Then here comes the union sequence:
$$\mathcal{X}_1\rightarrow \mathcal{X}_1 \cup \mathcal{X}_2 \leftarrow \mathcal{X}_2 \rightarrow \dots \leftarrow \mathcal{X}_n$$ 
If two samples are sparse, the intersection sequence is not useful, because they are unlikely to intersect very much. 

In this work, the greedy matroid algorithm we developed for finding the homology basis of a given point cloud is a new version of topological bootstrapping to tackle the above problems. Instead of computing Rips persistence diagram or persistence barcode on sub-samples, we compute induced maps from sub-samples to full point cloud and update the homology basis iteratively. The homology basis we computed could comprehensively describe the persistence homology of the given full point cloud. 

%% file: 03background.tex
\section{Background}\label{chap:background}
In this section, we review the concepts, definitions, and propositions used in this work.
For definitions of homology, and computation of induced map, we refer to Algebraic topology\cite{MR1867354} and Parameterized topological data analysis\cite{nelson}. 
For descriptions of matroid and greedy algorithm, we refer to Matroids and the greedy algorithm\cite{edmonds1971matroids}. 
\subsection{Homology}
The input of our greedy matroid algorithm includes a hypothetical homology basis of the given point cloud and a set of induced maps computed from sub-samples to the full space.

Homology is a fundamental concept in algebraic topology that provides a way to study the shape and structure of a space. It associates algebraic objects, called homology groups, to spaces and maps between them. These groups capture information about the presence of holes, tunnels, or voids in a space at different dimensions. The idea behind homology is to capture the essential topological features of a space by considering cycles and boundaries. In simple terms, a cycle is a loop or a closed path in the space, while a boundary is an edge or a path that forms the boundary of a region. The boundary of a cycle is another cycle, and the homology groups measure the cycles that are not boundaries, thus capturing the ``holes'' in the space.

Homology is a functor, which is a mapping between categories. In the context of algebraic topology, it is a mapping between the category of topological spaces and the category of algebraic objects (usually groups). The functoriality of homology means that it preserves the structure and relationships between spaces and their maps. Specifically, if we have a continuous map between two spaces, $f: X \rightarrow Y$, the functorality of homology ensures that there is a corresponding map between their homology groups, denoted as $f: H(X) \rightarrow H_*(Y)$, where $*$ represents the dimension or degree of the homology groups. This functorial property of homology is powerful because it allows us to analyze the properties of spaces and maps by studying the homology groups and their relationships. It enables us to compare different spaces, detect topological invariants, and understand how spaces transform under continuous maps. 

Homology provides a systematic and algebraic way to study the shape and structure of spaces, and its functoriality allows us to analyze and compare spaces using the language of algebraic objects. 

\begin{definition}[Simplicial Complexes\cite{MR1867354}]
A simplicial complex can be described combinatorially as a set $\mathcal{X}_0$ of vertices together with sets $\mathcal{X}_n$ of $n$-simplices, which are $(n+1)$-element subsets of $\mathcal{X}_0$. The only requirement is that each $(k+1)$-element subset of the vertices of an $n$-simplex in $\mathcal{X}_n$ is a $k$-simplex in $\mathcal{X}_k$. From this combinatorial data a $\delta$-complex $\mathcal{X}$ can be constructed, once we choose a partial ordering of the vertices $\mathcal{X}_0$ that restricts to a linear ordering on the vertices of each simplex in $\mathcal{X}_n$. 
\end{definition}
Classical simplicial complexes include Vietoris-Rips complex \ref{v-rips}, Cech complex, and Delaunay complex. 
\begin{definition}[Chain Complex\cite{nelson}]
A chain complex is a sequence of vector spaces $\{C_k\}$, where $k = 0,1, \dots$ with boundary maps $\partial: C_k \rightarrow C_{k-1}$ with the property that $\partial_{k-1} \circ \partial_{k} = 0$ \\ In general, $k$ need not start at $0$, which implies $\partial_0 = 0$
$$0\leftarrow C_0 \xleftarrow{\partial_1} C_1 \xleftarrow{\partial_2} \dots C_{k-1} \xleftarrow{\partial_k} C_k \xleftarrow{\partial_{k+1}} \dots .$$
Elements of $C_k$ are referred to as $k$-chains, elements of $\operatorname{ker}\partial_k$ are referred to as cycles, and elements of $\operatorname{img}\partial_{k+1}$ are referred to as boundaries.
\end{definition}

A filtration is a nested sequence of spaces $$\mathcal{X}_0 \subseteq \mathcal{X}_1 \subseteq \dots \subseteq \mathcal{X}_n. $$

\begin{definition}[Persistence Homology of Filtrations\cite{nelson}]
The persistence homology of filtration studies how homology changes through the sequence of spaces 
$$\mathcal{H}_k(\mathcal{X}_0) \rightarrow \mathcal{H}_k(\mathcal{X}_1) \rightarrow \dots  \rightarrow \mathcal{H}_k(\mathcal{X}_n). $$
\end{definition}

\begin{definition}[Homology Revealing Basis\cite{nelson}]
A homology revealing basis for $\mathcal{C}_k$ is a pair $(\mathcal{B}_k, \mathcal{I}_k)$, where $\mathcal{B}_k$ is a basis for $\mathcal{C}_k$, and $\mathcal{I}_k$ is an index set such that $\{b_i \in \mathcal{B}_k\}_{i\in \mathcal{I}_k} \subseteq \mathcal{B}_k$ generates a basis for $\mathcal{H}_k(\mathcal{C}_*)$. Explicitly, a basis for $\mathcal{H}_k$ is $$\{[b_i]|b_i\in \mathcal{B}_k, i \in \mathcal{I}_k \}.$$
\end{definition}
We use the Reduction Algorithm\cite{nelson} for computing the homology-revealing basis. Since the algorithm has been embedded in BATS\cite{bats}, we call that function directly for computation. 
\begin{proposition}
Given a homology-revealing basis $(\mathcal{B}_k, \mathcal{I}_k)$ for $\mathcal{C}_k$, every homology class $[x] \in \mathcal{H}_k(\mathcal{C}_*)$ has a unique preferred representative. 
\end{proposition}

We visualized the homology changes through Vietoris-Rips persistence diagram.  The Vietoris-Rips persistence diagram on sub-samples of Figure-8 and Annulus are in Figure \ref{fig:fig_8_rips} and \ref{fig:ann_rips}. 

\begin{algorithm}
\caption{Computation of Induced Map\cite{nelson}} \label{alg:im}
\begin{algorithmic}
\Require Homology representative $\mathbf{x} = {U^C_{k}}_i$ in $H_k(C_*), U^D_k, R^D_{k+1}$, from reduction algorithm applied to $\partial_*^D$, with index set $\mathcal{I}^D_*$, with index set $\mathcal{I}^D_k$. Chain map $\mathcal{F}_k$ in original basis. 
\Ensure Induced map on homology, $\tilde{\mathcal{F}_k}_x$.
\State $y \leftarrow (U^D_k)^{-1}F_k$x
\State $n \leftarrow \operatorname{dim}D_k $
\State $\tilde\partial^D_{k+1} \leftarrow (U^d_k)^{-1}R_{k+1}$
\For{$j = n, n-1, \cdots, 1$}
\If{$y_j$ $\neq 0$ and $j$ is a pivot of column $i$ of $\tilde\partial^D_{k+1}$}
\State $\alpha \leftarrow \frac{y_j}{\tilde{\partial^D_{k+1}}_{j,i}}$
\State $y \leftarrow y - \alpha \tilde{\partial^D_{k+1}}_i$
\EndIf
\EndFor
\end{algorithmic}
\end{algorithm}
\begin{figure}
    \centering
    \begin{subfigure}{0.49\linewidth}
    \includegraphics[width=\textwidth]{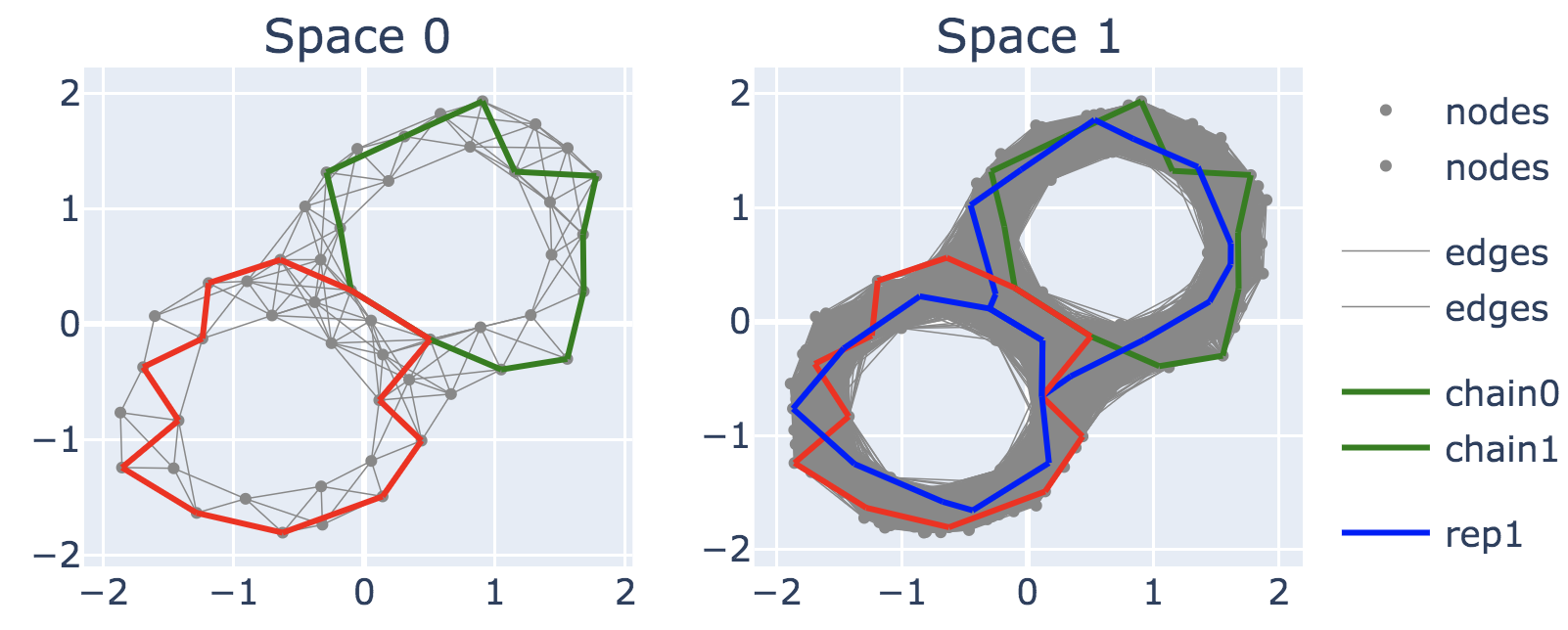}
    \caption{Induced map.}
    \end{subfigure}
    \begin{subfigure}{0.49\linewidth}
    \includegraphics[width=\textwidth]{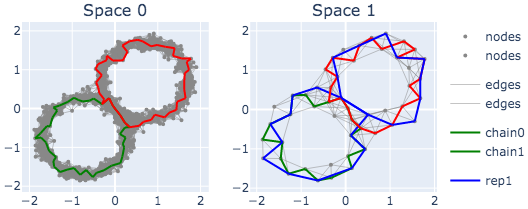}
    \caption{Projection map.}        
    \end{subfigure}
    \caption{Induced map and projection map of Figure-8.}
    \label{fig:im_pm}
\end{figure}

BATS\cite{bats} has embedded Algorithm \ref{alg:im} for computing induced map, we directly call it for computation. 
The reason of computing induced map, instead of projection map is that For matrices of induced maps, the column space of induced maps is the column space of homology of Space 1 in Figure \ref{fig:im_pm}. Space 1 is always different, because sub-samples are different.  If we want to apply Kruskal’s algorithm, we need to have these images to be in the same space.  Induced map is to compute from sub-sample to full space. Projection map is to compute from full space to sub-sample, it could be used while computing co-homology. The point is that the image of each single one of the maps should be a consistent space and the consistent space is the full space that has not been sampled from. Projection map means that there are more spaces in Space 0s than in Space 1s, so we cannot induce. But in our case, the Space 0 should be included in Space 1 in every single case, so we use the induced map.

\subsection{Greedy Algorithm}
The idea of sorting induced maps by how likely they could be the homology basis of full point cloud is greedy algorithm, i.e., Kruskal's algorithm. Kruskal's algorithm is an algorithm in which we make the optimal step at each stage in order to find the global optimum. A classical usage of Kruskal’s algorithm is finding the minimum spanning tree of a graph. The averaged run time complexity of classical Kruskal's algorithm is $O(m \operatorname{log} n)$, where $m$ is the number of edges, and $n$ is the number of vertices. Since we are dealing with matroids instead of graphs, in Section \ref{chap:alg}, we adapt the idea of Kruskal's algorithm to our case. 
\begin{definition}
A matroid $M = (E, I)$ where $E$ is a finite set, and $I$ is a collection of subsets of $E$, satisfies these axioms:
\begin{itemize}
    \item $\emptyset \in I$,
    \item If $A\in I$, and $B\subset A$, then $B\in I$, and 
    \item If $A, B \in I$ and $|A|>|B|$, $ \exists x\in A$ such that $B\cup x \in I$.
\end{itemize}
\end{definition}

\subsection{LU Decomposition}
The approach of our choice for checking whether a vector representation of induced map is in the span of homology basis is LU decomposition with rows and columns pivoting. We manually implemented this algorithm to adapt it to our case.  

Why use rows and columns pivoting? The induced maps could be
$\begin{bmatrix}
0 & 0 & 0 \\
0 & 1 & 0 \\
1 & 0 & 0 
\end{bmatrix}$ or 
$\begin{bmatrix}
0 & 1 & 0 \\
0 & 1 & 0 \\
0 & 0 & 0 
\end{bmatrix}$. If we only do rows pivoting, or columns pivoting, or neither of them, we may not form a valid basis for linear system to determine whether we should include the vector of induced map or not. 

What if forward or backward substitution fail, and/or the linear system is not solvable? The forward or backward substitution may fail because of the inconsistent shape of inputs which means that the number of columns of homology basis and the dimension of induced map are different. Our solution is returning false, and adding that vector to the basis of homology. If the linear system is solvable which means that the induced map is in the span of homology basis, then we don’t include it and eliminate it from consideration. If the linear system is not solvable which means that the induced map is not in the span of homology basis, then we add that vector to the basis of homology.

What if the inputs are sparse matrices? Usually the homology basis is low dimensional, so we exclude sparse matrix for consideration at this point. 

%% file: 04gma.tex
\section{Greedy Matroid Algorithm}\label{chap:alg}
We developed a Greedy Matroid Algorithm \ref{alg:gma}
to find the comprehensive homology basis of a space. 
The input of the algorithm is a list of induced maps. 
We first sorted the list based on how likely the vector of an induced map is in the span of homology basis of the given space by Algorithm \ref{alg:imr_w}. 
Our selection of criteria is computing the expected rank of a list of induced maps. 
The list of ranks represents the list of weights of induced maps. 
We furthermore created a dictionary to map each weight to the corresponding induced map and sorted the induced maps in terms of their weights in decreasing order. 
Then we initialized the induced map with the highest weight as the hypothetical homology basis. The reason is we want to initialize a homology basis that contains as much information about the full space as possible. Their rank/weight to some extent represents the amount of information they contain.  

Second, we update the basis of homology by iteratively looping into the list of induced maps and adding the induced maps that were not in the span of homology basis to the basis. The approach we choose to do this is LU decomposition with rows and columns pivoting. In practice, we adapt the classical LU decomposition to accommodate our case in the way mentioned in section \ref{chap:background}. As a result, if the linear system is not solvable or returns a $False$, which means that the vector of induced map is not in the span of homology basis, then we include that vector in the basis. If the linear system is solvable which means the vector of induced map is in the span of the basis of homology, then we do not include it in the basis and eliminate it from consideration. Notice that once we add a vector into the homology basis the dimension of the basis will increase by one. As a result, we should extend the dimension of the next vector of induced map each time accordingly to ensure the linear system is valid.

The best case scenario of greedy matroid algorithm \ref{alg:gma} is $O(log(n))$, the worst case scenario is $O(n^2)$, and the average case scenario is $O(n \cdot log(n))$.  

\begin{figure}
    \centering
    \includegraphics[width=0.23\linewidth]{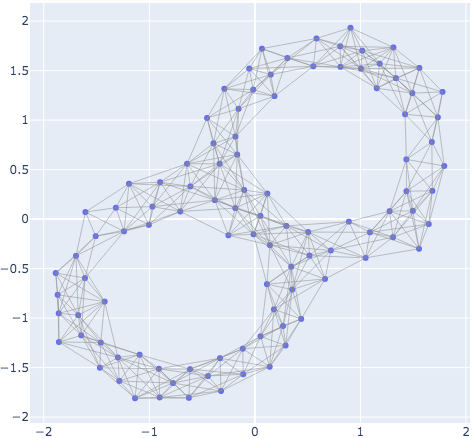}
    \includegraphics[width=0.23\linewidth]{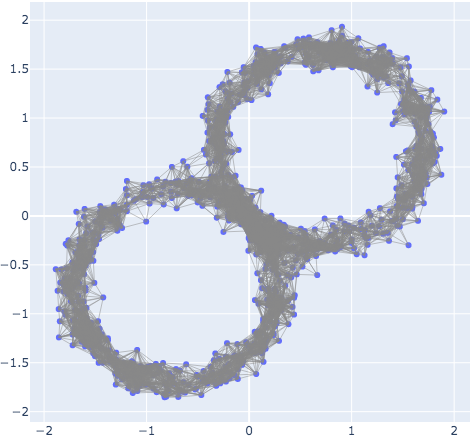}
    \includegraphics[width=0.23\linewidth]{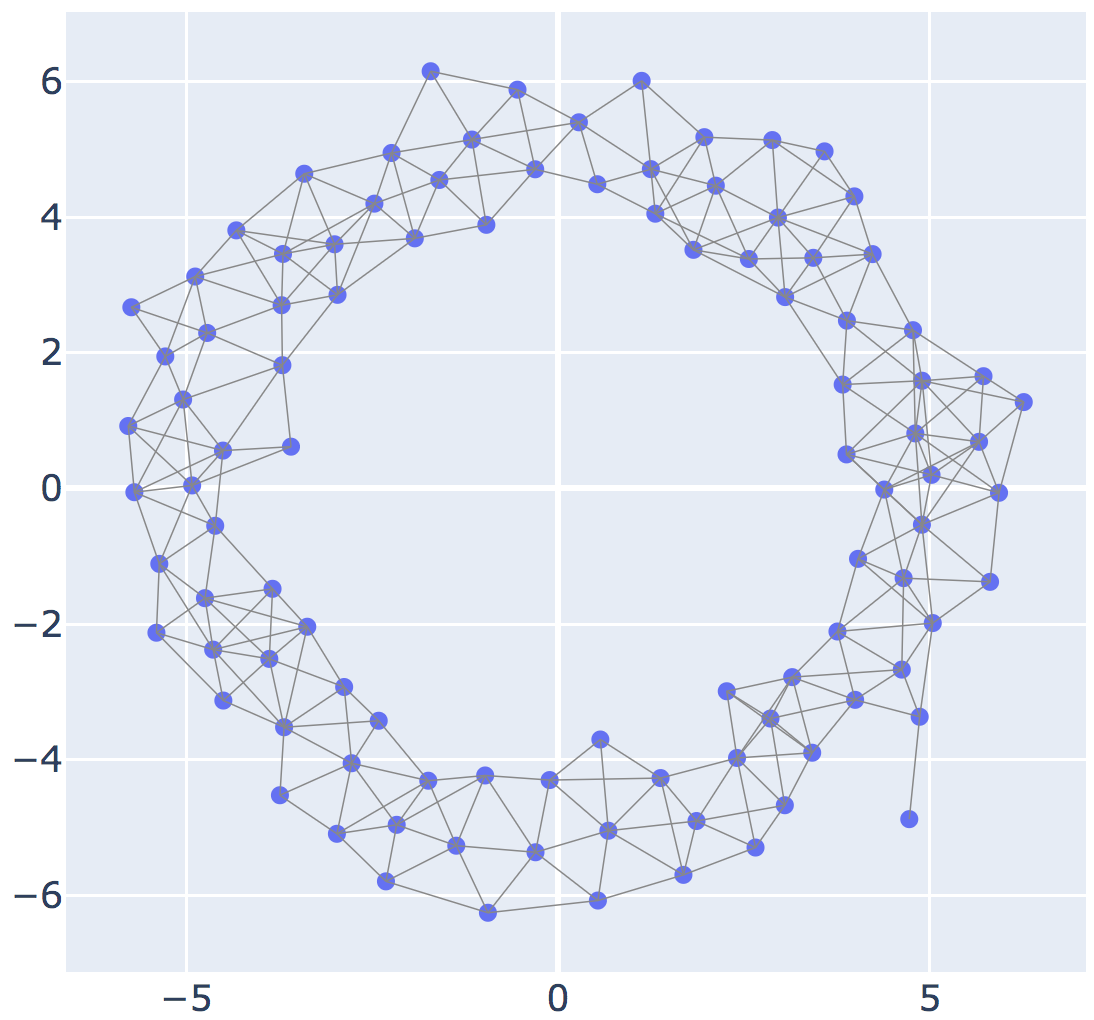}
    \includegraphics[width=0.23\linewidth]{figs/intro/annulus.png}
    \caption{Vietoris-Rips complexes of both sub-samples and full samples. The threshold of our choice are $0.25 +
    2*\text{Hausdorff distance}$ respectively.}
    \label{fig:fig_8_rc} 
\end{figure}

\begin{algorithm}
\caption{Reduction Algorithm with formation of weights}\label{alg:imr_w}
\begin{algorithmic}[1]
\Require A list of induced maps $\mathcal{IM}$.
\Ensure A list of reduced $\mathcal{IM}$, and a list of corresponding weights $\mathcal{W}$.
\For{$i = 1, \cdots, n$} 
\For{$j = 2, \cdots, n$}
\If{$\mathcal{IM}_{i,j} \neg 0$}
\State $\mathcal{IM}_{i,j} \leftarrow 1$
\EndIf
\EndFor
\EndFor
\State Initialize weights $\mathcal{W} = []$
\For{$d = 1, \cdots, l_\mathcal{IM}$}
\State Initialize $\mathcal{L} \leftarrow [1, \cdots, l_{\mathcal{IM}_d}]$
\For{$i = 1,\cdots, l_{\mathcal{IM}_d}$}
\State $j \leftarrow i + 1$
\While{$j <l_{\mathcal{IM}_d}$}
\If{$\mathcal{IM}_{d,i} = \mathcal{IM}_{d,j}$}
\State $\mathcal{L}\leftarrow\mathcal{L}$ remove $j$
\EndIf
\EndWhile 
\EndFor
\State $\mathcal{W}\leftarrow \mathcal{W};l_\mathcal{L}$
\EndFor
\end{algorithmic}
\end{algorithm}
\begin{algorithm}
\caption{Greedy Matroid Algorithm}\label{alg:gma}
\begin{algorithmic}[1]
\Require A list of reduced induced maps $\mathcal{IM}$, and a list of corresponding weights $\mathcal{W}$.
\Ensure A hypothetical homology basis, $\mathcal{H}$.
\State Initialize a dictionary $\mathcal{D}$, where keys $K\leftarrow \mathcal{IM}$ and values $V\leftarrow\mathcal{W}$ correspondingly.
\State Sort $\mathcal{D}$ in terms of $V$.
\State Initialize $\mathcal{H}\leftarrow K_1$, $\mathcal{B}\leftarrow K_{2:}$.
\State Initialize $\mathbf{n_{zeros}}\leftarrow 0$.
\State Solve $\mathcal{H}\mathbf{x} = \mathcal{B}_{1,1}$ for $\mathbf{x}$ by LU decomposition with rows and columns pivoting.
\If{$\mathbf{x} =$ False {\bf or} the linear system is unsolvable}
\State $\mathcal{H}\leftarrow\mathcal{H};\mathcal{B}_{1,1}$, $\mathcal{B}_{1,2}\leftarrow\mathcal{B}_{1,2};0$ or $\mathcal{B}_{2,1}\leftarrow\mathcal{B}_{2,1};0$, 
\State $\mathbf{n_{zeros}}\leftarrow\mathbf{n_{zeros}}+1$
\EndIf
\For{$i = 1, \cdots, l_\mathcal{B}$} 
\For{$j = 2, \cdots, l_{\mathcal{B}_i}$}
\State Initialize $\mathbf{b} \leftarrow \mathcal{B}_{i,j}$
\State $\mathbf{b}\leftarrow\mathbf{b};0\times \mathbf{n_{zeros}}$
\State Solve $\mathcal{H}\mathbf{x} = \mathbf{b}$ for $\mathbf{x}$ by LU decomposition with rows and columns pivoting
\If{$\mathbf{x}$ = False {\bf or} the linear system is unsolvable}
\State $\mathcal{H} \leftarrow \mathcal{H};\mathcal{B}_{i,j}$
\State $\mathbf{n_{zeros}} \leftarrow \mathbf{n_{zeros}} + 1$
\EndIf
\EndFor
\EndFor
\end{algorithmic}
\end{algorithm}

%% file: 05_1experiment.tex
\section{Experiments}\label{chap:experiments}
We give an example of finding the homology basis of a noisy construction using Greedy Matroid Algorithm \ref{alg:gma}. The constructions are noisy Figure-8 and noisy Annulus. The size of full point cloud is $1000$. The sub-sample sizes are $20$, $50$, $100$, $300$, $500$, and $800$. For each sub-sample size, we randomly generated ten sub-samples from the full point cloud and computed their induced maps.  

\subsection{Induced Maps of Figure-8}
The computation of induced maps follows this procedure. We first computed the Vietoris-Rips complex of sub-samples and full point cloud (Figure \ref{fig:fig_8_rc}).
Then we converted it to reduced chain complex and finally computed induced map on it. The results of induced maps for Figure-8 are in Table \ref{tab:fig_8_800_im}. 
To give an example, while sample size is $50$, the homology values are $H_0=1$ and $H_1=3$, then there is one connected component and three loops.  
We collected induced maps in Table \ref{tab:fig_8_800_im} to a matroid $\mathbf{IM_{fig8-50}}$ as the input of our greedy matroid algorithm. All inputs are summarized in Table \ref{tab:f8_im_all}.

We convert lists of induced maps of each index to matrices, where row vectors in a matrix are elements of induced maps and column vectors are elements of input for greedy matroid algorithm. For empty lists we use zeros to represent. 
\begin{table}[!htb]
\footnotesize
    \begin{tabular*}{\columnwidth}{@{\extracolsep{\fill}}ccccc}
    \hline
        Indices&Subsample size&TB Homology&$H_1$& $\mathbf{H}$\\\hline 
        0-9&20&$(1,0,...)$&$\mathbf{IM_{fig8-20}}$&$\mathbf{H_{20}}$\\
        10-19&50&$(1,3,...)$&$\mathbf{IM_{fig8-50}}$&$\mathbf{H_{50}}$\\
        20-29&100&$(2,1,...)$&$\mathbf{IM_{fig8-100}}$&$\mathbf{H_{100}}$\\
        30-39&300&$(6,3,...)$&$\mathbf{IM_{fig8-300}}$&$\mathbf{H_{300}}$\\
        40-49&500&$(8,7,...)$&$\mathbf{IM_{fig8-500}}$&$\mathbf{H_{500}}$\\
        50-59&800&$(12,7,...)$&$\mathbf{IM_{fig8-800}}$&$\mathbf{H_{800}}$\\\hline 
    \end{tabular*}
    \caption{Figure-8. Induced maps of different sample sizes. Sub-samples were randomly sampled.}\label{tab:f8_im_all}
\end{table}
\begin{table}[!htb]
\footnotesize
    \begin{tabular*}{.5\columnwidth}{@{\extracolsep{\fill}}ccc}
        \hline
        index   & size  &  $H_1$\\\hline 
        0 & 20   & [] \\
        1 & 20   & [] \\
        2 & 20   & [] \\
        3 & 20   & [] \\
        4 & 20   & [] \\
        5 & 20   & [] \\
        6 & 20   & [] \\
        7 & 20   & [] \\
        8 & 20   & [] \\
        9 & 20   & [] \\
    \end{tabular*}
    \label{tab:fig_8_20_im}
    \begin{tabular*}{.5\columnwidth}{@{\extracolsep{\fill}}ccc}
    \hline
    index & size &  $H_1$\\\hline 
        10 & 50   & [[0], [0], [1]]\\
        11 & 50   & [[0, 0], [0, 0], [1, 0]]\\
        12 & 50   & [[0], [0], [2]] \\
        13 & 50   & [[0], [0], [2]] \\
        14 & 50   & [[0, 0], [0, 0], [0, 1]]\\
        15 & 50   & [[0, 0, 0], [0, 0, 0], [2, 2, 0]]\\
        16 & 50   & [[0, 0], [0, 0], [0, 1]]\\
        17 & 50   & [[0], [0], [1]]\\
        18 & 50   & [[0, 0], [0, 0], [2, 0]]\\
        19 & 50   & [[], [], []]\\
    \end{tabular*}
    \label{tab:fig_8_50_im}
    \begin{tabular*}{.5\columnwidth}{@{\extracolsep{\fill}}ccc}
    \hline
     index & size  & $H_1$ \\\hline 
        20 & 100   & [[0, 2]] \\
        21 & 100   & [[0, 2]] \\
        22 & 100   & [[0, 2]] \\
        23 & 100   & [[0, 1]] \\
        24 & 100   & [[1, 0, 0]] \\
        25 & 100   & [[2, 0]] \\
        26 & 100   & [[0]] \\
        27 & 100   & [[0, 2]] \\
        28 & 100   & [[0, 1, 0]] \\
        29 & 100   & [[0, 2]] \\
    \end{tabular*}
    \label{tab:fig_8_100_im}
    \begin{tabular*}{.5\columnwidth}{@{\extracolsep{\fill}}ccc}
     \hline
     index & size  & $H_1$\\\hline 
        30 & 300   &  [[1, 0, 0],[0, 0, 1],[2, 1, 0]]\\
        31 & 300   &  [[0, 1], [2, 0], [0, 0]] \\
        32 & 300   &  [[0, 1, 0],[0, 0, 2],[0, 0, 0]]\\
        33 & 300   &  [[0, 0, 2, 0],[0, 1, 0, 0],[0, 0, 0, 0]]\\
        34 & 300   &  [[0, 2, 0],[2, 0, 0],[0, 0, 0]]\\
        35 & 300   &  [[0, 2], [1, 0], [0, 0]] \\
        36 & 300   &  [[2, 0], [0, 2], [0, 0]] \\
        37 & 300   &  [[2, 0], [0, 2], [0, 0]] \\
        38 & 300   &  [[0, 2, 0, 0],[1, 0, 0, 0],[0, 0, 0, 0]]\\
        39 & 300   &  [[0, 1, 0, 0],[1, 0, 0, 0],[0, 0, 0, 0]]\\
    \end{tabular*}
    \label{tab:fig_8_300_im}
    \begin{tabular*}{\columnwidth}{@{\extracolsep{\fill}}ccc}
        \hline
        index & size & $H_1$ \\
        \hline 
        40 & 500 &[[0, 0, 0, 0, 0, 0, 1], [0, 1, 0, 0, 0, 0, 0], [0, 0, 0, 0, 1, 0, 0], [0, 1, 1, 0, 0, 0, 0], \\
        &  &[0, 0, 0, 0, 0, 1, 0], [0, 0, 0, 2, 0, 2, 0], [1, 2, 0, 1, 0, 0, 0]] \\
        41 & 500 &[[0, 0, 0, 0], [1, 0, 2, 0], [0, 2, 0, 0], [0, 0, 0, 2], [0, 0, 0, 0], [0, 0, 1, 0], [0, 0, 0, 0]] \\
        42 & 500 &[[0, 0, 0, 0], [0, 0, 2, 0], [0, 0, 0, 2], [2, 2, 0, 0], [0, 0, 0, 0], [0, 1, 0, 0], [0, 0, 0, 0]] \\
        43 & 500 &[[0, 0, 0], [0, 1, 0], [2, 0, 0], [0, 0, 2], [0, 0, 0], [0, 0, 0], [0, 0, 0]] \\
        44 & 500 &[[0, 0, 0, 0], [0, 1, 0, 0], [2, 0, 0, 0], [0, 0, 1, 0], [0, 0, 0, 0], [0, 0, 0, 0], [0, 0, 0, 0]] \\
        45 & 500 &[[0, 0, 0], [0, 1, 0], [2, 0, 0], [0, 0, 0], [0, 0, 0], [0, 0, 0], [0, 0, 0]] \\
        46 & 500 &[[0, 0, 0, 0], [0, 2, 0, 1], [0, 0, 1, 0], [1, 0, 0, 0], [0, 0, 0, 0], [0, 0, 0, 2], [0, 0, 0, 0]] \\
        47 & 500 &[[0, 0, 0, 0], [0, 2, 0, 0], [1, 0, 0, 0], [0, 0, 2, 2], [0, 0, 0, 0], [0, 0, 0, 0], [0, 0, 0, 0]] \\
        48 & 500 &[[0, 0, 0, 0, 0], [0, 1, 0, 0, 0], [1, 0, 0, 0, 0], [0, 0, 1, 0, 0], \\
        & &[0, 0, 0, 0, 0], [0, 0, 0, 0, 0], [0, 0, 0, 0, 0]] \\
        49 & 500 &[[0, 0, 0, 0], [1, 2, 0, 0], [0, 0, 2, 0], [0, 0, 0, 2], [0, 0, 0, 0], [2, 0, 0, 0], [0, 0, 0, 0]] 
    \end{tabular*}
    \label{tab:fig_8_500_im}
    \begin{tabular*}{\columnwidth}{@{\extracolsep{\fill}}ccc}
        \hline
        index& size & $H_1$ \\\hline
        50 & 800 &[[1, 2, 1, 1, 1, 0, 1], [0, 0, 0, 1, 0, 0, 0], [1, 2, 0, 0, 1, 0, 0], [0, 0, 0, 0, 0, 0, 2],\\
        & & [0, 0, 0, 0, 0, 1, 0], [0, 2, 0, 0, 0, 0, 0], [0, 0, 0, 0, 2, 2, 0]]\\      
        51 & 800 &[[1, 0, 2, 2], [1, 2, 2, 2], [1, 0, 0, 0], [0, 0, 0, 0], [0, 0, 0, 0], [0, 0, 0, 1], [0, 0, 0, 0]]\\
        52 & 800 &[[0, 0, 1, 1], [0, 2, 0, 0], [2, 0, 0, 1], [0, 0, 0, 0], [0, 0, 0, 0], [2, 0, 0, 0], [0, 0, 0, 0]]\\
        53 & 800 &[[0, 0, 2, 0, 0], [0, 2, 2, 0, 0], [2, 1, 0, 0, 2], [0, 0, 0, 0, 2], \\
        & & [0, 0, 0, 0, 0], [2, 1, 0, 1, 2], [0, 0, 0, 0, 0]]\\
        54 & 800 &[[2, 0, 1, 0, 1], [0, 2, 0, 0, 0], [0, 0, 1, 2, 1], [0, 0, 0, 0, 0], \\
        & & [0, 0, 0, 0, 0],[0, 0, 0, 2, 0], [0, 0, 0, 0, 0]]\\
        55 & 800 &[[2, 1, 2, 0, 0, 0, 1], [0, 0, 2, 0, 0, 0, 0], [2, 0, 0, 0, 0, 0, 0], [0, 0, 0, 0, 0, 0, 2], \\
        & & [0, 0, 0, 0, 0, 1, 0], [2, 0, 0, 1, 2, 0, 0], [0, 0, 0, 0, 2, 2, 0]],\\
        56 & 800 &[[1, 1, 0, 0], [0, 0, 1, 0], [1, 0, 0, 2], [0, 0, 0, 0], [0, 0, 0, 0], [0, 0, 0, 2], [0, 0, 0, 0]]\\  
        57 & 800 &[[1, 1, 0, 0], [0, 0, 1, 0], [1, 0, 0, 2], [0, 0, 0, 0], \\
        & & [0, 0, 0, 0], [0, 0, 0, 2], [0, 0, 0, 0]]\\
        58 & 800 &[[0, 0, 2, 0, 0], [0, 2, 2, 0, 0], [2, 1, 0, 0, 2], [0, 0, 0, 0, 2],\\
        & &[0, 0, 0, 0, 0], [2, 1, 0, 1, 2], [0, 0, 0, 0, 0]]\\ 
        59 & 800 &[[1, 2, 0, 0], [0, 0, 1, 0],[1, 0, 0, 2], [0, 0, 0, 0], [0, 0, 0, 0], [0, 0, 0, 2], [0, 0, 0, 0]]
    \end{tabular*}
    \caption{Induced maps of Figure-8 from sub-samples of difference sizes to the full point cloud.}
    \label{tab:fig_8_800_im}
\end{table}
\clearpage

$\mathbf{IM_{fig8-20}}=\begin{Bmatrix}
\begin{bmatrix}&\end{bmatrix},
\begin{bmatrix}&\end{bmatrix},
\begin{bmatrix}&\end{bmatrix},
\begin{bmatrix}&\end{bmatrix},
\begin{bmatrix}&\end{bmatrix},
\begin{bmatrix}&\end{bmatrix},
\begin{bmatrix}&\end{bmatrix},
\begin{bmatrix}&\end{bmatrix},
\begin{bmatrix}&\end{bmatrix},
\begin{bmatrix}&\end{bmatrix}
\end{Bmatrix}$.

$\mathbf{IM_{fig8-50}}=\begin{Bmatrix}
\begin{bmatrix}
&0&\\ 
&0&\\ 
&1&\end{bmatrix}, 
\begin{bmatrix}
&0 &0&\\
&0 &0&\\
&1 &0& \end{bmatrix}, 
\begin{bmatrix}
&0 &\\
&0 &\\
&2& \end{bmatrix},
\begin{bmatrix}
&0& \\
&0& \\
&2& \end{bmatrix},
\begin{bmatrix}
&0 &0& \\
&0 &0& \\
&0 &1& \end{bmatrix},\\
\begin{bmatrix}
&0 &0 &0&\\ 
&0 &0 &0& \\
&2 &2 &0& \end{bmatrix},
\begin{bmatrix}
&0& 0& \\
&0& 0& \\
&0& 1& \end{bmatrix},
\begin{bmatrix}
&0& \\
&0& \\
&1& \end{bmatrix},
\begin{bmatrix}
&0& 0 &\\
&0& 0 &\\
&2& 0& \end{bmatrix},
\begin{bmatrix}
&0& \\
&0& \\
&0& \end{bmatrix}
\end{Bmatrix}$.

$\mathbf{IM_{fig8-100}}=\begin{Bmatrix}
    \begin{bmatrix}&0 & 2 &\end{bmatrix}, 
     \begin{bmatrix}&0 & 2 &\end{bmatrix},
     \begin{bmatrix}&0 & 2 &\end{bmatrix},
     \begin{bmatrix}&0 & 1 &\end{bmatrix}, 
     \begin{bmatrix}&1 & 0 & 0 &\end{bmatrix},\\
     \begin{bmatrix}&2 & 0 &\end{bmatrix}, 
     \begin{bmatrix}&0 &\end{bmatrix}, 
     \begin{bmatrix}&0 & 2 &\end{bmatrix}, 
     \begin{bmatrix}&0 &1 & 0 &\end{bmatrix},
     \begin{bmatrix}&0 &2 &\end{bmatrix}
\end{Bmatrix}$.

$\mathbf{IM_{fig8-300}}=
\begin{Bmatrix}
\begin{bmatrix}
&1 &0 &0& \\
&0 &0 &1& \\
&2 &1 &0&
\end{bmatrix}, 
\begin{bmatrix}
&0 &1& \\
&2 &0& \\
&0 &0&
\end{bmatrix},
\begin{bmatrix}
&0 &1 &0 &\\
&0 &0 &2&\\
&0 &0 &0&
\end{bmatrix}, \\
\begin{bmatrix}
&0 &0 &2 &0& \\
&0 &1 &0 &0& \\
&0 &0 &0 &0&
\end{bmatrix}, 
\begin{bmatrix}
&0 &2 &0& \\
&2 &0 &0& \\
&0 &0 &0&
\end{bmatrix},
\begin{bmatrix}
&0& 2 &\\
&1& 0 &\\
&0& 0&
\end{bmatrix},
\begin{bmatrix}
&2 &0&\\
&0 &2&\\
&0 &0&\end{bmatrix},\\
\begin{bmatrix}
&2 &0& \\
&0 &2& \\
&0 &0&
\end{bmatrix}, 
\begin{bmatrix}
&0& 2& 0& 0& \\
&1& 0& 0& 0& \\
&0& 0& 0& 0&
\end{bmatrix},
\begin{bmatrix}
&0 &1 &0 &0 & \\
&1 &0 &0 &0 & \\
&0 &0 &0 &0 &
\end{bmatrix}
\end{Bmatrix}$.

$\mathbf{IM_{fig8-500}}=\begin{Bmatrix}
\begin{bmatrix}
&0& 0& 0& 0& 0& 0& 1&\\
&0& 1& 0& 0& 0& 0& 0&\\
&0& 0& 0& 0& 1& 0& 0&\\
&0& 1& 1& 0& 0& 0& 0&\\
&0& 0& 0& 0& 0& 1& 0&\\
&0& 0& 0& 2& 0& 2& 0&\\
&1& 2& 0& 1& 0& 0& 0&
\end{bmatrix},
\begin{bmatrix}
&0 &0 &0 &0 & \\
&1 &0 &2 &0 & \\
&0 &2 &0 &0 & \\
&0 &0 &0 &2 & \\
&0 &0 &0 &0 & \\
&0 &0 &1 &0 & \\
&0 &0 &0 &0 &
\end{bmatrix},
\begin{bmatrix}
&0& 0& 0& 0&\\
&0& 0& 2& 0&\\
&0& 0& 0& 2&\\
&2& 2& 0& 0&\\
&0& 0& 0& 0& \\
&0& 1& 0& 0&\\ 
&0& 0& 0& 0&
\end{bmatrix}, \\
\begin{bmatrix}
&0 &0 &0& \\
&0 &1 &0& \\
&2 &0 &0& \\
&0 &0 &2& \\
&0 &0 &0& \\
&0 &0 &0& \\
&0 &0 &0&
\end{bmatrix},
\begin{bmatrix}
&0& 0& 0& 0 &\\
&0& 1& 0& 0 &\\
&2& 0& 0& 0 &\\
&0& 0& 1& 0 &\\
&0& 0& 0& 0 &\\
&0& 0& 0& 0 &\\
&0& 0& 0& 0&
\end{bmatrix},
\begin{bmatrix}
&0& 0& 0 &\\
&0& 1& 0 &\\
&2& 0& 0 &\\
&0& 0& 0&\\
&0& 0& 0&\\
&0& 0& 0&\\
&0& 0& 0&
\end{bmatrix},
\begin{bmatrix}
&0& 0& 0& 0& \\
&0& 2& 0& 1& \\
&0& 0& 1& 0& \\
&1& 0& 0& 0& \\
&0& 0& 0& 0& \\
&0& 0& 0& 2& \\
&0& 0& 0& 0&
\end{bmatrix},\\
\begin{bmatrix}
&0& 0& 0& 0& \\
&0& 2& 0& 0&\\
&1& 0& 0& 0&\\
&0& 0& 2& 2&\\
&0& 0& 0& 0& \\
&0& 0& 0& 0& \\
&0& 0& 0& 0&
\end{bmatrix}, 
\begin{bmatrix}
&0& 0& 0& 0 &0& \\
&0& 1& 0& 0 &0& \\
&1& 0& 0& 0 &0& \\
&0& 0& 1& 0 &0& \\
&0& 0& 0& 0 &0& \\
&0& 0& 0& 0 &0& \\
&0& 0& 0& 0 &0&
\end{bmatrix},
\begin{bmatrix}
&0& 0& 0& 0 &\\
&1& 2& 0& 0 &\\
&0& 0& 2& 0 &\\
&0& 0& 0& 2 &\\
&0& 0& 0& 0 &\\
&2& 0& 0& 0 &\\
&0& 0& 0& 0&
\end{bmatrix}
\end{Bmatrix}$

$\mathbf{IM_{fig8-800}}=\begin{Bmatrix}
\begin{bmatrix}
&0& 0& 0& 0& 0& 0& 1&\\
&0& 1& 0& 0& 0& 0& 0&\\
&0& 0& 0& 0& 1& 0& 0&\\
&0& 1& 1& 0& 0& 0& 0&\\
&0& 0& 0& 0& 0& 1& 0&\\
&0& 0& 0& 2& 0& 2& 0&\\
&1& 2& 0& 1& 0& 0& 0&\end{bmatrix},
\begin{bmatrix}
&0& 0& 0& 0&\\
&1& 0& 2& 0&\\
&0& 2& 0& 0&\\
&0& 0& 0& 2&\\
&0& 0& 0& 0&\\
&0& 0& 1& 0&\\
&0& 0& 0& 0&\end{bmatrix},
\begin{bmatrix}
&0& 0& 0& 0&\\
&0& 0& 2& 0&\\
&0& 0& 0& 2&\\
&2& 2& 0& 0&\\
&0& 0& 0& 0&\\
&0& 1& 0& 0&\\
&0& 0& 0& 0&\end{bmatrix},\\
\begin{bmatrix}
&0& 0& 0&\\
&0& 1& 0&\\
&2& 0& 0&\\
&0& 0& 2&\\
&0& 0& 0&\\
&0& 0& 0&\\
&0& 0& 0&\end{bmatrix},
\begin{bmatrix}
&0& 0& 0& 0&\\
&0& 1& 0& 0&\\
&2& 0& 0& 0&\\
&0& 0& 1& 0&\\
&0& 0& 0& 0&\\
&0& 0& 0& 0&\\
&0& 0& 0& 0&\end{bmatrix},
\begin{bmatrix}
&0& 0& 0&\\
&0& 1& 0&\\
&2& 0& 0&\\
&0& 0& 0&\\
&0& 0& 0&\\
&0& 0& 0&\\
&0& 0& 0&\end{bmatrix},
\begin{bmatrix}
&0& 0& 0& 0&\\
&0& 2& 0& 1&\\
&0& 0& 1& 0&\\
&1& 0& 0& 0&\\
&0& 0& 0& 0&\\
&0& 0& 0& 2&\\
&0& 0& 0& 0&\end{bmatrix},\\
\begin{bmatrix}
&0& 0& 0& 0&\\
&0& 2& 0& 0&\\
&1& 0& 0& 0&\\
&0& 0& 2& 2&\\
&0& 0& 0& 0&\\
&0& 0& 0& 0&\\
&0& 0& 0& 0&\end{bmatrix},
\begin{bmatrix}
&0& 0& 0& 0& 0&\\
&0& 1& 0& 0& 0&\\
&1& 0& 0& 0& 0&\\
&0& 0& 1& 0& 0&\\
&0& 0& 0& 0& 0&\\
&0& 0& 0& 0& 0&\\
&0& 0& 0& 0& 0&\end{bmatrix},
\begin{bmatrix}
&0& 0& 0& 0&\\
&1& 2& 0& 0&\\
&0& 0& 2& 0&\\
&0& 0& 0& 2&\\
&0& 0& 0& 0&\\
&2& 0& 0& 0&\\
&0& 0& 0& 0&\end{bmatrix}.
\end{Bmatrix}$
\subsection{Homology Bases of Figure-8}
The estimated matroids of homology basis of noisy Figure-8 computed by Greedy Matroid Algorithm \ref{alg:gma} based on sub-samples of different sizes are below.

$\mathbf{H_{20}}=
\begin{bmatrix}
&0&
\end{bmatrix}$
$\mathbf{H_{50}}=
\begin{bmatrix}
&0&0&1&
\end{bmatrix}$
$\mathbf{H_{100}}=
\begin{bmatrix}
&1&
\end{bmatrix}$
$\mathbf{H_{300}}=
\begin{bmatrix}
&1 & 0 & 1 &\\ 
&0 & 0 & 1 &\\
&0 & 1 & 0 &\\
&1 & 0 & 0 &
\end{bmatrix}$

$\mathbf{H_{500}}=
\begin{bmatrix}
&0 & 0 & 0 & 0 & 0 & 0 & 1& \\ 
&0 & 1 & 0 & 1 & 0 & 0 & 1& \\
&0 & 0 & 0 & 1 & 0 & 0 & 0& \\
&0 & 0 & 0 & 0 & 0 & 1 & 1& \\ 
&0 & 0 & 1 & 0 & 0 & 0 & 0& \\
&0 & 0 & 0 & 0 & 1 & 1 & 0& \\ 
&1 & 0 & 0 & 0 & 0 & 0 & 0& \\
&0 & 1 & 0 & 0 & 0 & 0 & 0& \\
&0 & 1 & 0 & 0 & 0 & 1 & 0& \\
&0 & 0 & 0 & 1 & 0 & 1 & 0& 
\end{bmatrix}$
$\mathbf{H_{800}}=
\begin{bmatrix}
&1&0 &1 &0 &0 &0 &0&\\
&1&0 &1 &0 &0 &1 &0&\\
&1 &0 &0 &0 &0 &0 &0&\\
&1 &1 &0 &0 &0 &0 &0&\\
&1 &0 &1 &0 &0 &0 &1&\\
& 0 &0 &0 &0 &1 &0 &1&\\
& 1 &0 &0 &1 &0 &0 &0&\\
& 0 &0 &0 &0 &0 &1 &0&\\
& 0 &0 &0 &0 &0 &1 &1&\\
& 0 &0 &1 &0 &0 &1 &0&\\
& 0 &1 &1 &0 &0 &1 &0&\\
& 0 &0 &1 &1 &0 &1 &0&\\
& 1 &1 &1 &0 &0 &0 &0&\\
& 0 &1 &0 &0 &0 &0 &0&\\
& 1 &1 &0 &0 &0 &1 &0&
\end{bmatrix}$

We now show that the estimated homology bases above may or may not be a basis of Figure-8.
\begin{definition}[\cite{siu1998introduction}]\label{def_cycle}
A matroid $\mathbf{M}$ consists of a non-empty set $\mathbf{E}$, and a collection $\mathbf{C}$ of non-empty subsets of $\mathbf{E}$ (called cycles) satisfying the following properties:
\begin{itemize}
    \item $\mathbf{C}(i)$ no cycle properly contains another cycle;
    \item $\mathbf{C}(ii)$ if $\mathbf{C_1}$ and $\mathbf{C_2}$ are two distinct cycles each containing an element $e$, then exists a cycle in $\mathbf{C_1} \cup \mathbf{C_2}$ that does not contain ${e}$.
\end{itemize}
\end{definition}
From Definition \ref{def_cycle}, we could derive the following property: 
\begin{proposition}[Basis of Figure-8]\label{prop:bo8}
If there are two bases $\mathbf{A}$ and $\mathbf{B}$ in the estimated homology bases $\mathbf{H}$, such that, $d\cdot \mathbf{A} = 0$, $e\cdot \mathbf{B} = 0$, and there is a common vector $h \in \mathbf{A}$ and $h \in \mathbf{B}$, then by Definition \ref{def_cycle} there exists a construction of Figure-8 in $\mathbf{H}$. 
\end{proposition}
\begin{algorithm}
\caption{Check Figure-8 Construction}\label{alg:cfig8}
\begin{algorithmic}[1]
\Require $\mathbf{H}$
\Ensure There exists or does not exist a Figure-8 construction in $\mathbf{H}$. 
\For{$\mathbf{X},\mathbf{Y} \subset \mathbf{H}$ and $\mathbf{X},\mathbf{Y} \notin \emptyset$}
\If{$\mathbf{X}\neq\mathbf{Y}$ and $\mathbf{i} \in \mathbf{X}$ and $\mathbf{i} \in \mathbf{Y}$}
\State{Solve $\mathbf{X}\mathbf{a} = \mathbf{0}$ and $\mathbf{Y}\mathbf{c} = \mathbf{0}$ for $\mathbf{a}$ and $\mathbf{c}$ by LU decomposition with rows and columns pivoting.}
\EndIf
\EndFor
\end{algorithmic}
\end{algorithm}
\begin{algorithm}
\caption{Check Annulus Construction}\label{alg:cdependency}
\begin{algorithmic}[1]
\Require $\mathbf{H}$
\Ensure There exists or does not exist an Annulus construction in $\mathbf{H}$.
\For{$\mathbf{X} \subset \mathbf{H}$ and $\mathbf{X} \neq \emptyset$}
\State{Solve $\mathbf{X}\mathbf{a} = \mathbf{0}$ for $\mathbf{a}$ by LU decomposition with rows and columns pivoting.}
\EndFor
\end{algorithmic}
\end{algorithm}
Since each row represents an element of induced map, here we are checking the row space. 
Intuitively, $\mathbf{H_{20}}^\top$, $\mathbf{H_{50}}^\top$, and $\mathbf{H_{100}}^\top$ does not satisfy the Proposition \ref{prop:bo8}. 
$\mathbf{H_{300}}^\top$ contains two sub-spaces
$\begin{bmatrix}
    &1&1&0&\\
    &0&0&0&\\
    &1&0&1&
\end{bmatrix}$
and itself
$\begin{bmatrix}
    &1&0&1&0&\\
    &0&0&0&1&\\
    &1&1&0&0&
\end{bmatrix}$.
The two matrices could form a cycle and they have common vectors 
$\begin{Bmatrix}
 \begin{bmatrix}
    &1&\\
    &0&\\
    &1&\end{bmatrix},
    \begin{bmatrix}
    &0&\\
    &0&\\
    &1&\end{bmatrix},
    \begin{bmatrix}
    &1&\\
    &0&\\
    &0&\end{bmatrix}   
\end{Bmatrix}$. 
Therefore, by Proposition \ref{prop:bo8}, $\mathbf{H_{300}}$ could form a construction of Figure-8, even this is a special case.
$\mathbf{H_{500}}^\top$ contains 
$\begin{bmatrix}
    &0&0&0&0&\\
    &0&1&0&1&\\
    &0&0&0&0&\\
    &0&1&1&0&\\
    &0&0&0&0&\\
    &0&0&0&0&\\
    &1&1&0&0&\\
\end{bmatrix}$ and 
$\begin{bmatrix}
    &0&0&0&0&\\
    &0&0&1&1&\\
    &0&0&0&0&\\
    &0&0&0&0&\\
    &0&0&0&0&\\
    &0&1&0&1&\\
    &1&1&0&0&\\
\end{bmatrix}$ subspaces. Both of the two contains the common vectors
$\begin{Bmatrix}
\begin{bmatrix}
    &0&\\
    &0&\\
    &0&\\
    &0&\\
    &0&\\
    &0&\\
    &1&\\
\end{bmatrix},
\begin{bmatrix}
    &0&\\
    &1&\\
    &0&\\
    &0&\\
    &0&\\
    &0&\\
    &0&\\
\end{bmatrix}
\end{Bmatrix}$. 
Therefore, by Proposition \ref{prop:bo8}, $\mathbf{H_{500}}$ could form a construction of Figure-8.
$\mathbf{H_{800}}^\top$ contains $\begin{bmatrix}
    &1&1&0&\\
    &0&0&0&\\
    &1&1&0&\\
    &0&0&0&\\
    &0&0&0&\\
    &0&1&1&\\
    &0&0&0&\\
\end{bmatrix}$ and 
$\begin{bmatrix}
    &1&1&0&\\
    &1&1&0&\\
    &0&0&0&\\
    &0&0&0&\\
    &0&0&0&\\
    &0&1&1&\\
    &0&0&0&\\
\end{bmatrix}$ sub-spaces. They contains the common vector
$\begin{Bmatrix}
\begin{bmatrix}
    &0&\\
    &0&\\
    &0&\\
    &0&\\
    &0&\\
    &1&\\
    &0&\\
\end{bmatrix}\end{Bmatrix}$. Therefore, by Proposition \ref{prop:bo8}, $\mathbf{H_{800}}$ could form a construction of Figure-8. This could be done by Algorithm \ref{alg:cfig8}.

\subsection{Induced Maps of Noisy Annulus}
\begin{table}[!htb]
\footnotesize
    \begin{tabular*}{\columnwidth}{@{\extracolsep{\fill}}ccccc}
    \hline
        Indices&Subsample size&IM Homology&$H_1$& $\mathbf{H}$\\\hline 
        0-9  &20 &$(1, 0,...)$&$\mathbf{IM_{annulus-20}}$&$\mathbf{H_{annulus-20}}$\\
        10-19&50 &$(1, 1,...)$&$\mathbf{IM_{annulus-50}}$&$\mathbf{H_{annulus-50}}$\\
        20-29&100&$(1, 5,...)$&$\mathbf{IM_{annulus-100}}$&$\mathbf{H_{annulus-100}}$\\
        30-39&300&$(19,5,...)$&$\mathbf{IM_{annulus-300}}$&$\mathbf{H_{annulus-300}}$\\
        40-49&500&$(2, 6,...)$&$\mathbf{IM_{annulus-500}}$&$\mathbf{H_{annulus-500}}$\\
        50-59&800&$(2, 6,...)$&$\mathbf{IM_{annulus-800}}$&$\mathbf{H_{annulus-800}}$\\\hline 
    \end{tabular*}
    \caption{Annulus. Induced maps of different sample sizes. Sub-samples were randomly sampled.}\label{tab:ann_im_all}
\end{table}
\begin{table}[!htb]
\footnotesize
    \begin{tabular*}{.48\columnwidth}{@{\extracolsep{\fill}}ccc}
        \hline 
        index & size & $H_1$ \\
        \hline 
        0:20 & 20 & []\\
        20:40 & 20 & []\\
        40:60 & 20 & []\\
        60:80 & 20 & []\\
        80:100 & 20 & []\\
        120:140 & 20 & []\\
        140:160 & 20 & []\\
        160:180 & 20 & []\\
        180:200 & 20 & []\\
        200:220 & 20 & []\\
    \end{tabular*} 
    \label{tab:annulus_20_im}
    \begin{tabular*}{.48\columnwidth}{@{\extracolsep{\fill}}ccc}
        \hline 
        index & size & $H_1$\\
        \hline 
        0:50 & 50 & [[0, 0, 0, 2, 0, 0]]\\
        50:100 & 50 & [[0, 0]]\\
        100:150 & 50 & [[0, 0]]\\
        150:200 & 50 & [[]]\\
        200:250 & 50 & [[0]]\\
        250:300 & 50 & [[0]]\\
        300:350 & 50 & [[0]]\\
        350:400 & 50 & [[]]\\
        400:450 & 50 &[[0]]\\
        450:500 & 50 &[[0]]\\
    \end{tabular*}
    \label{tab:annulus_50_im}
    \centering
    \begin{tabular*}{\columnwidth}{@{\extracolsep{\fill}}ccc}
        \hline 
        index & size & $H_1$ \\\hline 
        0:100 & 100 &[[0, 0, 0, 0, 2, 0, 0, 0, 0, 1],[0, 0, 0, 0, 1, 0, 0, 0, 0, 0],[0, 1, 0, 0, 0, 0, 0, 0, 0, 0],\\
        & &[0, 0, 0, 0, 0, 2, 0, 0, 0, 0],[0, 0, 0, 0, 0, 2, 0, 0, 0, 0]]\\
        100:200 & 100 & [[0, 1, 0], [0, 2, 0], [1, 0, 0], [0, 0, 0], [0, 0, 0]]\\
        200:300 & 100 & [[2], [1], [2], [0], [0]]\\
        300:400 & 100 & [[1], [2], [1], [0], [0]]\\
        400:500 & 100 & [[1], [2], [1], [0], [0]]\\
        500:600 & 100 & [[1], [2], [1], [0], [0]]\\
        600:700 & 100 & [[1], [2], [1], [0], [0]]\\
        700:800 & 100 & [[2], [1], [2], [0], [0]]\\
        800:900 & 100 & [[1], [2], [1], [0], [0]]\\
        900:1000 & 100 & [[1], [2], [1], [0], [0]]\\
    \end{tabular*}
    \label{tab:annulus_100_im}
    \begin{tabular*}{\columnwidth}{@{\extracolsep{\fill}}ccc}
        \hline 
        index & size & $H_1$ \\\hline 
        0:300 & 300 &[[0, 0, 2, 0, 0], [0, 0, 0, 0, 0], [0, 0, 0, 0, 2], [0, 1, 0, 0, 0], [0, 0, 0, 1, 0]]\\
        100:400 & 300 &[[1, 0, 0, 0], [0, 0, 0, 0], [0, 0, 0, 0], [0, 0, 0, 0], [0, 0, 0, 0]]\\
        200:500 & 300 & [[2], [0], [0], [0], [0]]\\
        300:600 & 300 & [[1, 0, 0], [0, 0, 0], [0, 0, 0], [0, 0, 0], [0, 0, 0]]\\
        400:700 & 300 & [[0, 0, 2, 0], [0, 0, 0, 0], [0, 0, 0, 0], [0, 0, 0, 0], [0, 0, 0, 0]]\\
        500:800 & 300 & [[2], [0], [0], [0], [0]] \\
        600:900 & 300 & [[1, 0], [0, 0], [0, 0], [0, 0], [0, 0]]\\
        700:1000 & 300 & [[0, 1], [0, 0], [0, 0], [0, 0], [0, 0]]\\
        250:550 & 300 &[[1, 0], [0, 0], [0, 0], [0, 0], [0, 0]]\\
        550:850 & 300 &[[2], [0], [0], [0], [0]]\\
    \end{tabular*}
    \label{tab:annulus_300_im}
    \begin{tabular*}{\columnwidth}{@{\extracolsep{\fill}}ccc}
        \hline 
        index&size& $H_1$ \\\hline 
        0:500&500&[[0, 0, 0, 0, 0, 1], [2, 0, 0, 1, 0, 0], [1, 0, 0, 0, 0, 0],\\
        & & [1, 2, 2, 2, 2, 0], [0, 2, 0, 0, 2, 0], [0, 2, 0, 0, 0, 0]]\\
        100:600&500&[[0, 0], [1, 0], [2, 0], [2, 2], [0, 0], [0, 0]]\\
        200:700&500&[[0, 0], [1, 0], [2, 0], [2, 1], [0, 0], [0, 0]]\\
        300:800&500&[[0], [1], [2], [2], [0], [0]]\\
        400:900&500&[[0], [1], [2], [2], [0], [0]]\\
        500:1000&500&[[0], [2], [1], [1], [0], [0]]\\
        450:950&500&[[0], [2], [1], [1], [0], [0]]\\
        350:850&500&[[0], [1], [2], [2], [0], [0]]\\
        250:750&500&[[0], [2], [1], [1], [0], [0]]\\
        150:650&500&[[0, 0], [2, 0], [1, 0], [1, 1], [0, 0], [0, 0]]\\
    \end{tabular*}
    \label{tab:annulus_500_im}
    \centering
    \begin{tabular*}{\columnwidth}{@{\extracolsep{\fill}}ccc}
        \hline 
        index & size & $H_1$ \\\hline 
        0:800 & 800 &[[0, 0, 0, 0, 0, 1], [2, 0, 0, 1, 0, 0], [1, 0, 0, 0, 0, 0],\\
        & & [1, 2, 2, 2, 2, 0], [0, 2, 0, 0, 2, 0], [0, 2, 0, 0, 0, 0]]\\
        100:900 & 800 & [[0, 0], [1, 0], [2, 0], [2, 2], [0, 0], [0, 0]]\\
        200:1000 & 800 & [[0, 0], [1, 0], [2, 0], [2, 1], [0, 0], [0, 0]]\\
        50:850 & 800 &[[0, 0, 0, 0, 0, 0],[0, 0, 1, 1, 0, 0],[0, 0, 2, 2, 1, 1],\\
        & &[0, 0, 0, 2, 0, 0],[0, 0, 0, 0, 0, 0],[0, 0, 0, 0, 0, 0]]\\
        150:950 & 800 &[[0, 0], [2, 0], [1, 0], [1, 1], [0, 0], [0, 0]]\\
        20:820 & 800 &[[0, 0, 0, 0, 0],[0, 1, 0, 2, 0],[0, 2, 1, 1, 0],[0, 2, 0, 0, 2],[0, 0, 0, 0, 2],[0, 0, 0, 0, 0]]\\
        120:920 & 800 & [[0, 0], [1, 0], [2, 0], [2, 1], [0, 0], [0, 0]]\\
        130:930 & 800 & [[0, 0], [2, 0], [1, 0], [1, 1], [0, 0], [0, 0]]\\
        40:840 & 800 &[[0, 0, 0, 0, 0, 0], [0, 0, 0, 1, 0, 2], [0, 2, 0, 0, 1, 0], \\
        & &[0, 0, 1, 0, 0, 0], [0, 0, 0, 0, 0, 0], [0, 0, 0, 0, 0, 0]]\\
        140:940 & 800 &[[0, 0], [1, 1], [2, 2], [2, 0], [0, 0], [0, 0]]\\
    \end{tabular*}
    \caption{Annulus. Induced maps from sub-samples to full point cloud of size $1000$. Sub-samples were randomly chosen from the full point cloud.}
    \label{tab:annulus_800_im}
\end{table} 
\clearpage

$\mathbf{IM_{annulus-20}}=\begin{Bmatrix}
\begin{bmatrix}&\end{bmatrix},
\begin{bmatrix}&\end{bmatrix},
\begin{bmatrix}&\end{bmatrix},
\begin{bmatrix}&\end{bmatrix},
\begin{bmatrix}&\end{bmatrix},
\begin{bmatrix}&\end{bmatrix},
\begin{bmatrix}&\end{bmatrix},
\begin{bmatrix}&\end{bmatrix},
\begin{bmatrix}&\end{bmatrix}
\end{Bmatrix}$.

$\mathbf{IM_{annulus-50}}=\begin{Bmatrix}
\begin{bmatrix}&0& 0& 0& 2& 0& 0&\end{bmatrix}, 
\begin{bmatrix}&0& 0&\end{bmatrix}, 
\begin{bmatrix}&0& 0&\end{bmatrix}, \\
\begin{bmatrix}&\end{bmatrix}, 
\begin{bmatrix}&0&\end{bmatrix}, 
\begin{bmatrix}&0&\end{bmatrix}, 
\begin{bmatrix}&0&\end{bmatrix}, 
\begin{bmatrix}&\end{bmatrix}, 
\begin{bmatrix}&0&\end{bmatrix},
\begin{bmatrix}&0&\end{bmatrix}
\end{Bmatrix}$.

$\mathbf{IM_{annulus-100}}=\begin{Bmatrix}
\begin{bmatrix}
& 0 & 2 & 0 & 1 &\\
& 0 & 1 & 0 & 0 &\\
& 1 & 0 & 0 & 0 &\\
& 0 & 0 & 2 & 0 &\\
& 0 & 0 & 2 & 0 &\end{bmatrix}, 
\begin{bmatrix}
&0& 1& 0& \\
&0& 2& 0& \\
&1& 0& 0& \\
&0& 0& 0& \\
&0& 0& 0& \end{bmatrix},
\begin{bmatrix}
&2& \\
&1& \\
&2& \\
&0& \\
&0& \end{bmatrix}, 
\begin{bmatrix}
&1& \\
&2& \\
&1& \\
&0& \\
&0& \end{bmatrix}, \\
\begin{bmatrix}
&1& \\
&2& \\
&1& \\
&0& \\
&0& \end{bmatrix}, 
\begin{bmatrix}
 &1& \\
 &2& \\
 &1& \\
 &0& \\
 &0&\end{bmatrix},
\begin{bmatrix}
 &1& \\
 &2& \\
 &1& \\
 &0& \\
 &0&\end{bmatrix},
\begin{bmatrix}
 &2& \\
 &1& \\
 &2& \\
 &0& \\
 &0& \end{bmatrix},
\begin{bmatrix}
 &1& \\
 &2& \\
 &1& \\
 &0& \\
 &0&\end{bmatrix},
\begin{bmatrix}
 &1& \\
 &2& \\
 &1& \\
 &0& \\
 &0& \end{bmatrix}
\end{Bmatrix}$.

$\mathbf{IM_{annulus-300}}=\begin{Bmatrix}
\begin{bmatrix}
&0& 0& 2& 0& 0&\\
&0& 0& 0& 0& 0&\\
&0& 0& 0& 0& 2&\\
&0& 1& 0& 0& 0&\\
&0& 0& 0& 1& 0&\end{bmatrix}, 
\begin{bmatrix}
&1& 0& 0& 0& \\
&0& 0& 0& 0& \\
&0& 0& 0& 0& \\
&0& 0& 0& 0& \\
&0& 0& 0& 0&\end{bmatrix}, 
\begin{bmatrix}
&2& \\
&0& \\
&0& \\
&0& \\
&0&\end{bmatrix}, \\
\begin{bmatrix}
&1& 0& 0& \\
&0& 0& 0& \\
&0& 0& 0& \\
&0& 0& 0& \\
&0& 0& 0&\end{bmatrix}, 
\begin{bmatrix}
&0& 0& 2& 0& \\
&0& 0& 0& 0& \\
&0& 0& 0& 0& \\
&0& 0& 0& 0& \\
&0& 0& 0& 0&\end{bmatrix},
\begin{bmatrix}
&2& \\
&0& \\
&0& \\
&0& \\
&0&\end{bmatrix},\\
\begin{bmatrix}
&1& 0&\\
&0& 0&\\
&0& 0&\\
&0& 0&\\
&0& 0&\end{bmatrix}, 
\begin{bmatrix}
&0& 1& \\
&0& 0& \\
&0& 0& \\
&0& 0& \\
&0& 0&\end{bmatrix},
\begin{bmatrix}
&1& 0& \\
&0& 0& \\
&0& 0& \\
&0& 0& \\
&0& 0&\end{bmatrix}, 
\begin{bmatrix}
&2&\\
&0&\\
&0&\\
&0&\\
&0&\end{bmatrix}
\end{Bmatrix}$.

$\mathbf{IM_{annulus-500}}=\begin{Bmatrix}
\begin{bmatrix}
&0& 0& 0& 0& 0& 1&\\
&2& 0& 0& 1& 0& 0&\\
&1& 0& 0& 0& 0& 0&\\
&1& 2& 2& 2& 2& 0&\\
&0& 2& 0& 0& 2& 0&\\
&0& 2& 0& 0& 0& 0&\end{bmatrix},
\begin{bmatrix}
&0& 0&\\
&1& 0&\\
&2& 0&\\
&2& 2&\\
&0& 0&\\
&0& 0&\end{bmatrix},
\begin{bmatrix}
&0& 0&\\ 
&1& 0&\\ 
&2& 0&\\
&2& 1&\\
&0& 0&\\ 
&0& 0&\\
\end{bmatrix},
\begin{bmatrix}
&0& \\
&1& \\
&2& \\
&2& \\
&0& \\
&0&\end{bmatrix},\\
\begin{bmatrix}
&0& \\
&1& \\
&2& \\
&2& \\
&0& \\
&0&\end{bmatrix},
\begin{bmatrix}
&0& \\
&2& \\
&1& \\
&1& \\
&0& \\
&0&\end{bmatrix},
\begin{bmatrix}
&0& \\
&2& \\
&1& \\
&1& \\
&0& \\
&0&\end{bmatrix},
\begin{bmatrix}
&0& \\
&1& \\
&2& \\
&2& \\
&0& \\
&0&\end{bmatrix},
\begin{bmatrix}
&0& \\
&2&\\
&1& \\
&1&\\
&0& \\
&0&\end{bmatrix},
\begin{bmatrix}
&0& 0& \\
&2& 0& \\
&1& 0& \\
&1& 1& \\
&0& 0& \\
&0& 0&\end{bmatrix}
\end{Bmatrix}$.

$\mathbf{IM_{annulus-800}}=\begin{Bmatrix}
\begin{bmatrix}
&0& 0& 0& 0& 0& 1&\\
&2& 0& 0& 1& 0& 0&\\
&1& 0& 0& 0& 0& 0&\\
&1& 2& 2& 2& 2& 0&\\
&0& 2& 0& 0& 2& 0&\\
&0& 2& 0& 0& 0& 0&
\end{bmatrix},
\begin{bmatrix}
&0& 0&\\
&1& 0&\\
&2& 0&\\
&2& 2&\\
&0& 0&\\
&0& 0&
\end{bmatrix}, 
\begin{bmatrix}
&0& 0&\\
&1& 0&\\
&2& 0&\\
&2& 1&\\
&0& 0&\\
&0& 0&
\end{bmatrix}, \\
\begin{bmatrix}
&0& 0& 0& 0& 0& 0& \\
&0& 0& 1& 1& 0& 0&\\
&0& 0& 2& 2& 1& 1& \\
&0& 0& 0& 2& 0& 0&\\
&0& 0& 0& 0& 0& 0& \\
&0& 0& 0& 0& 0& 0&
\end{bmatrix}, 
\begin{bmatrix}
&0& 0& \\
&2& 0&\\
&1& 0&\\
&1& 1& \\
&0& 0&\\
&0& 0&
\end{bmatrix}, 
\begin{bmatrix}
&0& 0& 0& 0& 0& \\
&0& 1& 0& 2& 0& \\
&0& 2& 1& 1& 0&\\
 &0& 2& 0& 0& 2& \\
 &0& 0& 0& 0& 2& \\
 &0& 0& 0& 0& 0&
\end{bmatrix}, \\
\begin{bmatrix}
&0& 0& \\
&1& 0& \\
&2& 0& \\
&2& 1& \\
&0& 0& \\
&0& 0&
\end{bmatrix},
\begin{bmatrix}
&0& 0& \\
&2& 0& \\
&1& 0& \\
&1& 1& \\
&0& 0& \\
&0& 0&
\end{bmatrix}, 
\begin{bmatrix}
&0& 0& 0& 0& 0& 0&\\
&0& 0& 0& 1& 0& 2&\\
&0& 2& 0& 0& 1& 0&\\
&0& 0& 1& 0& 0& 0&\\
&0& 0& 0& 0& 0& 0&\\
&0& 0& 0& 0& 0& 0&
\end{bmatrix}, 
\begin{bmatrix}
&0& 0&\\
&1& 1&\\
&2& 2&\\
&2& 0&\\
&0& 0&\\
&0& 0&
\end{bmatrix}
\end{Bmatrix}$.
\subsection{Homology Basis of Annulus}
The estimated homology basis of noisy annulus computed by our algorithm is listed below.

$\mathbf{H_{annulus-20}}=\begin{bmatrix}&\end{bmatrix}$
$\mathbf{H_{annulus-50}}=\begin{bmatrix}&1&\end{bmatrix}$
$\mathbf{H_{annulus-100}}=\begin{bmatrix}
&0&0&1&0&0&\\
&1&1&0&0&0&\\
&0&0&0&1&1&\\
&1&0&0&0&0&\\
&1&1&1&0&0&
\end{bmatrix}$

$\mathbf{H_{annulus-300}}=\begin{bmatrix}
&0&0&0&1&0&\\
&1&0&0&0&0&\\
&0&0&0&0&1&\\
&0&0&1&0&0&
\end{bmatrix}$
$\mathbf{H_{annulus-500}}=\begin{bmatrix}
&0&1&1&1&0&0&\\
&0&0&0&1&1&1&\\
&0&0&0&1&0&0&\\
&0&1&0&1&0&0&\\
&0&0&0&1&1&0&\\
&1&0&0&0&0&0&
\end{bmatrix}$

$\mathbf{H_{annulus-800}}=\begin{bmatrix}
&0 &1 &1 &1 &0 &0&\\
&0 &0 &0 &1 &1 &1&\\
&0 &0 &0 &1 &0 &0&\\
&0 &1 &0 &1 &0 &0&\\
&0 &0 &0 &1 &1 &0&\\
&1 &0 &0 &0 &0 &0&\\
&0 &1 &1 &0 &0 &0&\\
&0 &0 &1 &0 &0 &0&\\
&0 &1 &0 &0 &0 &0&
\end{bmatrix}$

Similarly, From Definition \ref{def_cycle}, we could derive the following property for Annulus.
\begin{proposition}[Basis of Annulus]\label{prop:boa}
If there is a subspace $\mathbf{A}$ in the estimated homology space $\mathbf{H}$, such that, $d\cdot \mathbf{A} = 0$, then by Definition \ref{def_cycle} there exists a construction of Annulus in $\mathbf{H}$. 
\end{proposition}
Since each row represents an element of induced map, we are checking the row spaces. 
Intuitively, $\mathbf{H_{20}}^\top$ and $\mathbf{H_{50}}^\top$ does not satisfy the condition of the above property. 
$\mathbf{H_{100}}^\top$ contains a subspace
$\begin{bmatrix}
    &0&1&1&\\
    &0&1&1&\\
    &1&0&1&\\
    &0&0&0&\\
    &0&0&0&
\end{bmatrix}$. Therefore, by Proposition \ref{prop:boa}, $\mathbf{H_{100}}^\top$ could form a construction of Annulus.
Intuitively, $\mathbf{H_{300}}^\top$ is linearly independent, so it forms a Forest and doesn't form a cycle.  Therefore, $\mathbf{H_{300}}^\top$ cannot be a representation of Annulus. $\mathbf{H_{500}}^\top$ is in the same situation. 
$\mathbf{H_{800}}^\top$ contains a subspace 
$\begin{bmatrix}
    &0&0&0&0&0&\\
    &1&1&1&1&0&\\
    &1&0&1&0&1&\\
    &1&1&0&0&0&\\
    &0&0&0&0&0&\\
    &0&0&0&0&0&    
\end{bmatrix}$. Therefore, by Proposition \ref{prop:boa}, $\mathbf{H_{800}}^\top$ could form a construction of Annulus. This could be done by Algorithm \ref{alg:cdependency}.

%% file: 05_2discussion.tex
\section{Discussion}\label{chap:discussion}
\begin{table}[!htb]
\footnotesize
\centering
    \begin{tabular*}{\columnwidth}{@{\extracolsep{\fill}}cccc}
    \hline
    Indices&Subsample size&RC speed (seconds)&GMA speed (seconds)\\\hline
        0-9&20& 28.352442026138306   & 0.0018012523651123047\\
        10-19&50& 17.811870098114014 & 0.002874135971069336\\
        20-29&100& 8.287723064422607 & 0.0024347305297851562\\
        30-39&300& 2.054633855819702 & 0.004442691802978516\\
        40-49&500& 1.607776165008545 & 0.007197856903076172\\
        50-59&800& 0.7690119743347168& 0.008867025375366211\\\hline     
     \end{tabular*}
    \caption{Figure-8. Average speed of computation using Rips persistence diagram(RC) and greedy matroid algorithm(GMA).}\label{tab:fig8_speed}
    \begin{tabular*}{\columnwidth}{@{\extracolsep{\fill}}cccc}
    \hline
    Indices&Subsample size&RC speed (seconds)&GMA speed (seconds)\\\hline
        0-9&20&   56.045454263687134    & 0.0025789737701416016\\
        10-19&50& 15.729853868484497    & 0.0007250308990478516\\
        20-29&100& 6.515740871429443    & 0.0009210109710693359\\
        30-39&300& 0.5750489234924316   & 0.0013933181762695312\\
        40-49&500& 0.1607518196105957   & 0.00101470947265625  \\
        50-59&800& 0.05244708061218262  & 0.0011279582977294922\\\hline     
     \end{tabular*}
    \caption{Annulus. Average speed of computation using Rips persistence diagram(RC) and greedy matroid algorithm(GMA).}\label{tab:annulus_speed}
    \begin{tabular*}{\columnwidth}{@{\extracolsep{\fill}}ccccc}
    \hline
        Indices&Subsample size&Homology by TB&Homology by GMA&Target Homology\\\hline 
        0-9  &20 &$(1,0,...)$&$(0,0,...)$&$(1,2,...)$\\
        10-19&50 &$(1,3,...)$&$(1,0,...)$&$(1,2,...)$\\
        20-29&100&$(2,1,...)$&$(1,0,...)$&$(1,2,...)$\\
        30-39&300&$(6,3,...)$&$(1,2,...)$&$(1,2,...)$\\
        40-49&500&$(8,7,...)$&$(1,2,...)$&$(1,2,...)$\\
        50-59&800&$(12,7,...)$&$(1,2,...)$&$(1,2,...)$\\\hline 
    \end{tabular*}
    \caption{Figure-8. Persistence Homology computed by topological bootstrapping(TB) and greedy matroid algorithm(GMA).}\label{tab:f8_homology}
    \begin{tabular*}{\columnwidth}{@{\extracolsep{\fill}}ccccc}
    \hline
        Indices&Subsample size&Homology by TB&Homology by GMA&Target Homology\\\hline 
        0-9  &20 &$(1, 0,...)$&$(0,0,...)$&$(0,1,...)$\\
        10-19&50 &$(1, 1,...)$&$(1,0,...)$&$(0,1,...)$\\
        20-29&100&$(1, 5,...)$&$(0,1,...)$&$(0,1,...)$\\
        30-39&300&$(19,5,...)$&$(3,0,...)$&$(0,1,...)$\\
        40-49&500&$(2, 6,...)$&$(3,0,...)$&$(0,1,...)$\\
        50-59&800&$(2, 6,...)$&$(0,1,...)$&$(0,1,...)$\\\hline 
    \end{tabular*}
    \caption{Annulus. Persistence Homology computed by topological bootstrapping(TB) and greedy matroid algorithm(GMA).}\label{tab:annulus_homology}
\end{table}    
We computed the speed of computing persistence homology using Vietoris-Rips reduced Chain complex and greedy matroid algorithm. The results, Table \ref{tab:fig8_speed} and Table \ref{tab:annulus_speed}, show that our greedy matroid algorithm is faster than Rips persistence diagram. We compared the homology values, $H_0$ and $H_1$, computed by our greedy matroid algorithm and topological bootstrapping. The results, Table \ref{tab:f8_homology} and Table \ref{tab:annulus_homology}, show that homology values computed by our greedy matroid algorithm are more accurate than homology values computed by topological bootstrapping.

%% file: 06conclusion.tex
\section{Conclusion}\label{chap:conclusion}
The greedy matroid algorithm speeds up computing persistence homology. It provides an efficient way of computing the homology basis of sub-samples and inducing homology basis of the full space from sub-samples. Our algorithm is computationally more efficient than algorithms of computing persistence homology, i.e., Vietoris-Rips persistence diagram and persistence barcode. 

The accuracy of homology values, $H_0$ and $H_1$, computed by greedy matroid algorithm is higher than $H_0$ and $H_1$ computed by topological bootstrapping. Thus, our greedy matroid algorithm effectively addresses the ambiguity problem in topological bootstrapping.

Our algorithm gives a comprehensive indication of the basic construction of a given point cloud, which could not be achieved by computing Vietoris–Rips persistence diagram, persistence barcode, and topological bootstrapping. 

We will continue to test our method on three dimensional constructions. As we observed that the estimated bases of construction depend on the threshold of Vietoris-Rips complex, we will explore how to determine the optimum threshold. 
\section*{Acknowledgement} 
We thank the support of DARPA research.  

\clearpage

%% file: 07appendix.tex
\section{Appendix}\label{appendix}
The code is available at \url{https://github.com/tianyisuntt/sampling}.
\begin{figure*}[!b]
\begin{subfigure}[!b]{0.48\textwidth}
    \includegraphics[width=\linewidth]{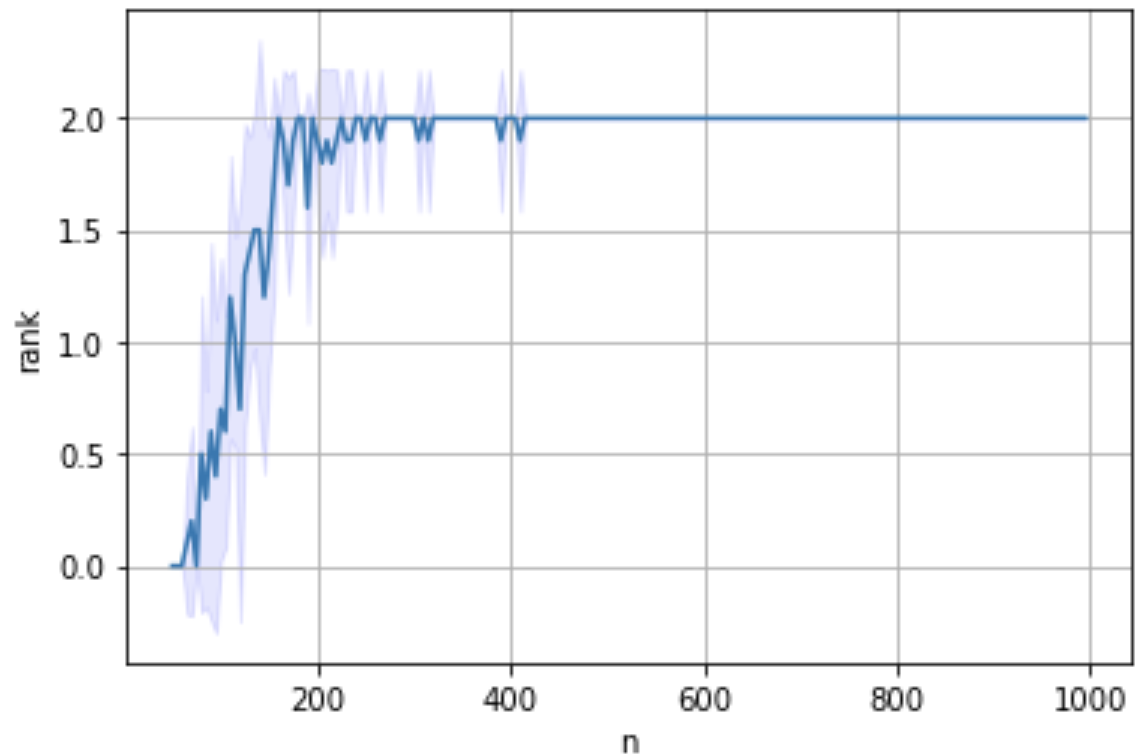}
    \caption{Expected rank of induced maps of Figure-8.}
\end{subfigure}
\begin{subfigure}[!b]{0.48\textwidth}
    \includegraphics[width=\linewidth]{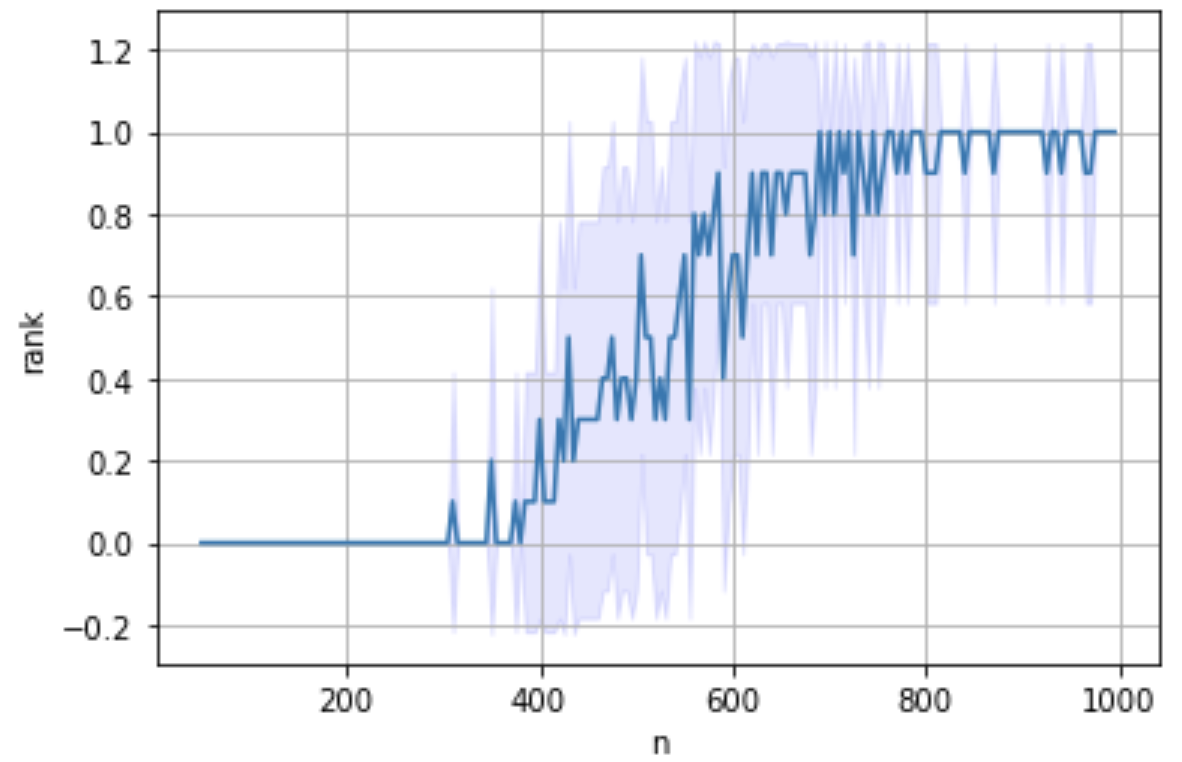}
    \caption{Expected rank of induced maps of Annulus.}
\end{subfigure}
    \caption{Rank statistics. $\mathbf{y}$-axis is the number of ranks and $\mathbf{x}$-axis is the number of samples.}
    \label{fig:ranks}
\end{figure*}
\subsection{Induced Maps of Figure-8}
\begin{figure}
\centering
    \includegraphics[width=0.48\linewidth]{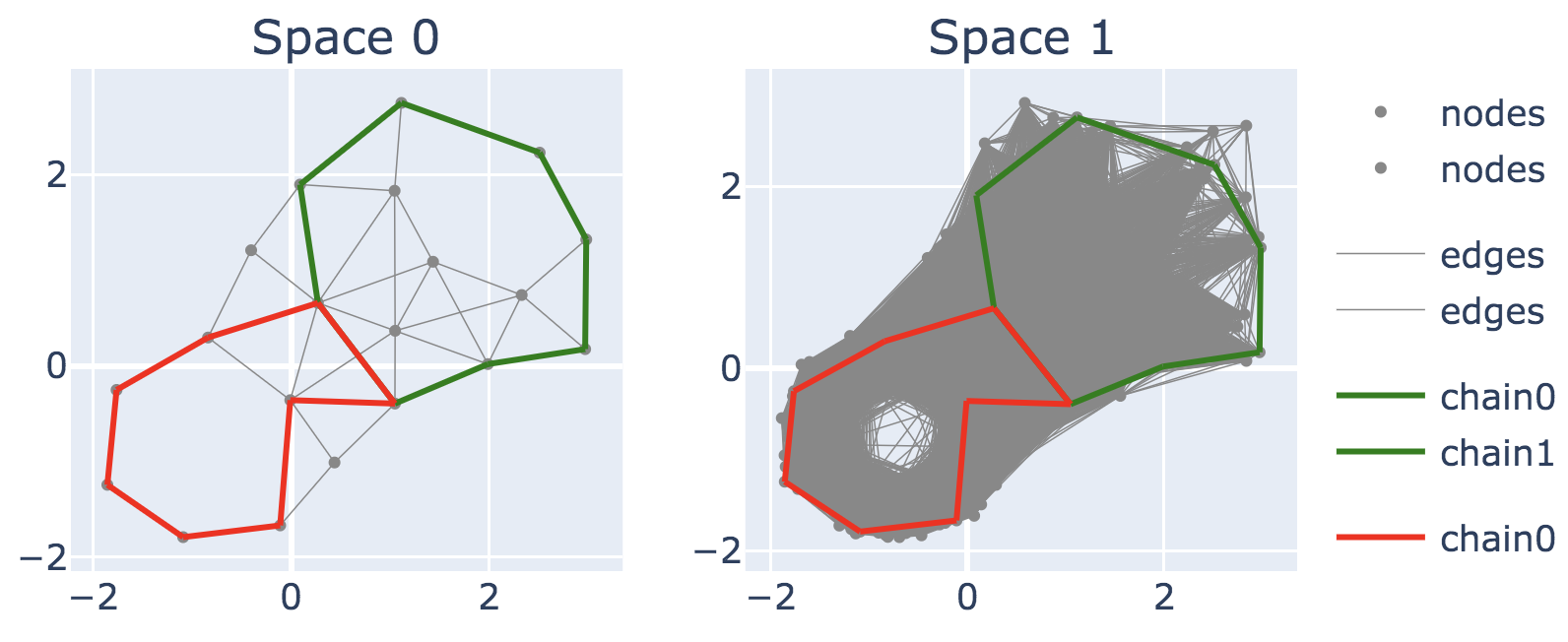}
    \includegraphics[width=0.48\linewidth]{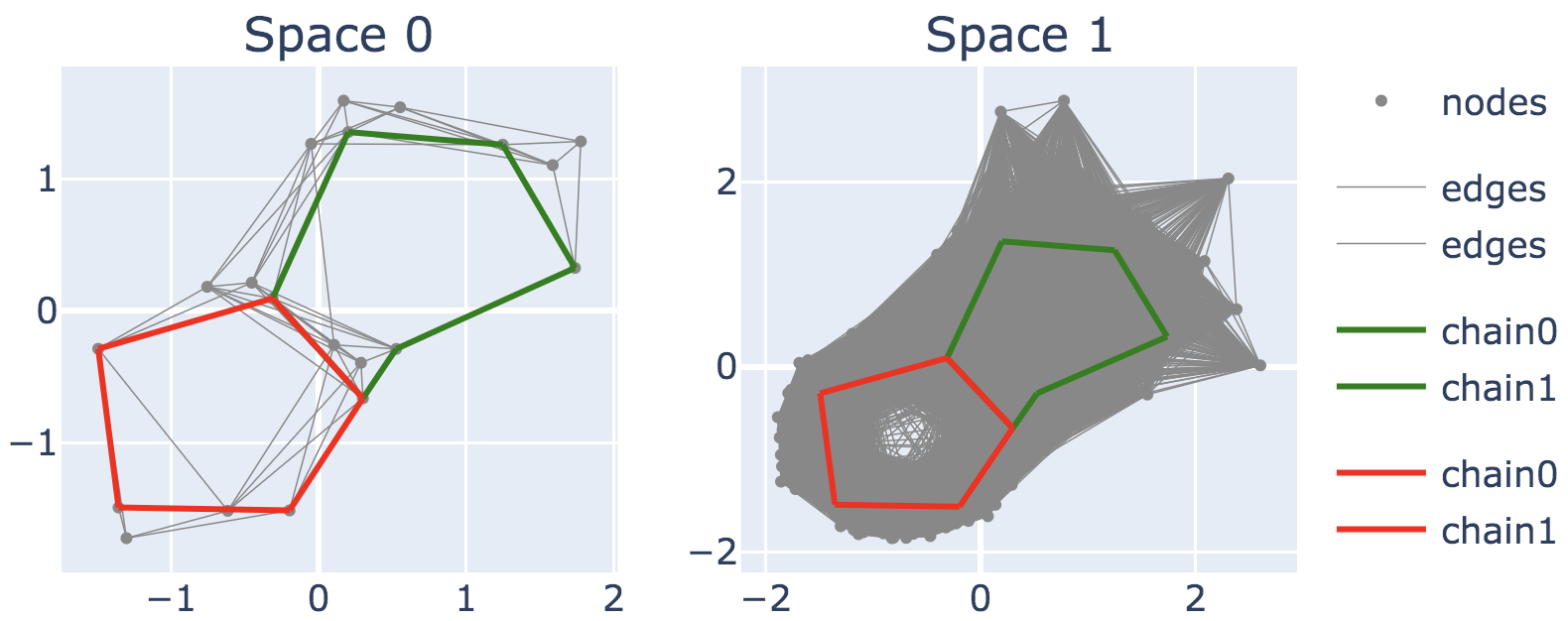}
    \includegraphics[width=0.48\linewidth]{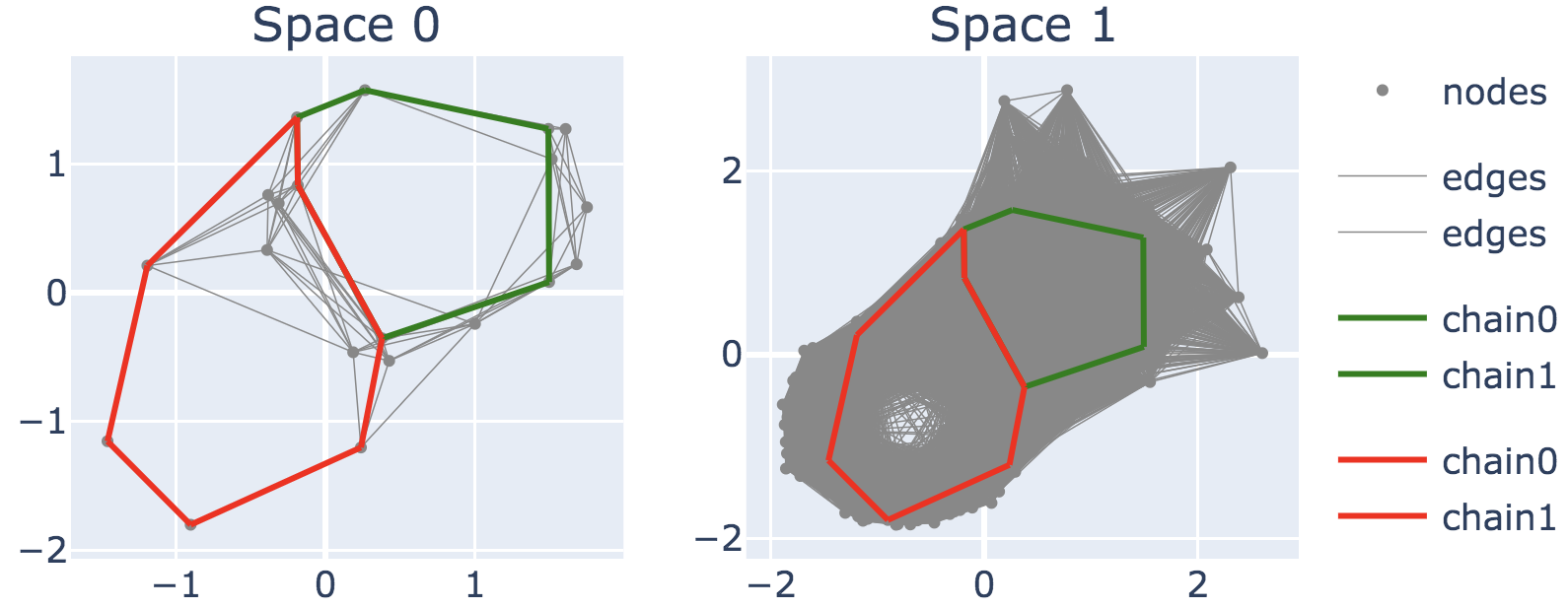}
    \includegraphics[width=0.48\linewidth]{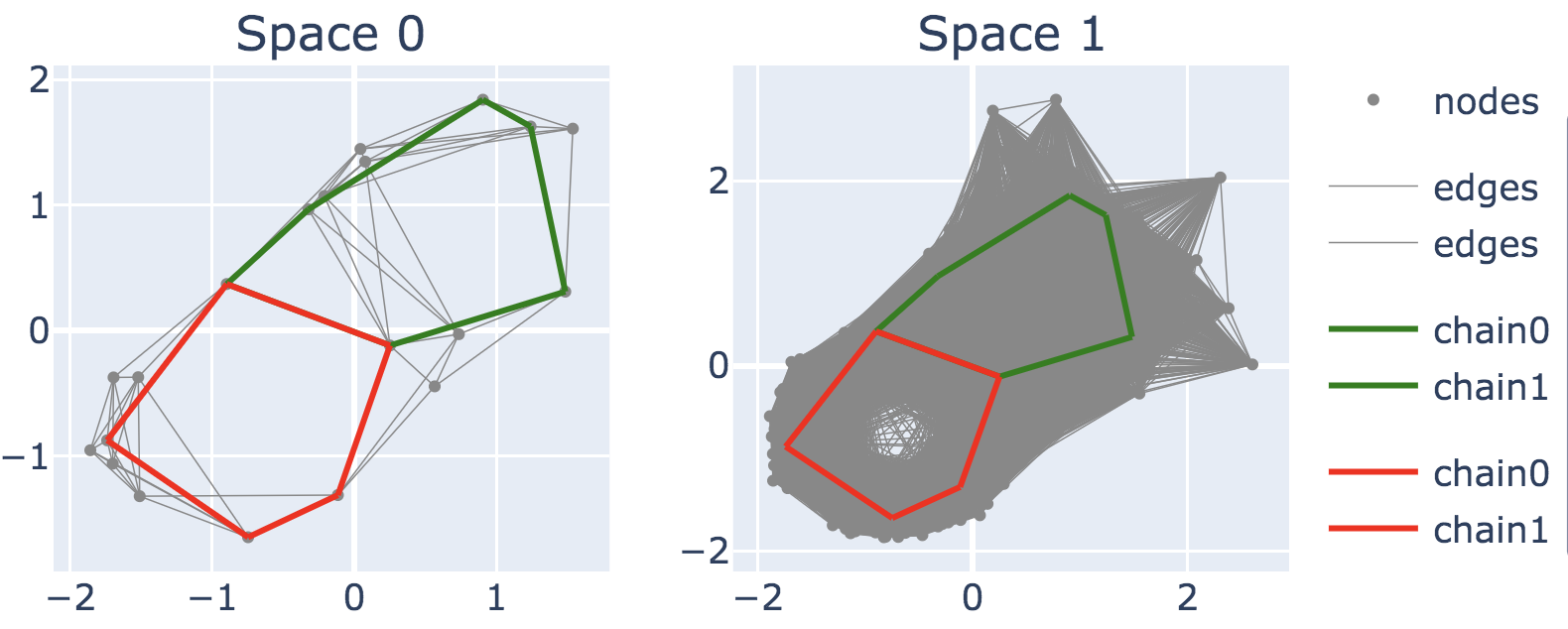}
    \includegraphics[width=0.48\linewidth]{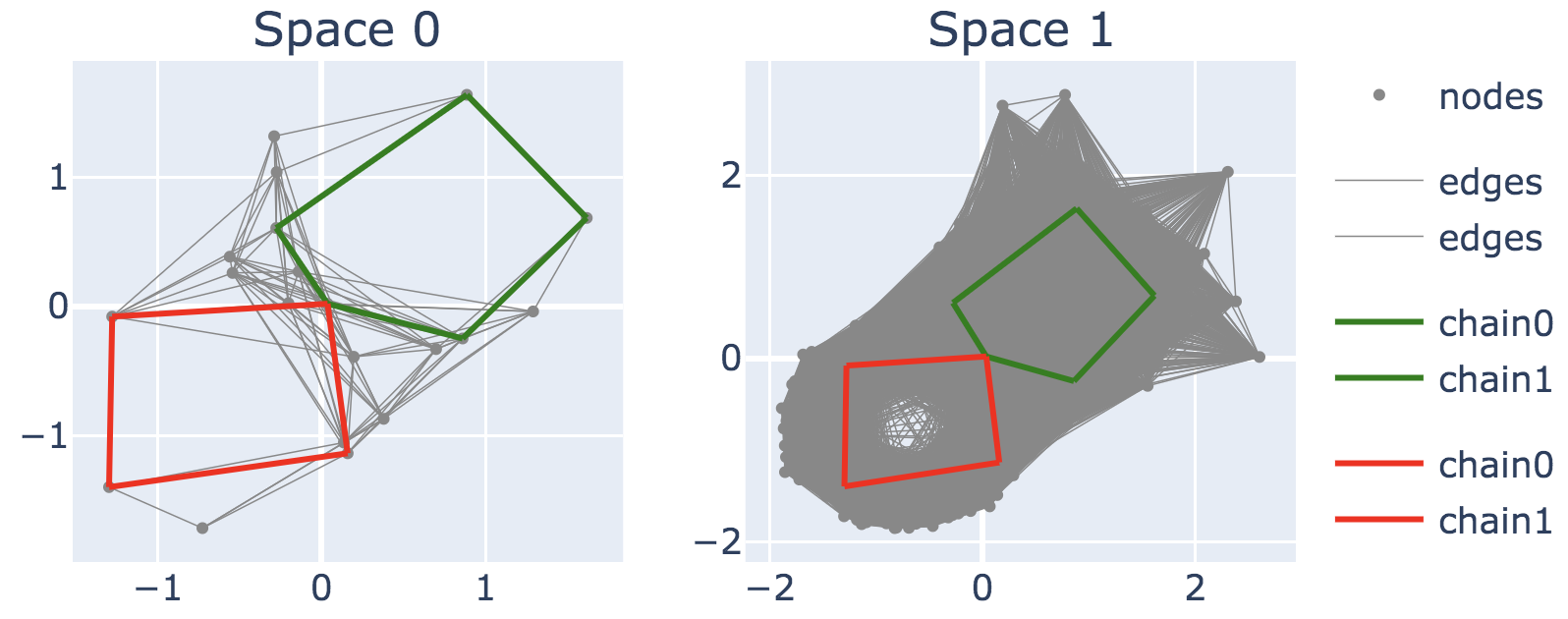}
    \includegraphics[width=0.48\linewidth]{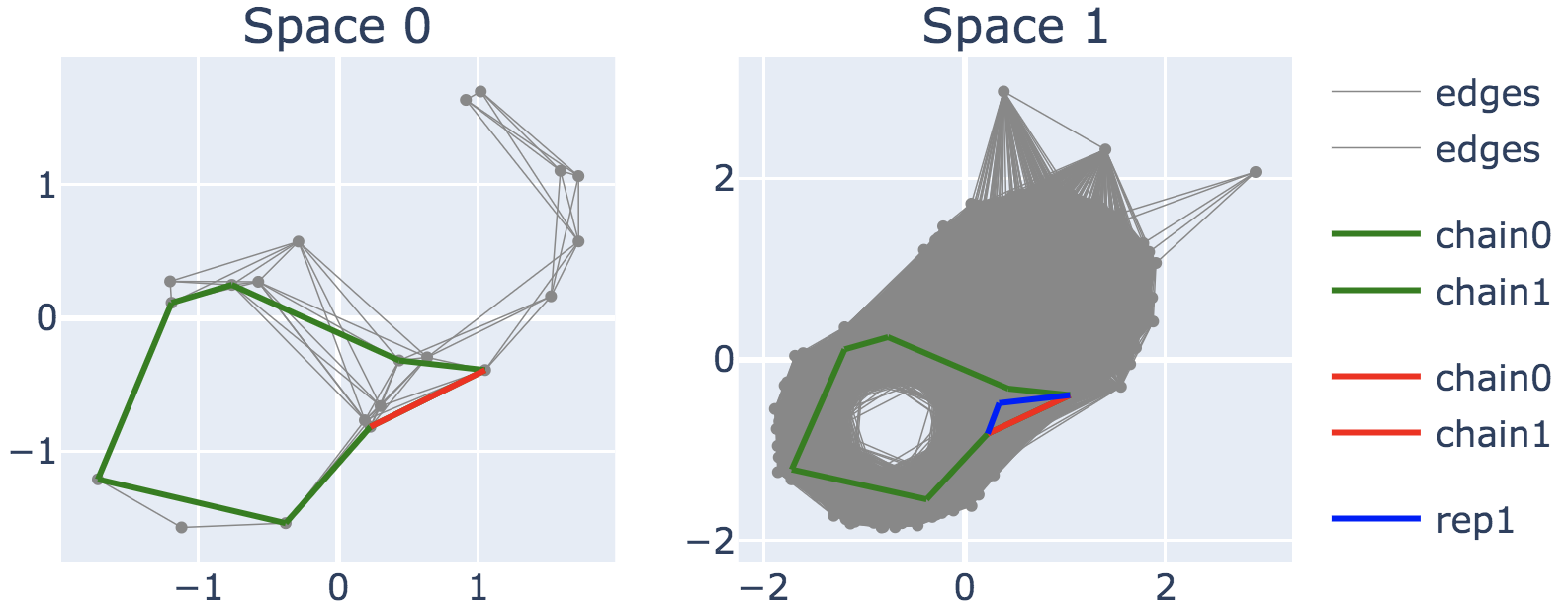}
    \includegraphics[width=0.48\linewidth]{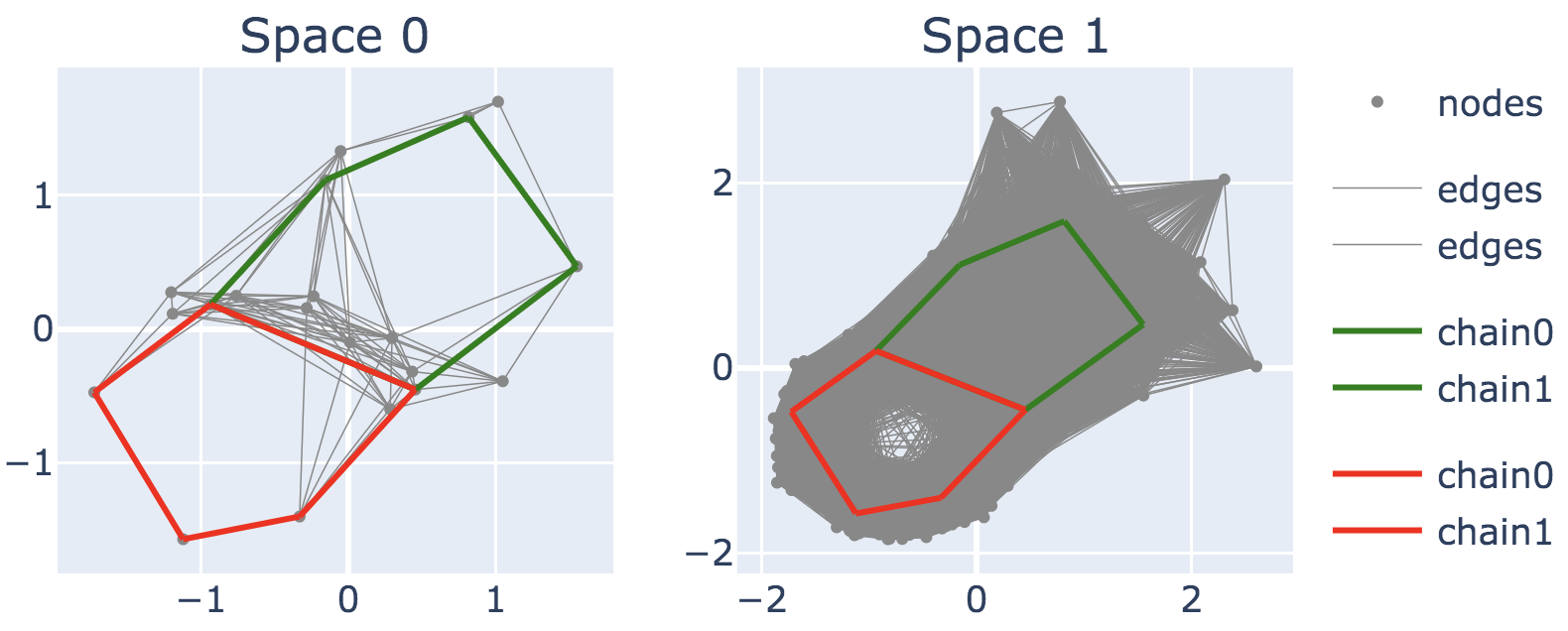}
    \includegraphics[width=0.48\linewidth]{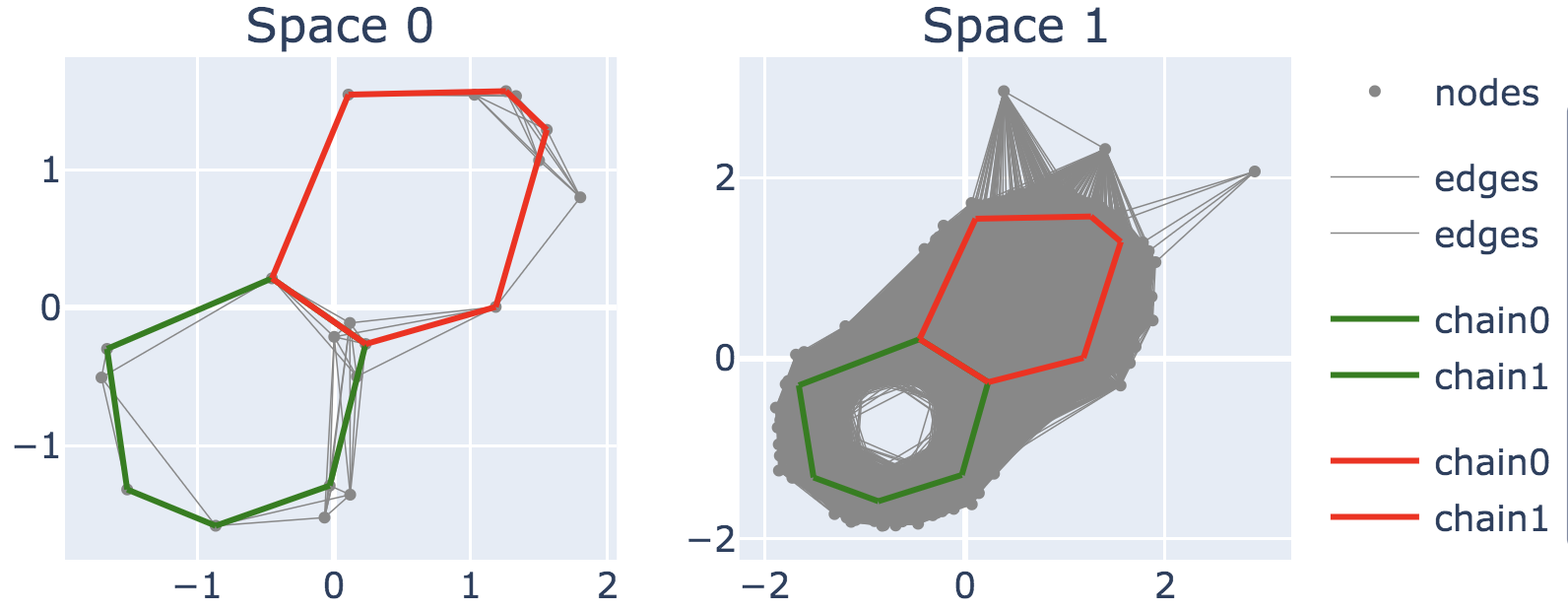}
    \includegraphics[width=0.48\linewidth]{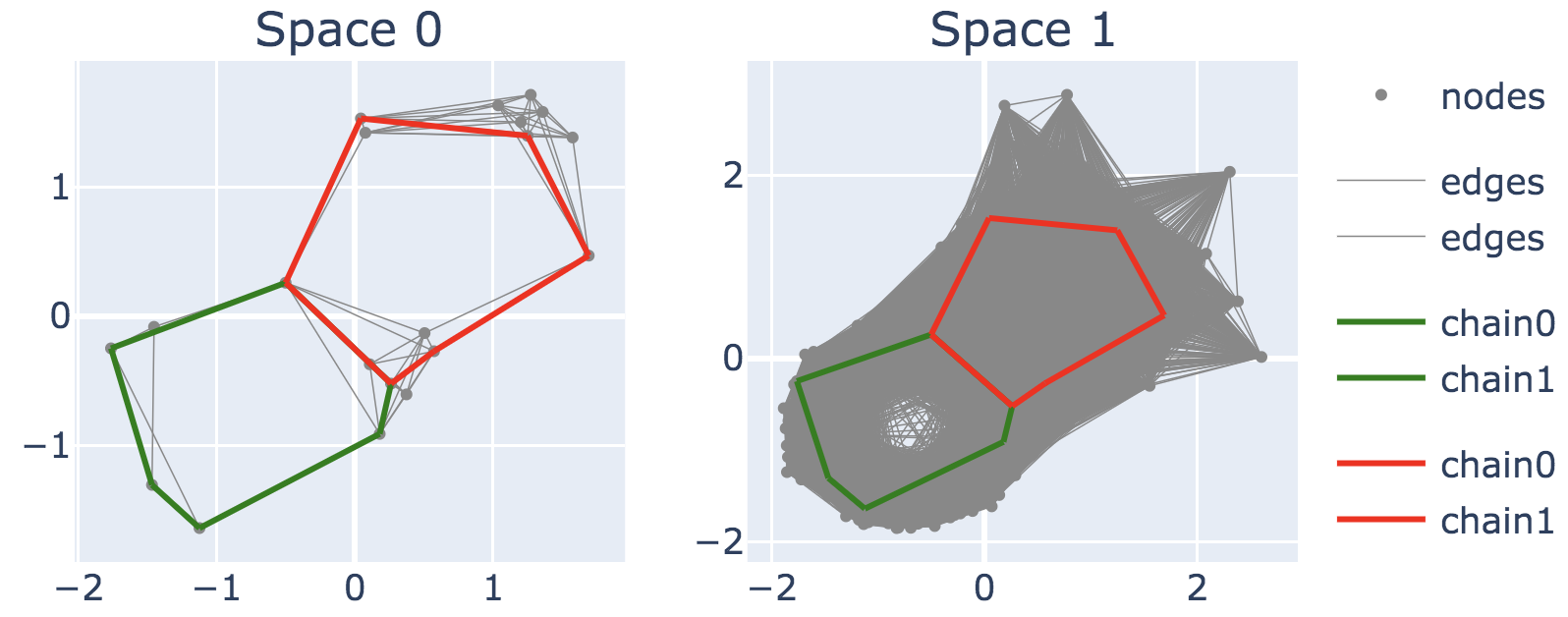}
    \includegraphics[width=0.48\linewidth]{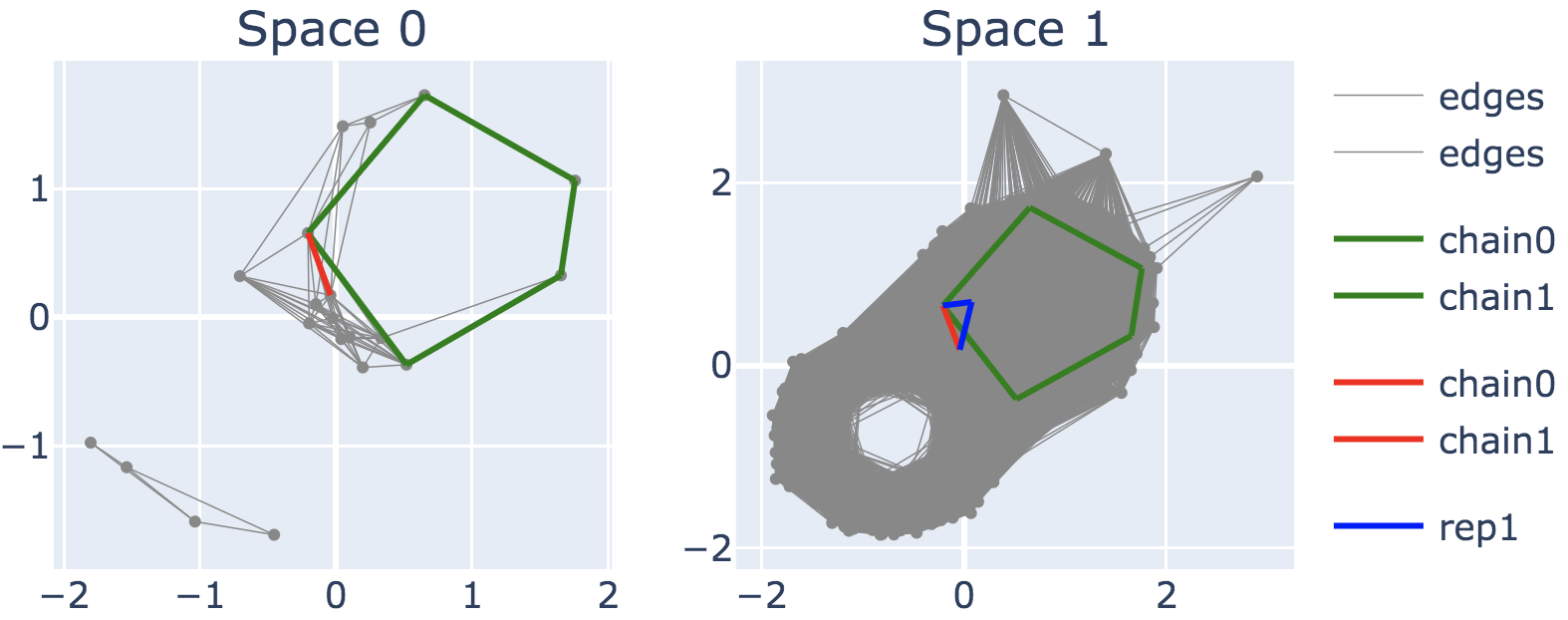}
    \caption{Visualization of induced maps of noisy Figure-8. Full size is $1000$. Sample size is $20$.}
    \label{fig:fig_8_20_im}
\end{figure}
\begin{figure}
\centering
    \includegraphics[width=0.48\linewidth]{figs/bad_fig8/50hd1_.png}
    \includegraphics[width=0.48\linewidth]{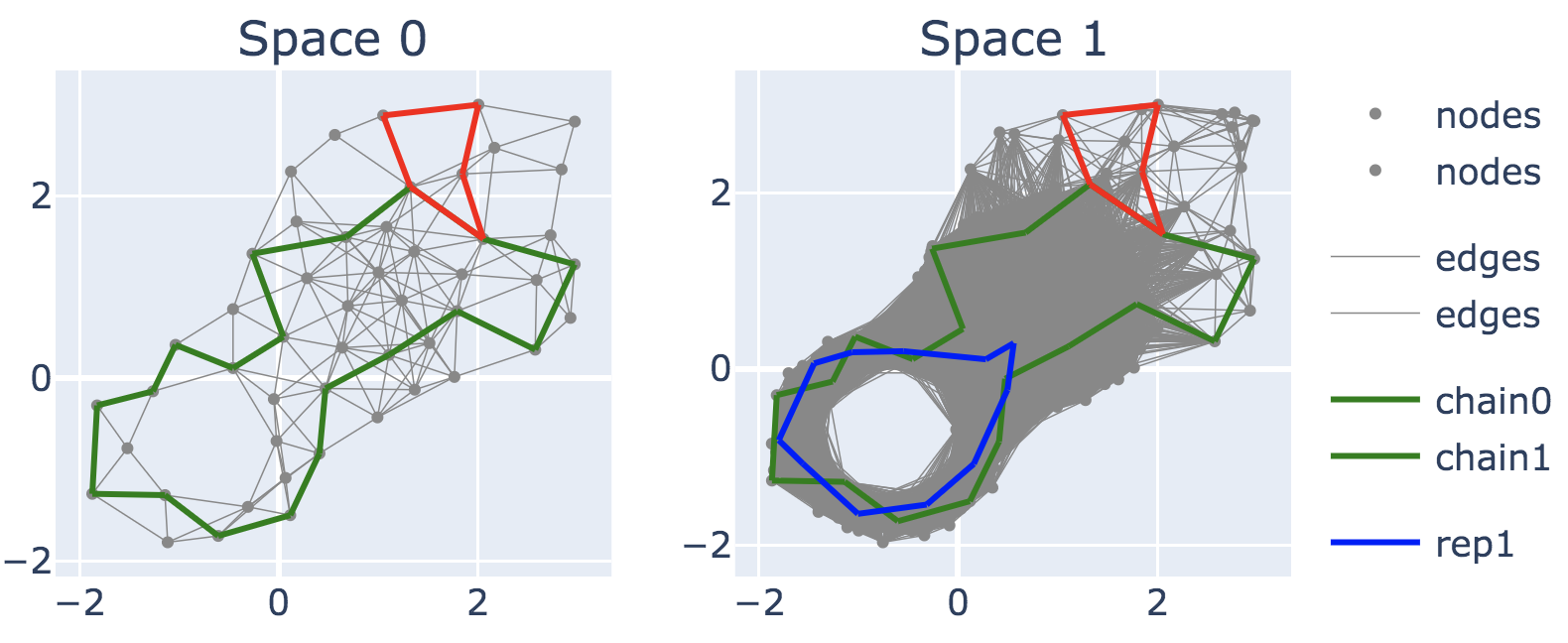}
    \includegraphics[width=0.48\linewidth]{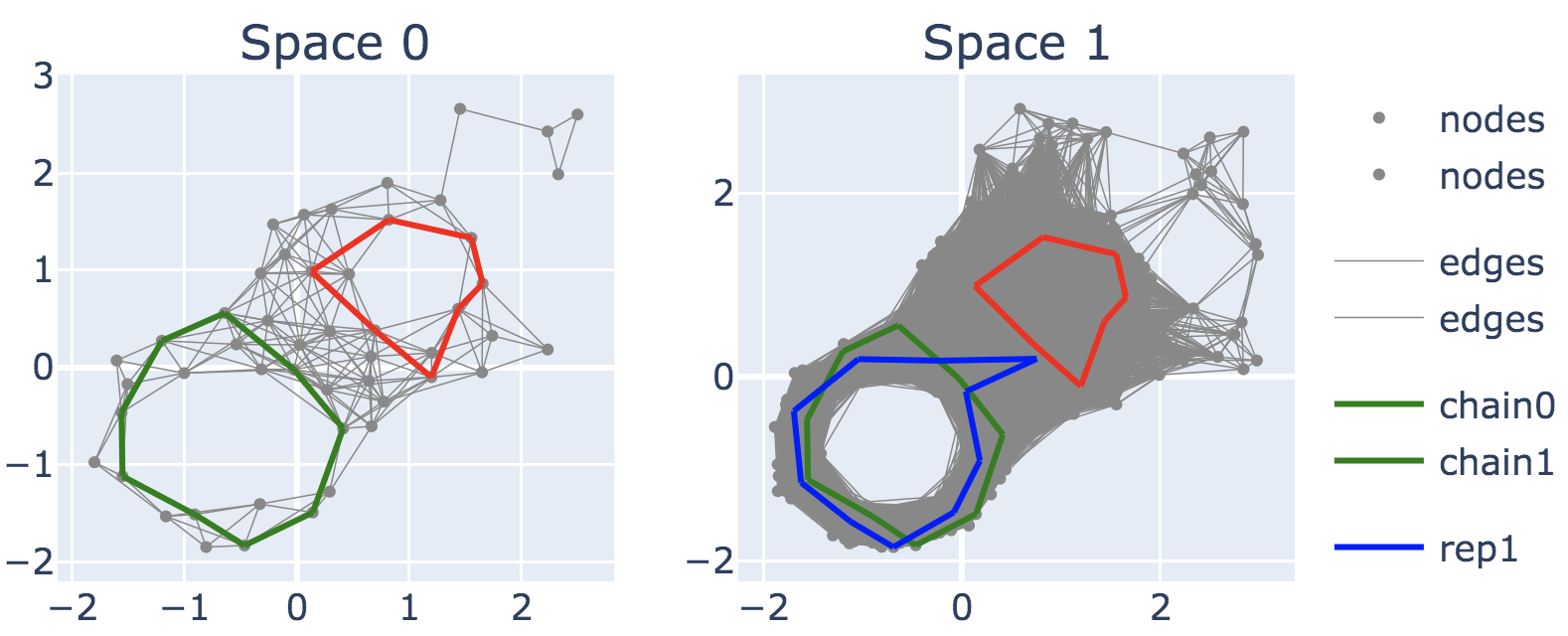}
    \includegraphics[width=0.48\linewidth]{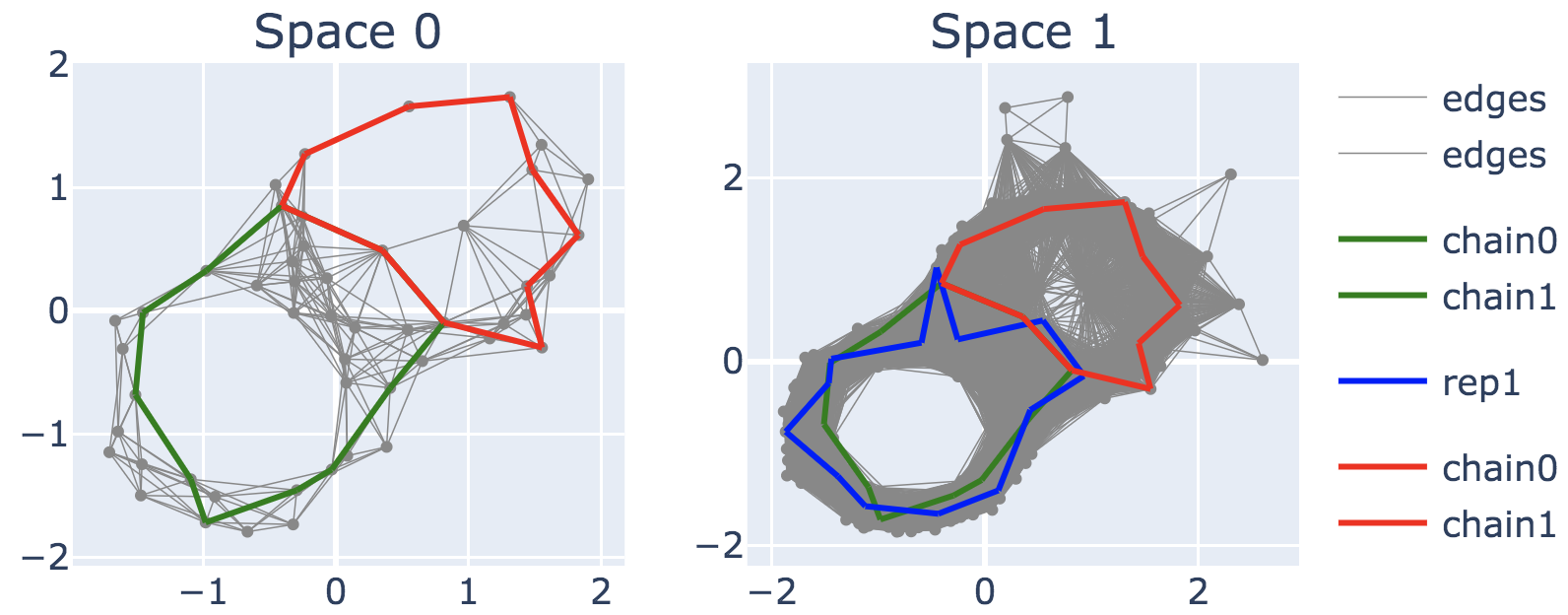}
    \includegraphics[width=0.48\linewidth]{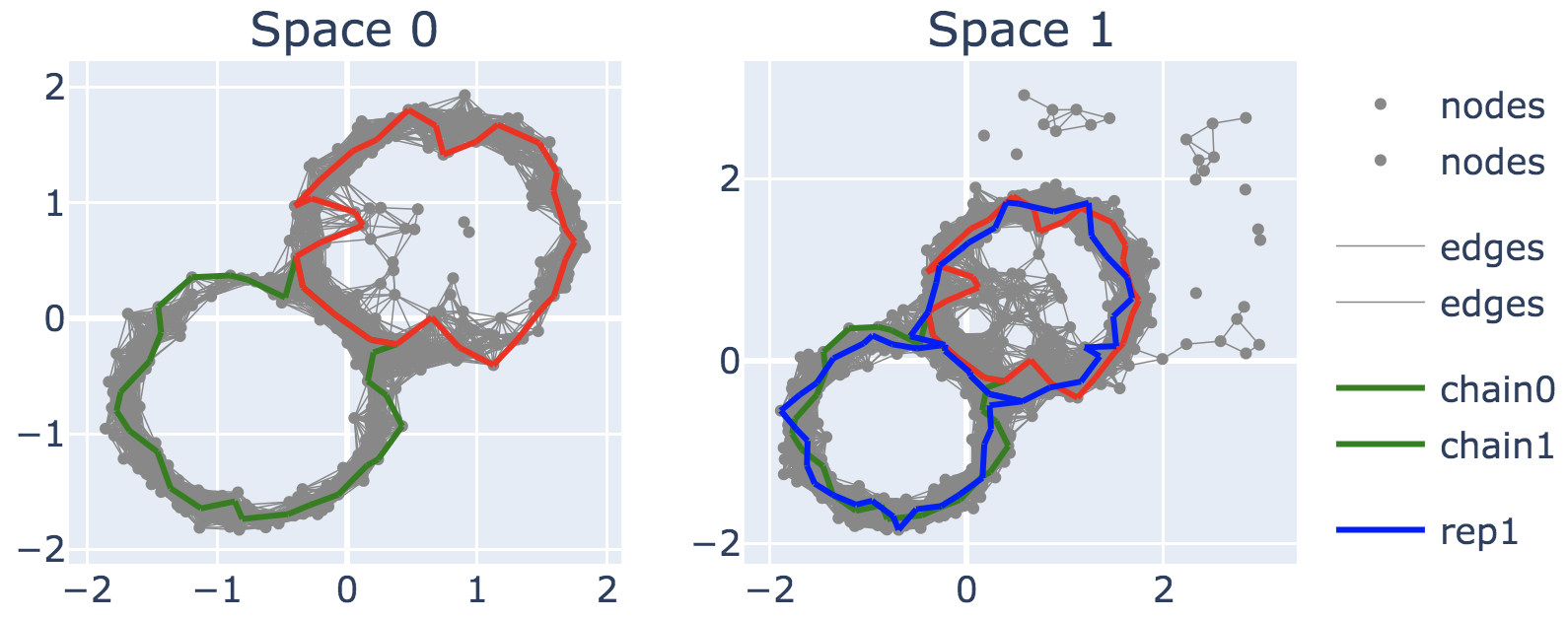}
    \includegraphics[width=0.48\linewidth]{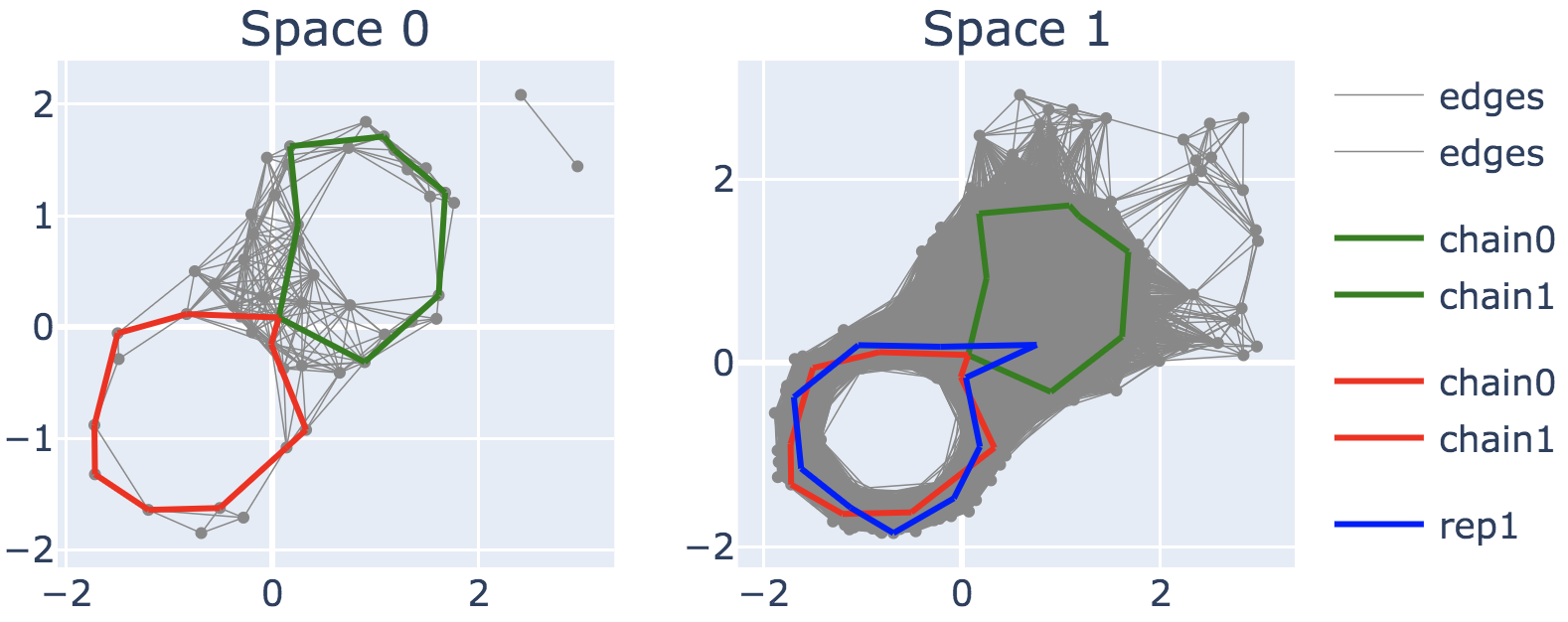}
    \includegraphics[width=0.48\linewidth]{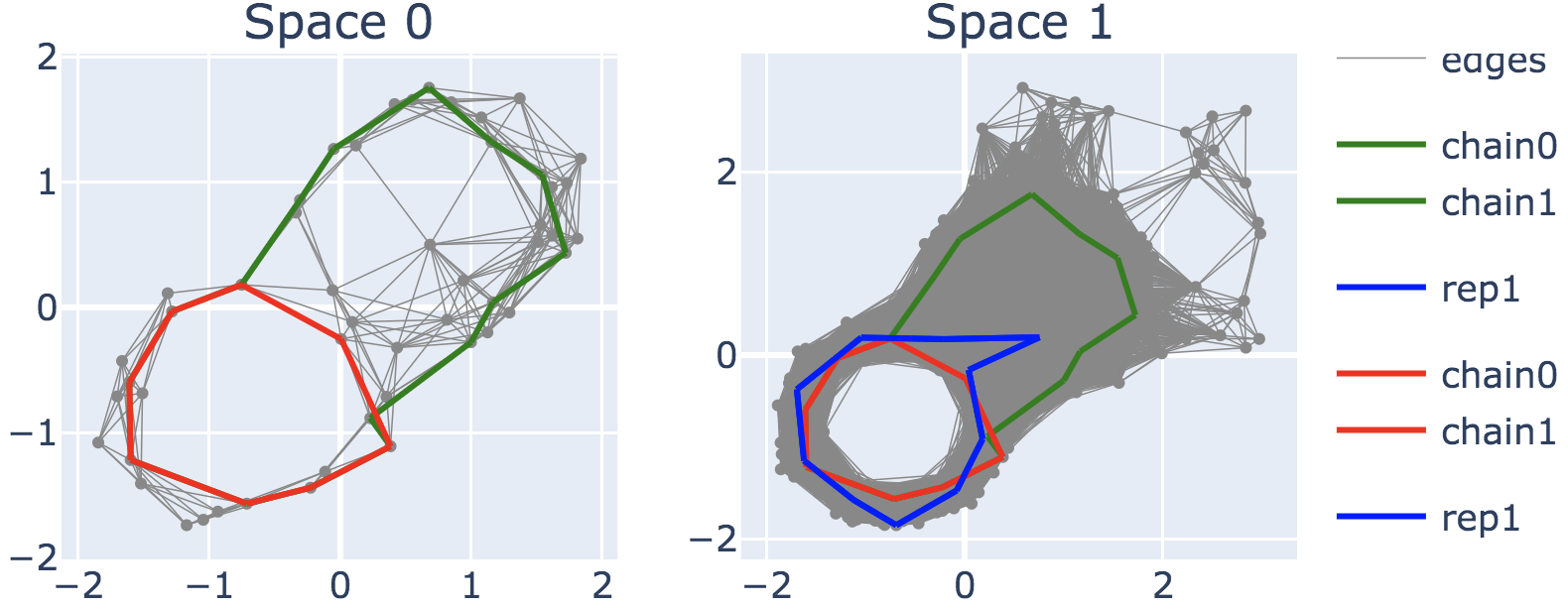}
    \includegraphics[width=0.48\linewidth]{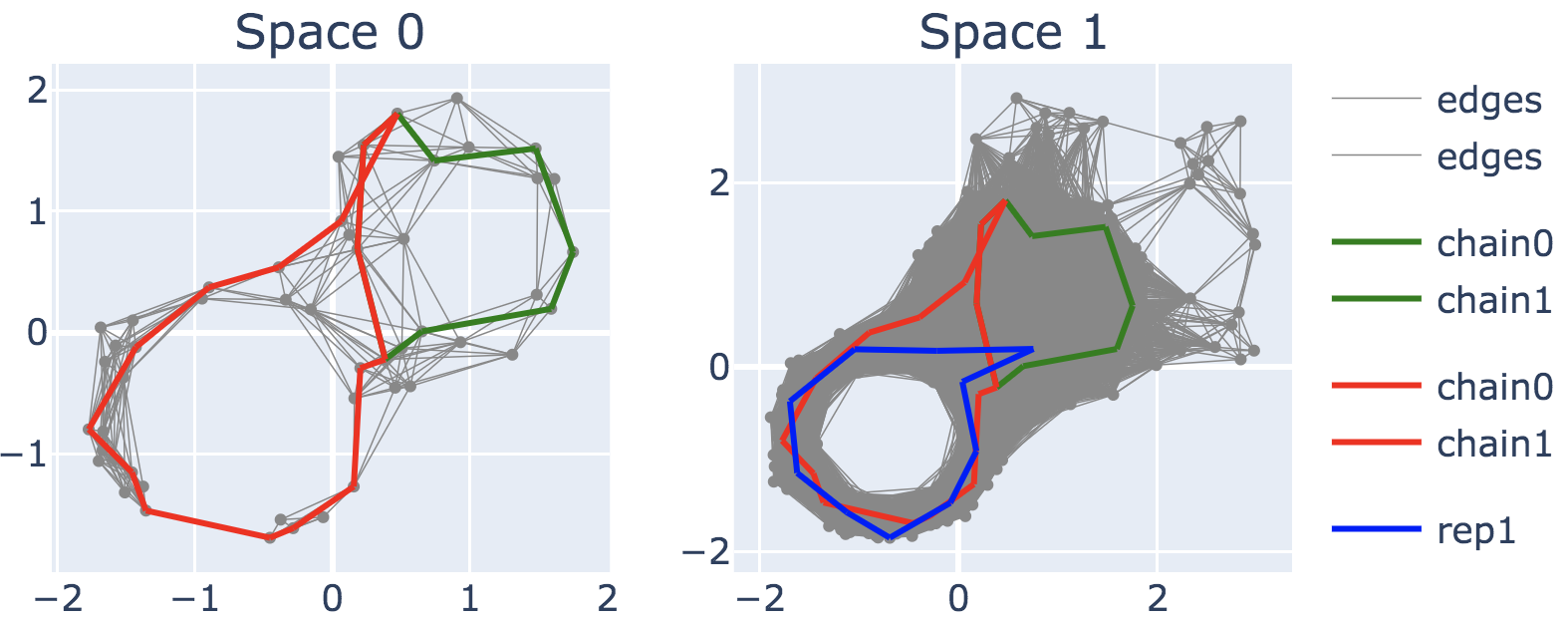}
    \includegraphics[width=0.48\linewidth]{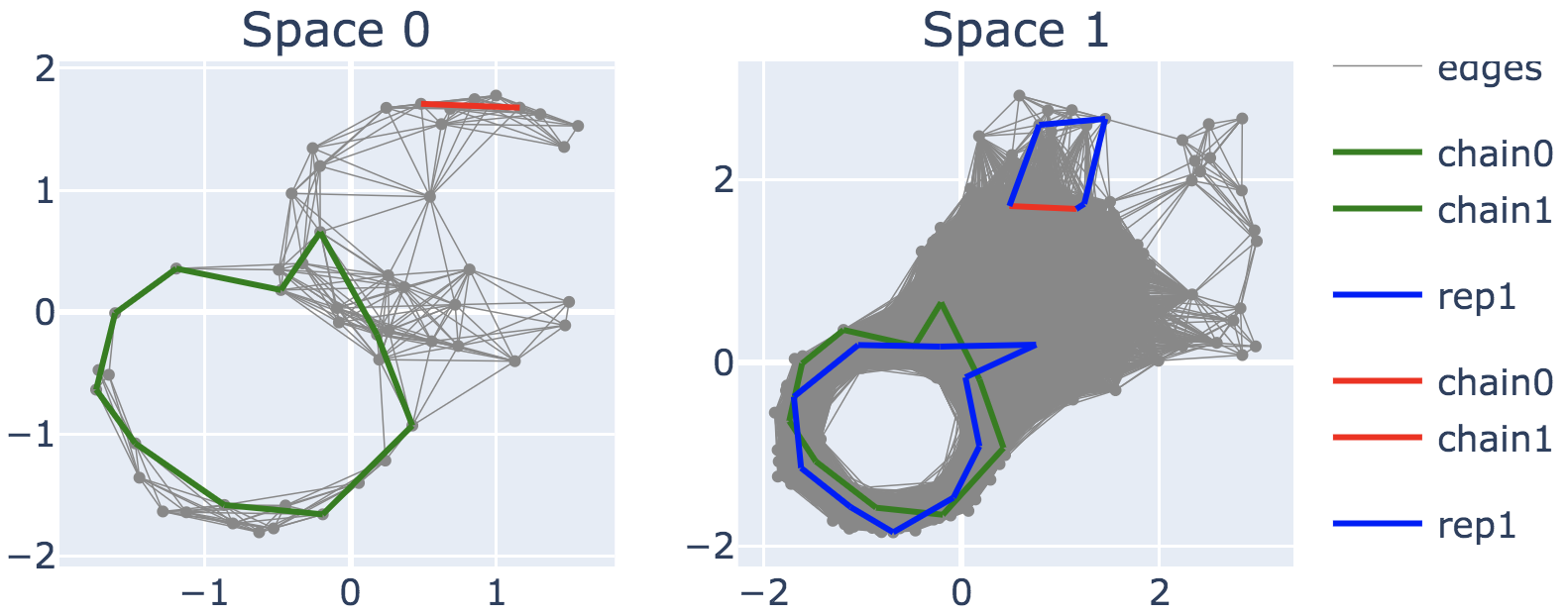}
    \includegraphics[width=0.48\linewidth]{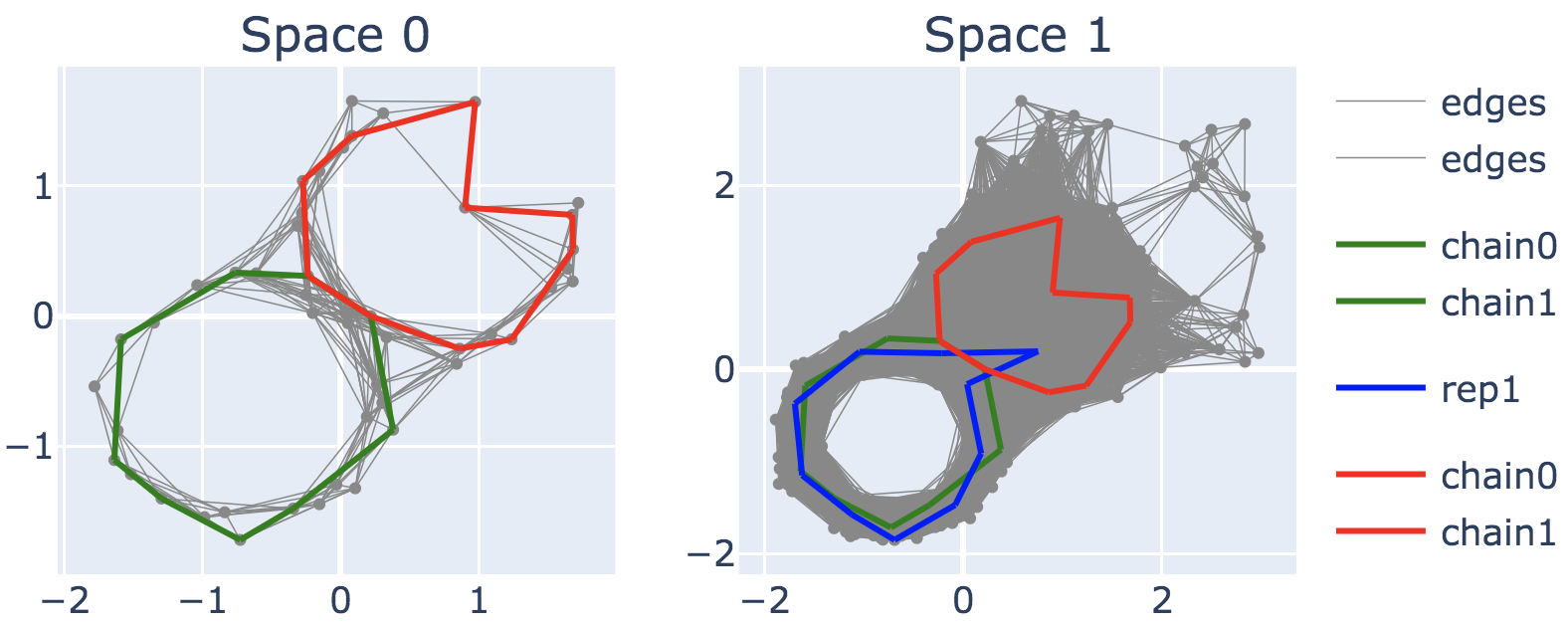}
    \caption{Visualization of induced maps of noisy Figure-8. Full size is $1000$. Sample size is $50$.}
    \label{fig:fig_8_50_im}
\end{figure}
\begin{figure}
\centering   
    \includegraphics[width=0.48\linewidth]{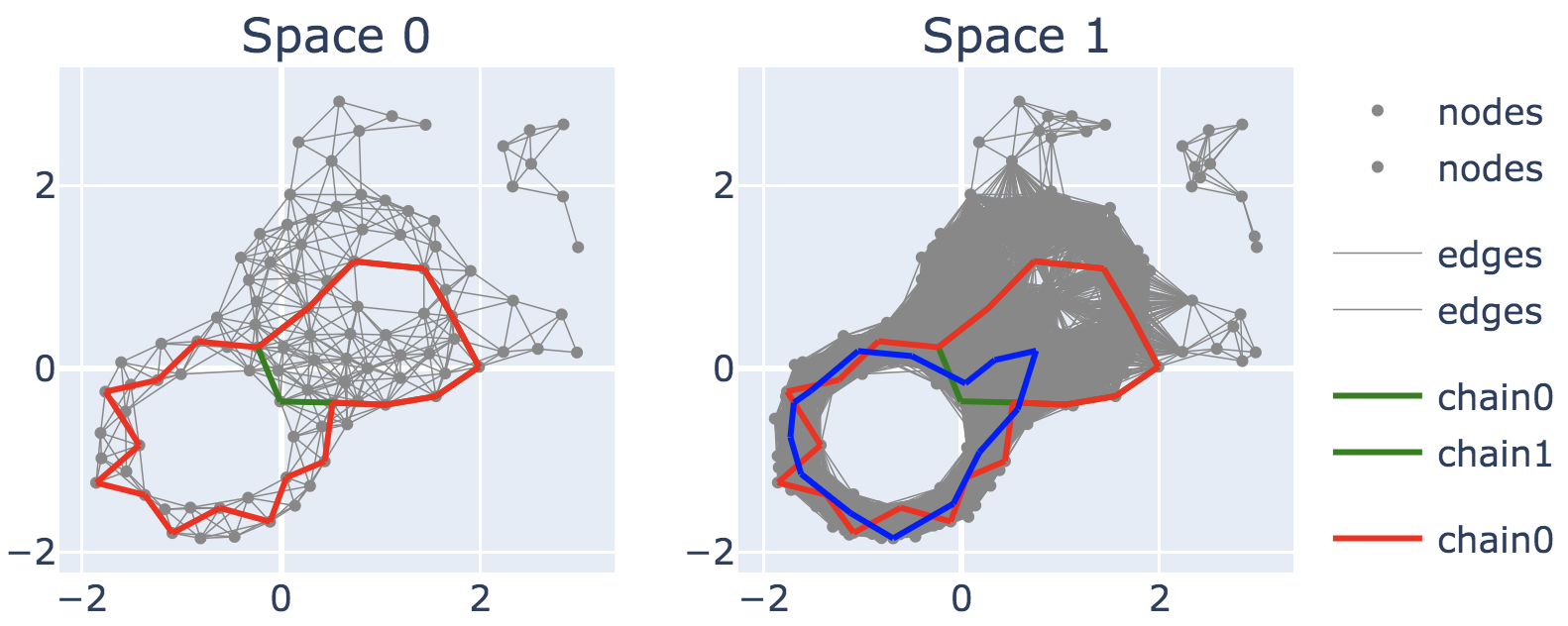}
    \includegraphics[width=0.48\linewidth]{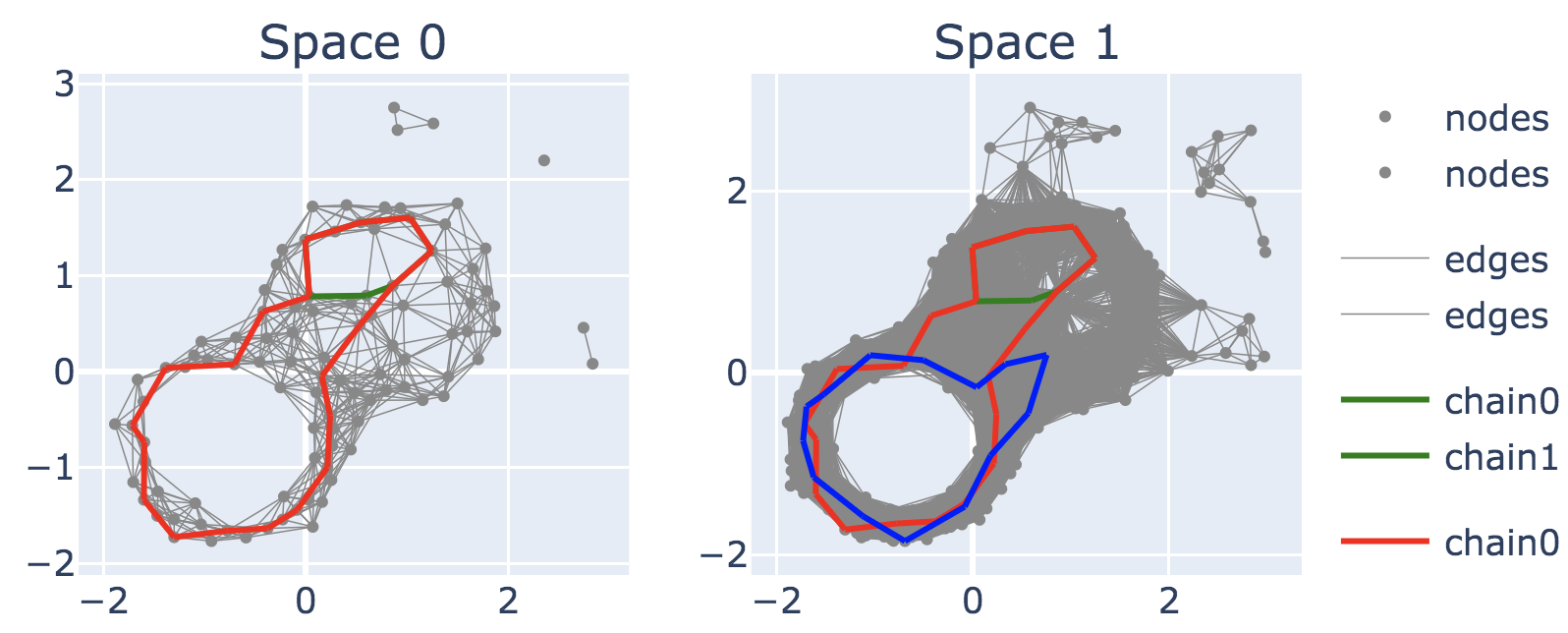}
    \includegraphics[width=0.48\linewidth]{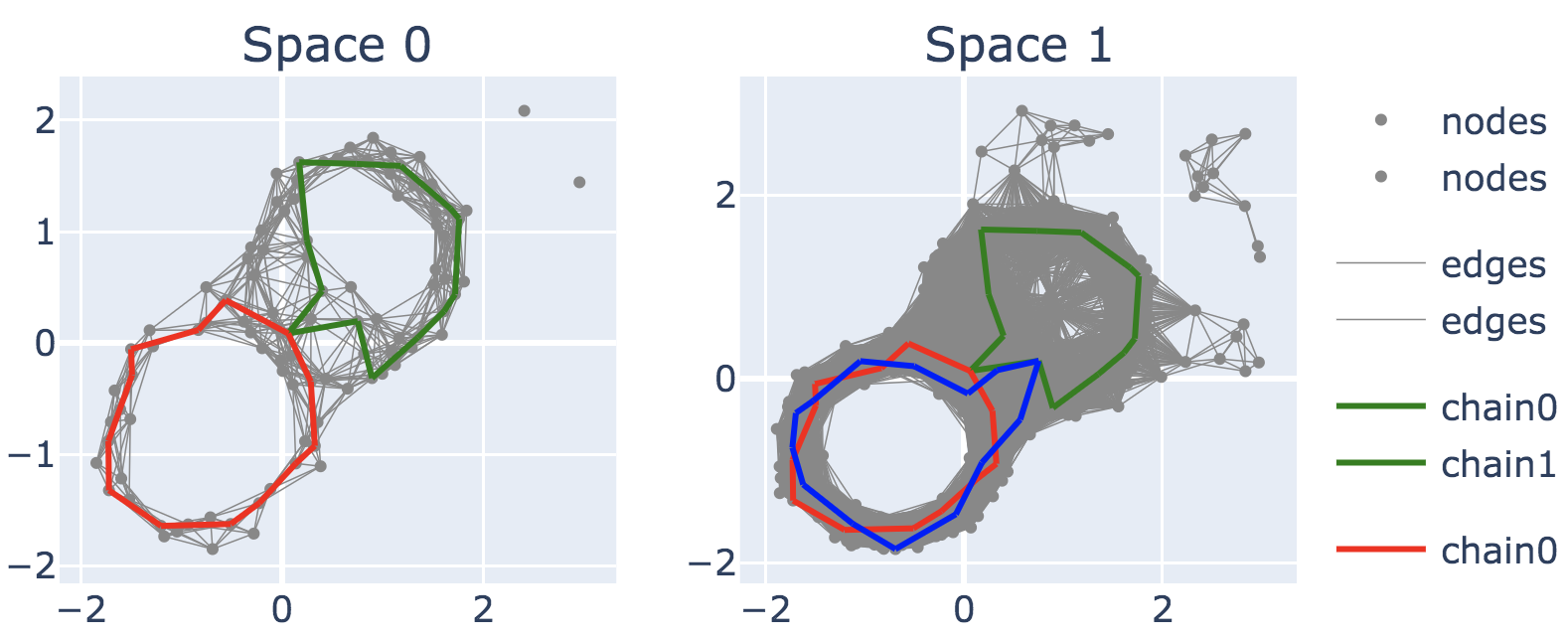}
    \includegraphics[width=0.48\linewidth]{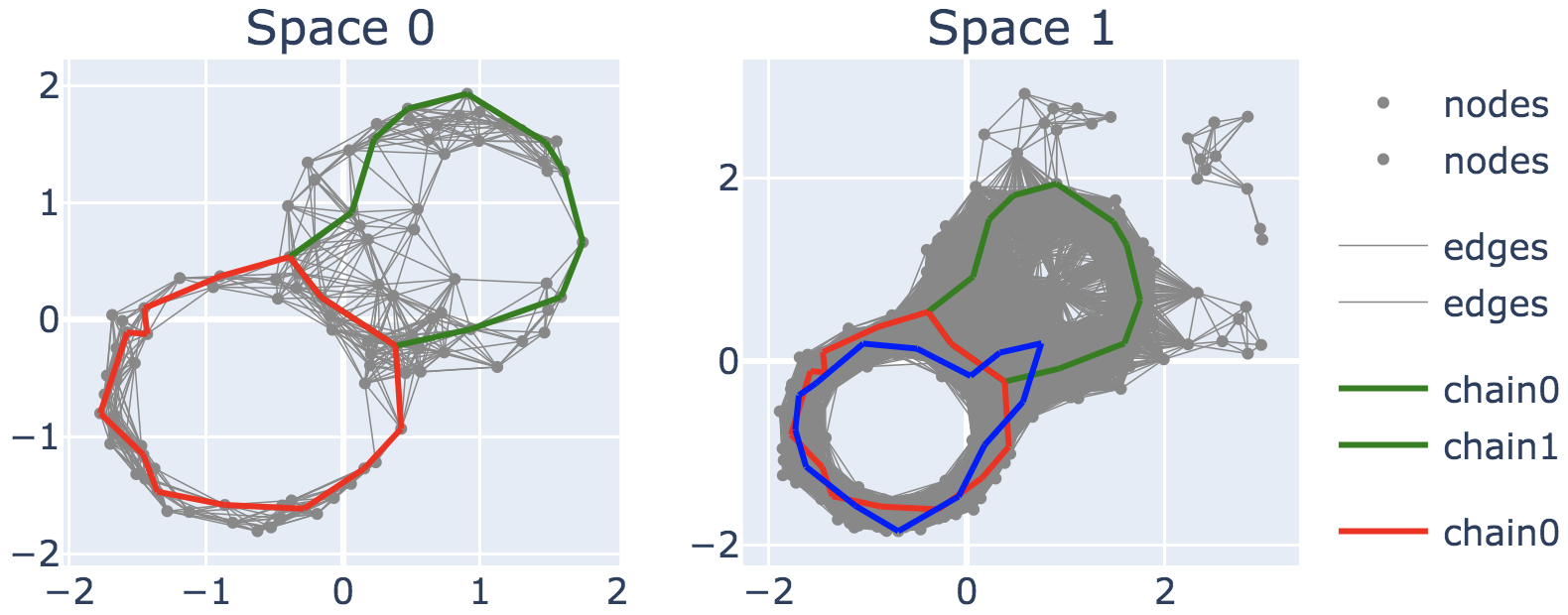}
    \includegraphics[width=0.48\linewidth]{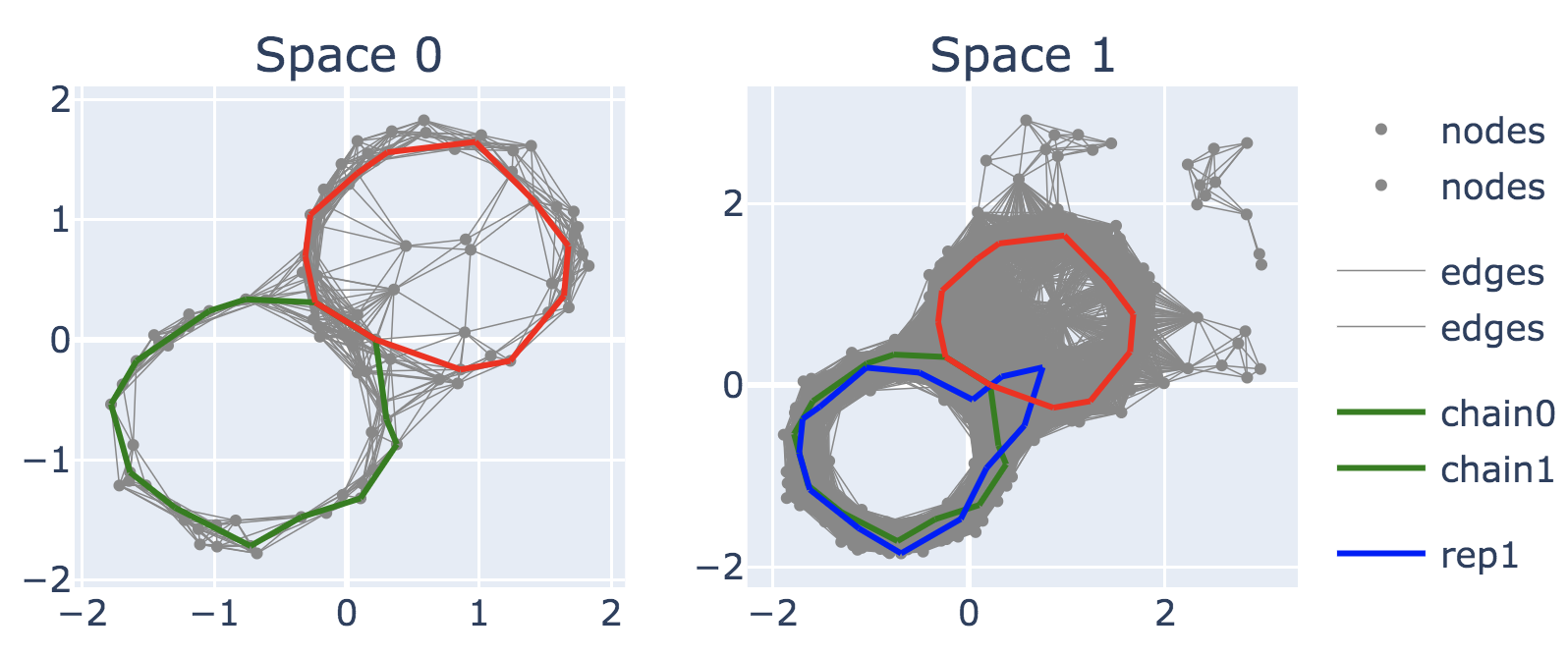}
    \includegraphics[width=0.48\linewidth]{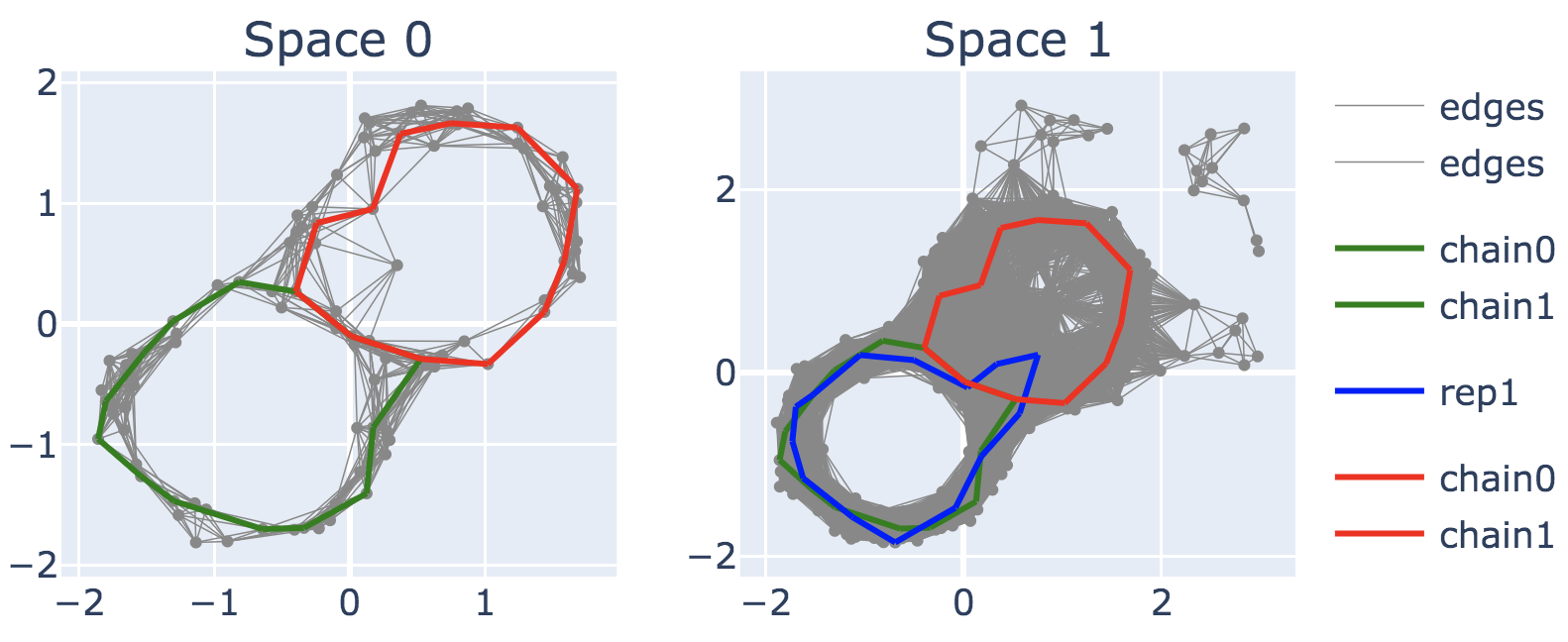}
    \includegraphics[width=0.48\linewidth]{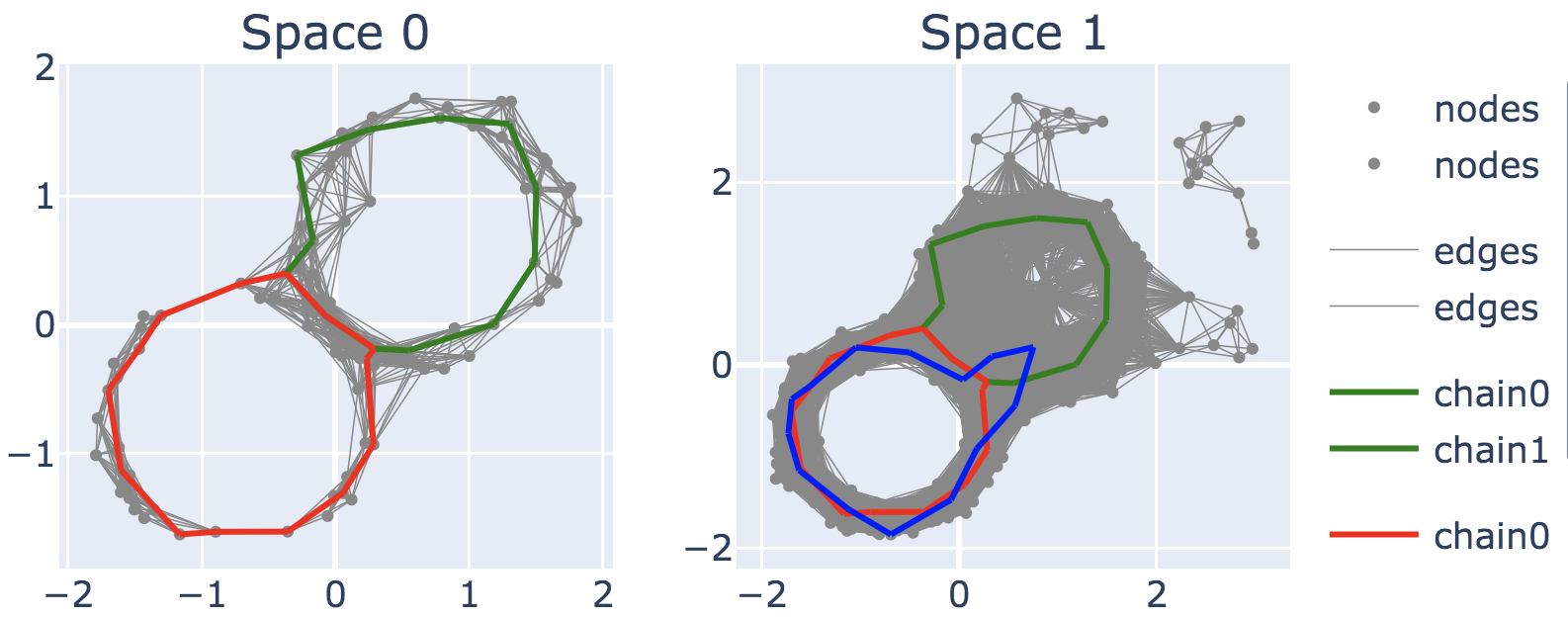}
    \includegraphics[width=0.48\linewidth]{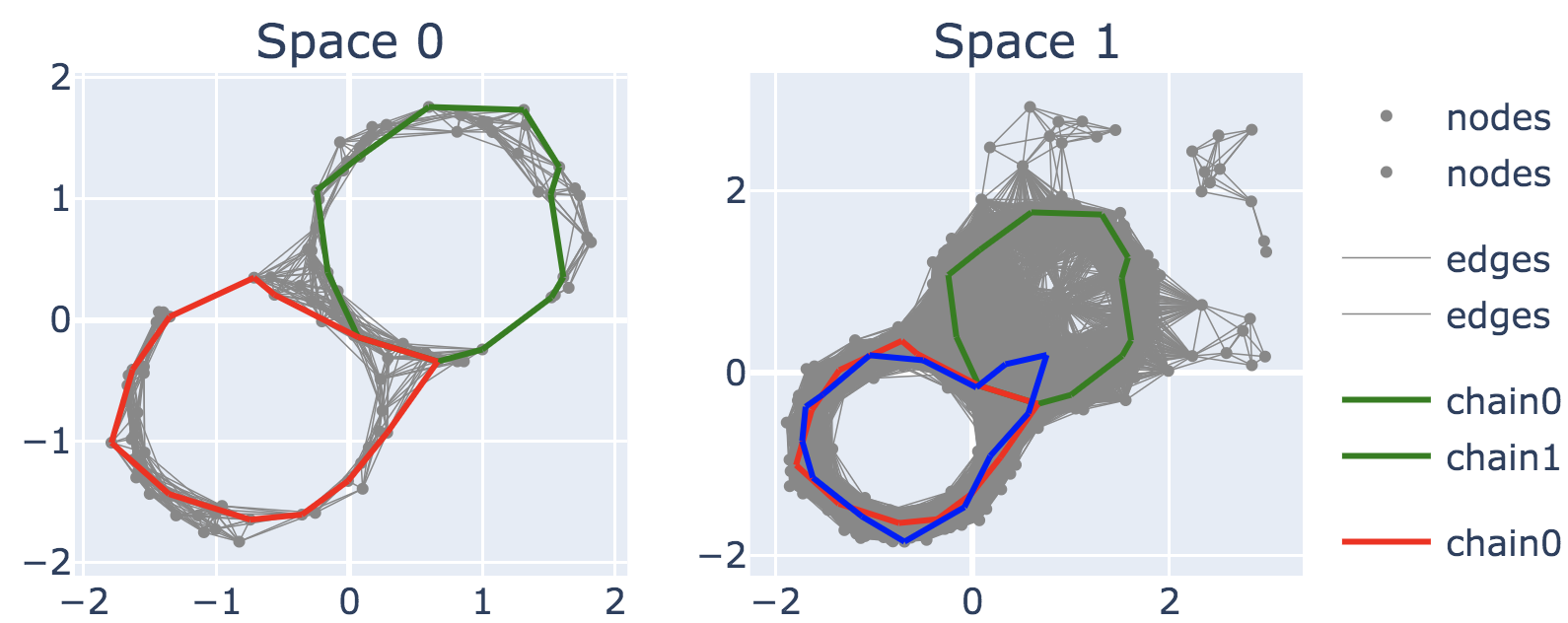}
    \includegraphics[width=0.48\linewidth]{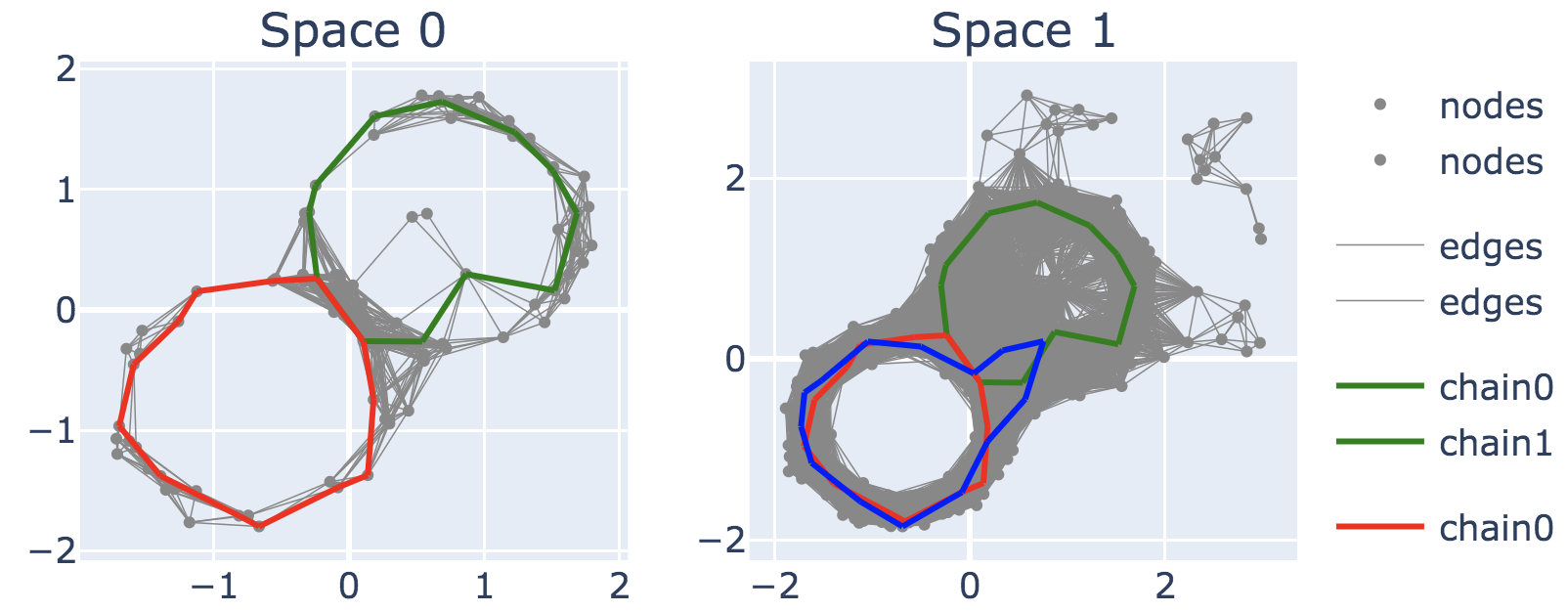}
    \includegraphics[width=0.48\linewidth]{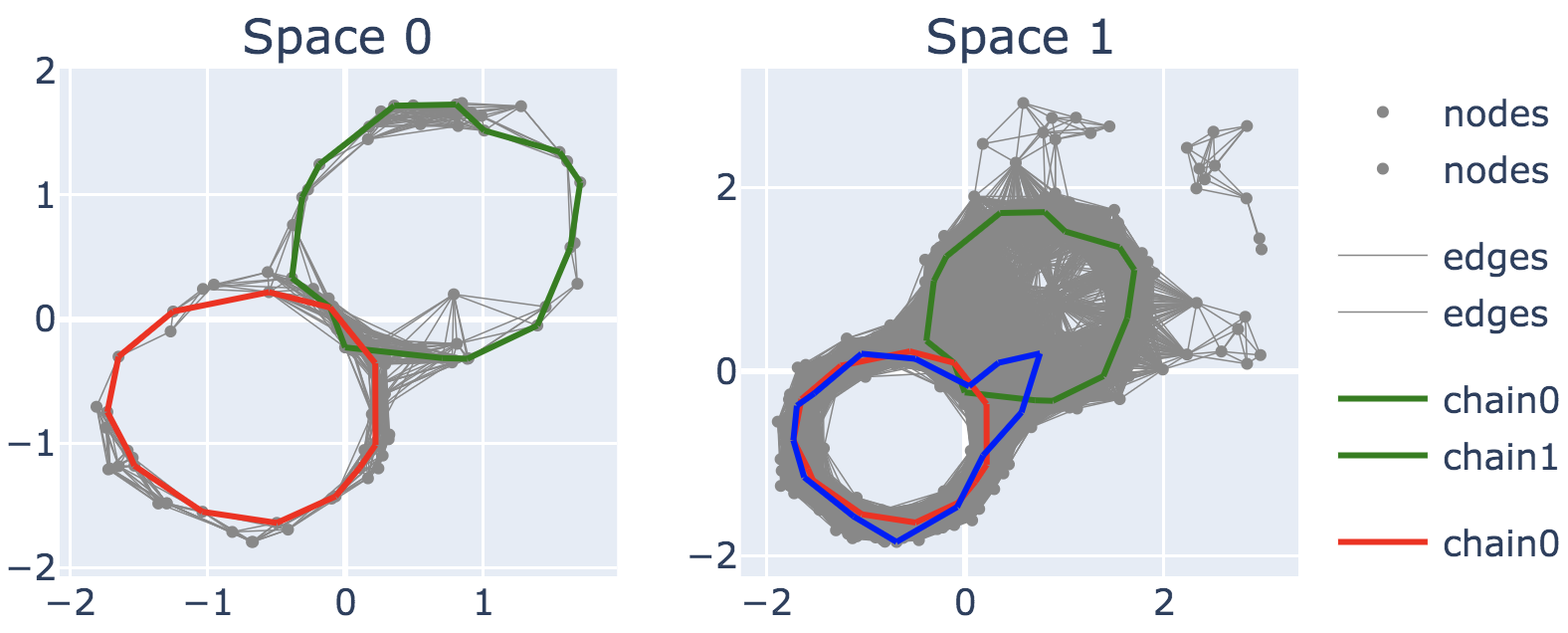}
    \caption{Visualization of induced maps of noisy Figure-8. Full size is $1000$. Sample size is $100$.}
    \label{fig:fig_8_100_im}
\end{figure}
\begin{figure}
\centering
    \includegraphics[width=0.48\linewidth]{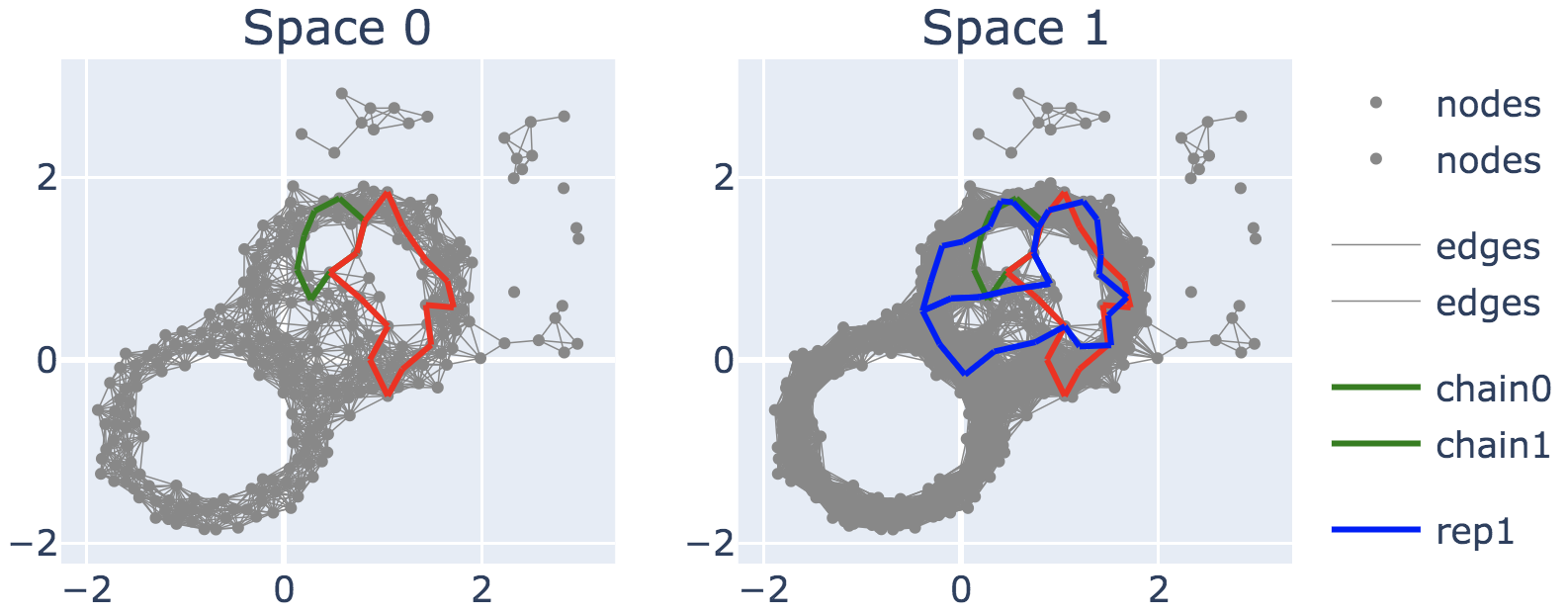}
    \includegraphics[width=0.48\linewidth]{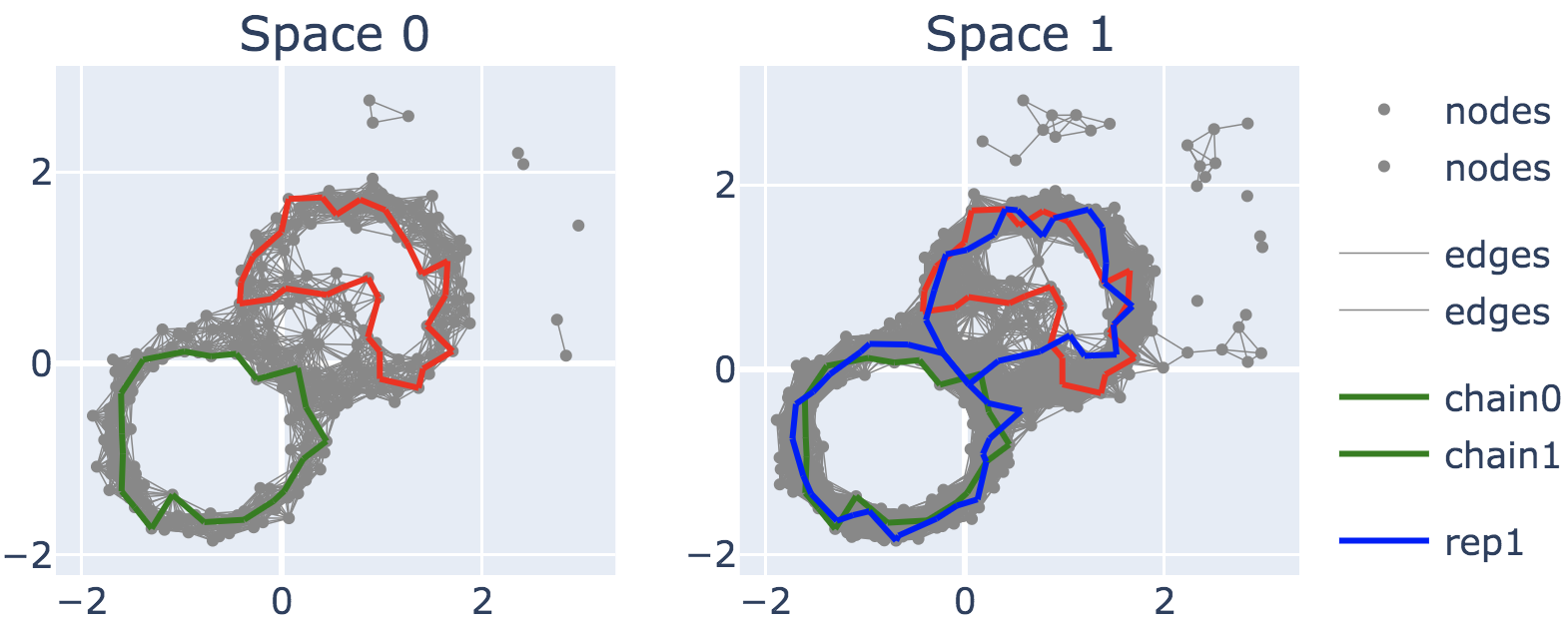}
    \includegraphics[width=0.48\linewidth]{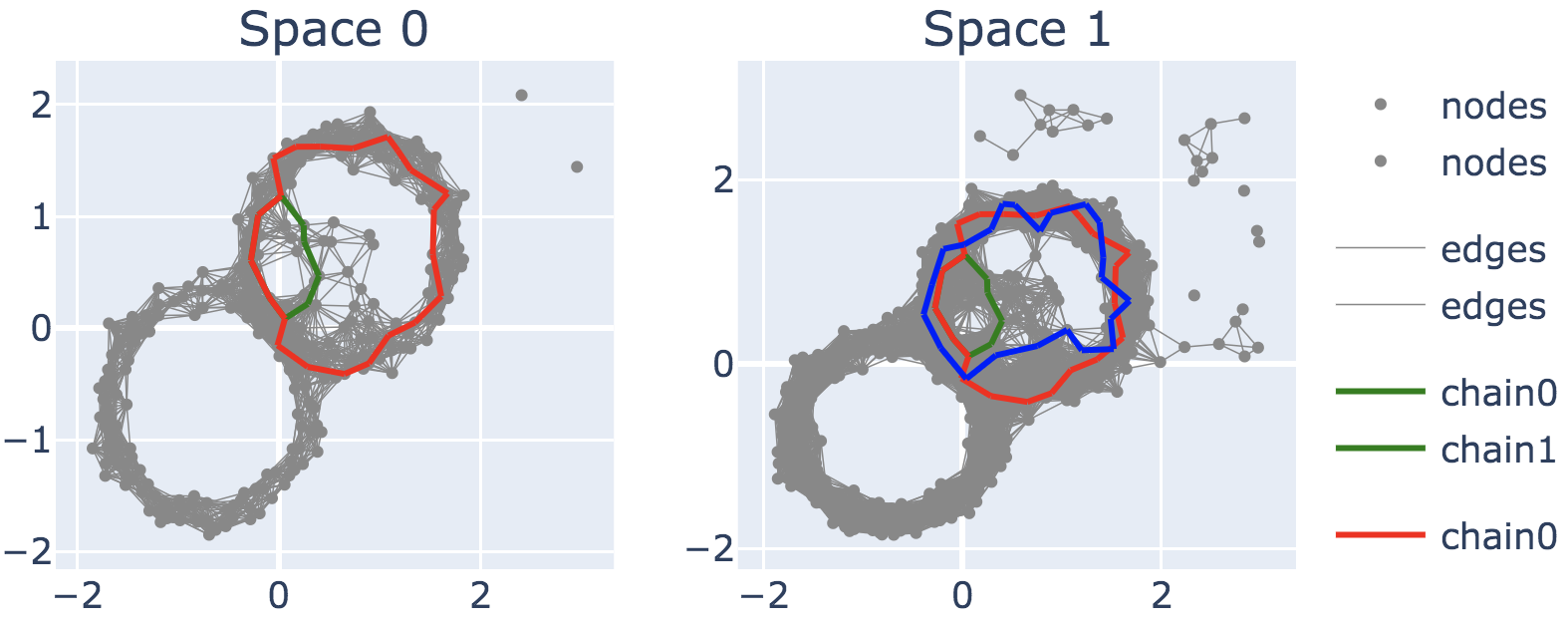}
    \includegraphics[width=0.48\linewidth]{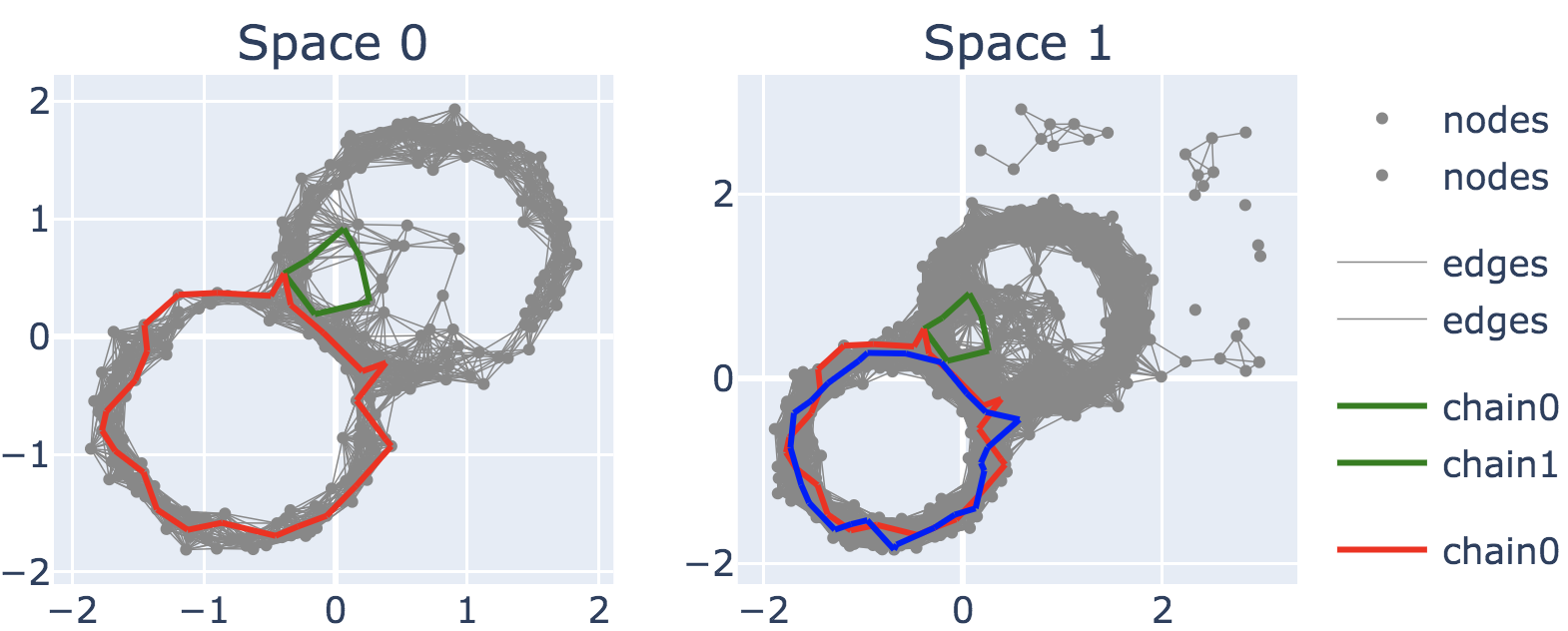}
    \includegraphics[width=0.48\linewidth]{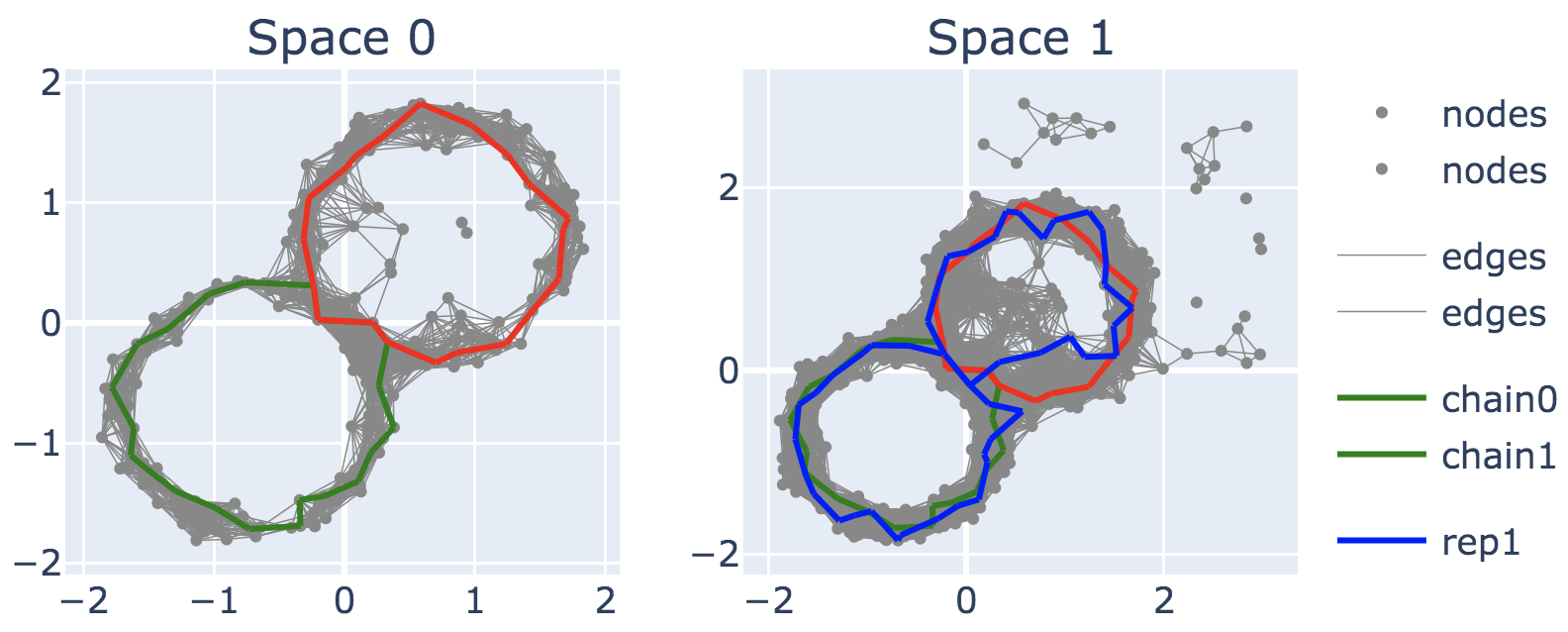}
    \includegraphics[width=0.48\linewidth]{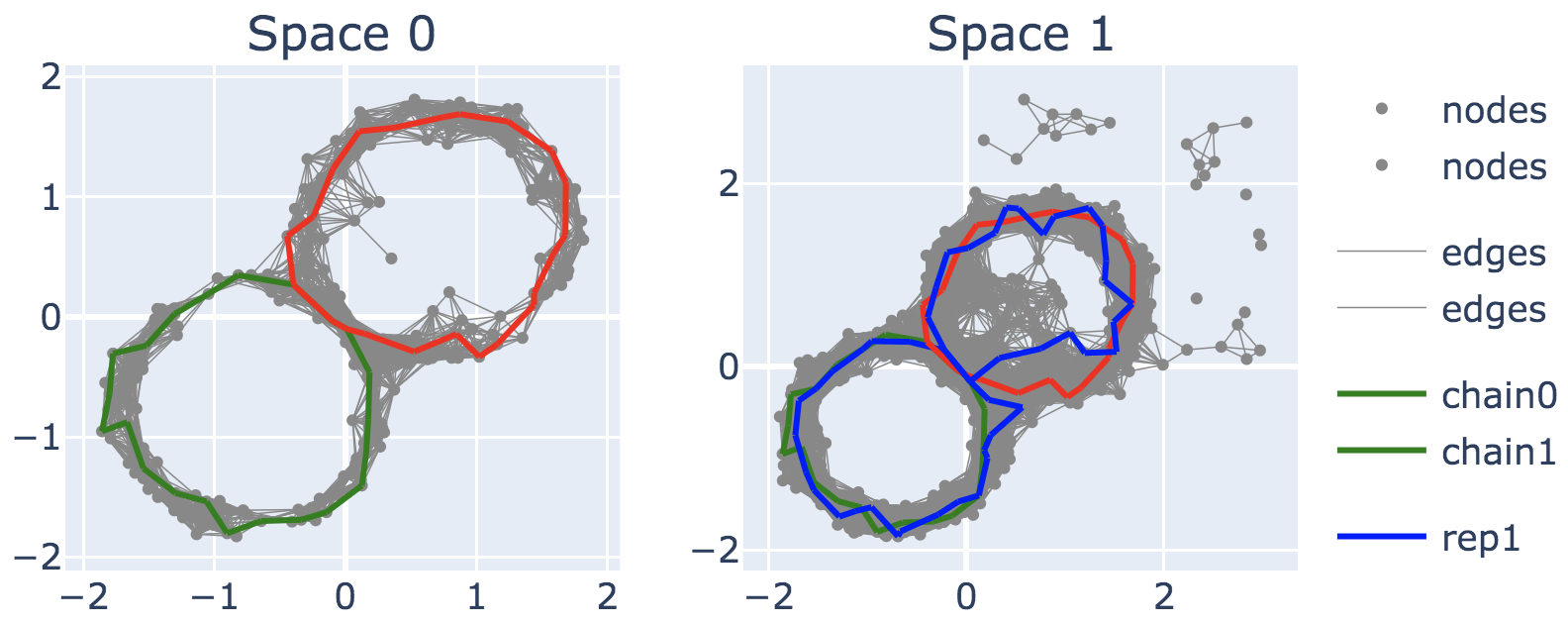}
    \includegraphics[width=0.48\linewidth]{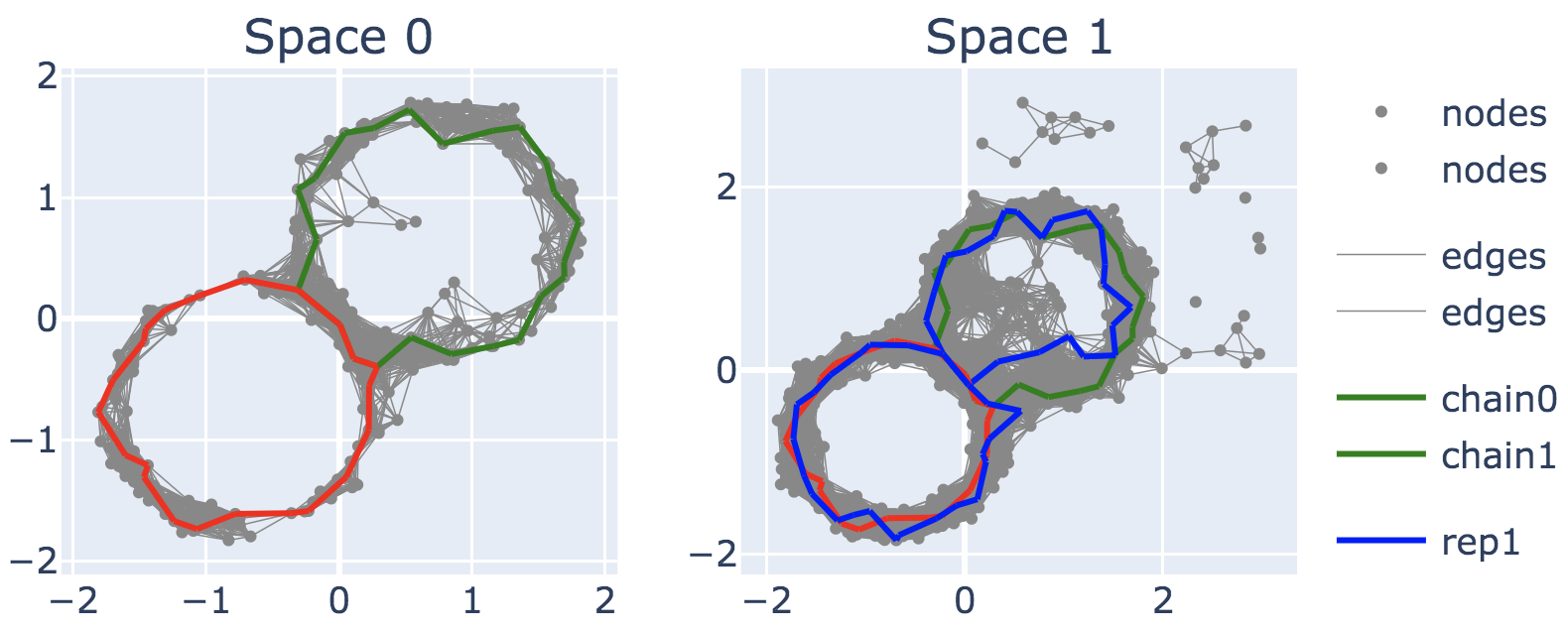}
    \includegraphics[width=0.48\linewidth]{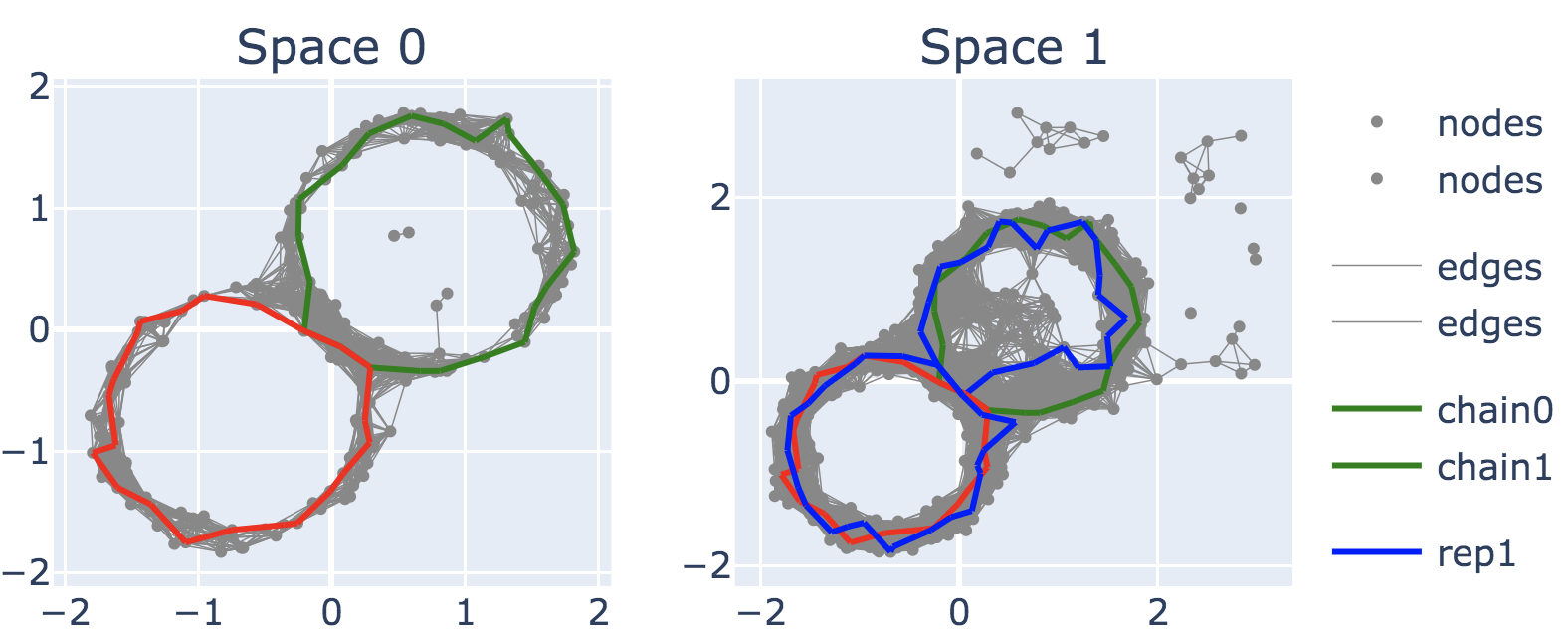}
    \includegraphics[width=0.48\linewidth]{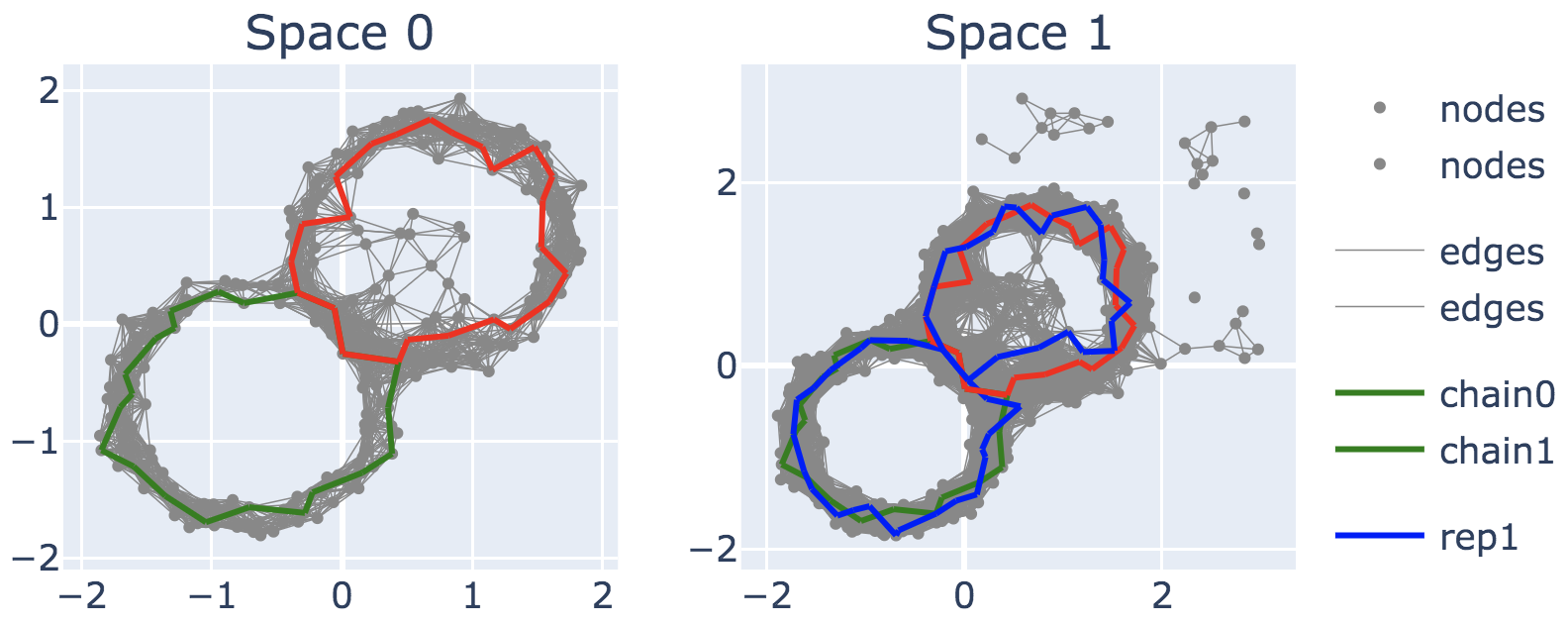}
    \includegraphics[width=0.48\linewidth]{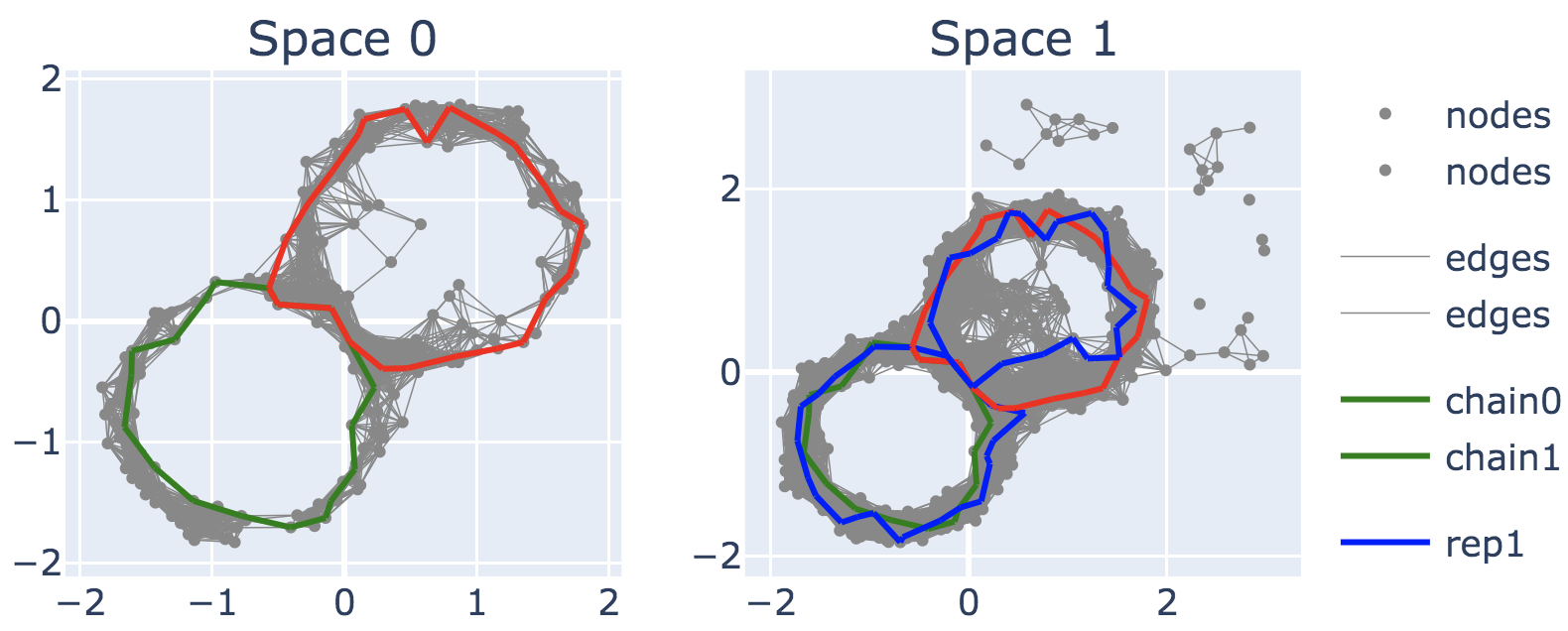}
    \caption{Visualization of induced maps of noisy Figure-8. Full size is $1000$. Sample size is $300$.}
    \label{fig:fig_8_300_im}
\end{figure}
\begin{figure}
\centering
    \includegraphics[width=0.48\linewidth]{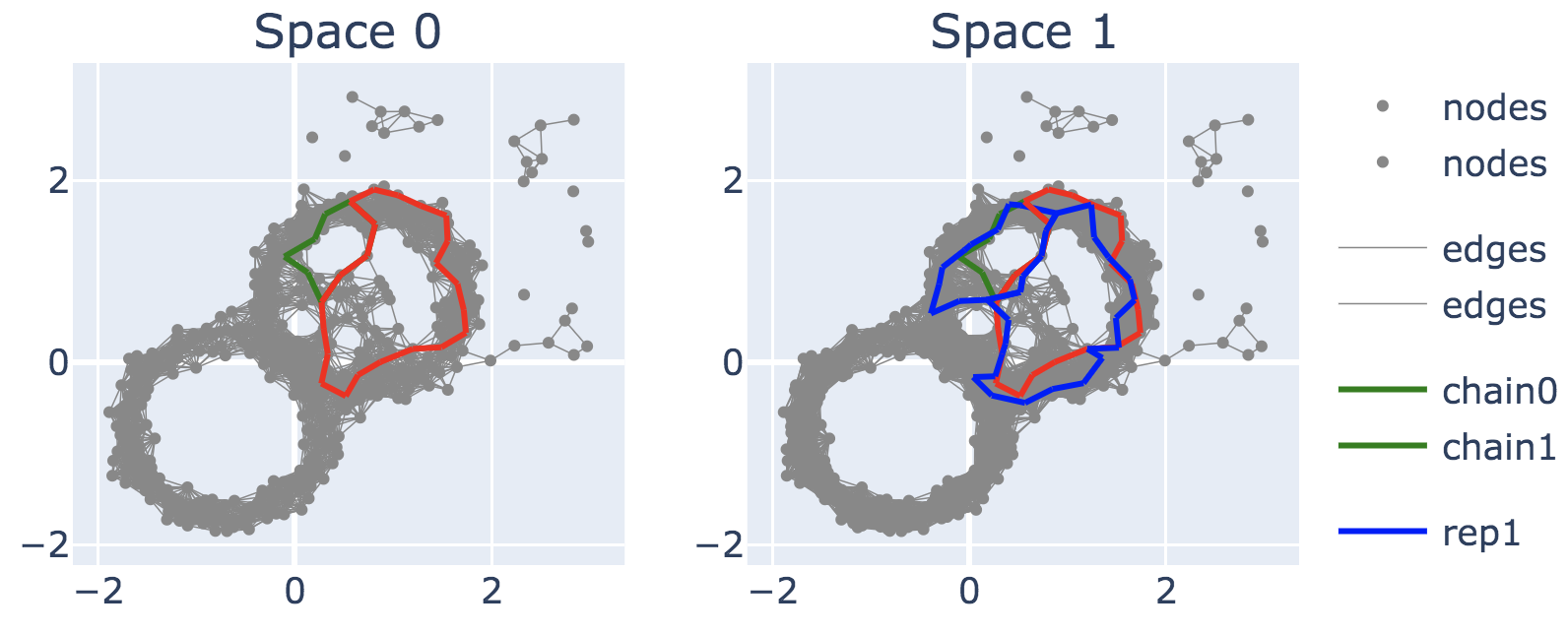}
    \includegraphics[width=0.48\linewidth]{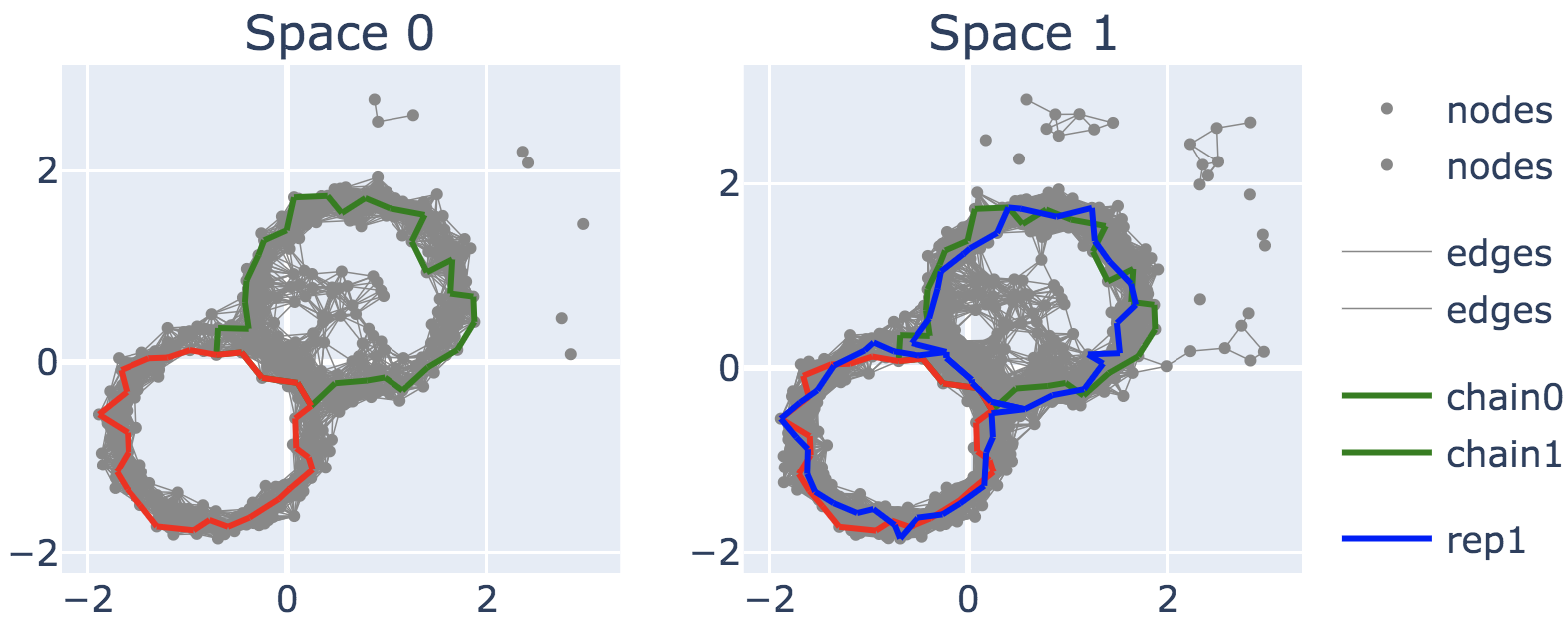}
    \includegraphics[width=0.48\linewidth]{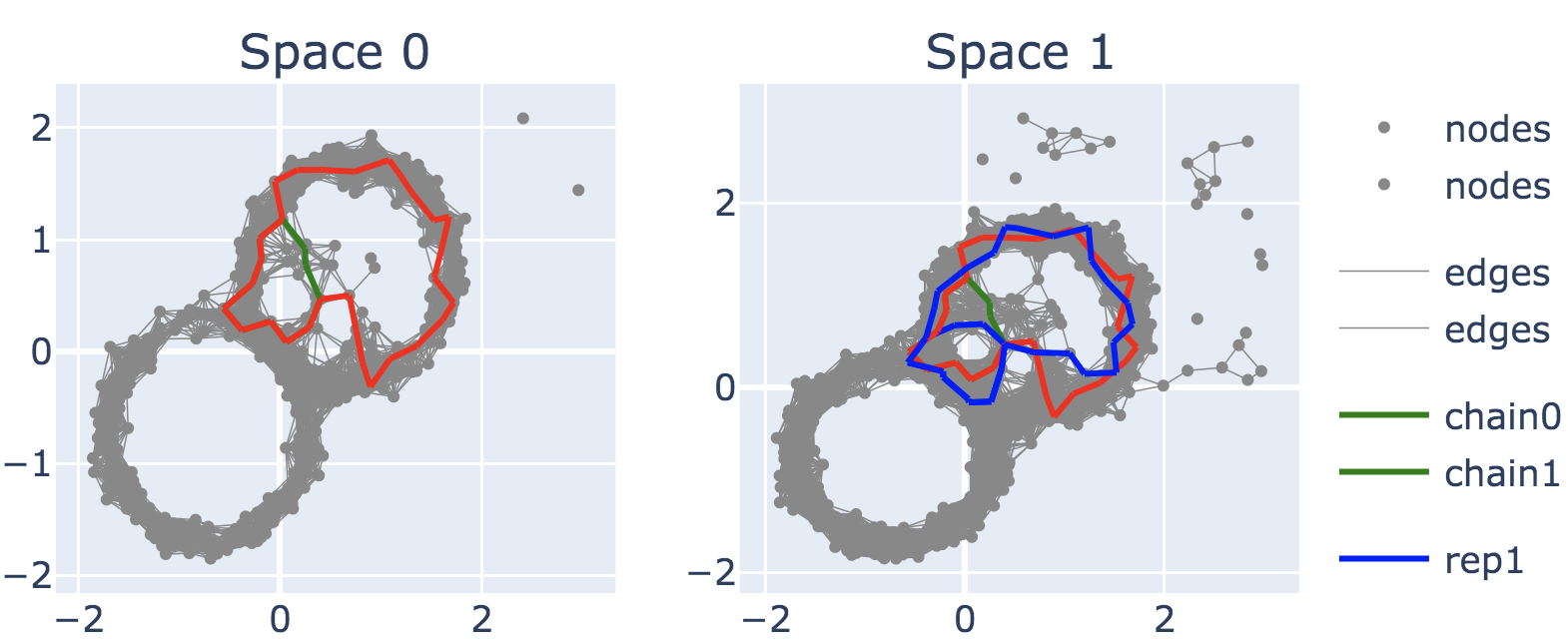}
    \includegraphics[width=0.48\linewidth]{figs/bad_fig8/500hd4.jpg}
    \includegraphics[width=0.48\linewidth]{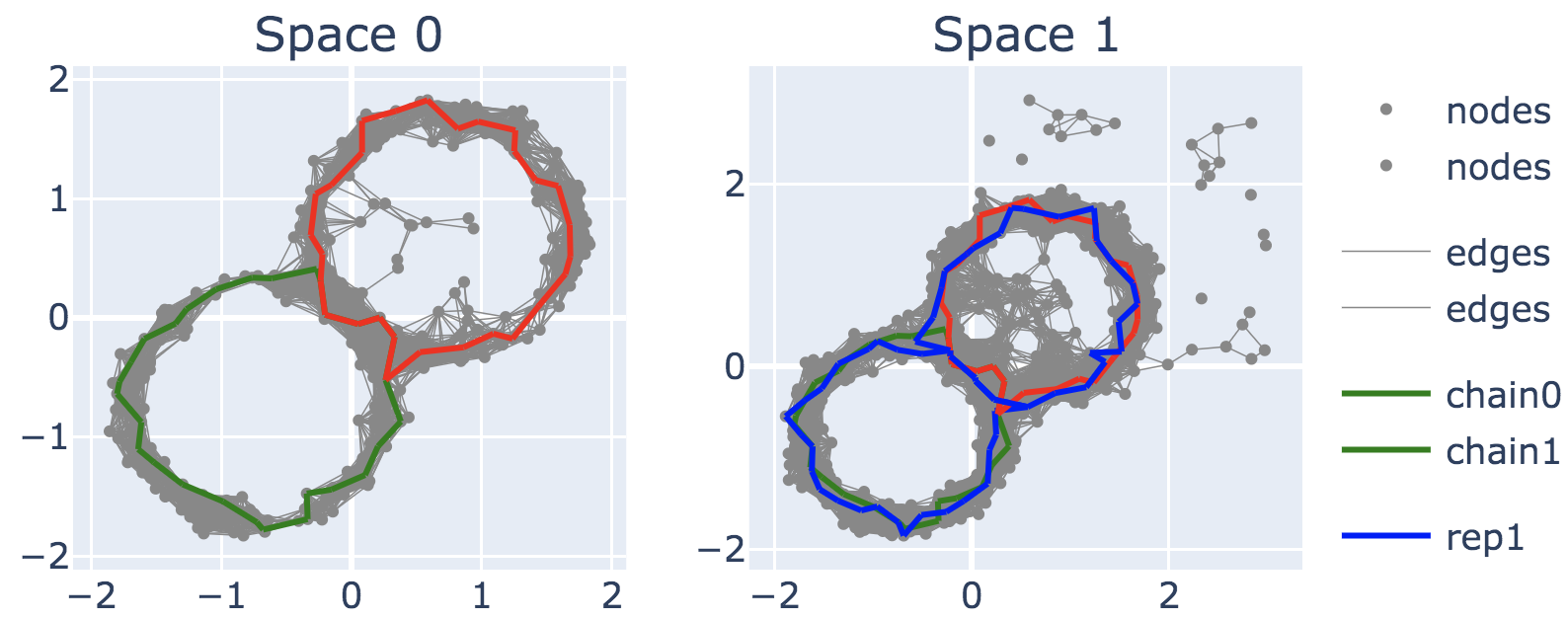}
    \includegraphics[width=0.48\linewidth]{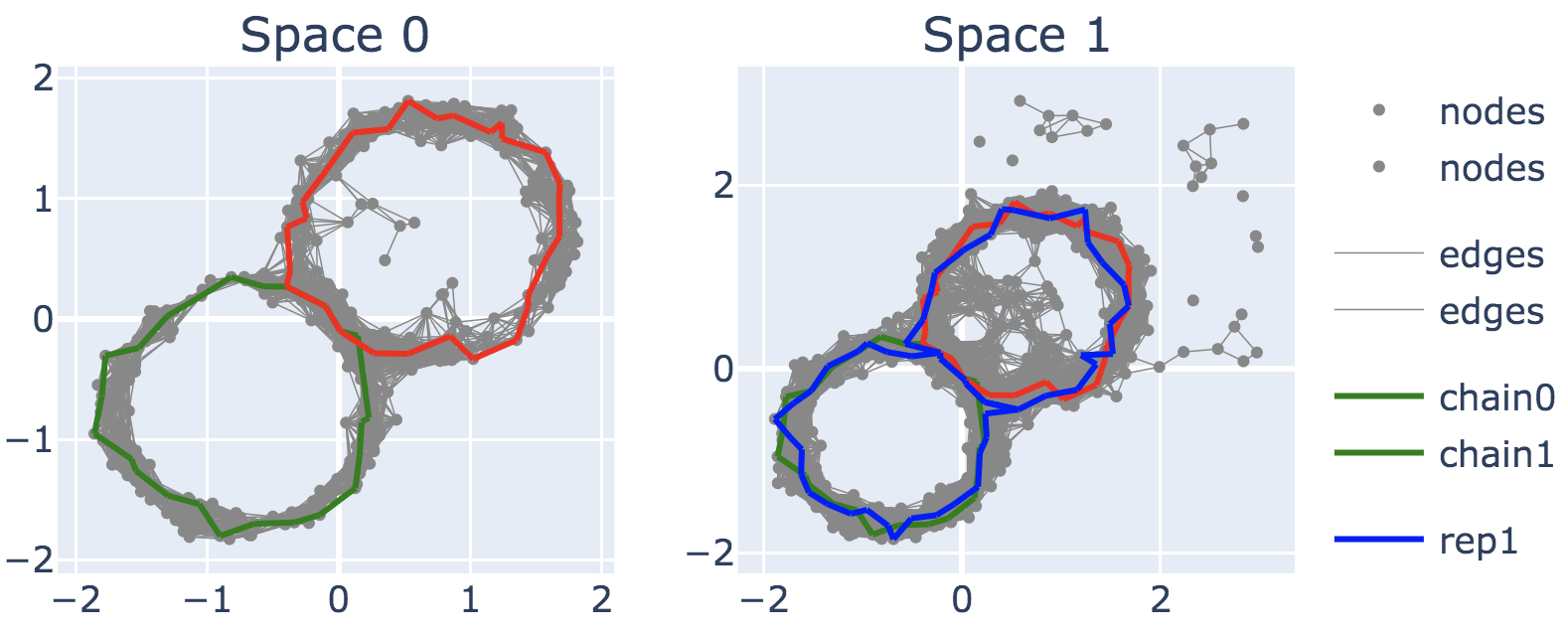}
    \includegraphics[width=0.48\linewidth]{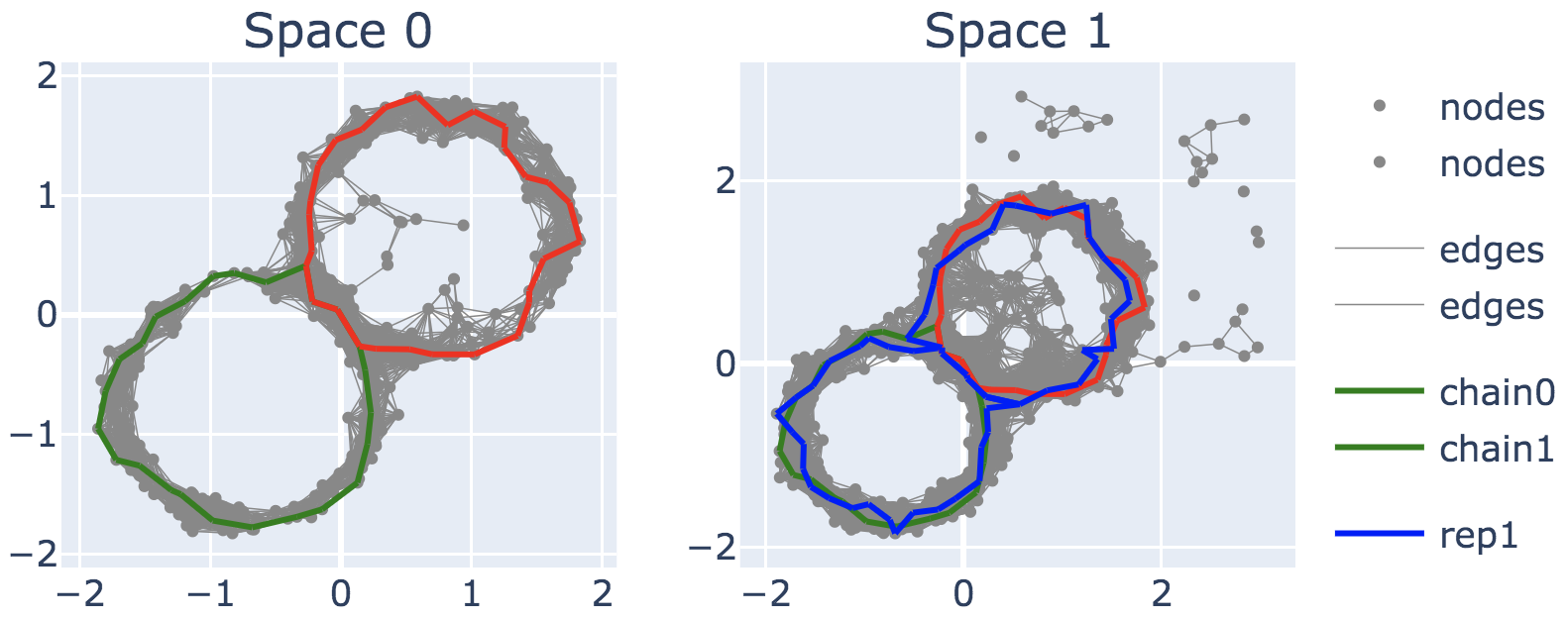}
    \includegraphics[width=0.48\linewidth]{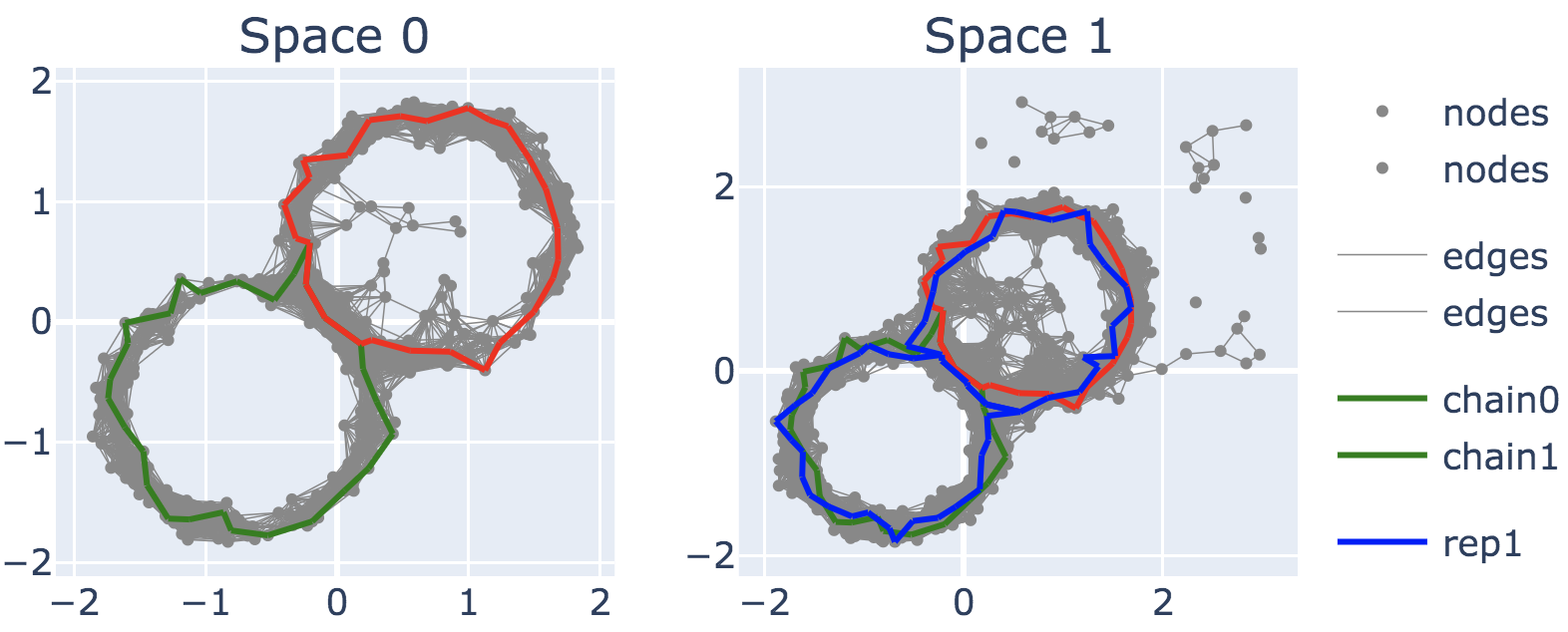}
    \includegraphics[width=0.48\linewidth]{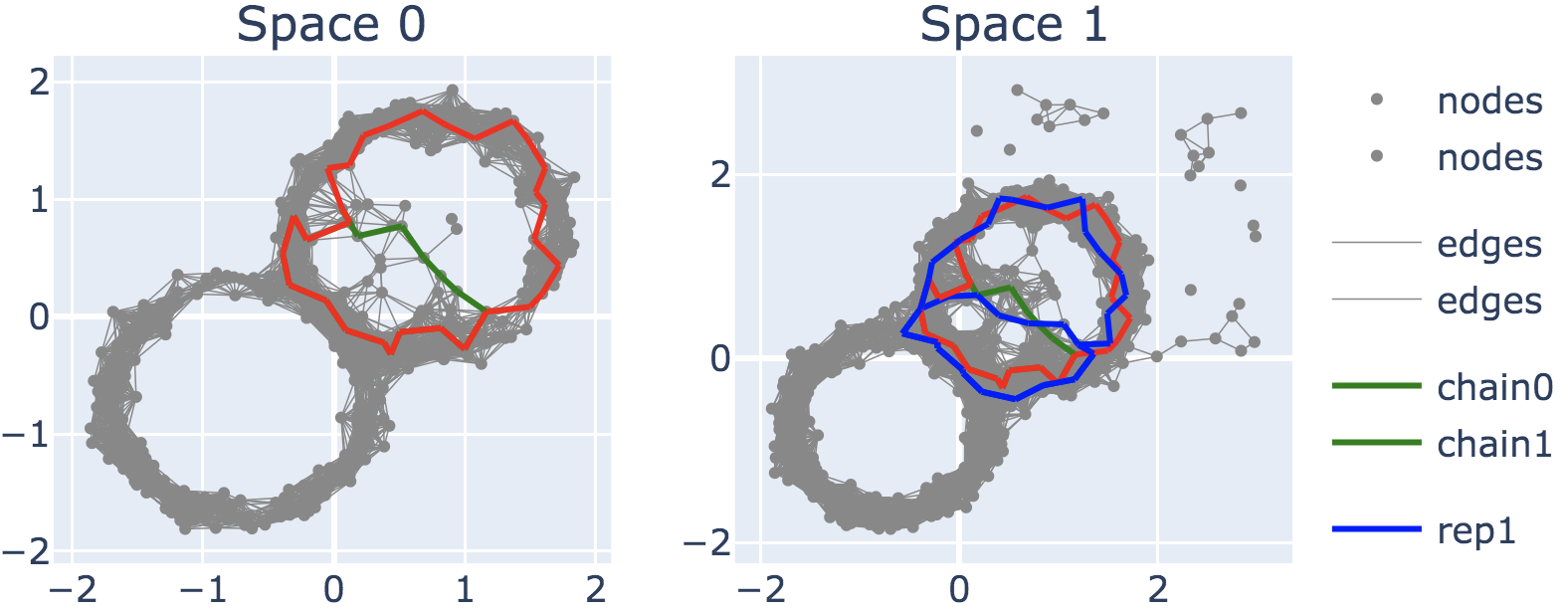}
    \includegraphics[width=0.48\linewidth]{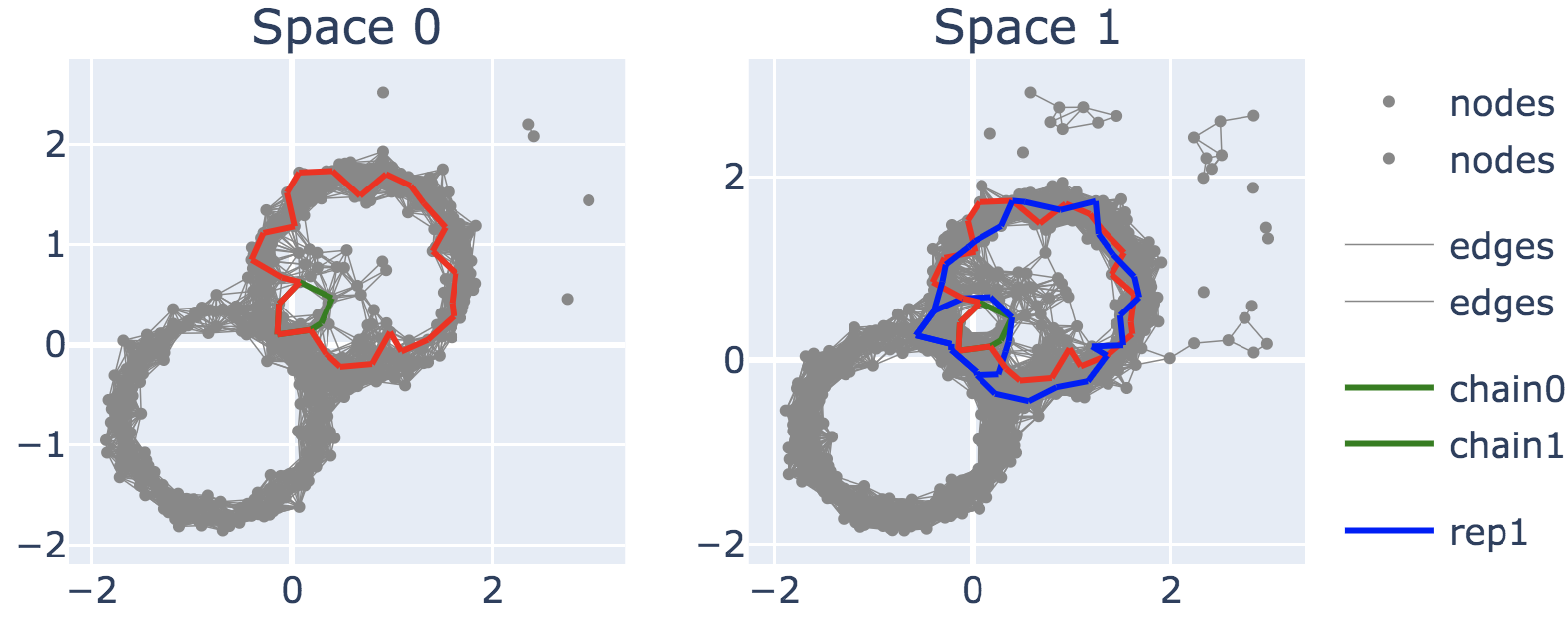}
    \caption{Visualization of induced maps of noisy Figure-8. Full size is $1000$. Sample size is $500$.}
    \label{fig:fig_8_500_im}
\end{figure}
\begin{figure}
\centering
    \includegraphics[width=0.48\linewidth]{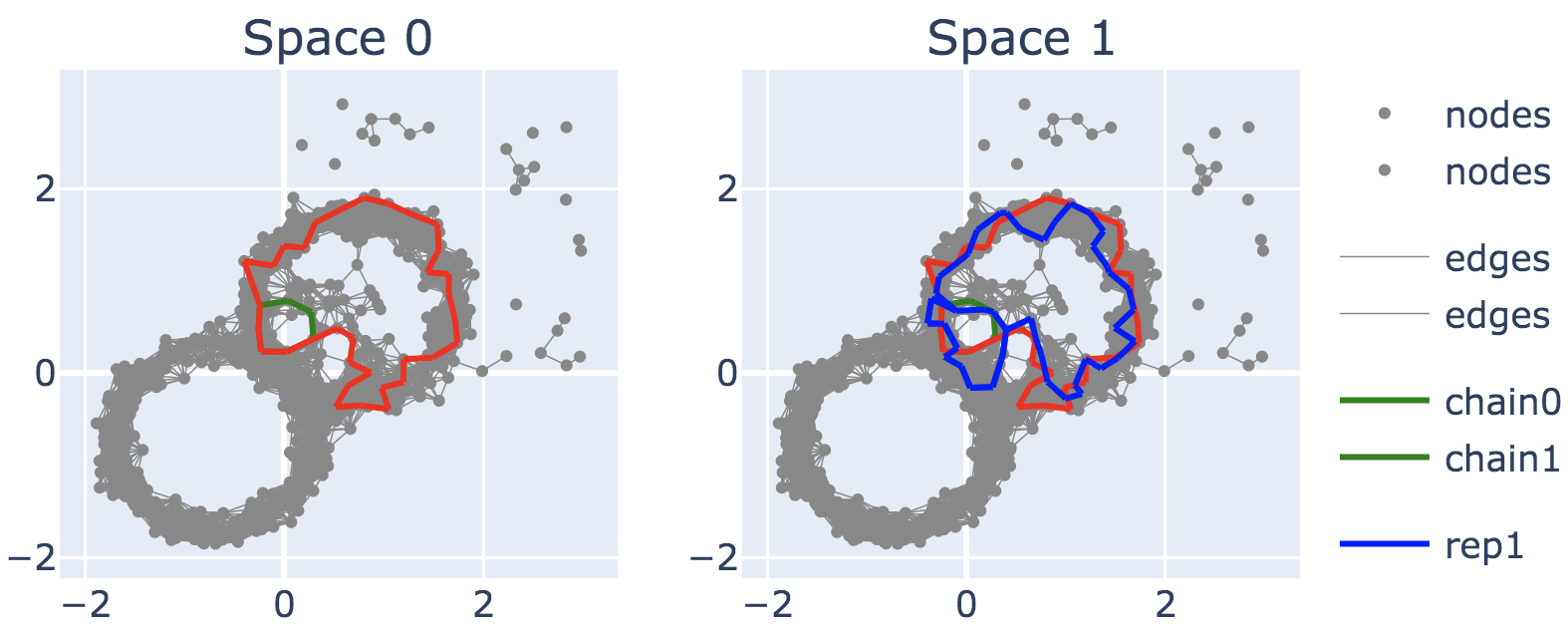}
    \includegraphics[width=0.48\linewidth]{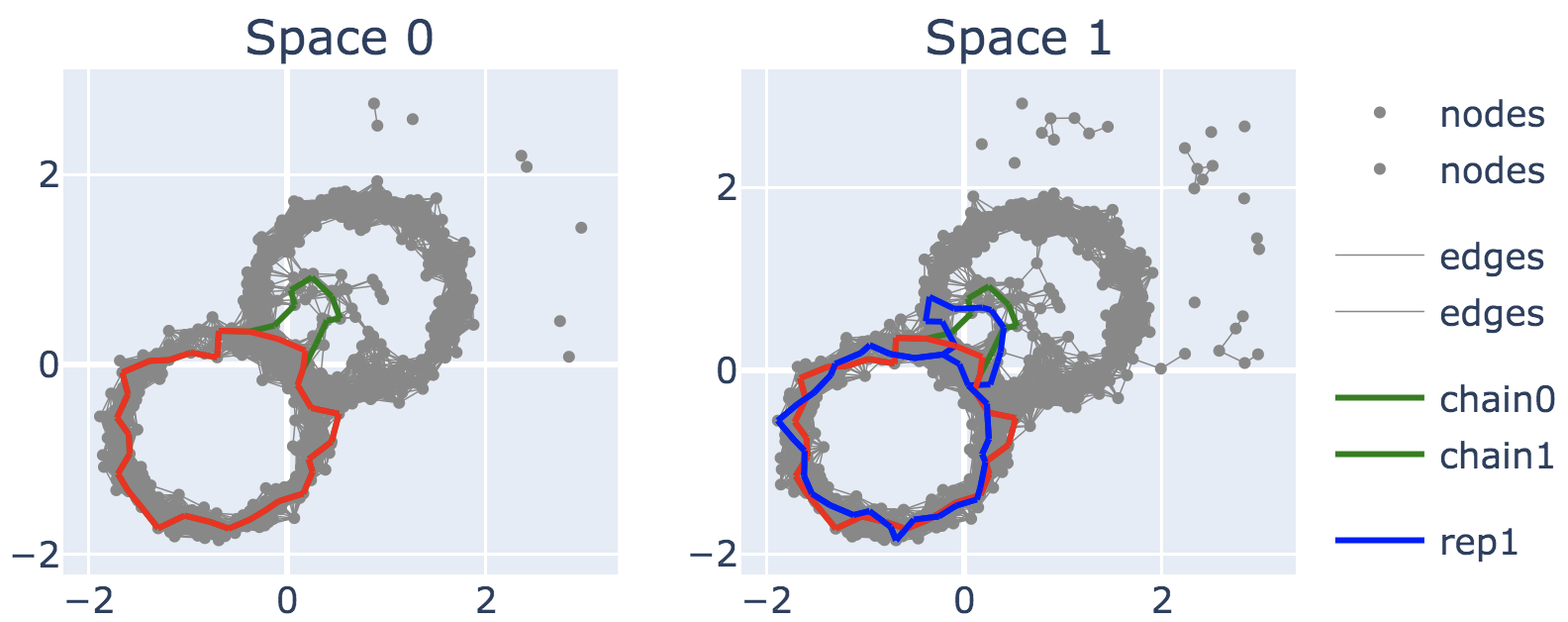}
    \includegraphics[width=0.48\linewidth]{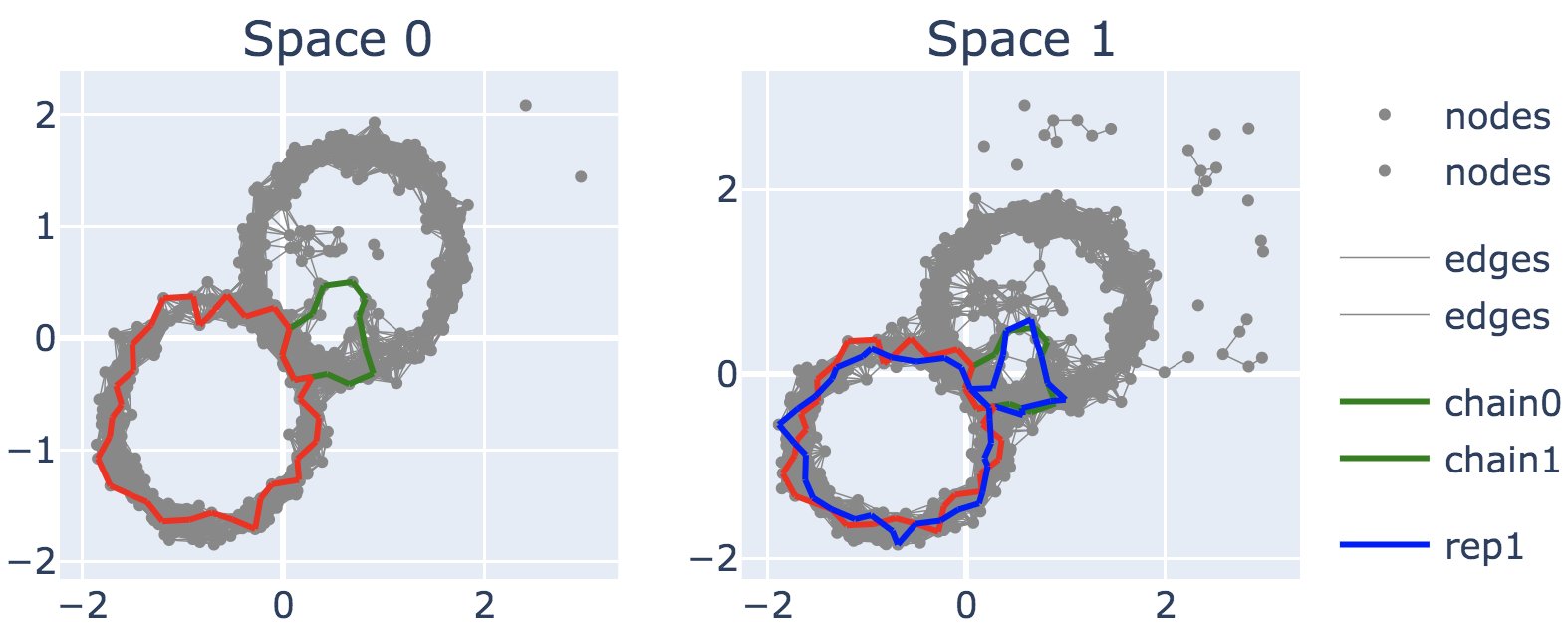}
    \includegraphics[width=0.48\linewidth]{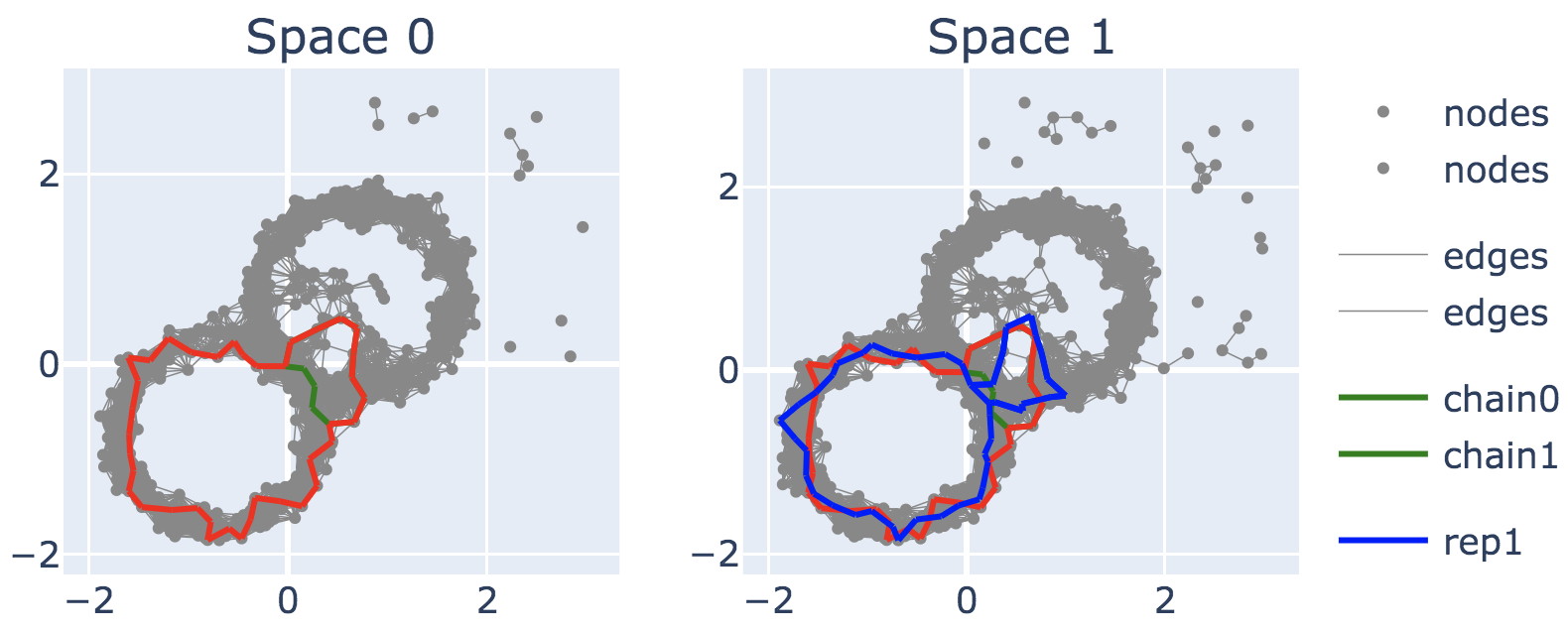}
    \includegraphics[width=0.48\linewidth]{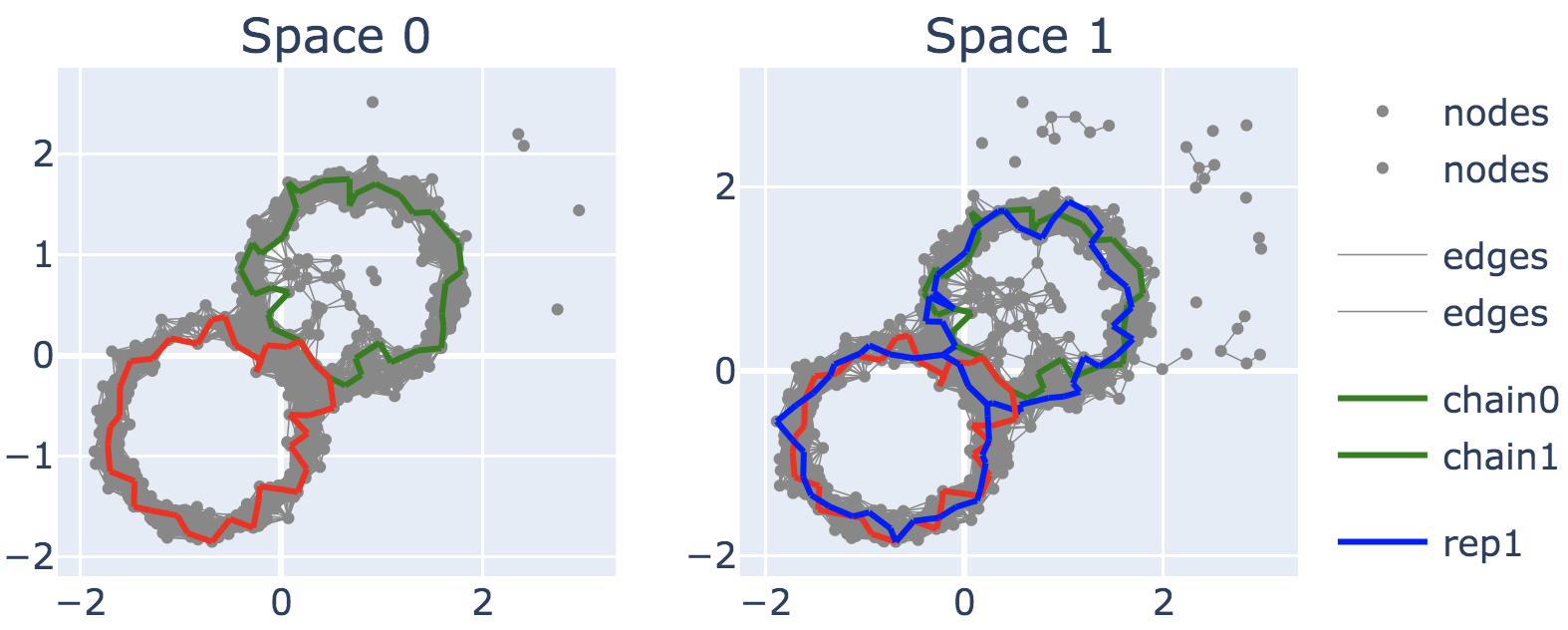}
    \includegraphics[width=0.48\linewidth]{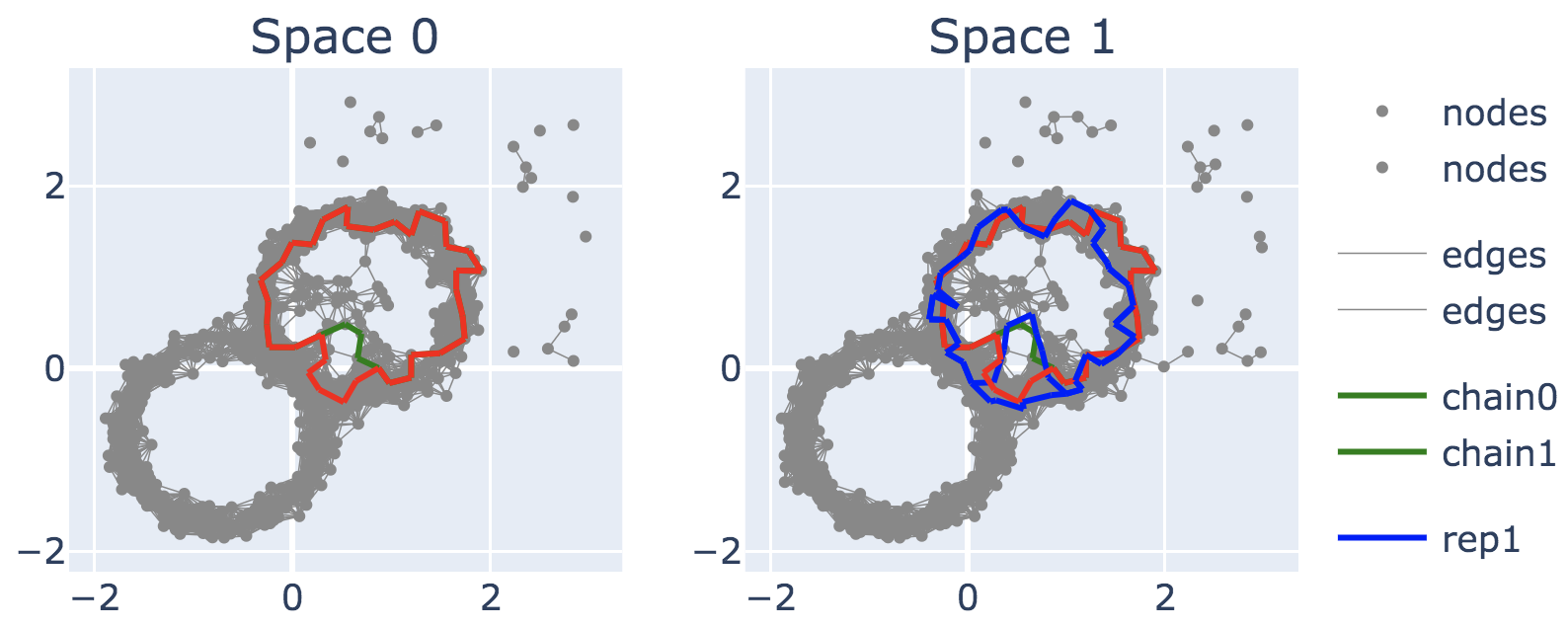}
    \includegraphics[width=0.48\linewidth]{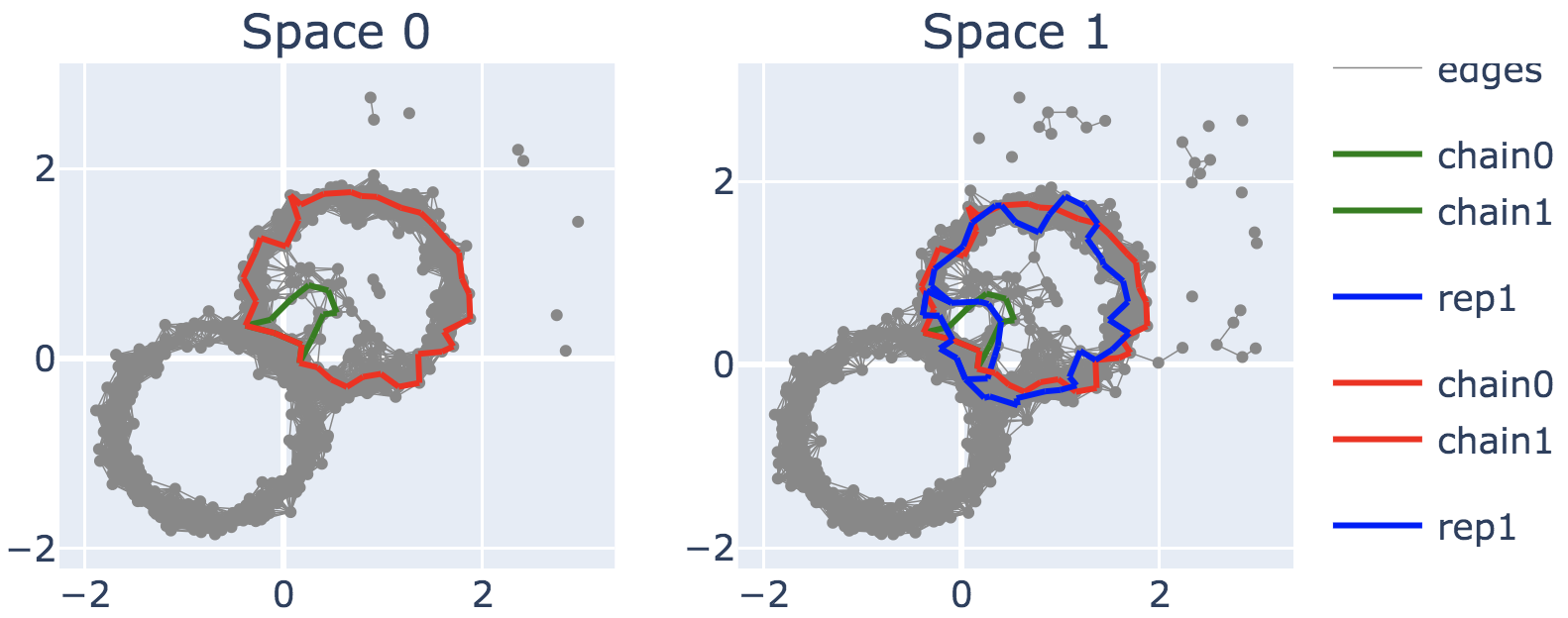}
    \includegraphics[width=0.48\linewidth]{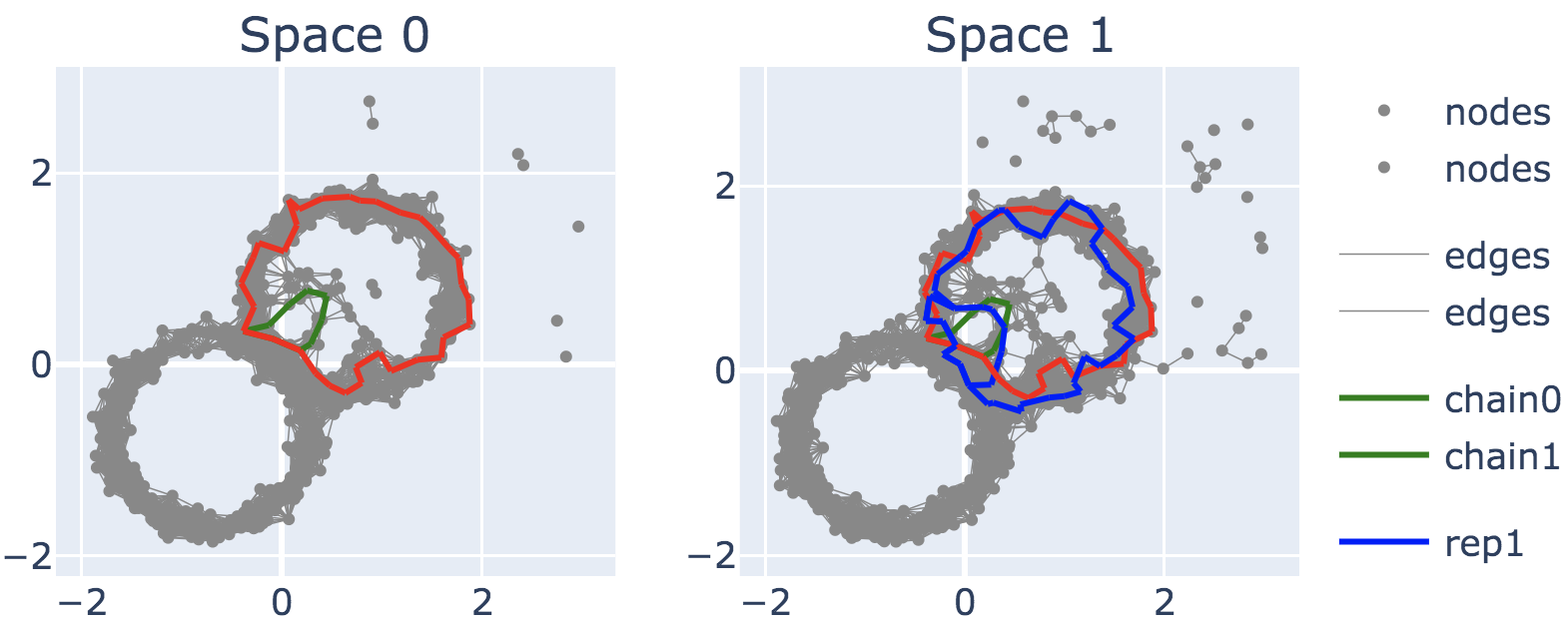}
    \includegraphics[width=0.48\linewidth]{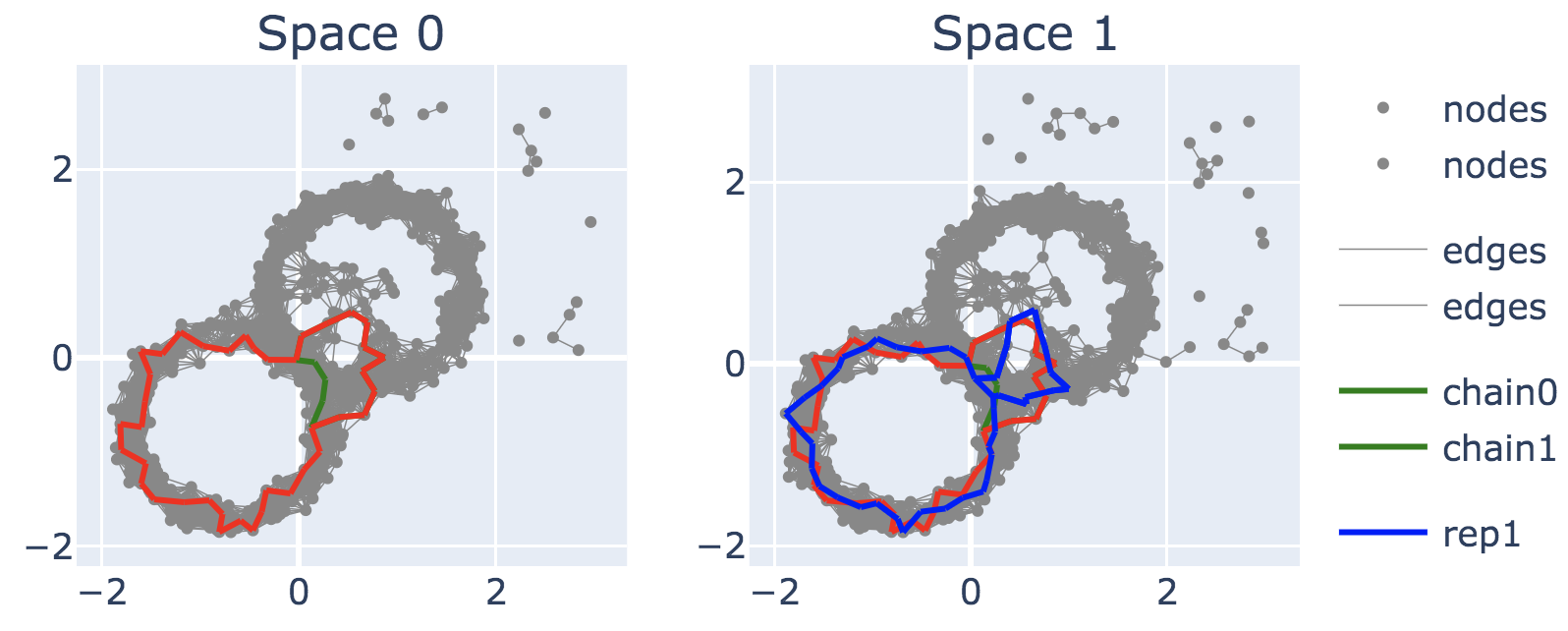}
    \includegraphics[width=0.48\linewidth]{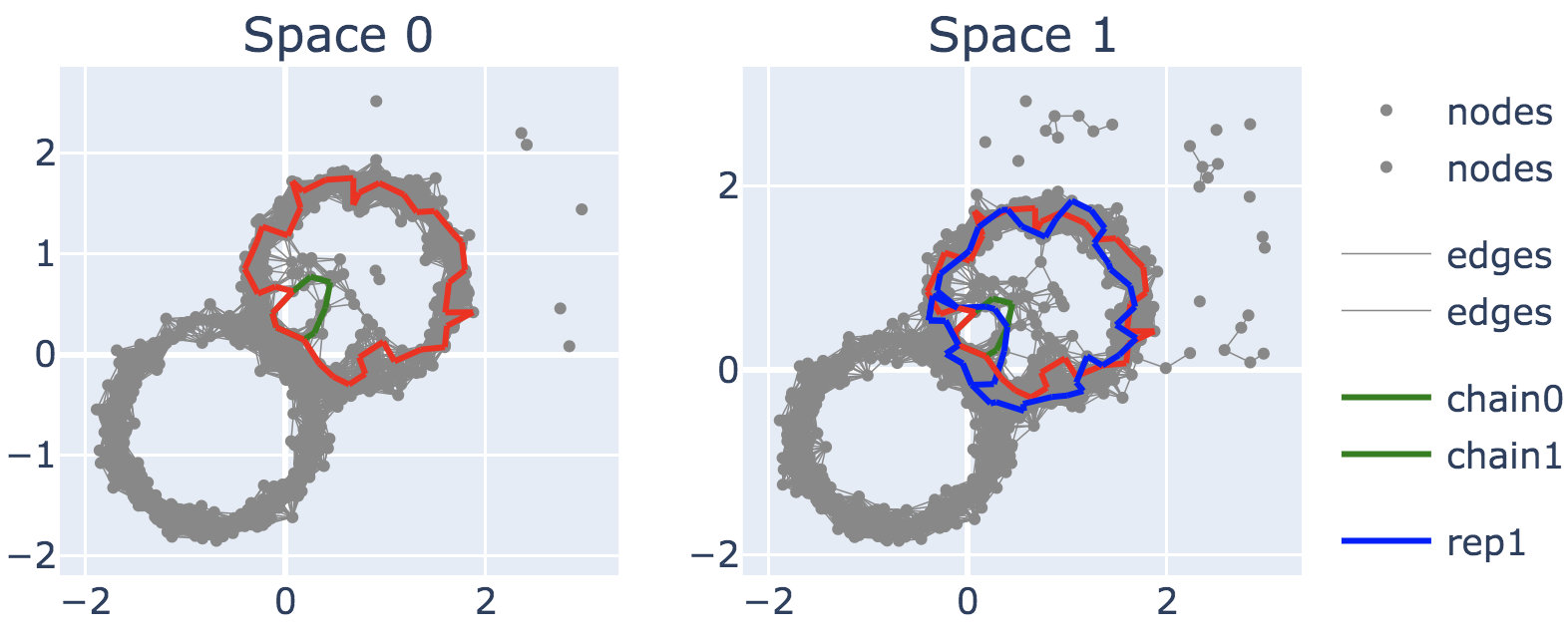}
    \caption{Visualization of induced maps of noisy Figure-8. Full size is $1000$. Sample size is $800$.}
    \label{fig:fig_8_800_im}
\end{figure}
\clearpage

\begin{figure}
\centering
\begin{subfigure}{\linewidth}
    \includegraphics[width=0.48\linewidth]{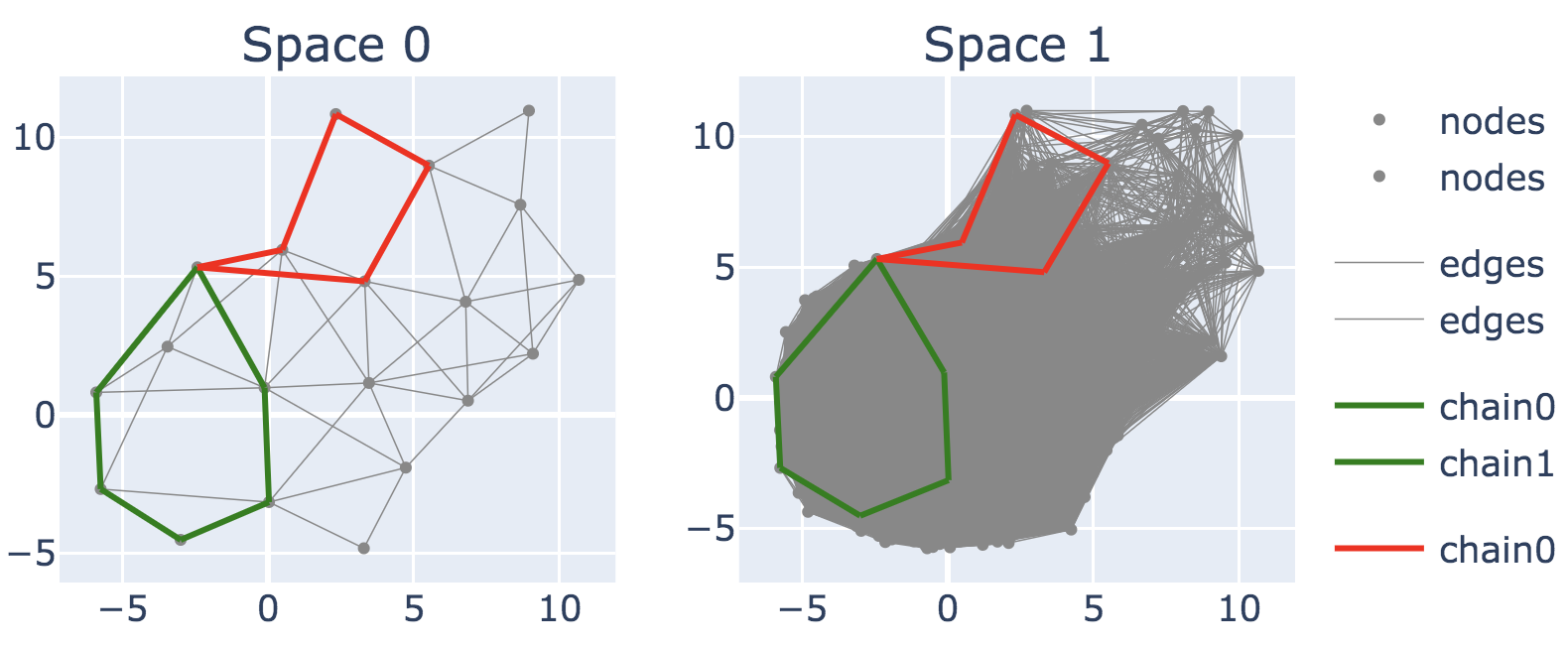}
    \includegraphics[width=0.48\linewidth]{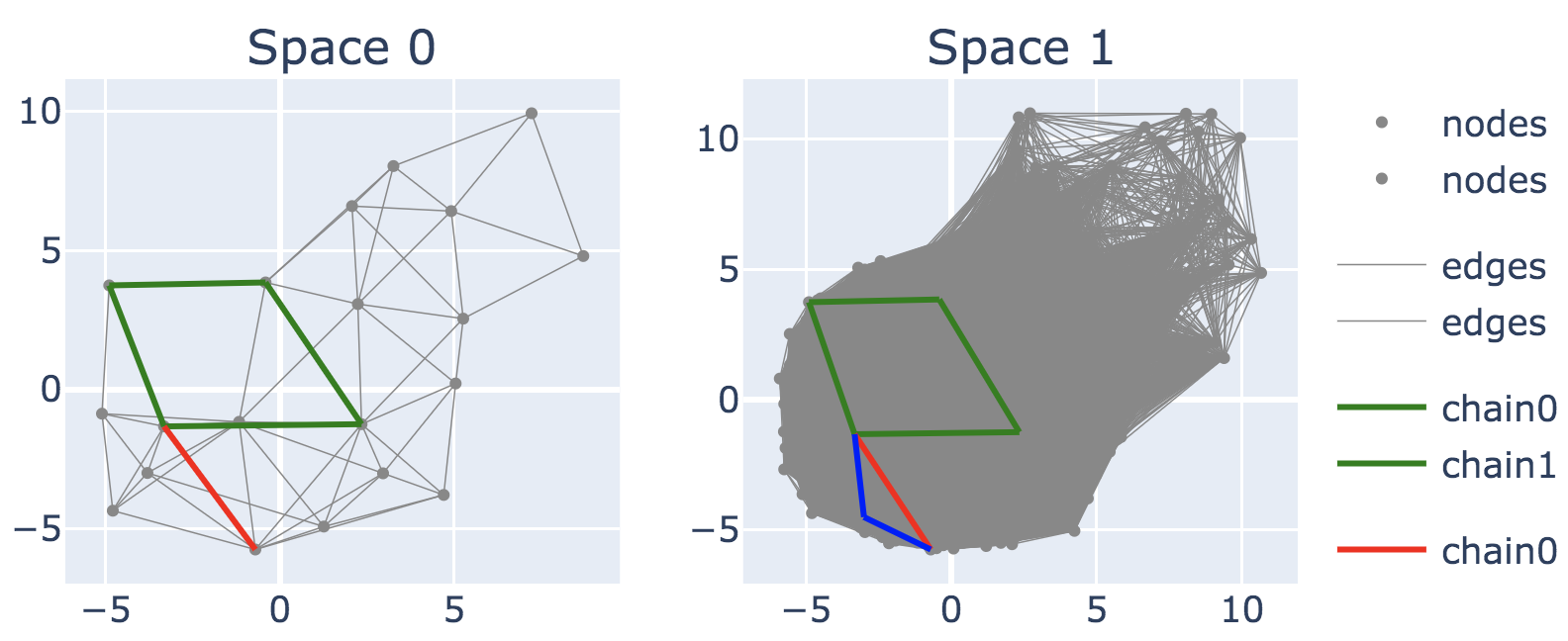} 
    \caption{Sample size is $20$. The threshold of Rips complex is $0.25 +\text{Hausdorff dist}*2$.}
    \label{fig:ann_20_im1}
    \end{subfigure}
\begin{subfigure}{\linewidth}
    \includegraphics[width=0.48\linewidth]{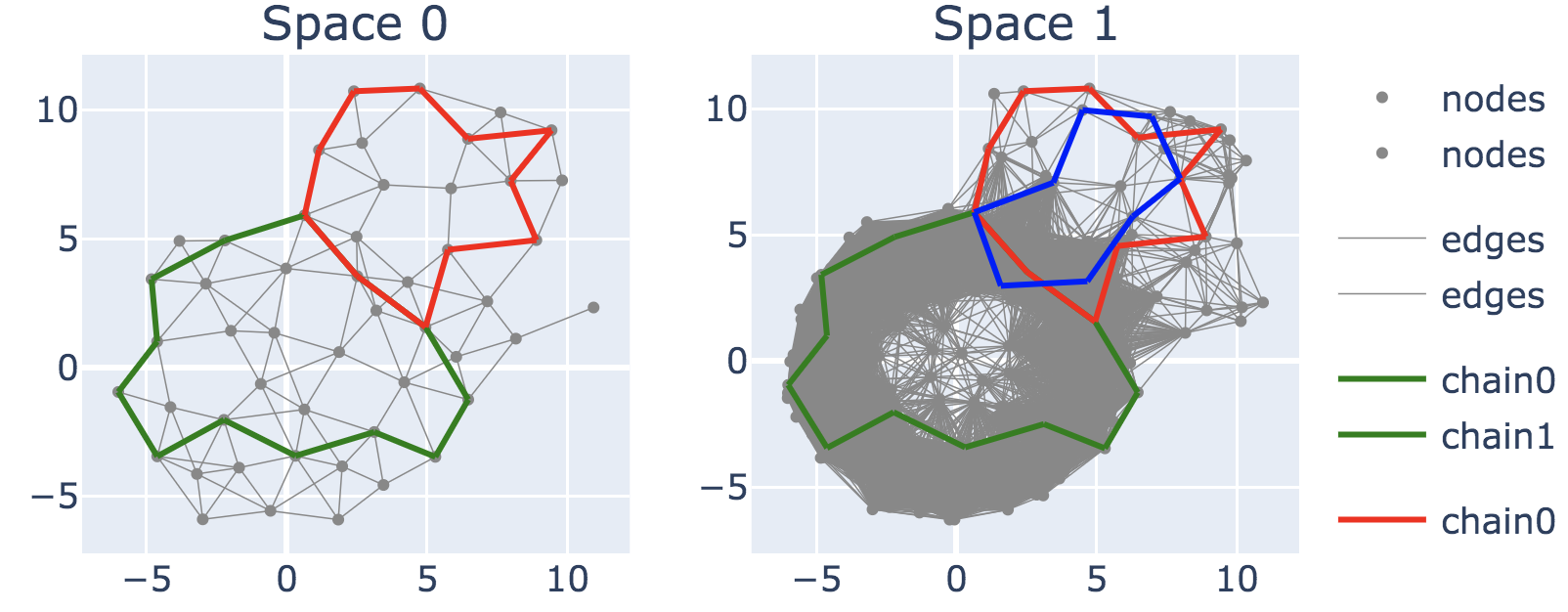}
    \includegraphics[width=0.48\linewidth]{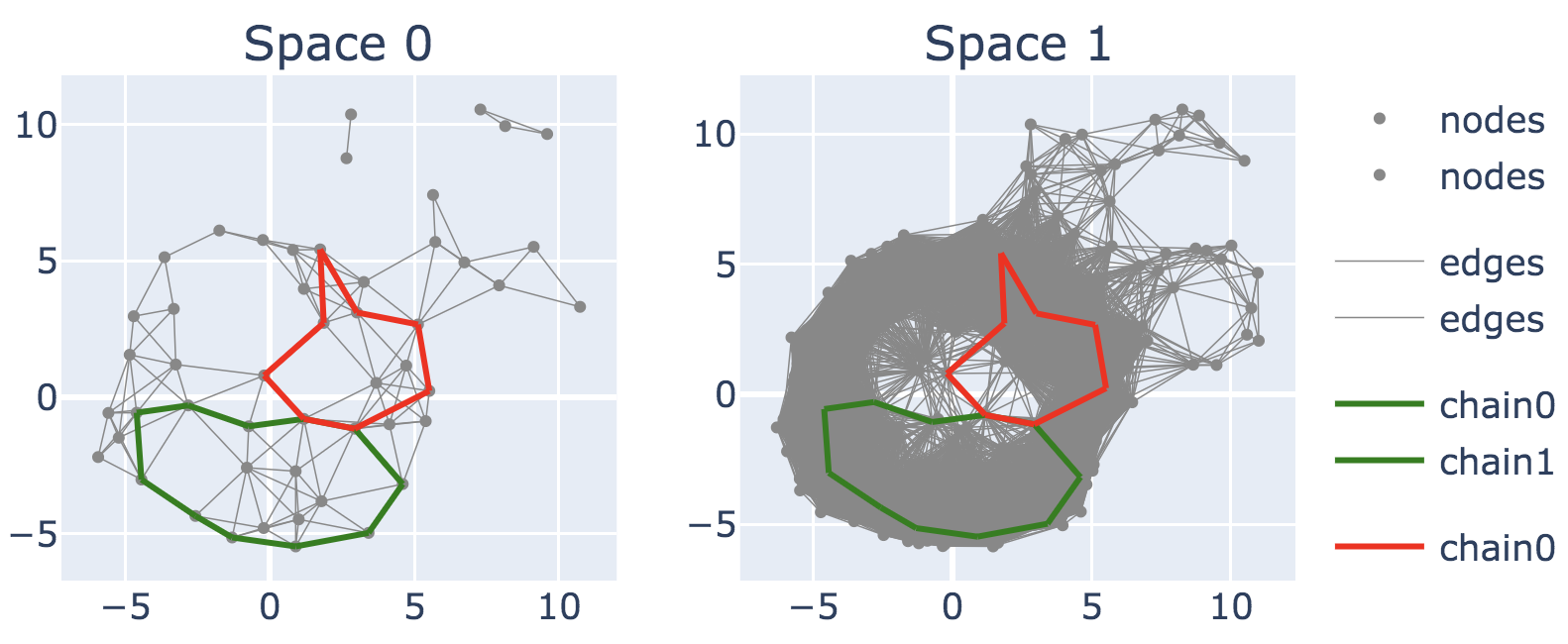}
    \caption{Sample size is $50$. The threshold of Rips complex is $0.25 +\text{Hausdorff dist}*2$.}
    \label{fig:ann_50_im1}
    \end{subfigure}
\begin{subfigure}{\linewidth}
    \includegraphics[width=0.48\linewidth]{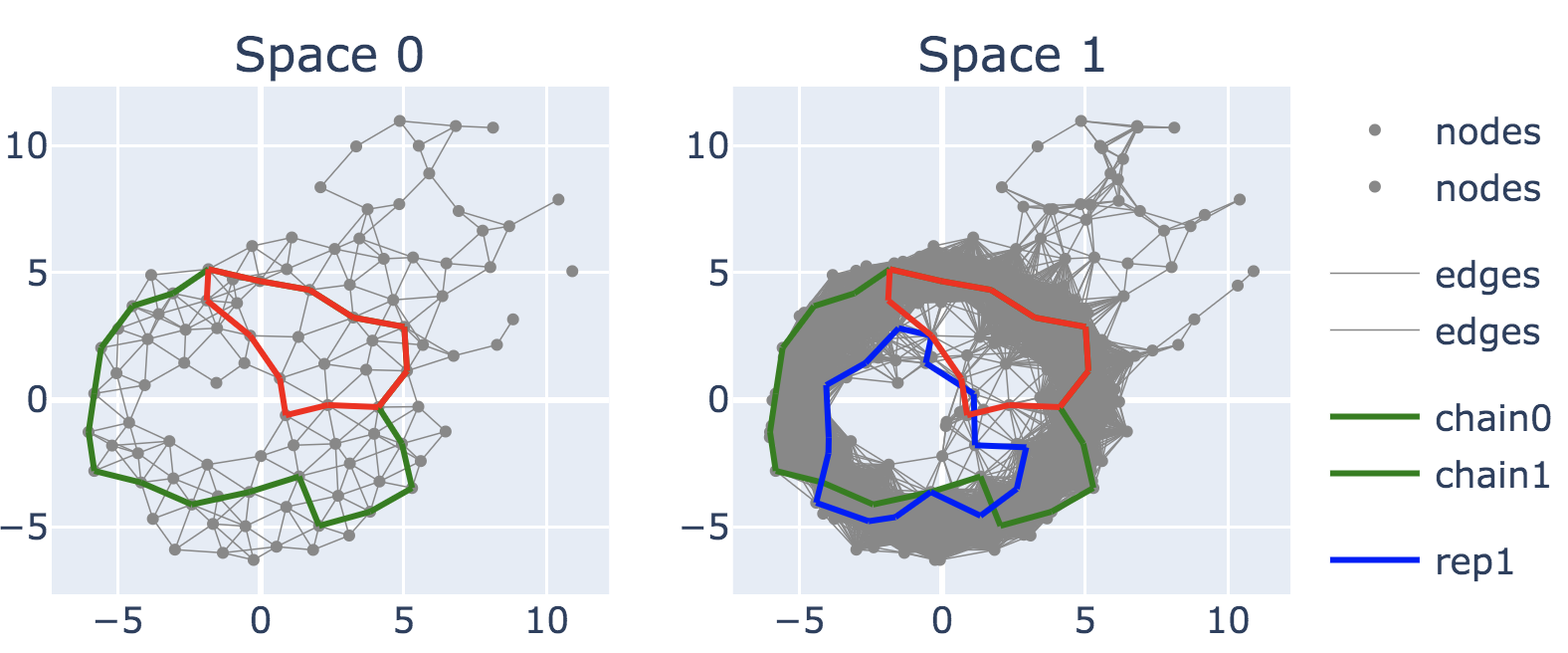}
    \includegraphics[width=0.48\linewidth]{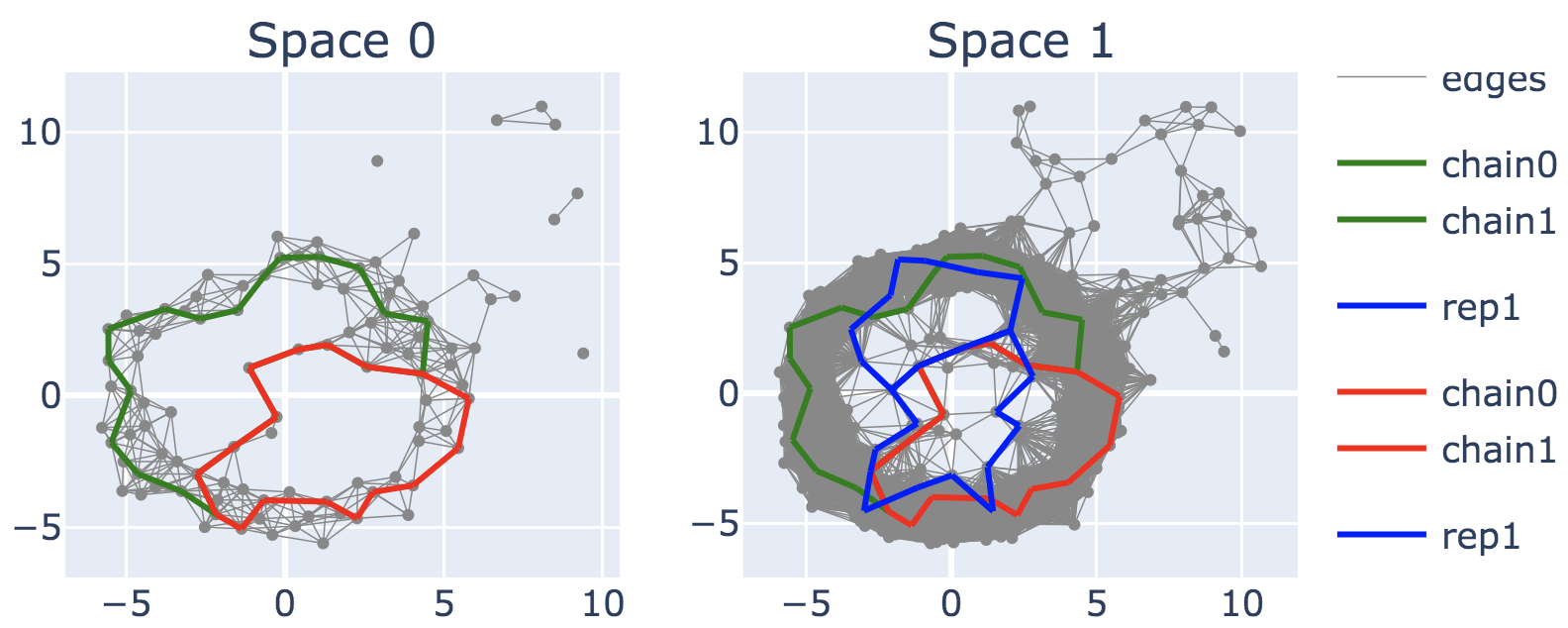}
    \caption{Sample size is $100$. The threshold of Rips complex is $0.25 +\text{Hausdorff dist}*2$.}
    \label{fig:ann_100_im1}
    \end{subfigure}
\begin{subfigure}{\linewidth}
    \includegraphics[width=0.48\linewidth]{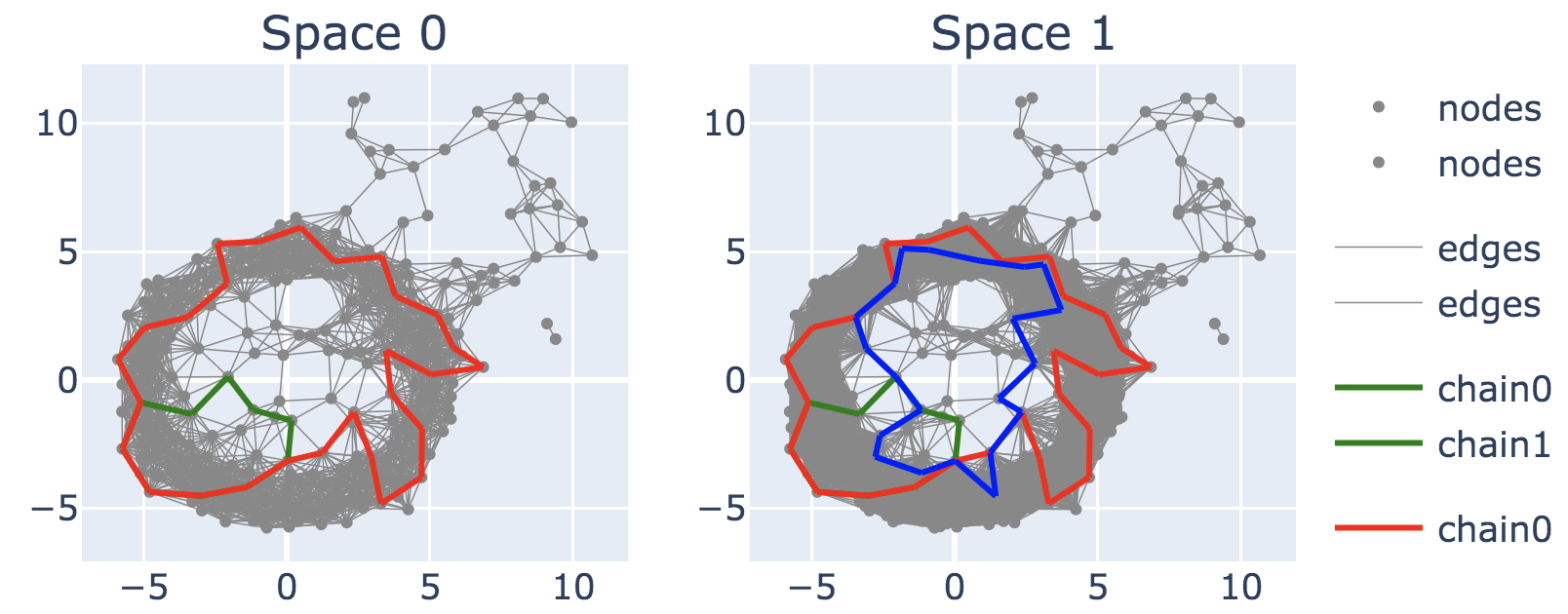}
    \includegraphics[width=0.48\linewidth]{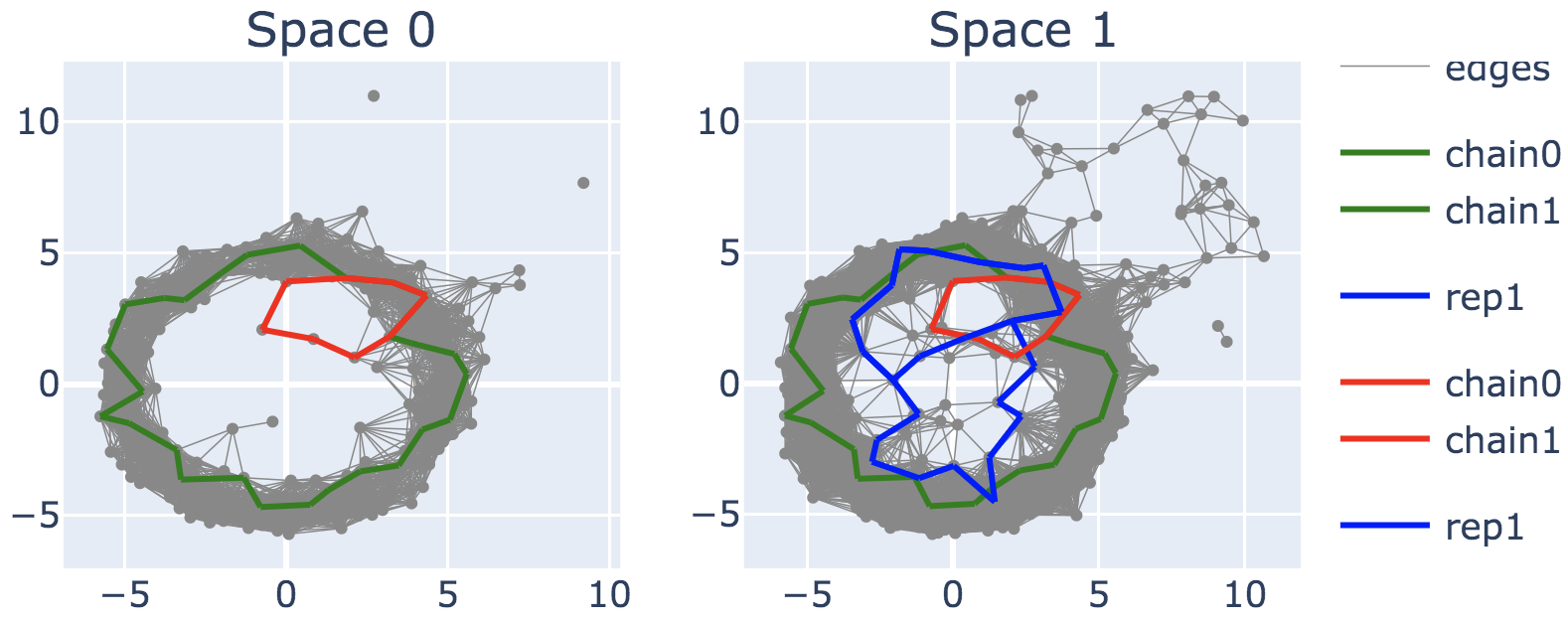} 
    \caption{Sample size is $300$. The threshold of Rips complex is $2$.}
    \label{fig:ann_300_im1}
    \end{subfigure}
\begin{subfigure}{\linewidth}
    \includegraphics[width=0.48\linewidth]{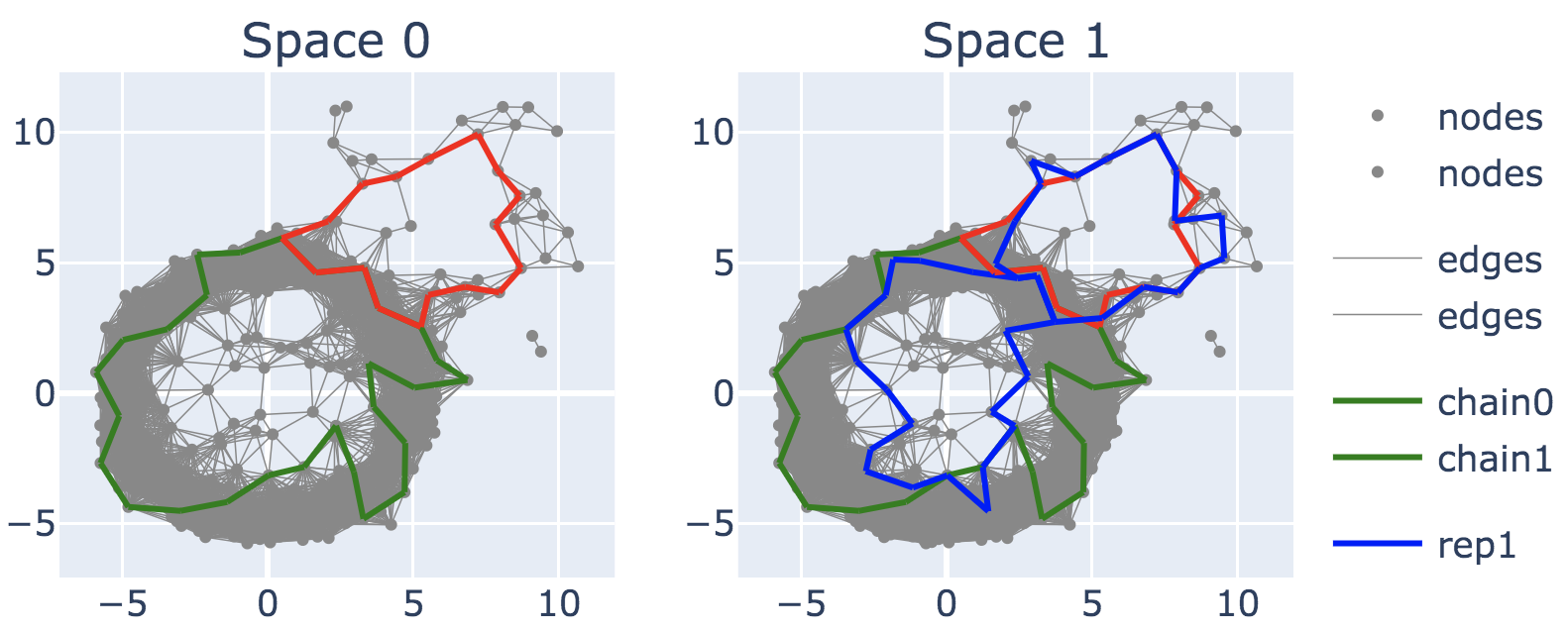}
    \includegraphics[width=0.48\linewidth]{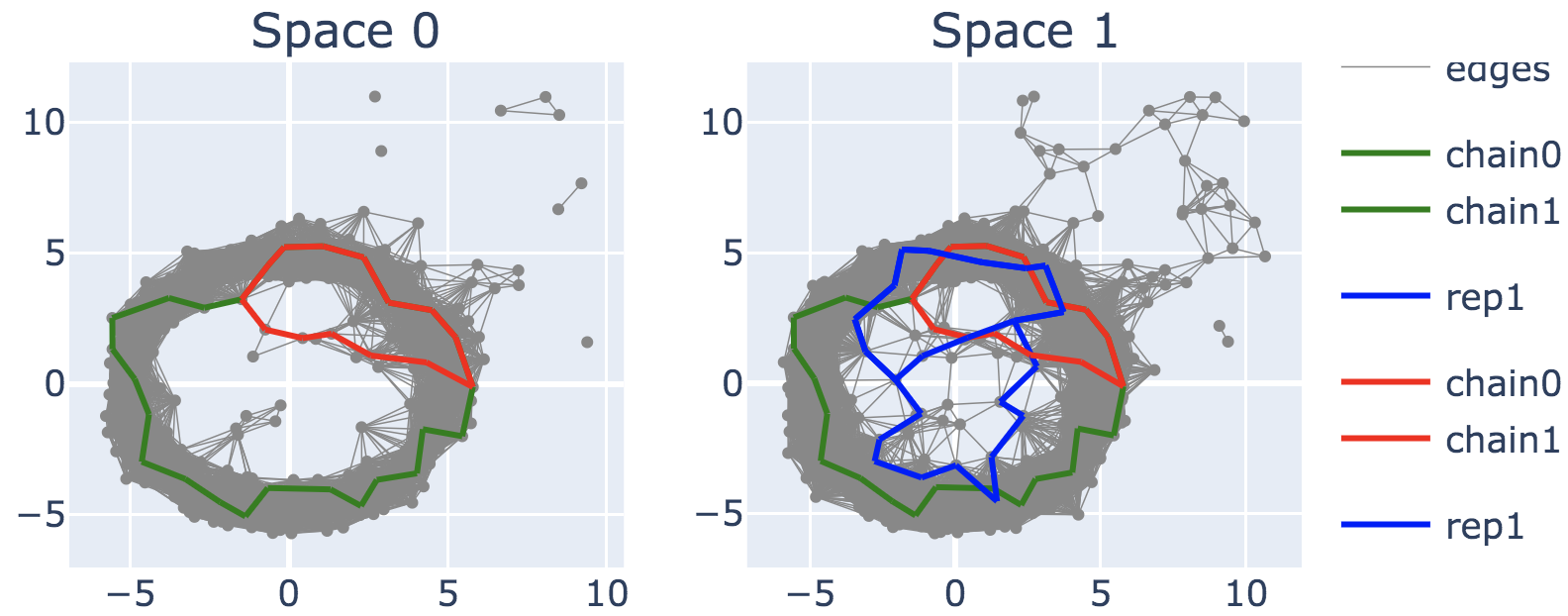}
    \caption{Sample size is $500$. The threshold of Rips complex is $2$.}
    \label{fig:ann_500_im1}
    \end{subfigure}
\begin{subfigure}{\linewidth}
    \includegraphics[width=0.48\linewidth]{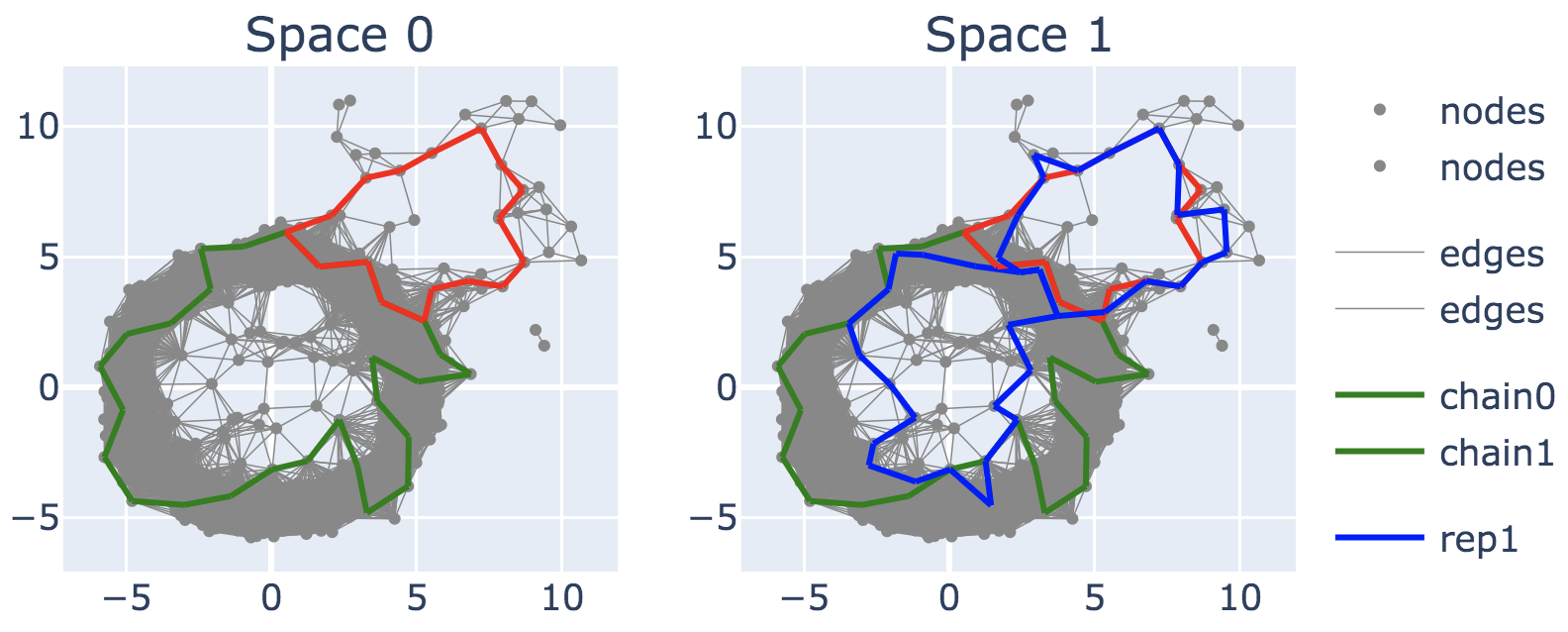}
    \includegraphics[width=0.48\linewidth]{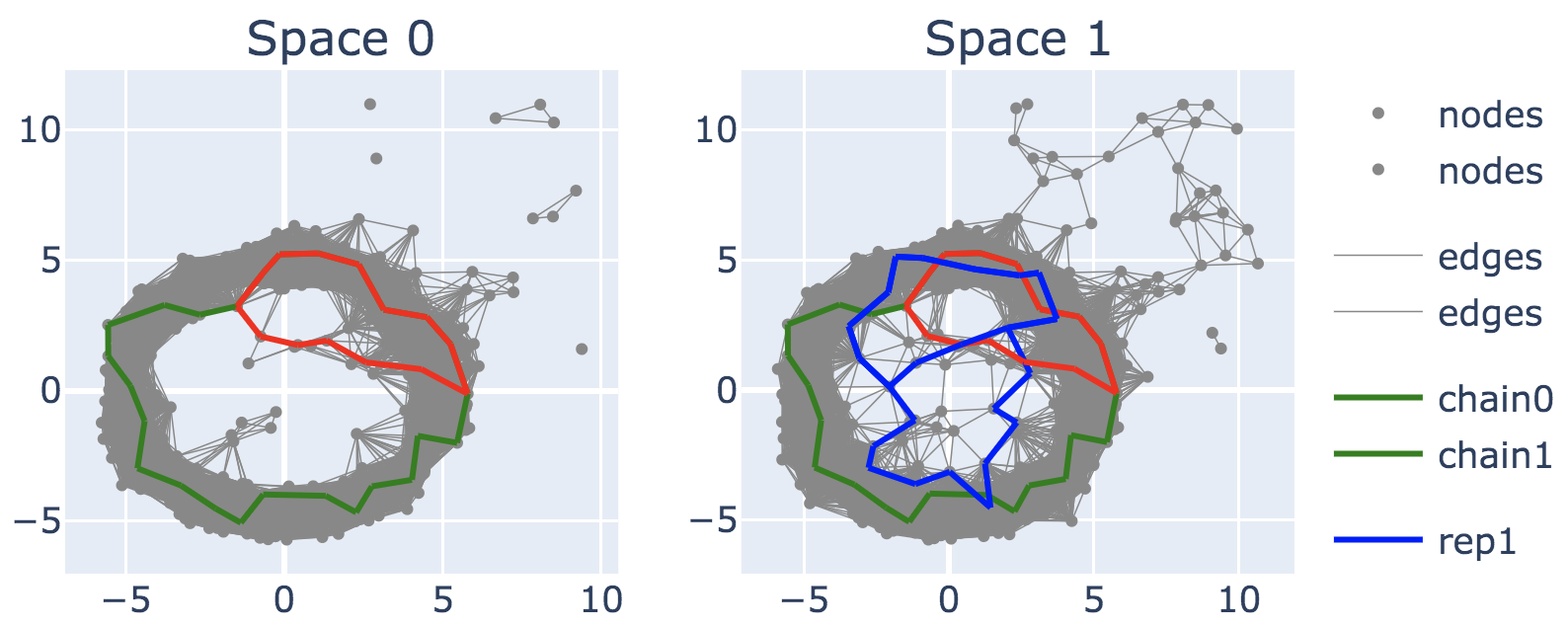}
    \caption{Sample size is $800$. The threshold of Rips complex is $2$.}
    \label{fig:ann_800_im1}
    \end{subfigure}
    \caption{Visualization of induced maps of annulus from sub-samples to full point cloud. Full size is $1000$.}
    \label{fig:ann_im1}
\end{figure}
\clearpage

\begin{figure}
\centering
    \includegraphics[width=0.48\linewidth]{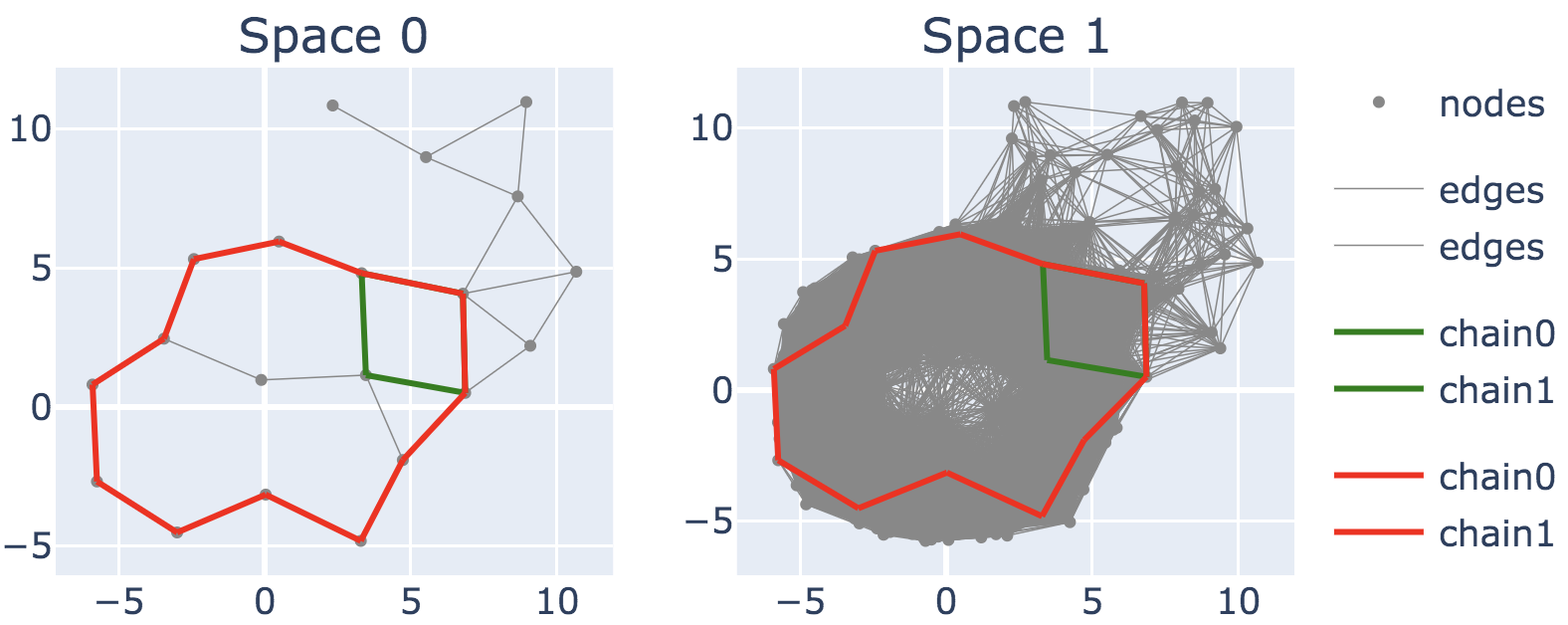}
    \includegraphics[width=0.48\linewidth]{figs/annulus/20hd1.png}
    \includegraphics[width=0.48\linewidth]{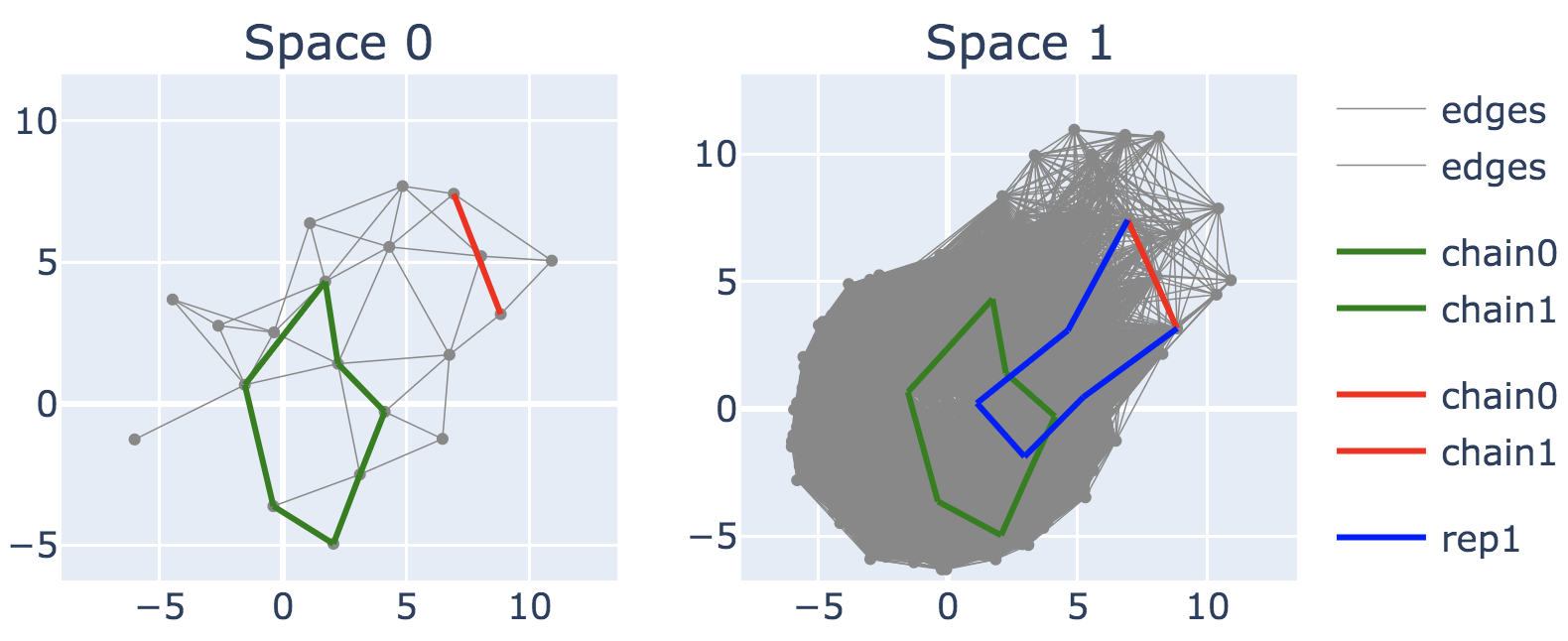}
    \includegraphics[width=0.48\linewidth]{figs/annulus/20hd2.png}
    \includegraphics[width=0.48\linewidth]{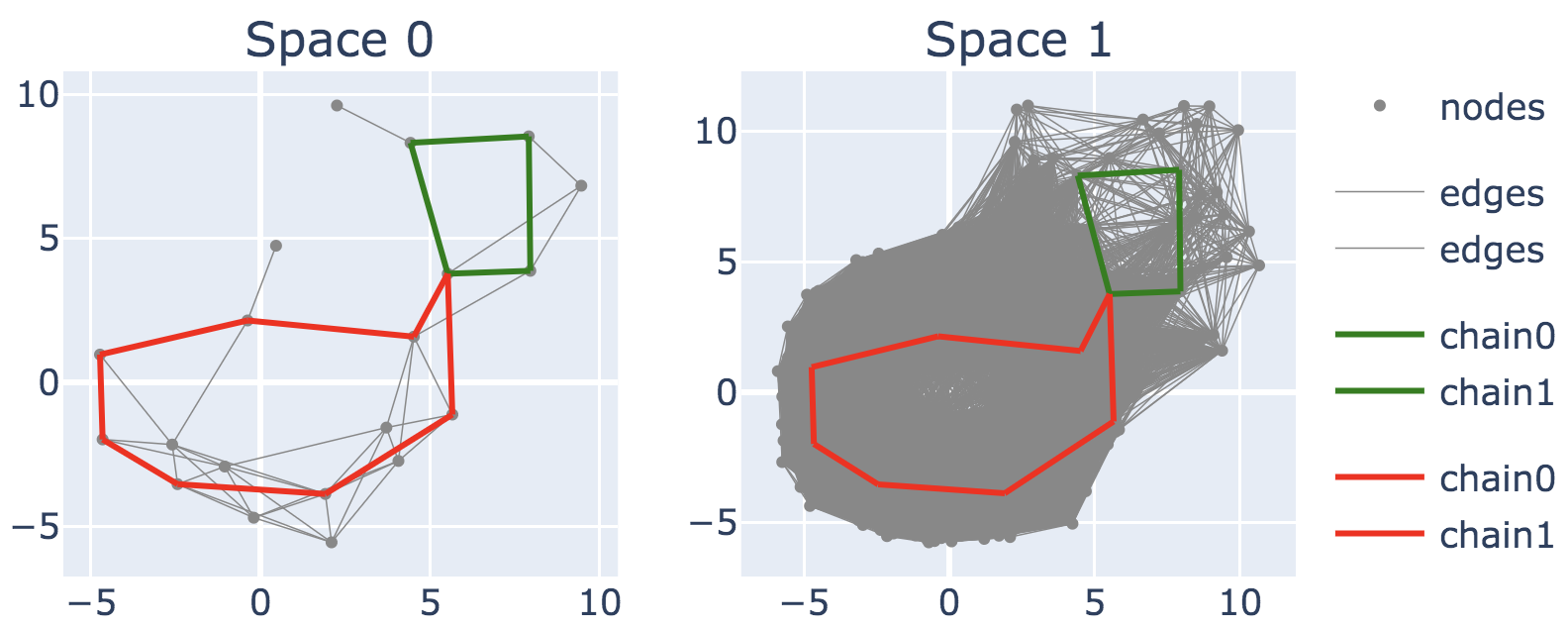}
    \includegraphics[width=0.48\linewidth]{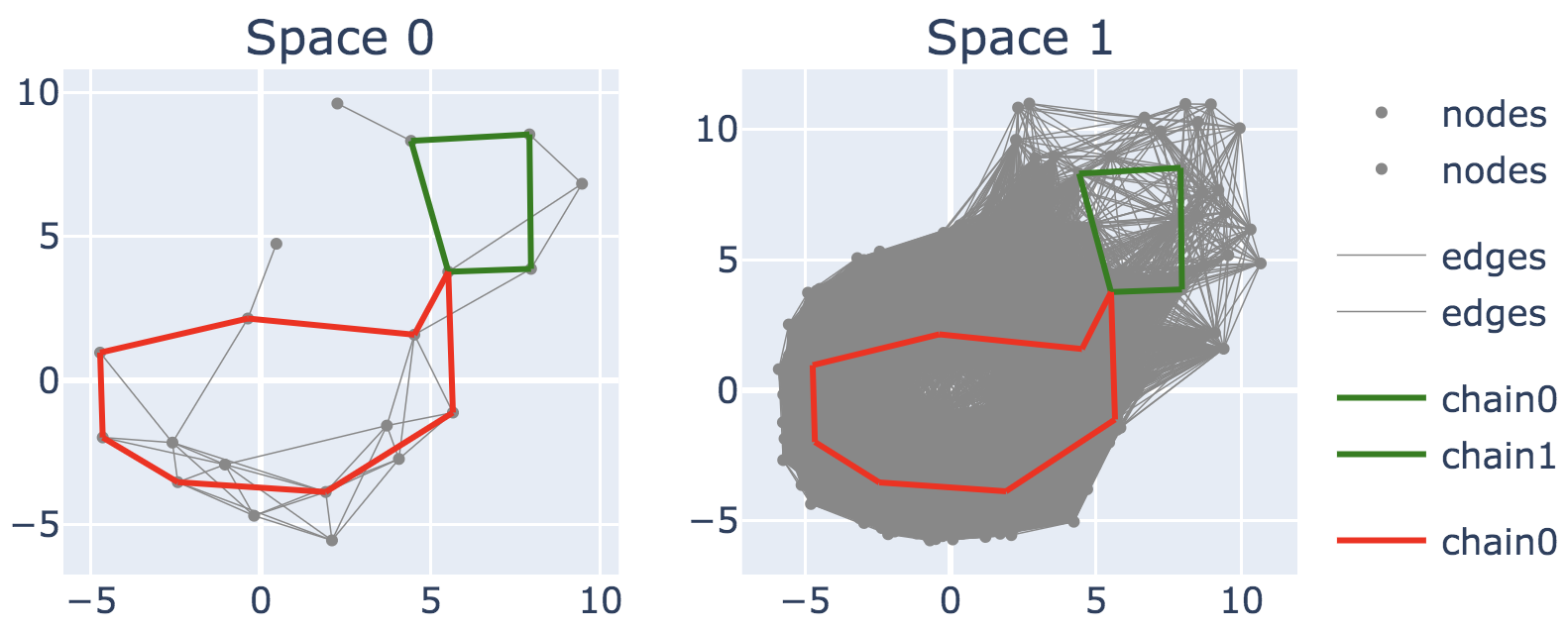}
    \includegraphics[width=0.48\linewidth]{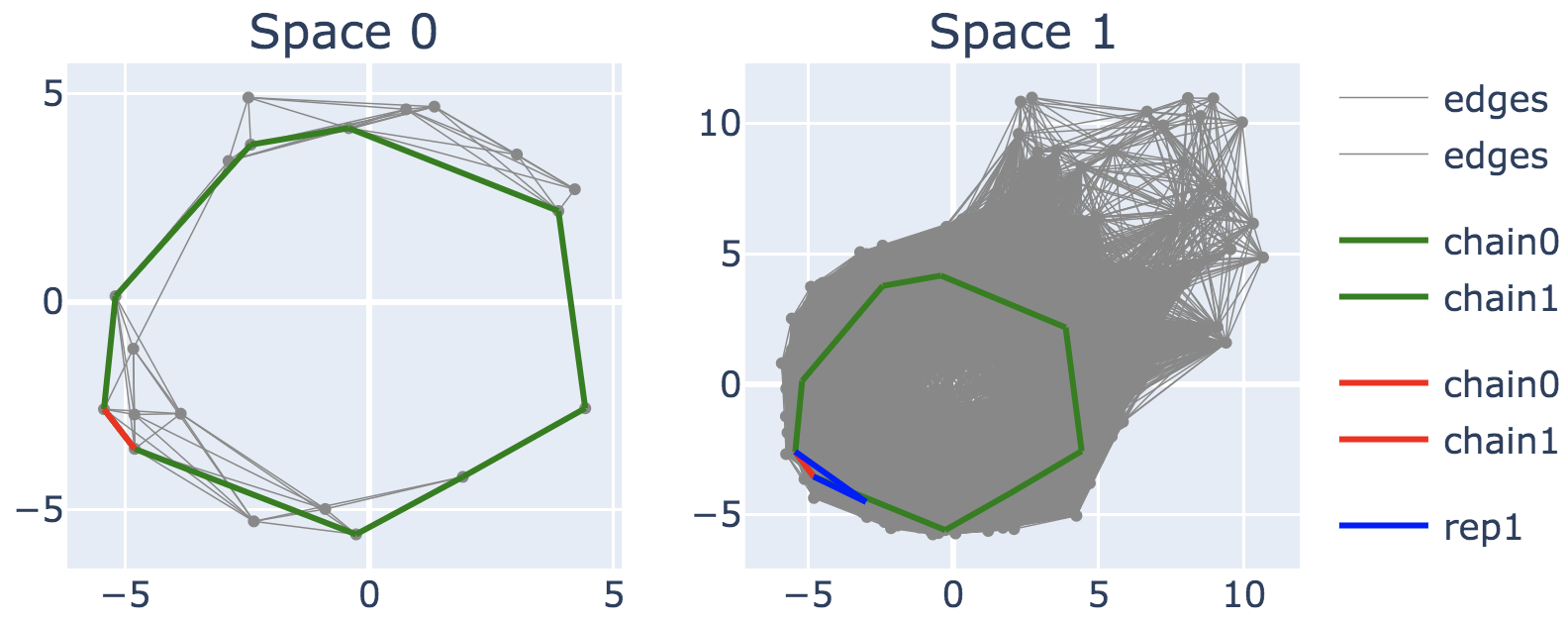}
    \includegraphics[width=0.48\linewidth]{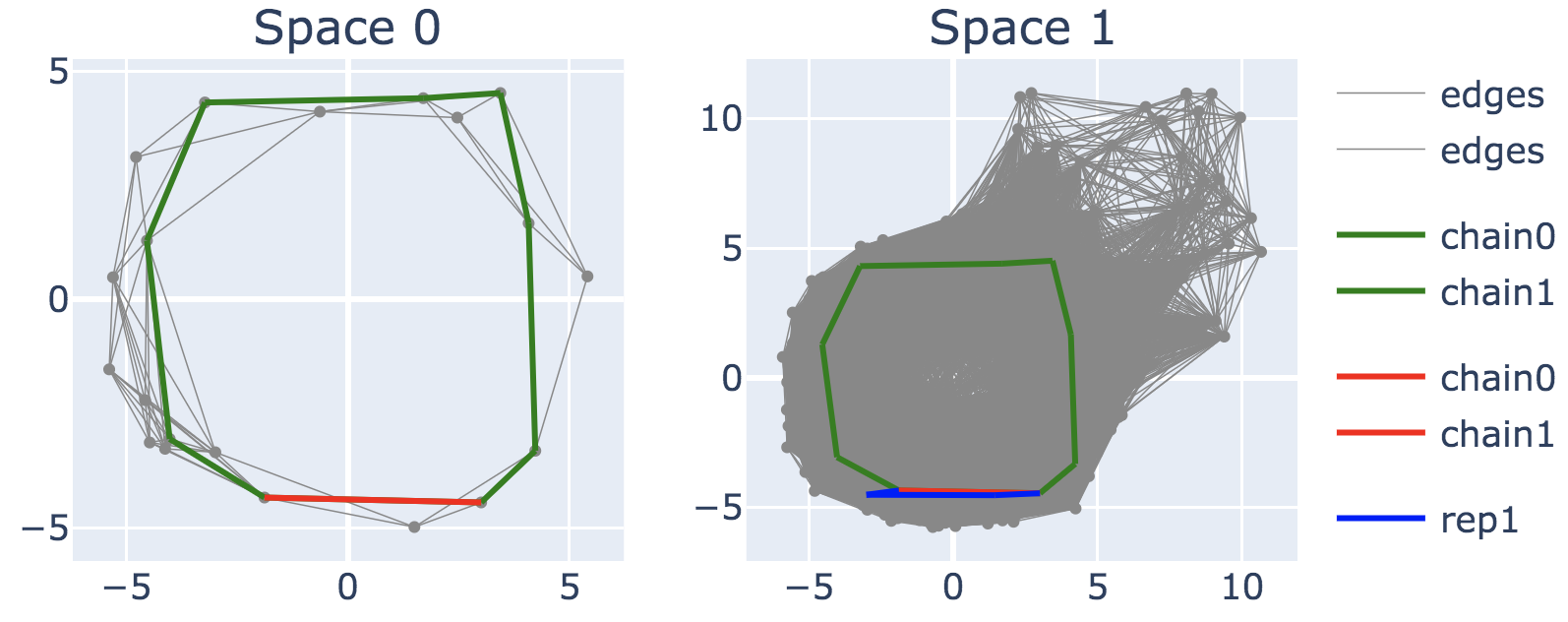}
    \includegraphics[width=0.48\linewidth]{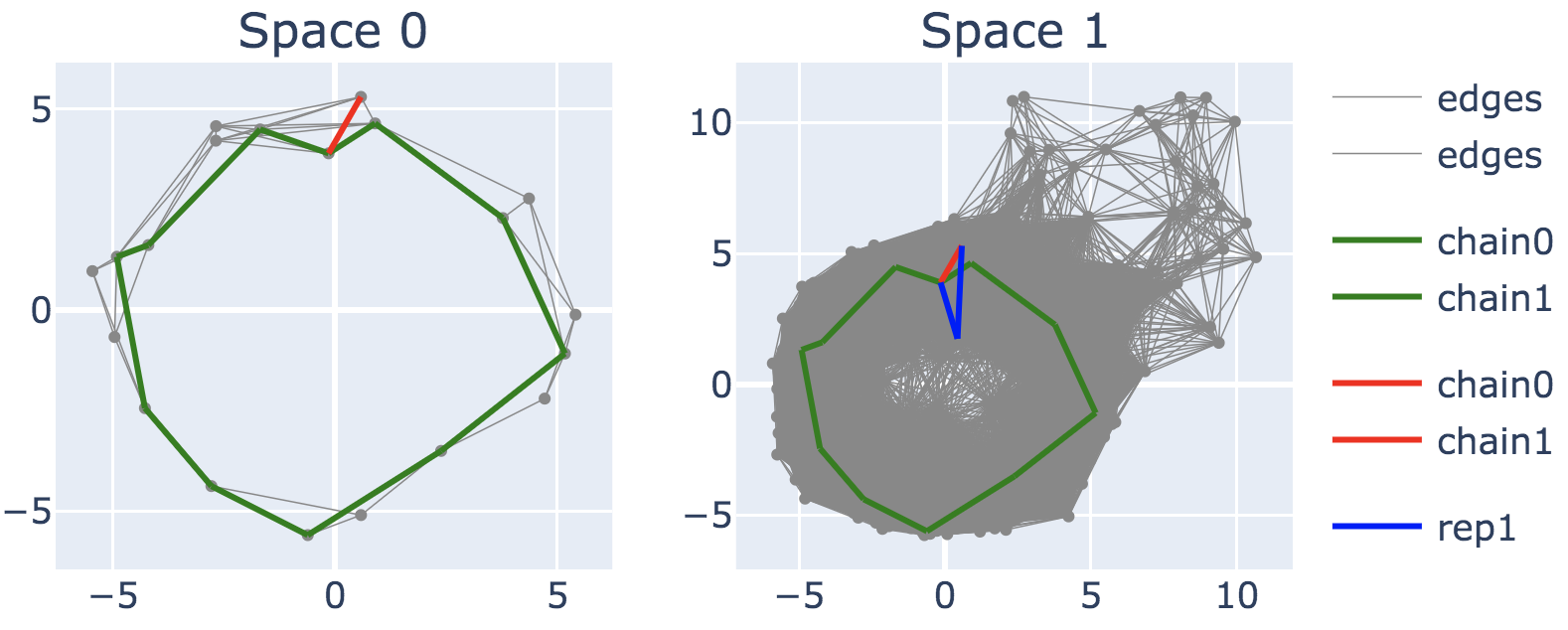}
    \includegraphics[width=0.48\linewidth]{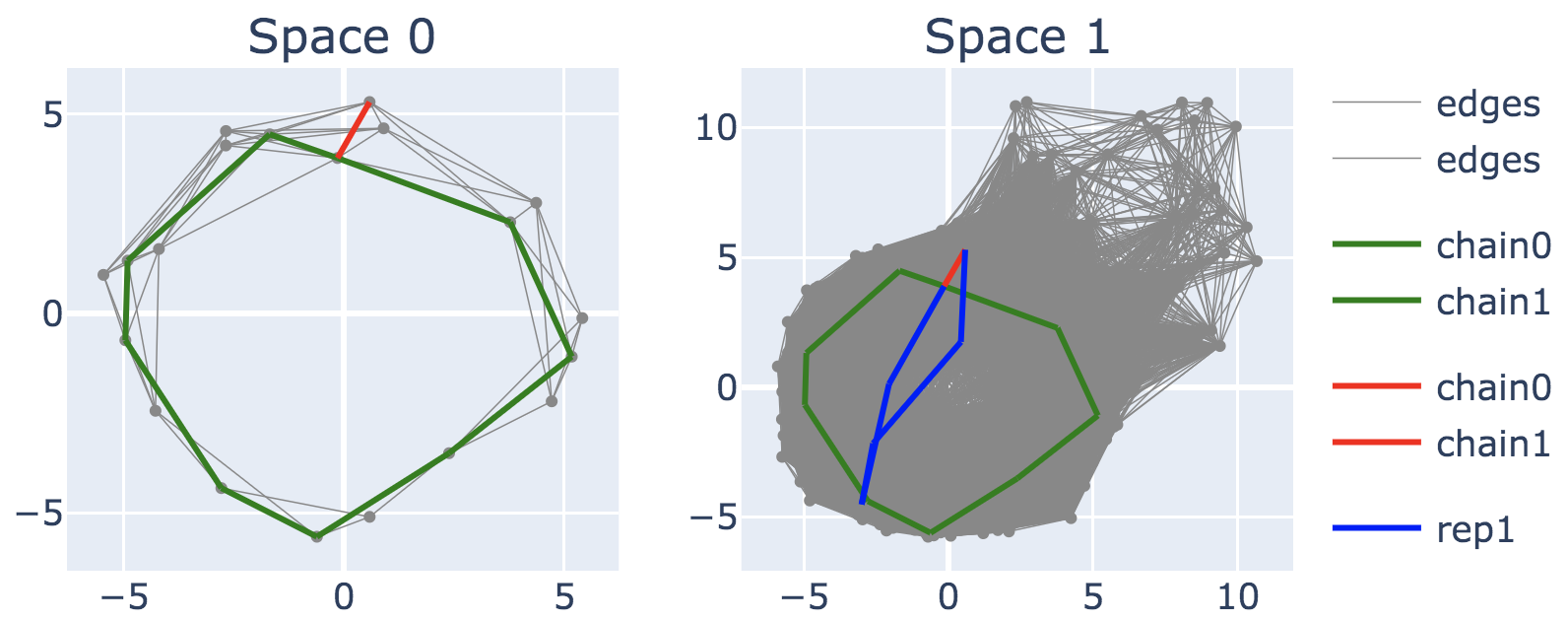}
    \caption{Induced maps of noisy Annulus. Full point cloud size is $1000$. Sample size is $20$.}
    \label{fig:annulus_20_im}
\end{figure}

\begin{figure}
\centering
    \includegraphics[width=0.48\linewidth]{figs/annulus/50hd1.png}
    \includegraphics[width=0.48\linewidth]{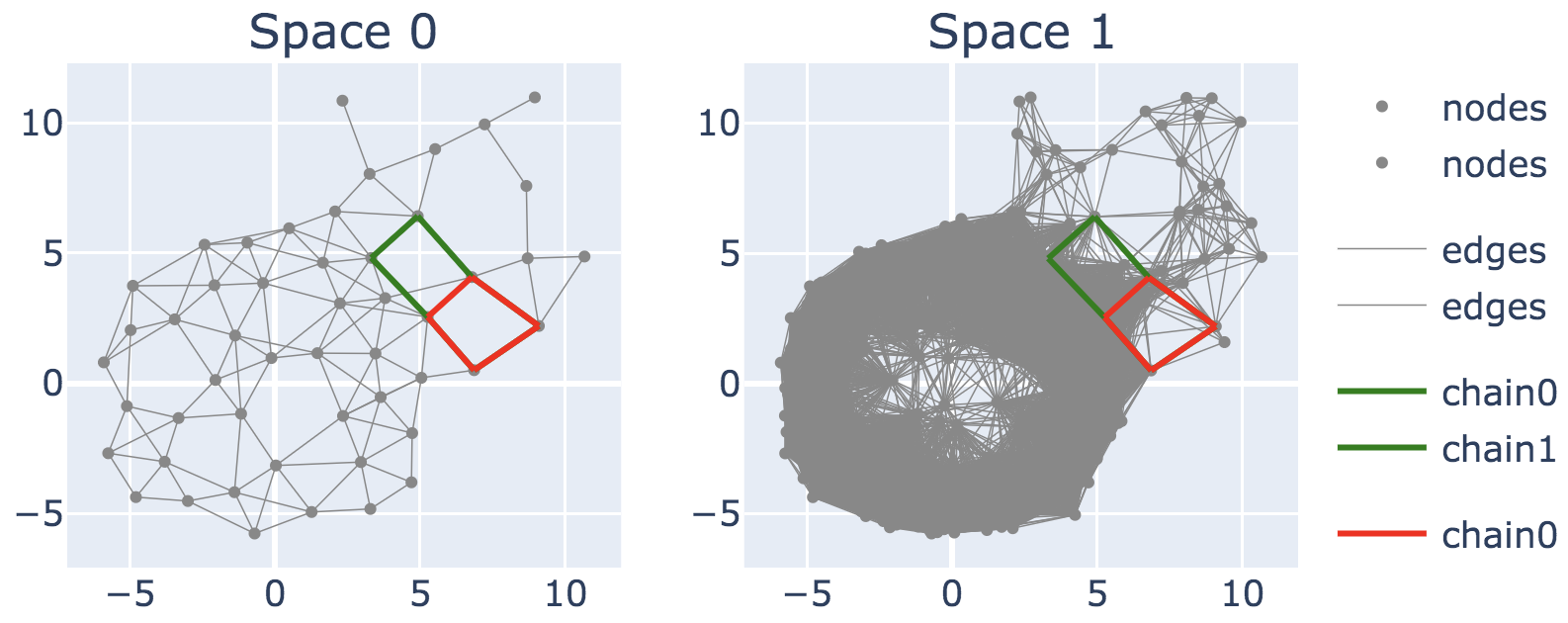}
    \includegraphics[width=0.48\linewidth]{figs/annulus/50hd2.png}
    \includegraphics[width=0.48\linewidth]{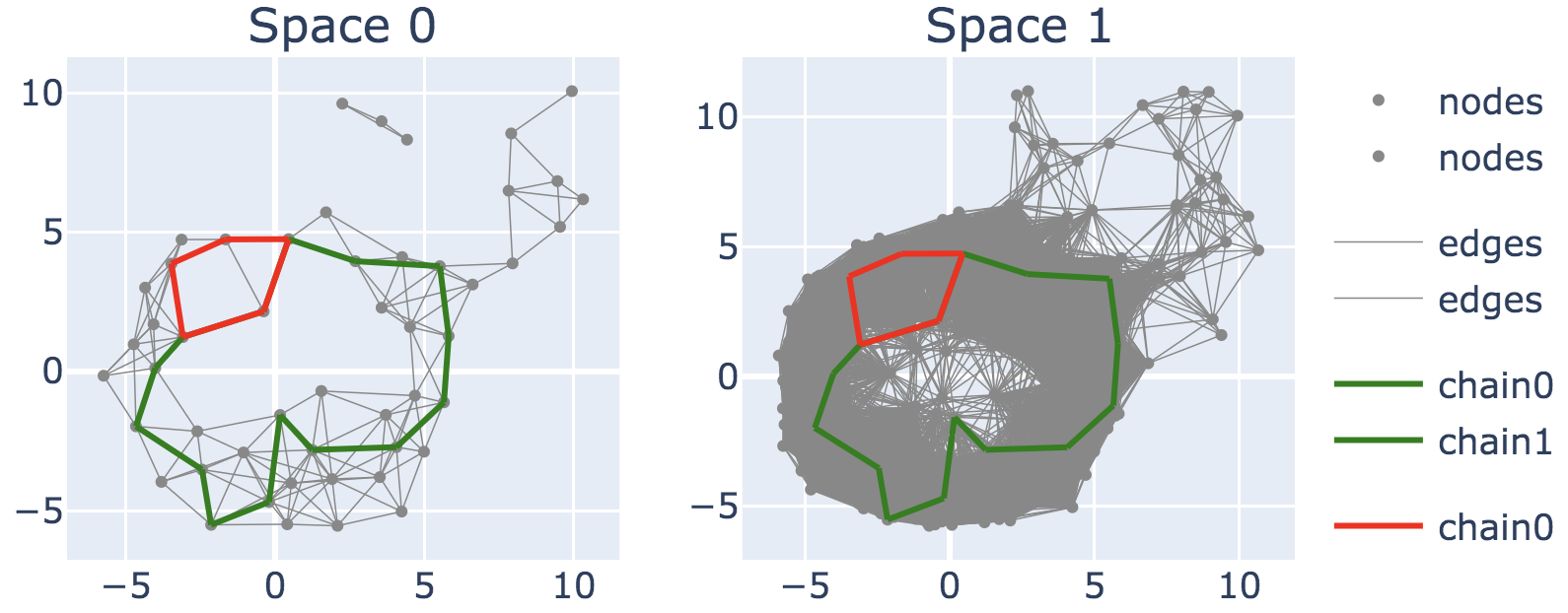}
    \includegraphics[width=0.48\linewidth]{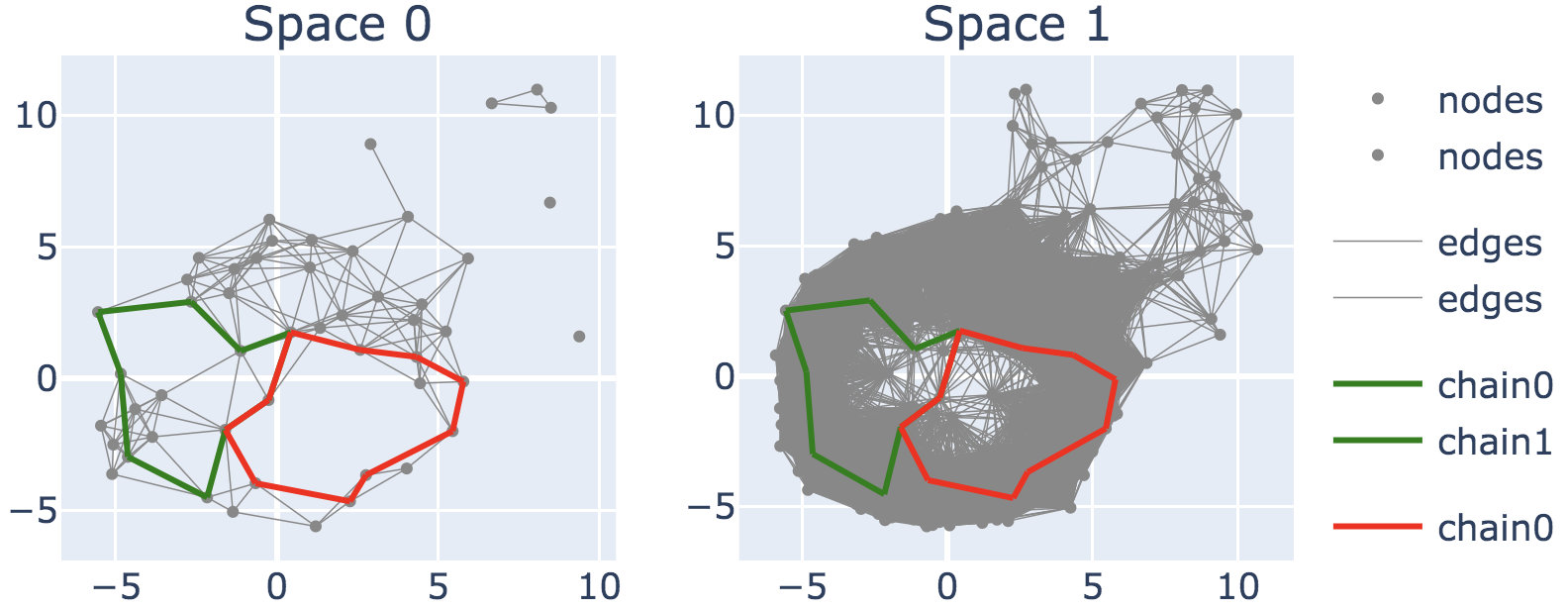}
    \includegraphics[width=0.48\linewidth]{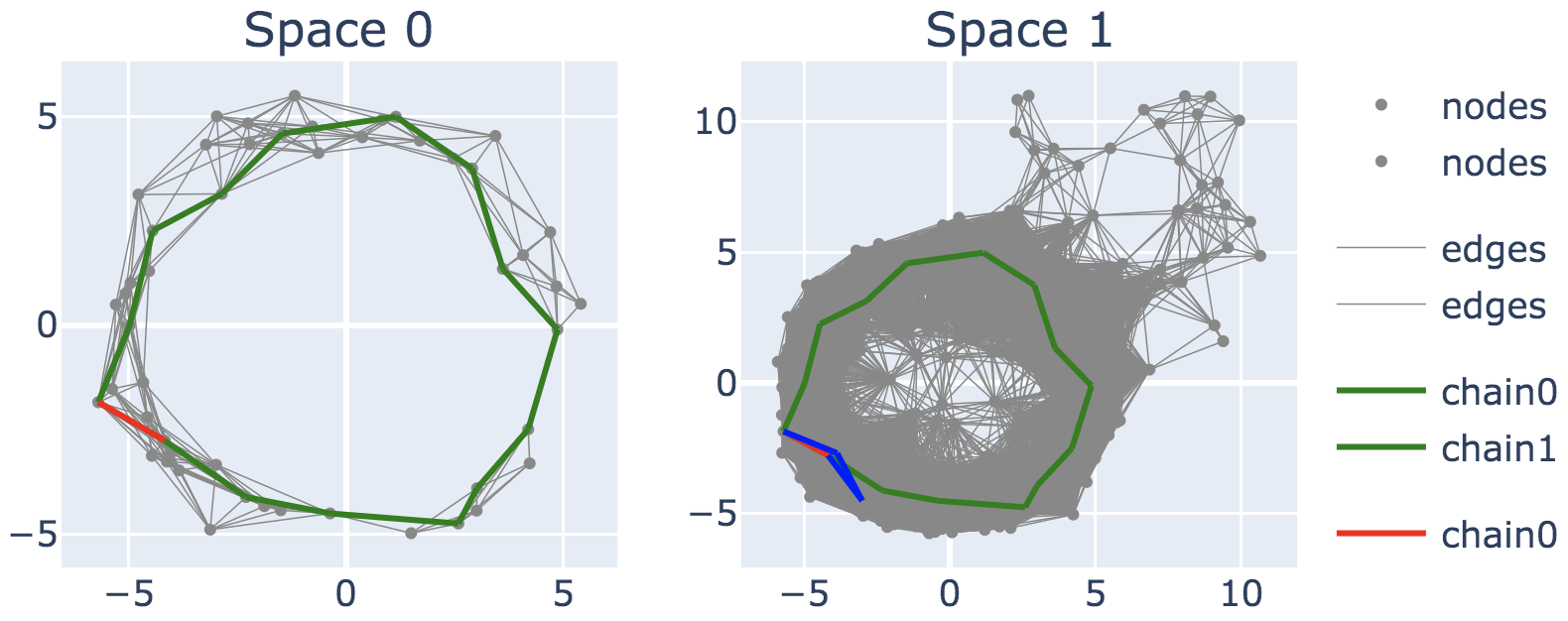}
    \includegraphics[width=0.48\linewidth]{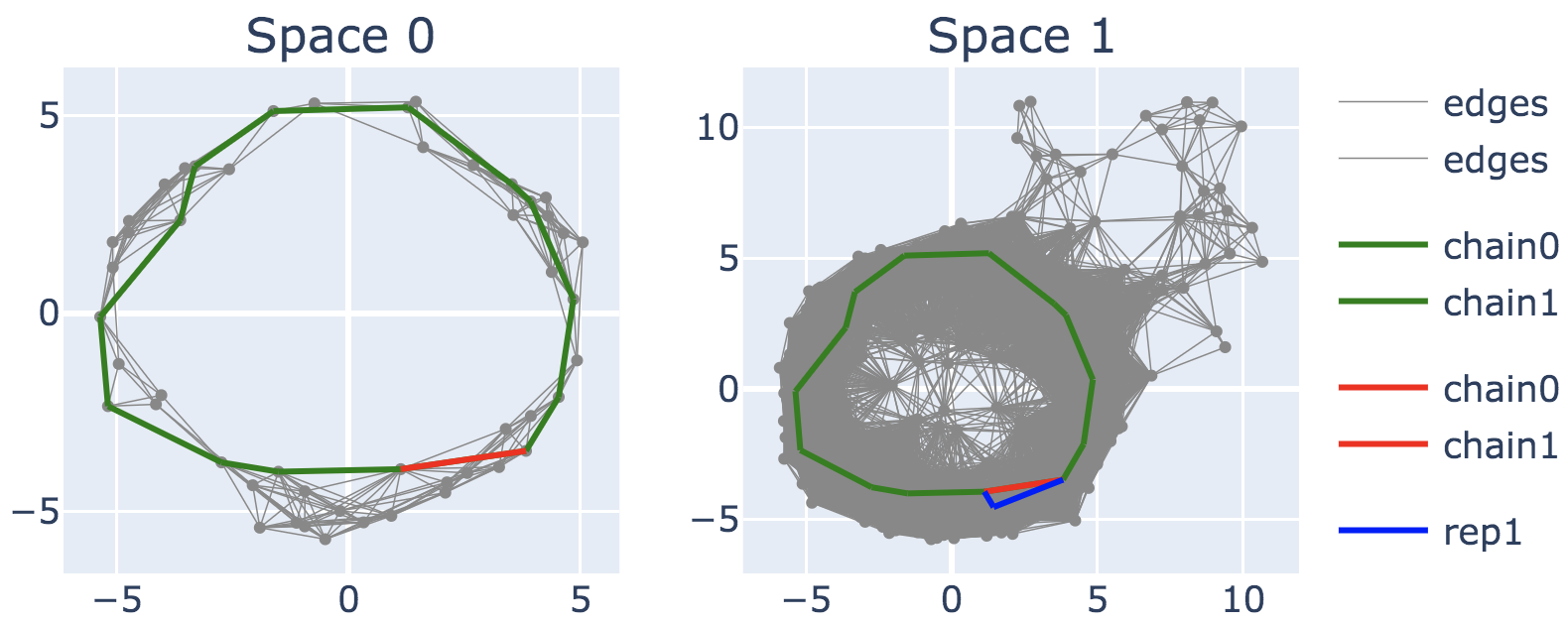}
    \includegraphics[width=0.48\linewidth]{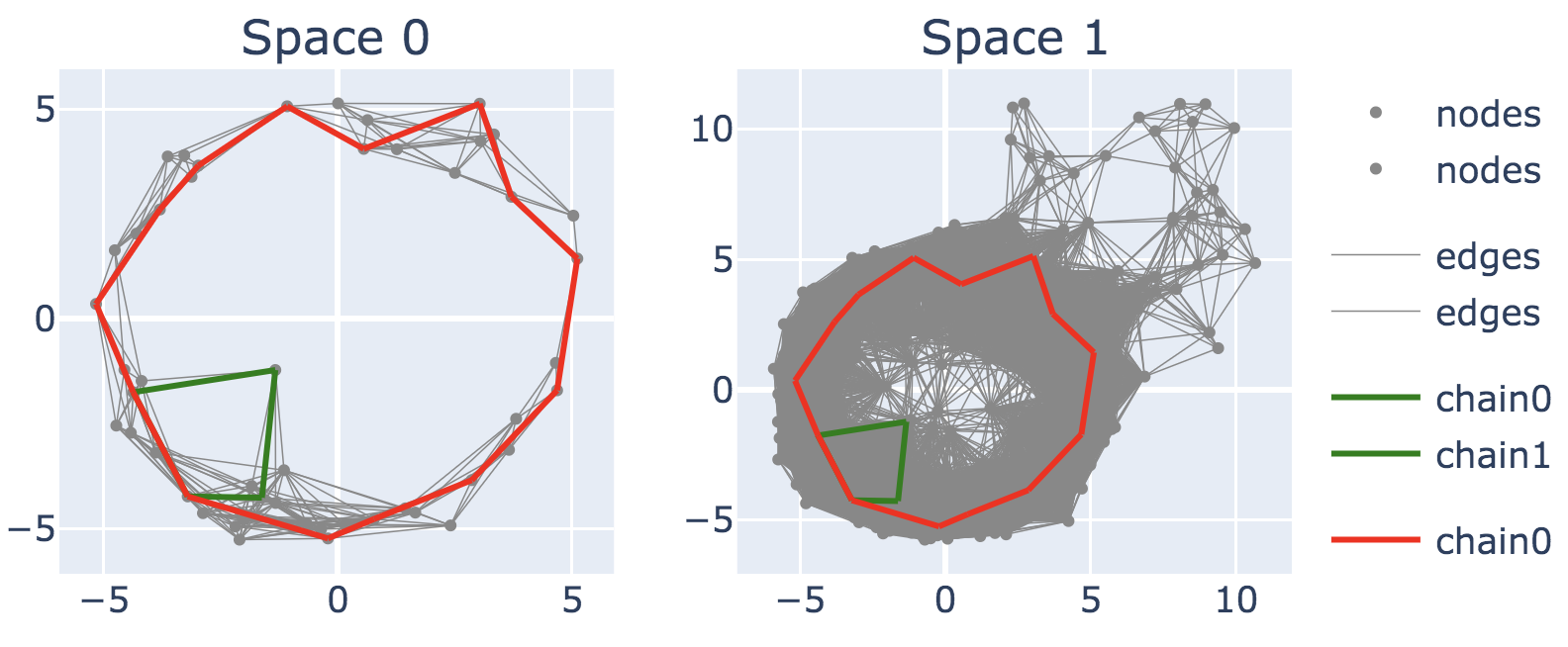}
    \includegraphics[width=0.48\linewidth]{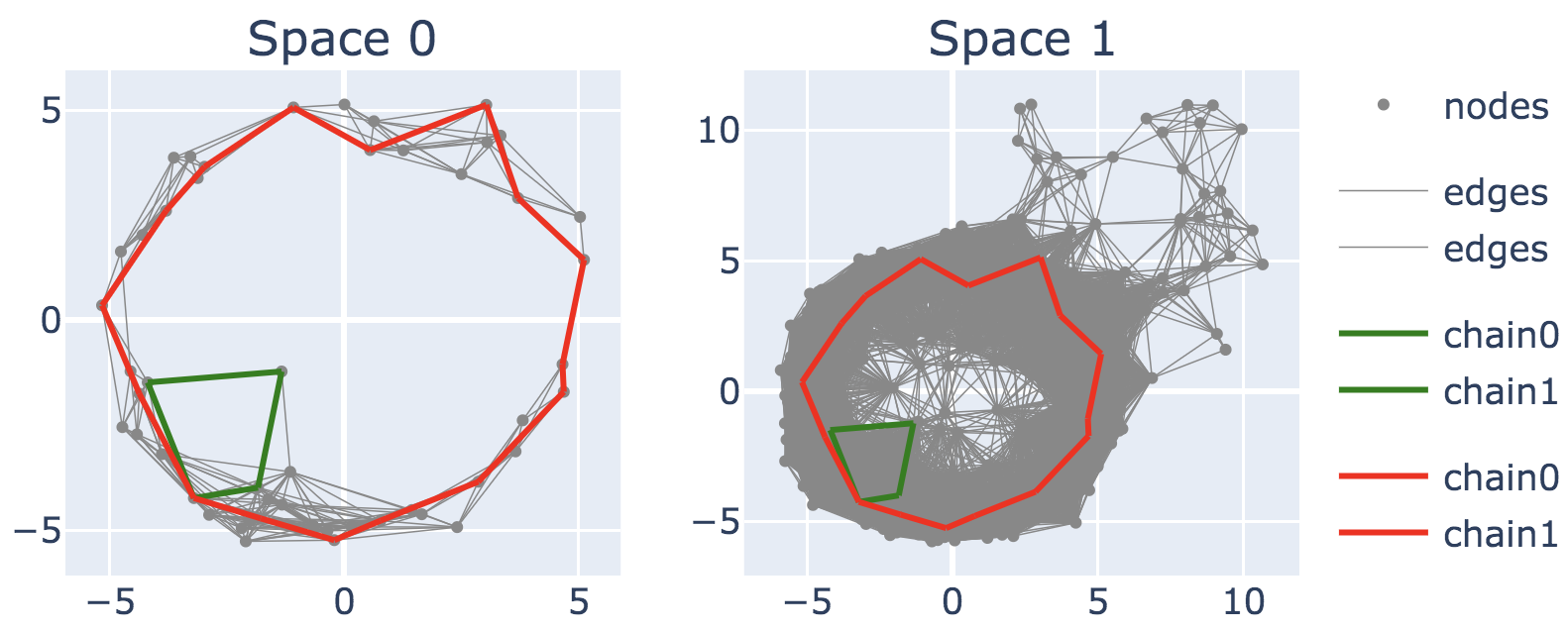}
    \includegraphics[width=0.48\linewidth]{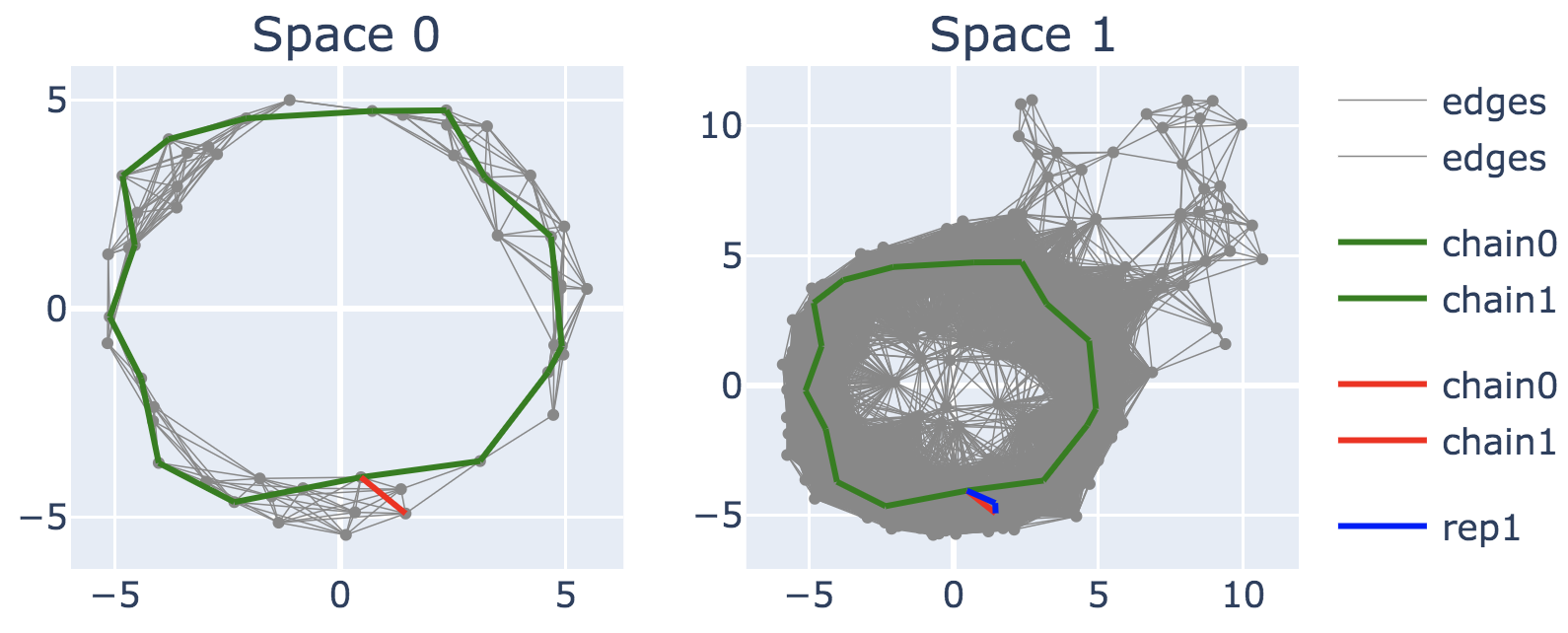}
    \caption{Induced maps of noisy Annulus. Full point cloud size is $1000$. Sample size is $50$.}
    \label{fig:annulus_50_im}
\end{figure}

\begin{figure}
\centering
    \includegraphics[width=0.48\linewidth]{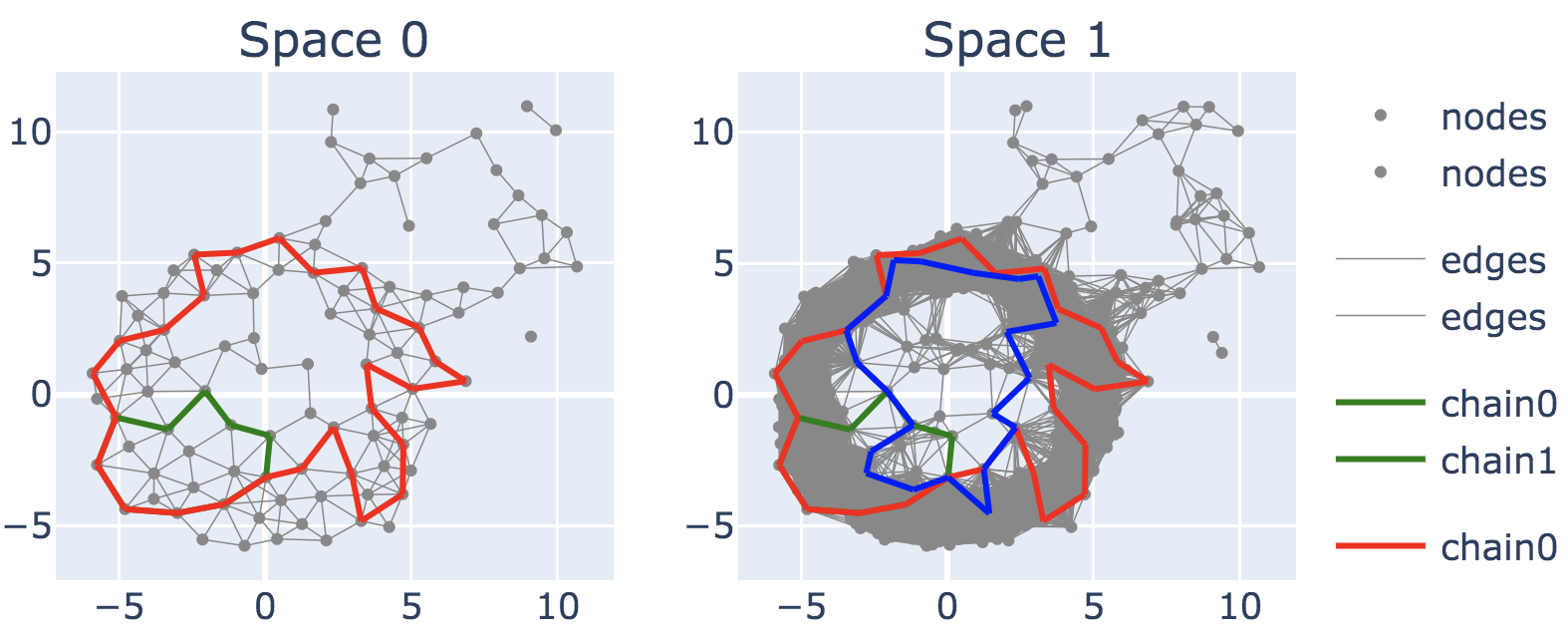}
    \includegraphics[width=0.48\linewidth]{figs/annulus/100hd1.png}
    \includegraphics[width=0.48\linewidth]{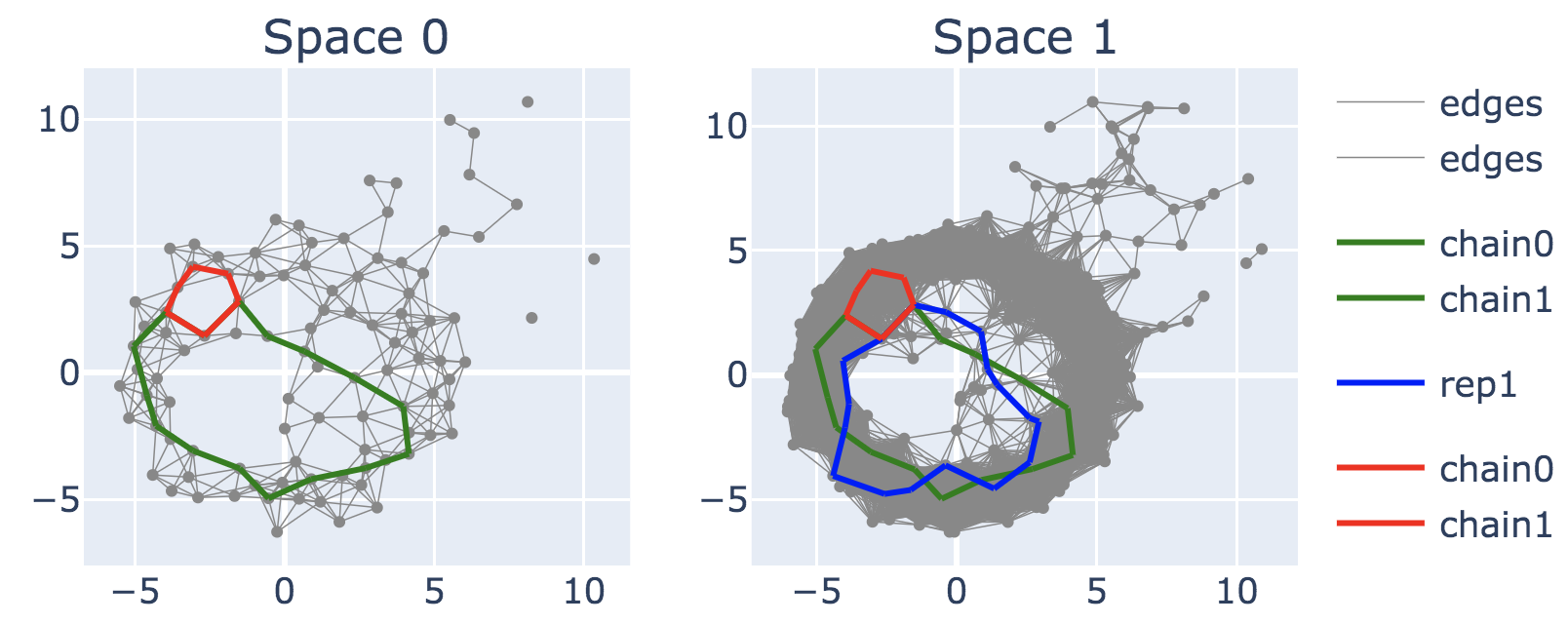}
    \includegraphics[width=0.48\linewidth]{figs/annulus/100hd2.png}
    \includegraphics[width=0.48\linewidth]{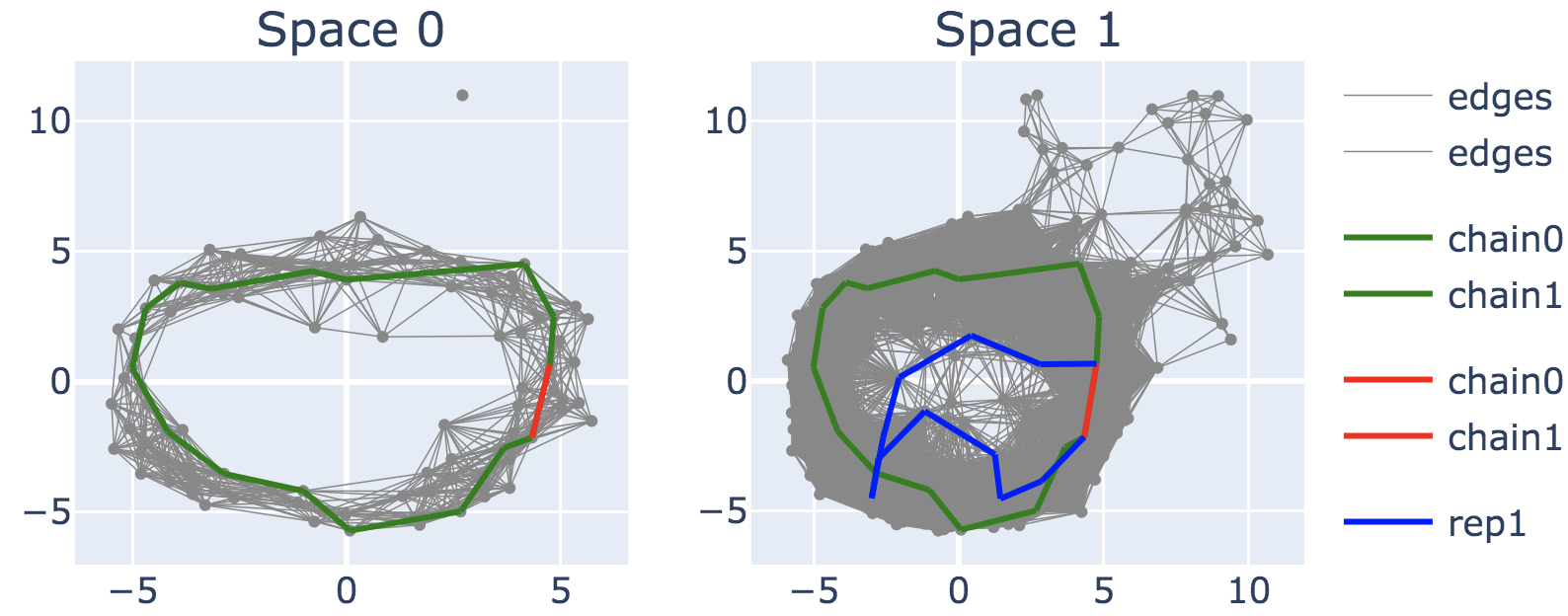}
    \includegraphics[width=0.48\linewidth]{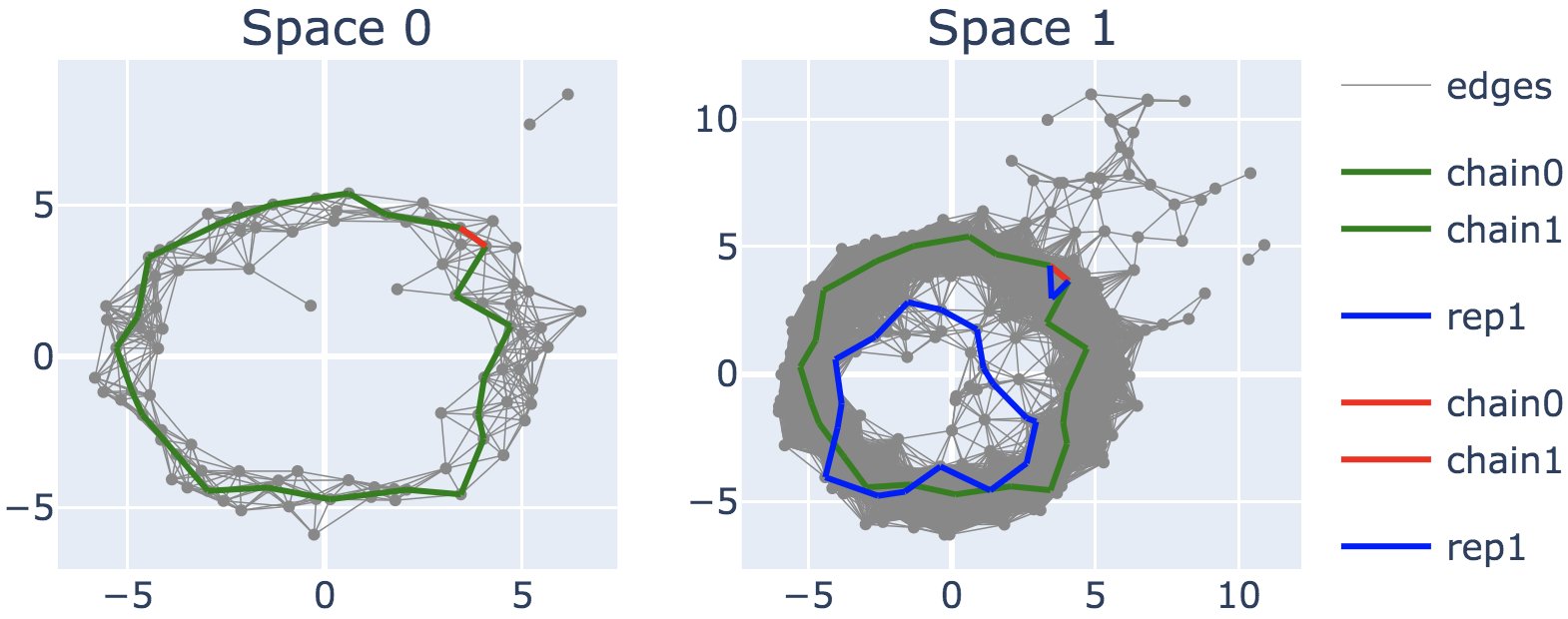}
    \includegraphics[width=0.48\linewidth]{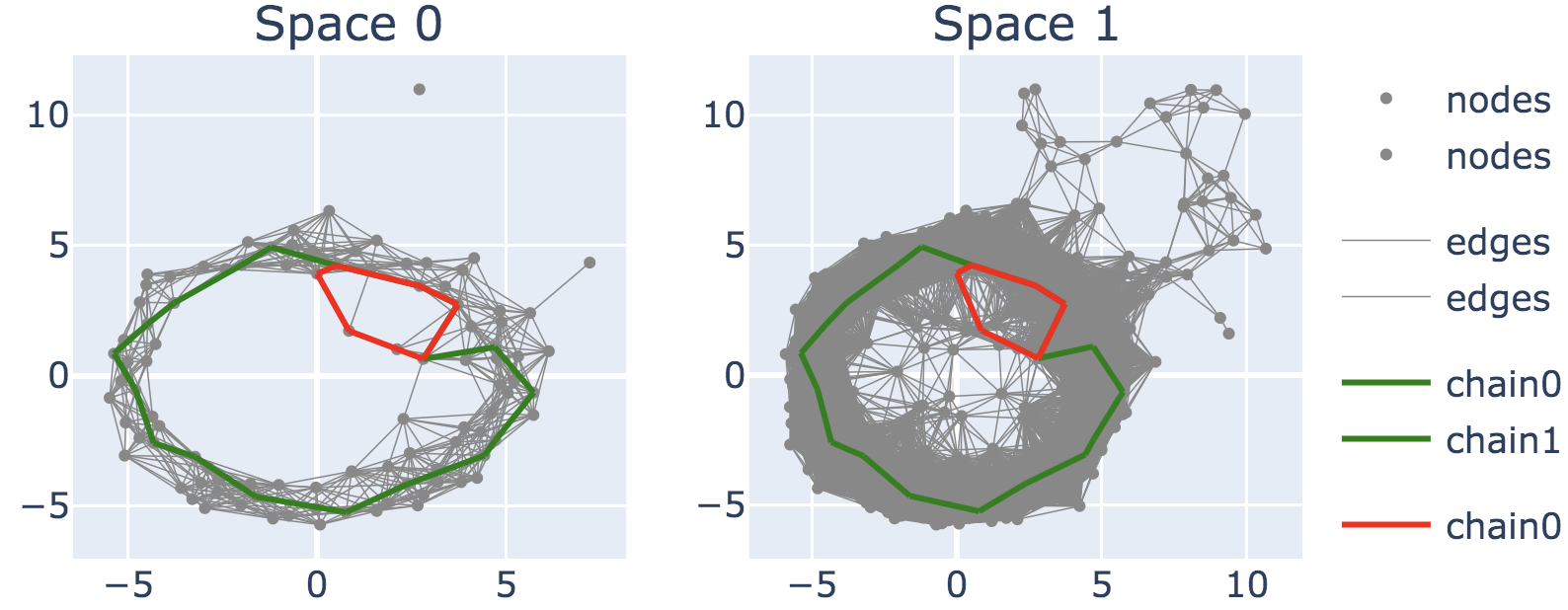}
    \includegraphics[width=0.48\linewidth]{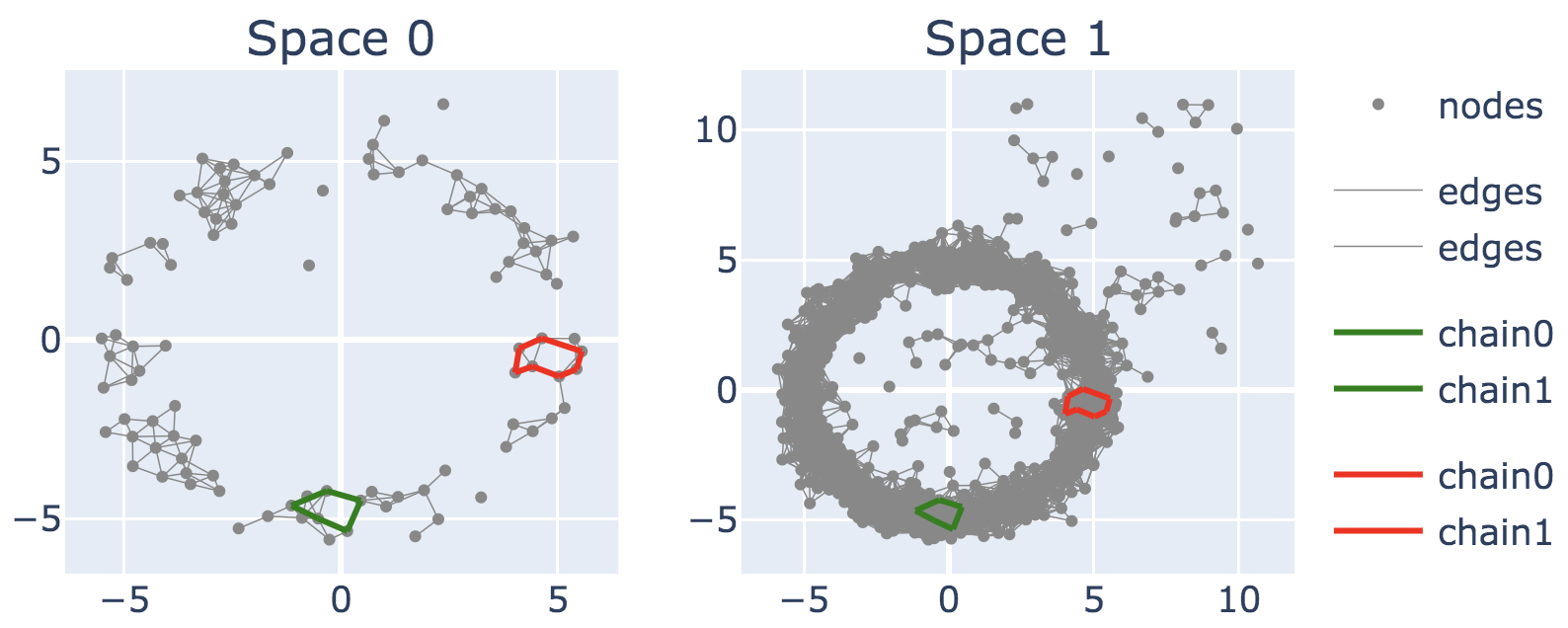}
    \includegraphics[width=0.48\linewidth]{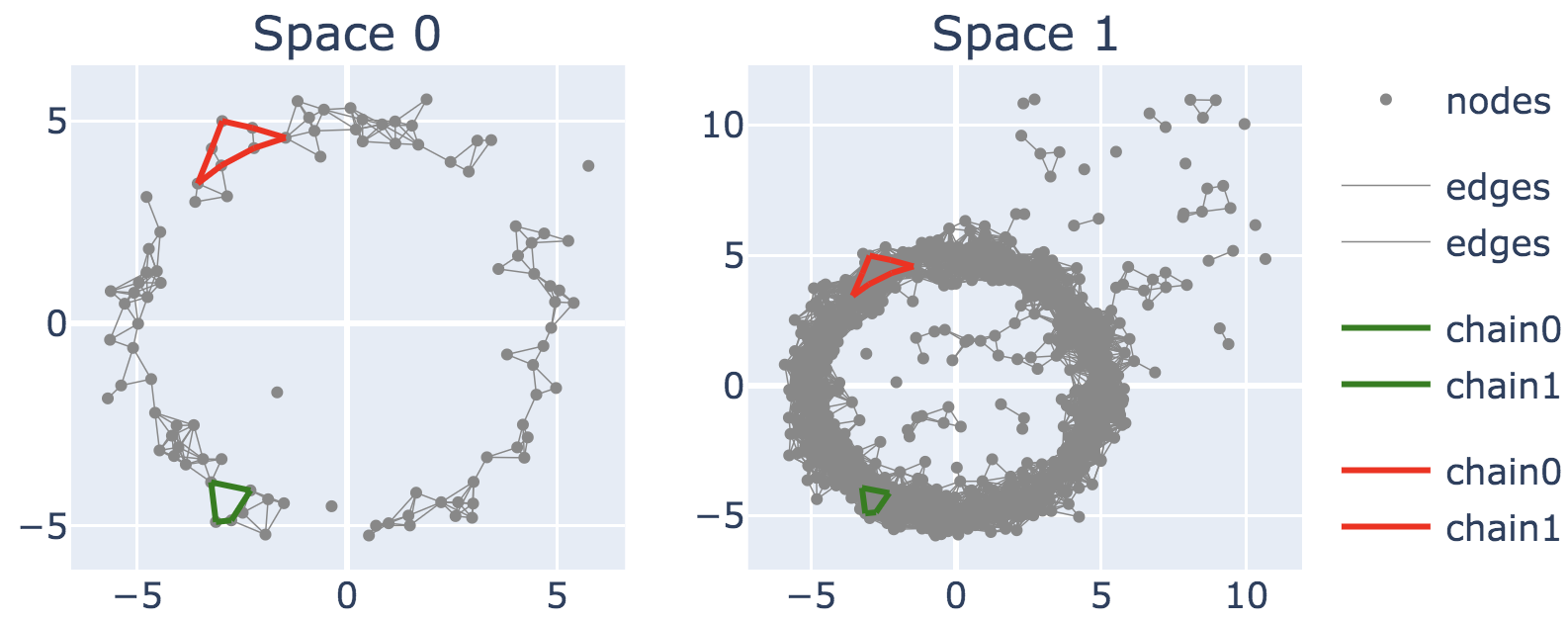}
    \includegraphics[width=0.48\linewidth]{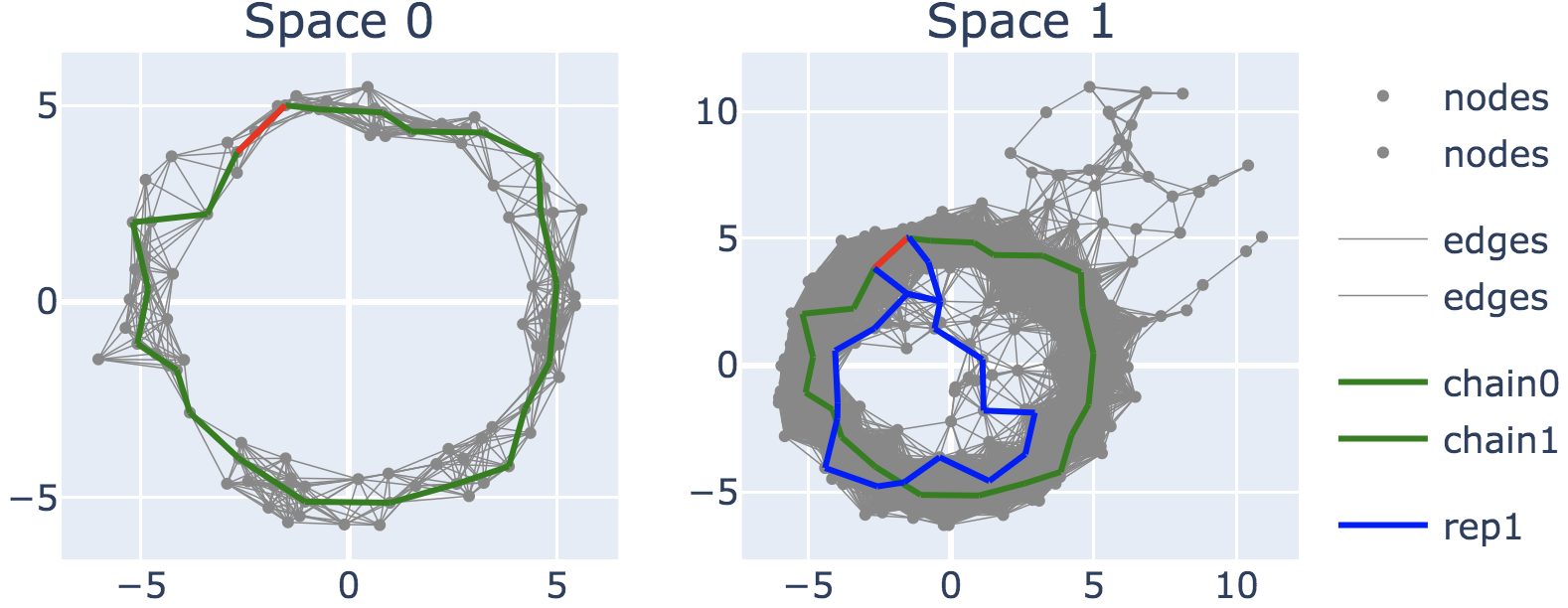}
    \caption{Induced maps of noisy Annulus. Full point cloud size is $1000$. Sample size is $100$.}
    \label{fig:annulus_100_im}
\end{figure}

\begin{figure}
\centering
    \includegraphics[width=0.48\linewidth]{figs/annulus/300hd1.png}
    \includegraphics[width=0.48\linewidth]{figs/annulus/300hd2.png}
    \includegraphics[width=0.48\linewidth]{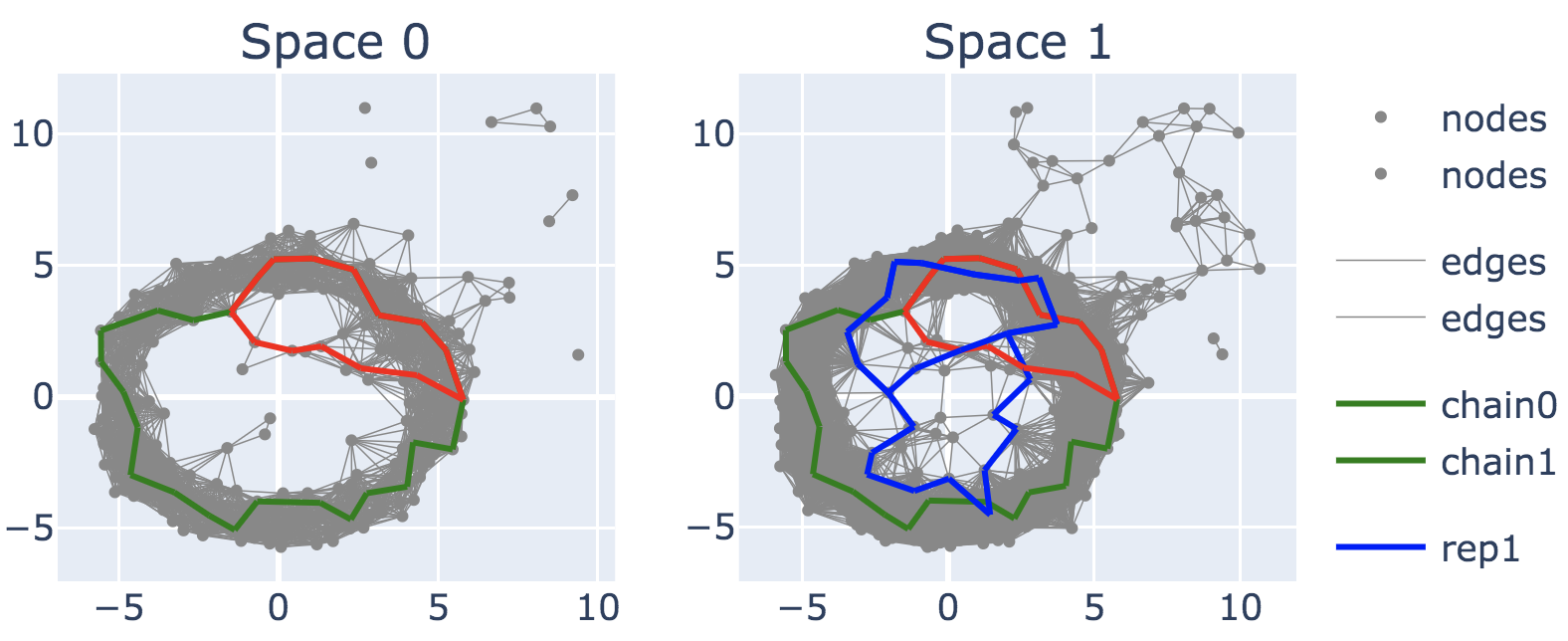}
    \includegraphics[width=0.48\linewidth]{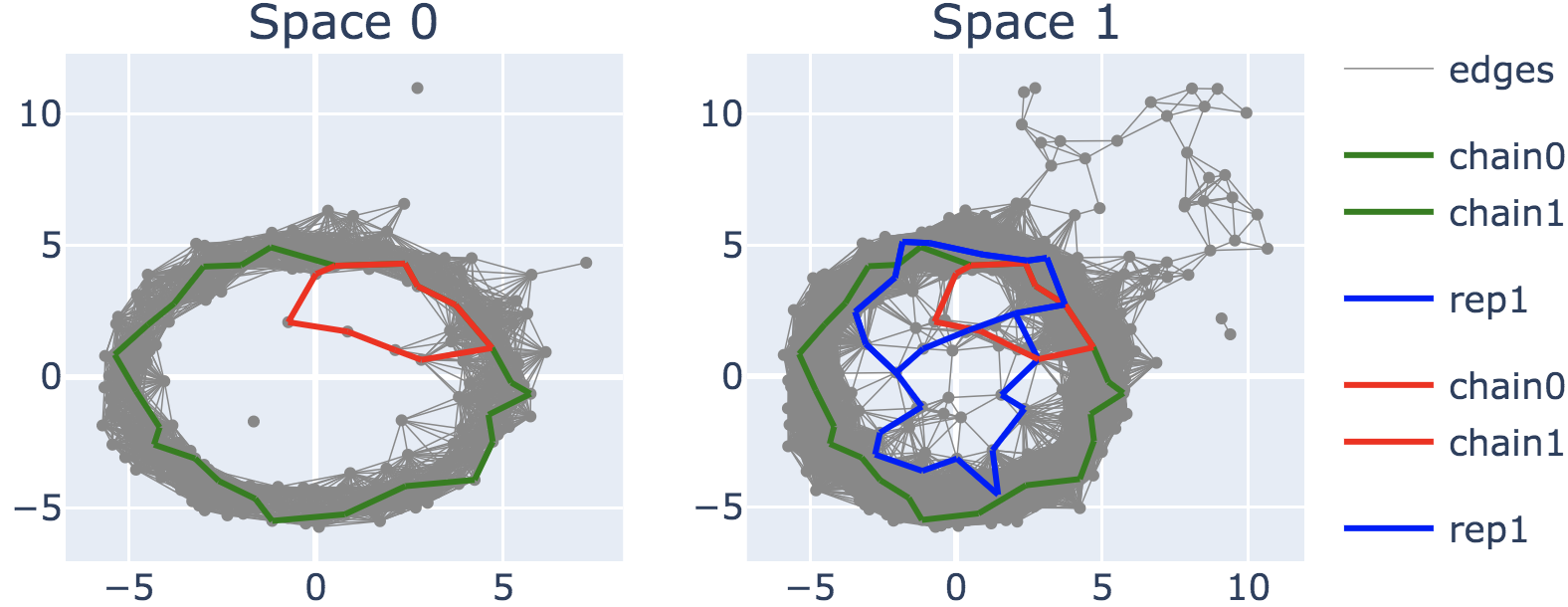}
    \includegraphics[width=0.48\linewidth]{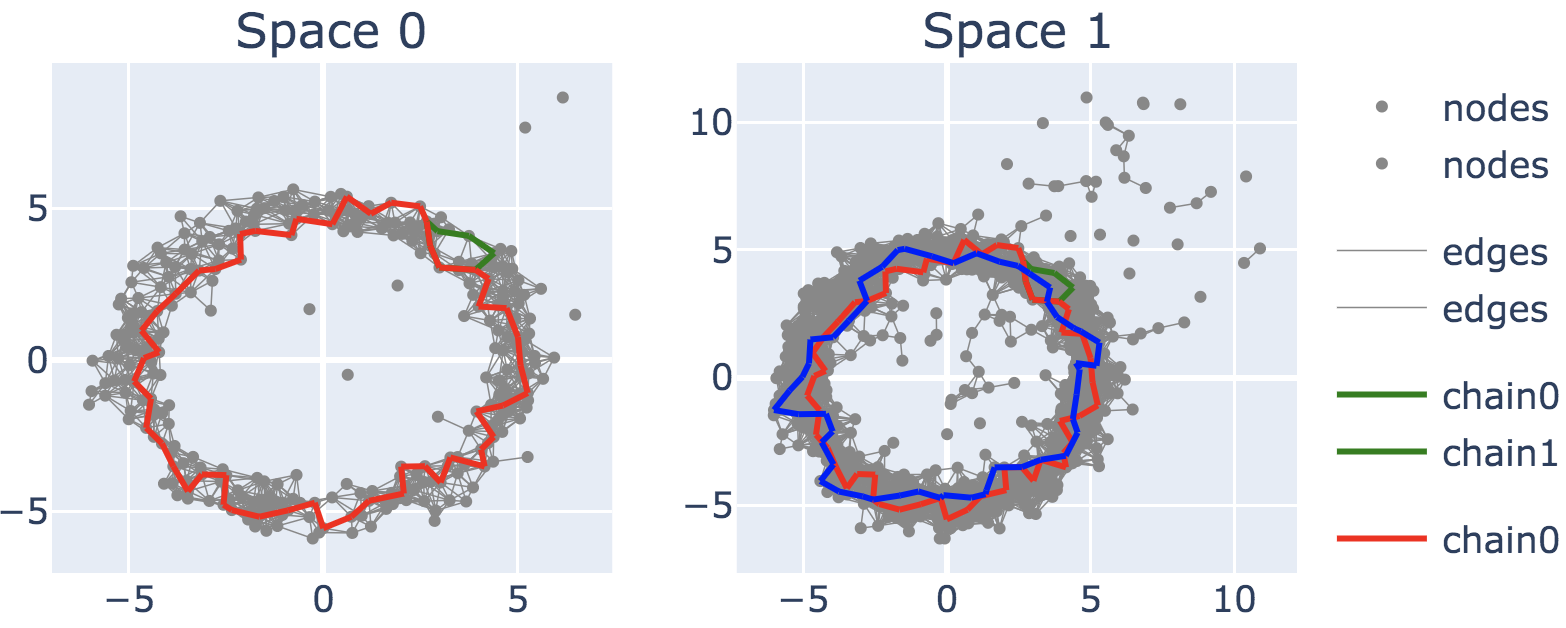}
    \includegraphics[width=0.48\linewidth]{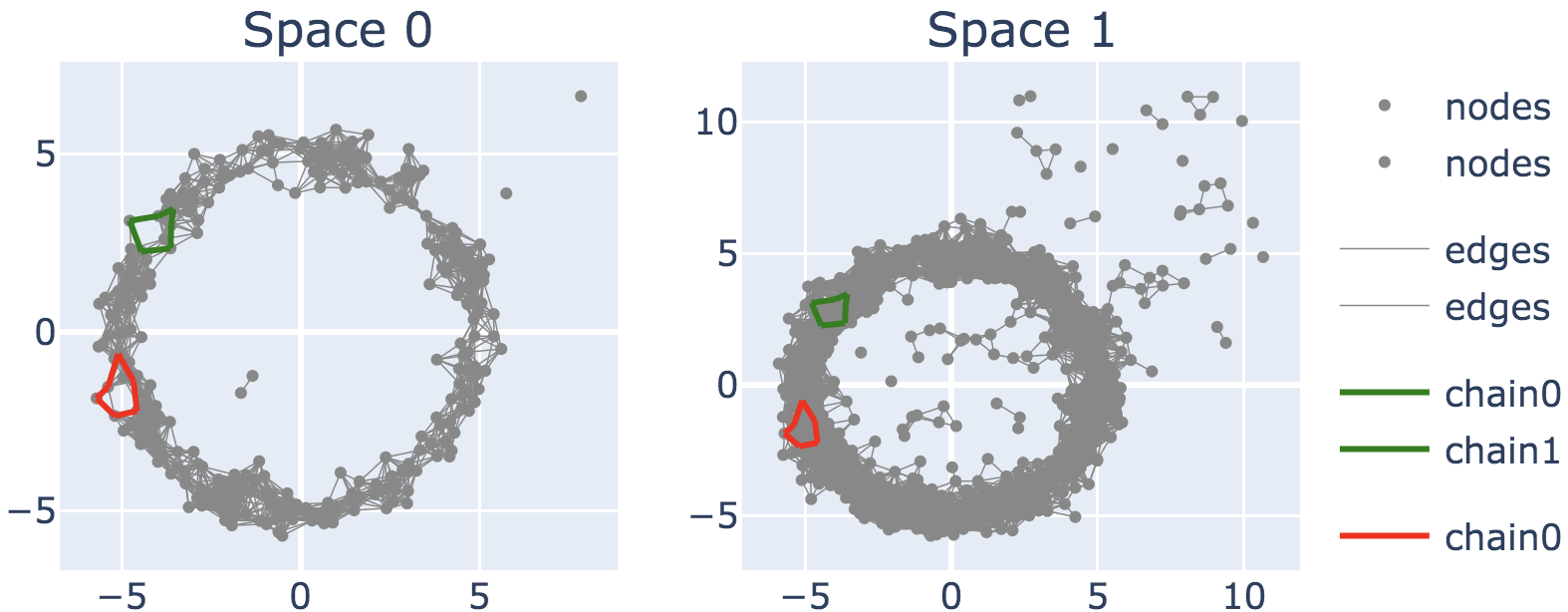}
    \includegraphics[width=0.48\linewidth]{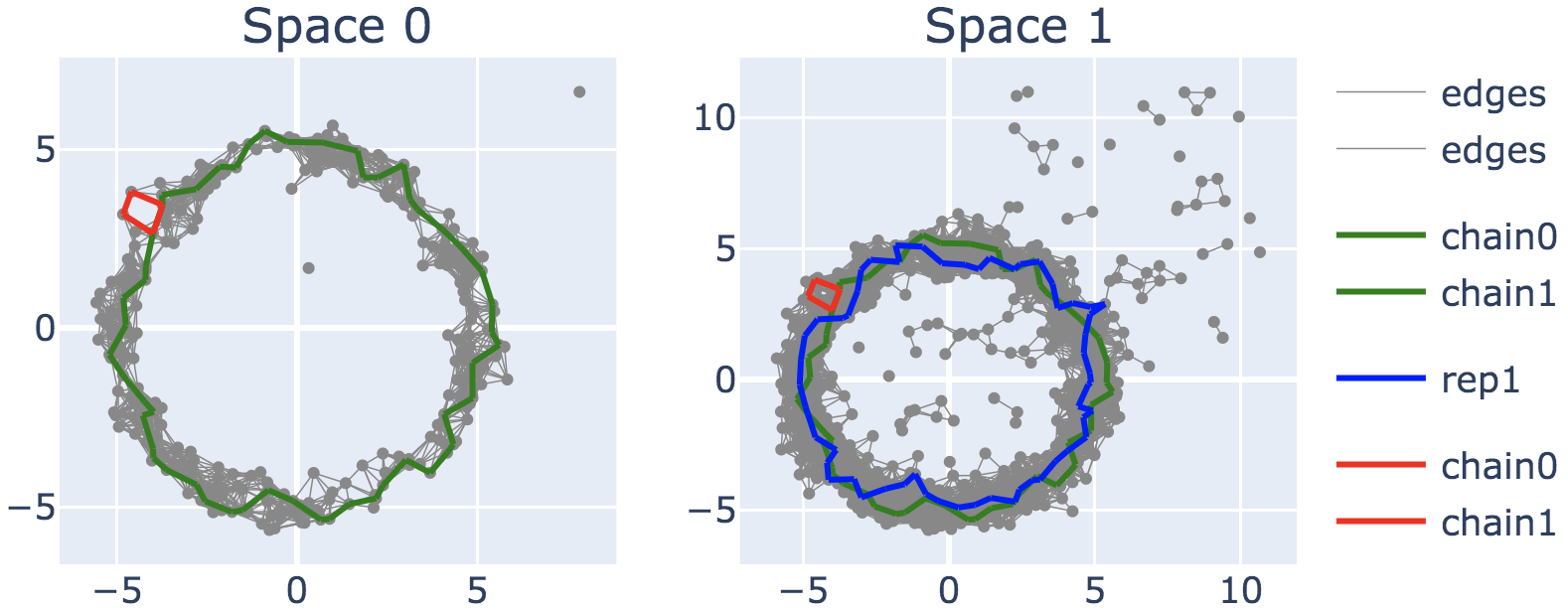}
    \includegraphics[width=0.48\linewidth]{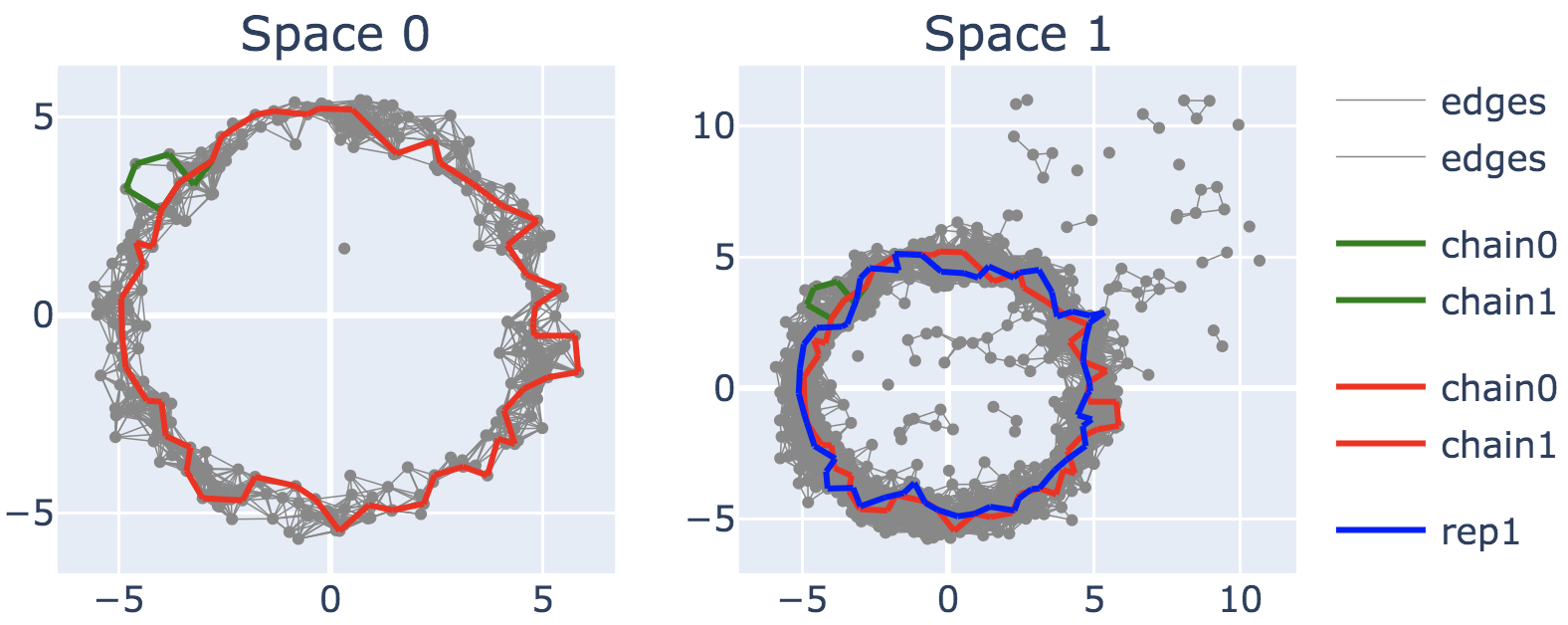}
    \includegraphics[width=0.48\linewidth]{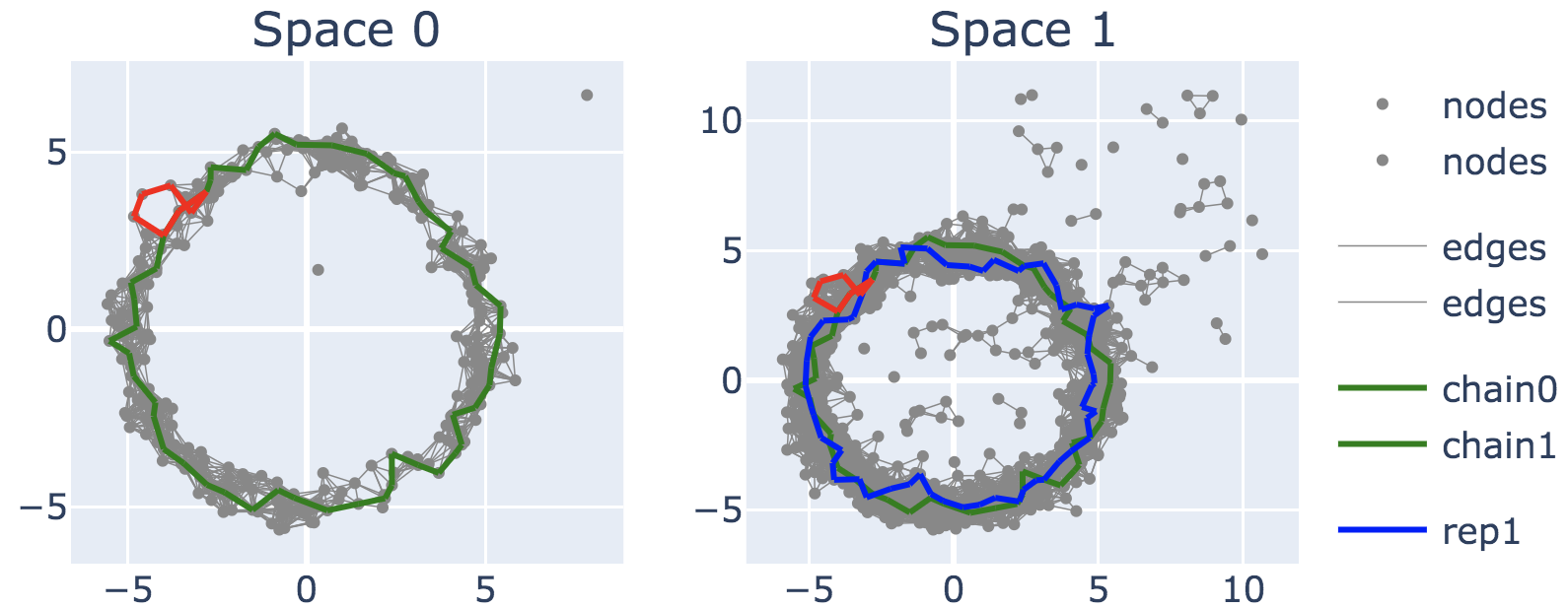}
    \includegraphics[width=0.48\linewidth]{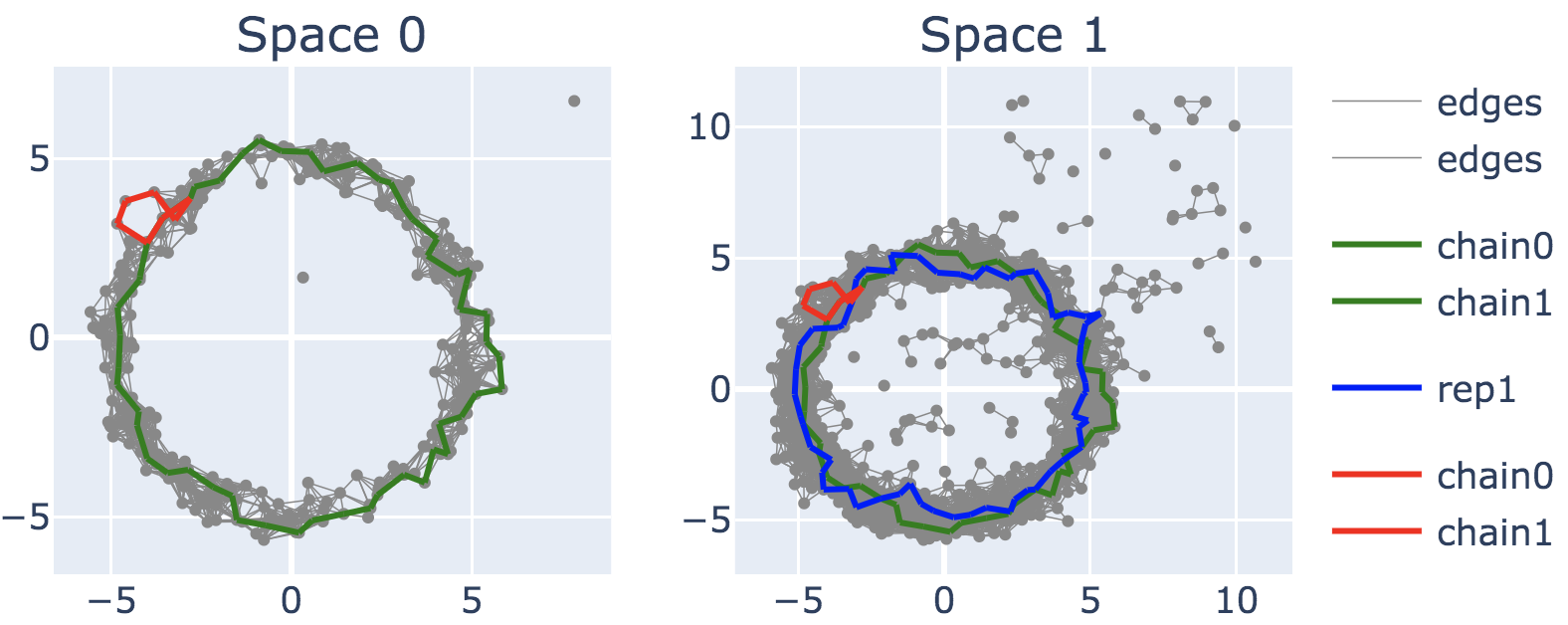}
    \caption{Induced maps of noisy Annulus. Full point cloud size is $1000$. Sample size is $300$.}
    \label{fig:annulus_300_im}
\end{figure}

\begin{figure}
\centering
    \includegraphics[width=0.48\linewidth]{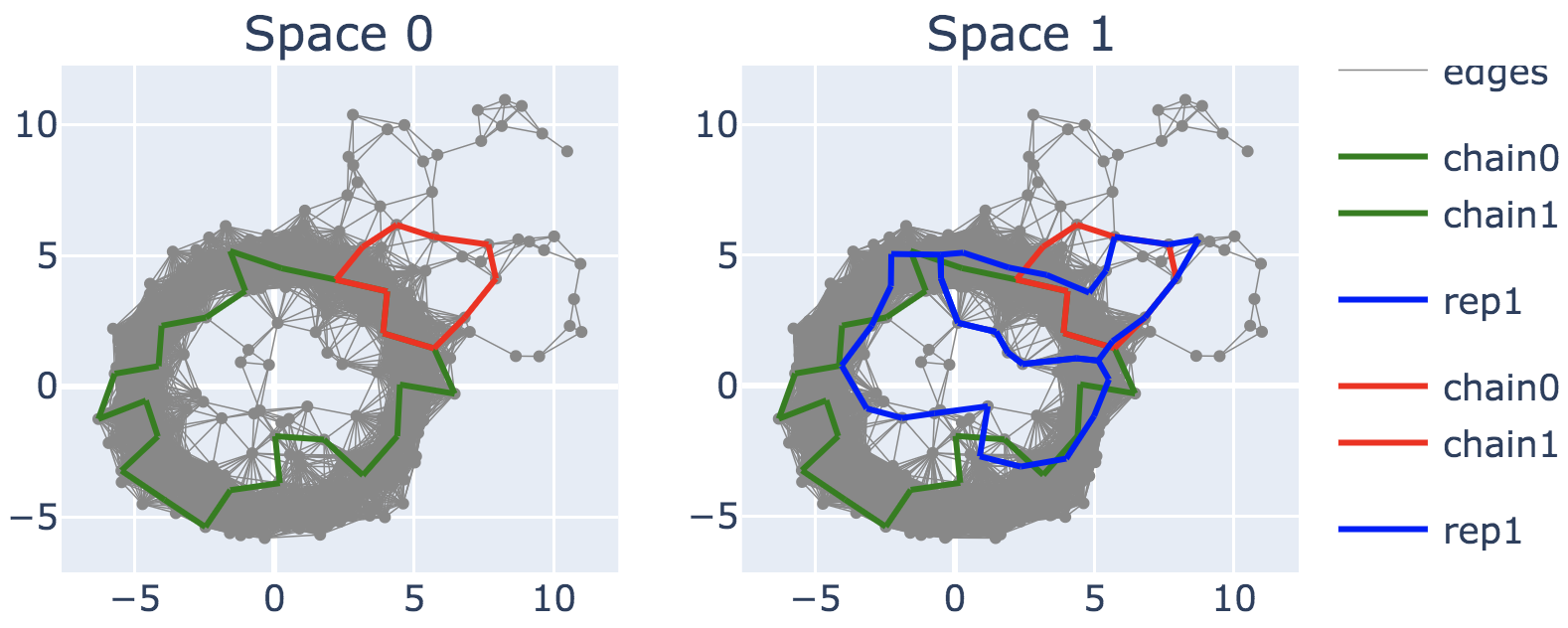}
    \includegraphics[width=0.48\linewidth]{figs/annulus/500hd2.png}
    \includegraphics[width=0.48\linewidth]{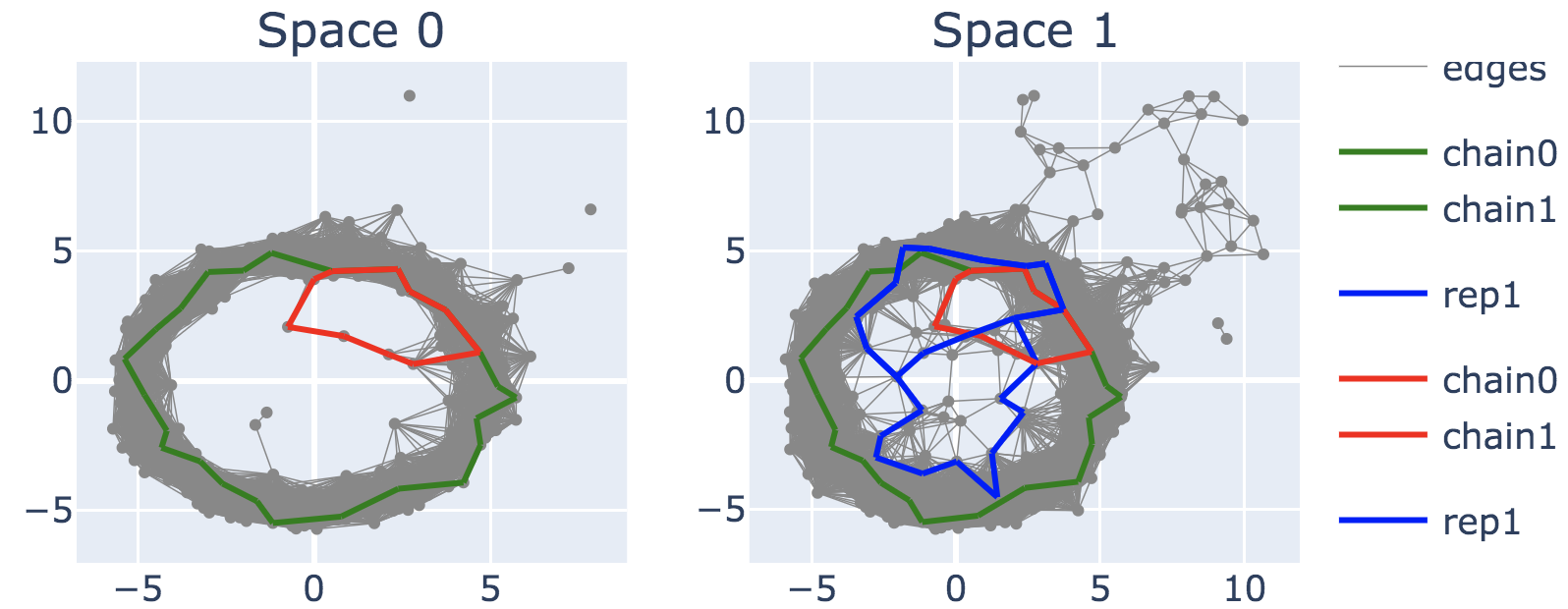}
    \includegraphics[width=0.48\linewidth]{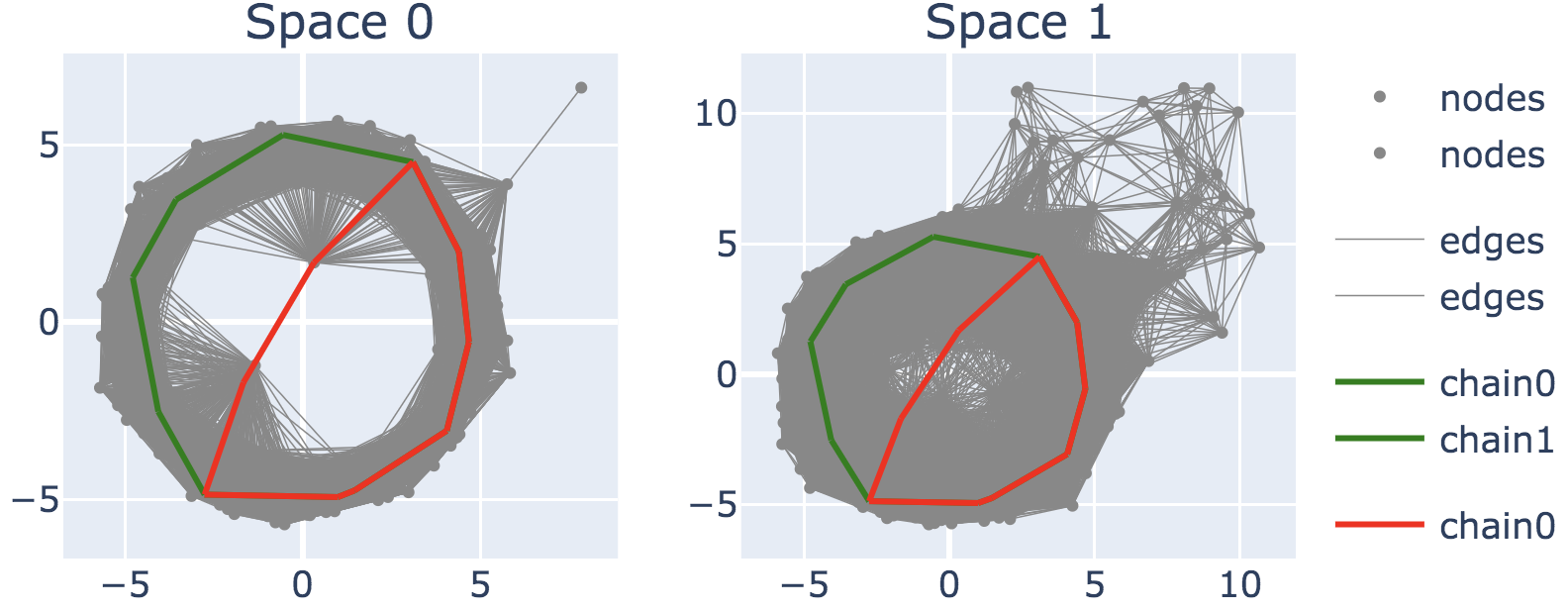}
    \includegraphics[width=0.48\linewidth]{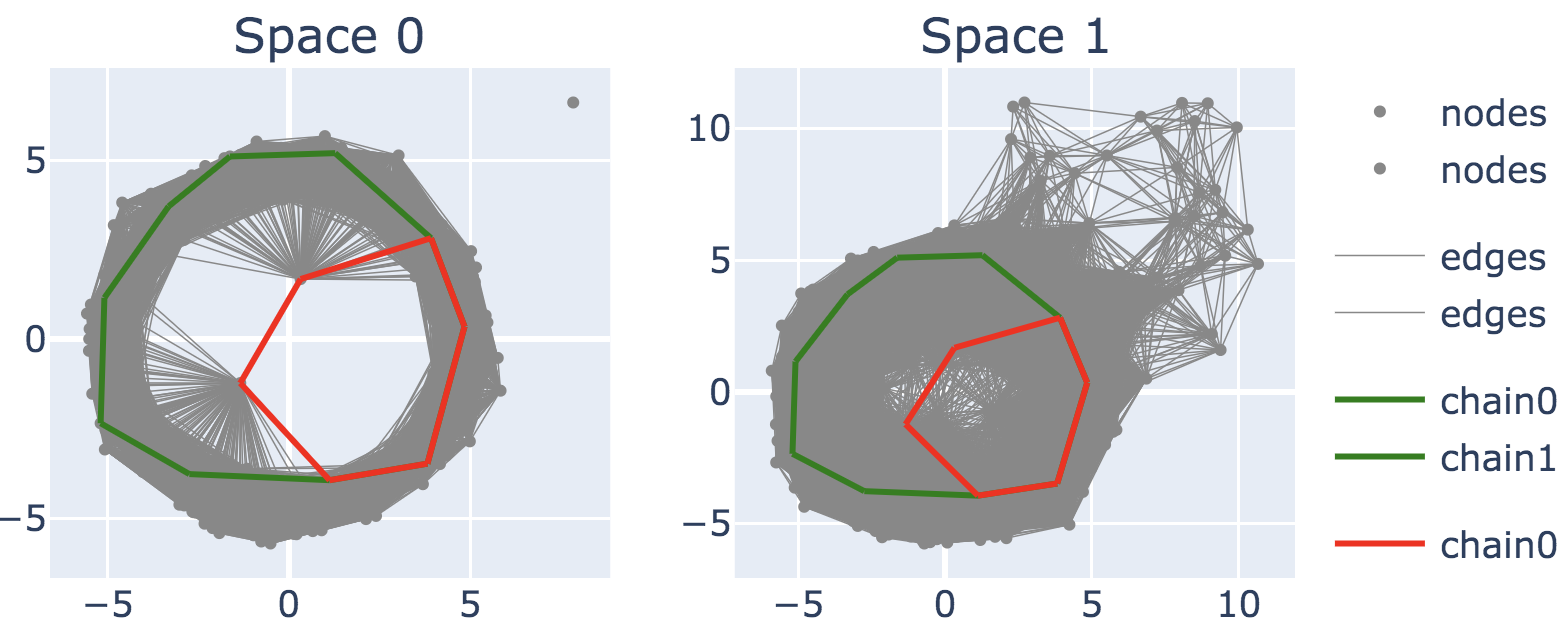}
    \includegraphics[width=0.48\linewidth]{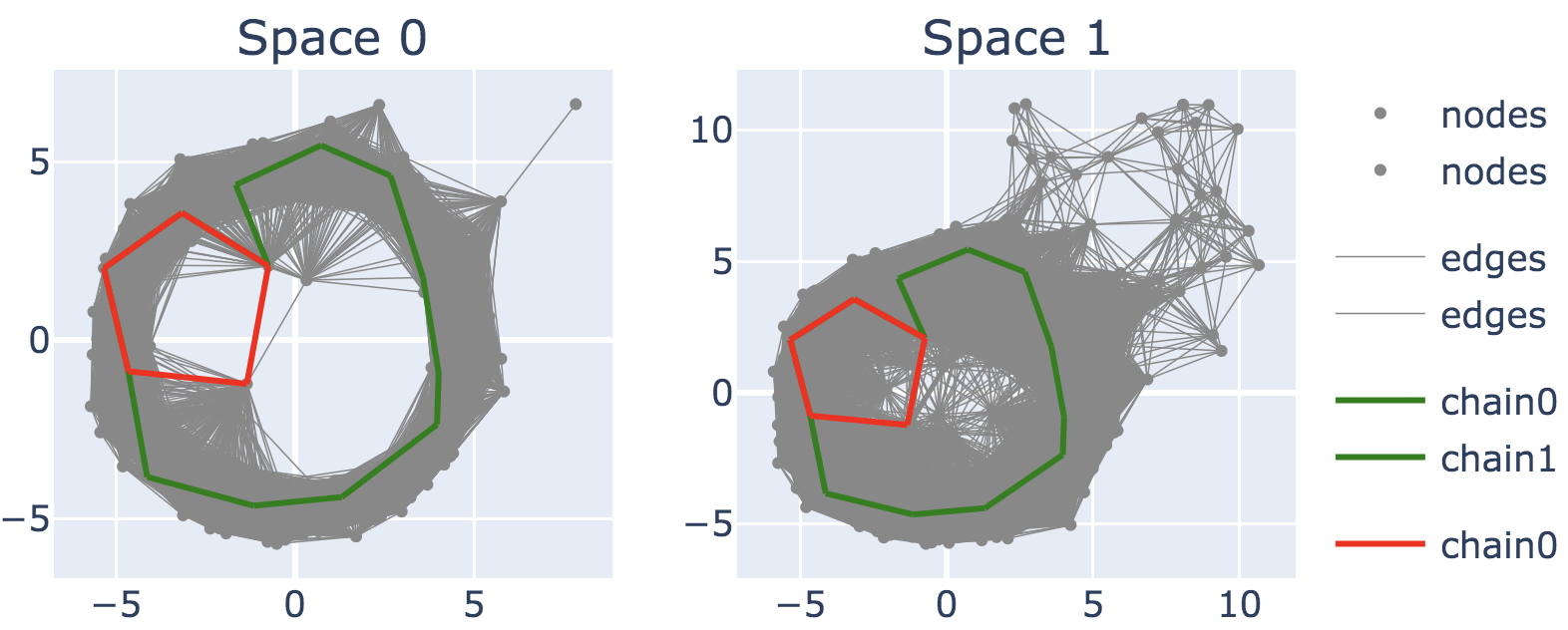}
    \includegraphics[width=0.48\linewidth]{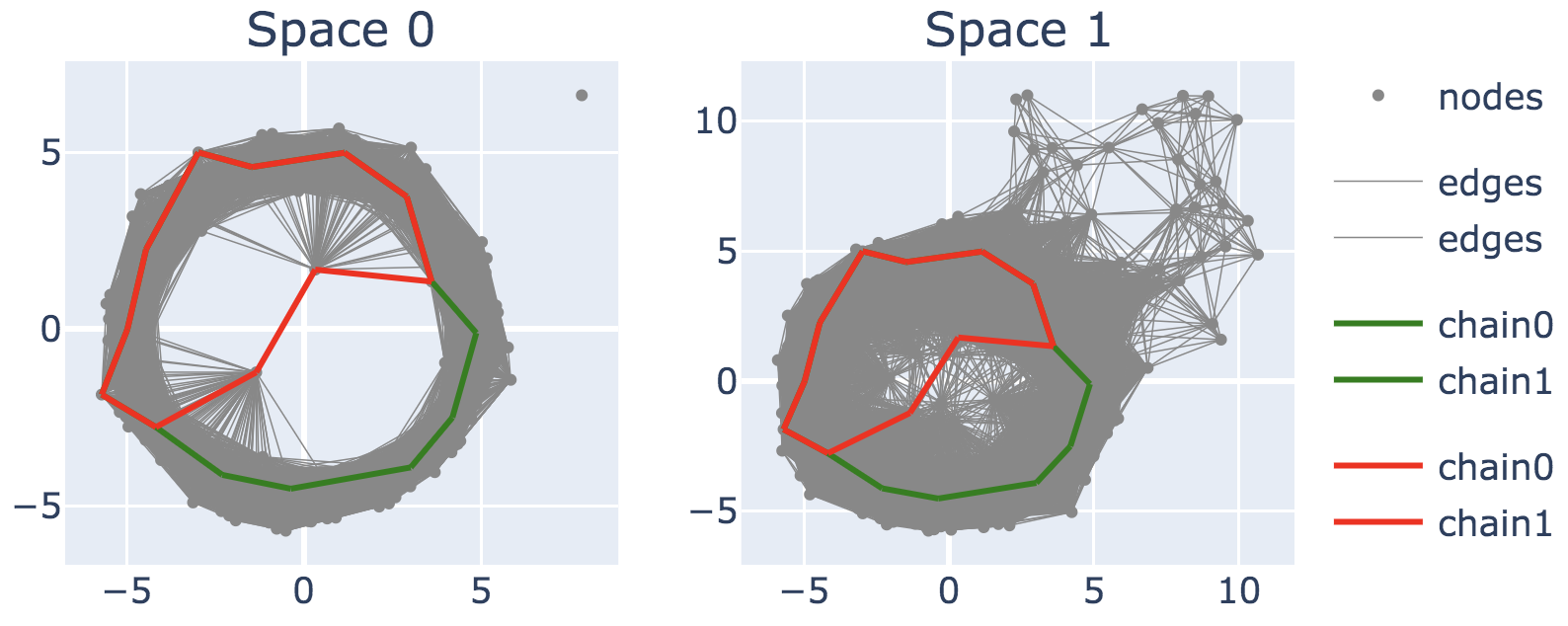}
    \includegraphics[width=0.48\linewidth]{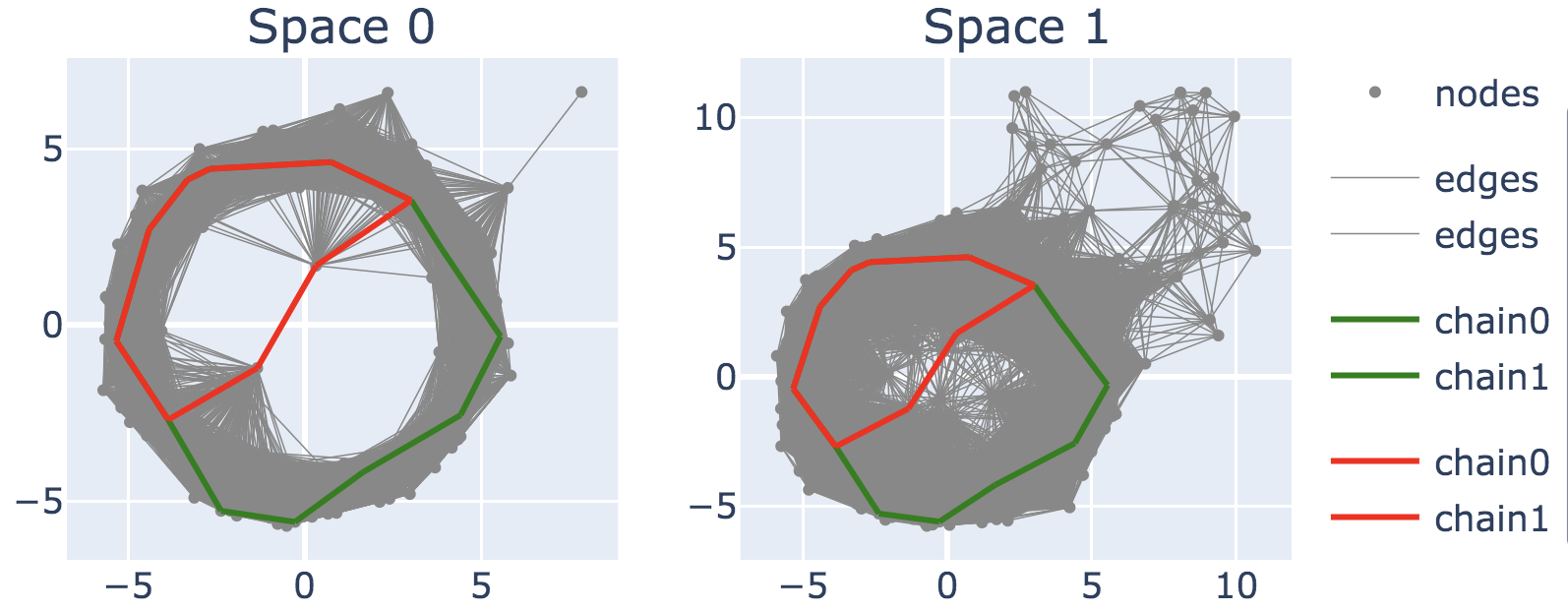}
    \includegraphics[width=0.48\linewidth]{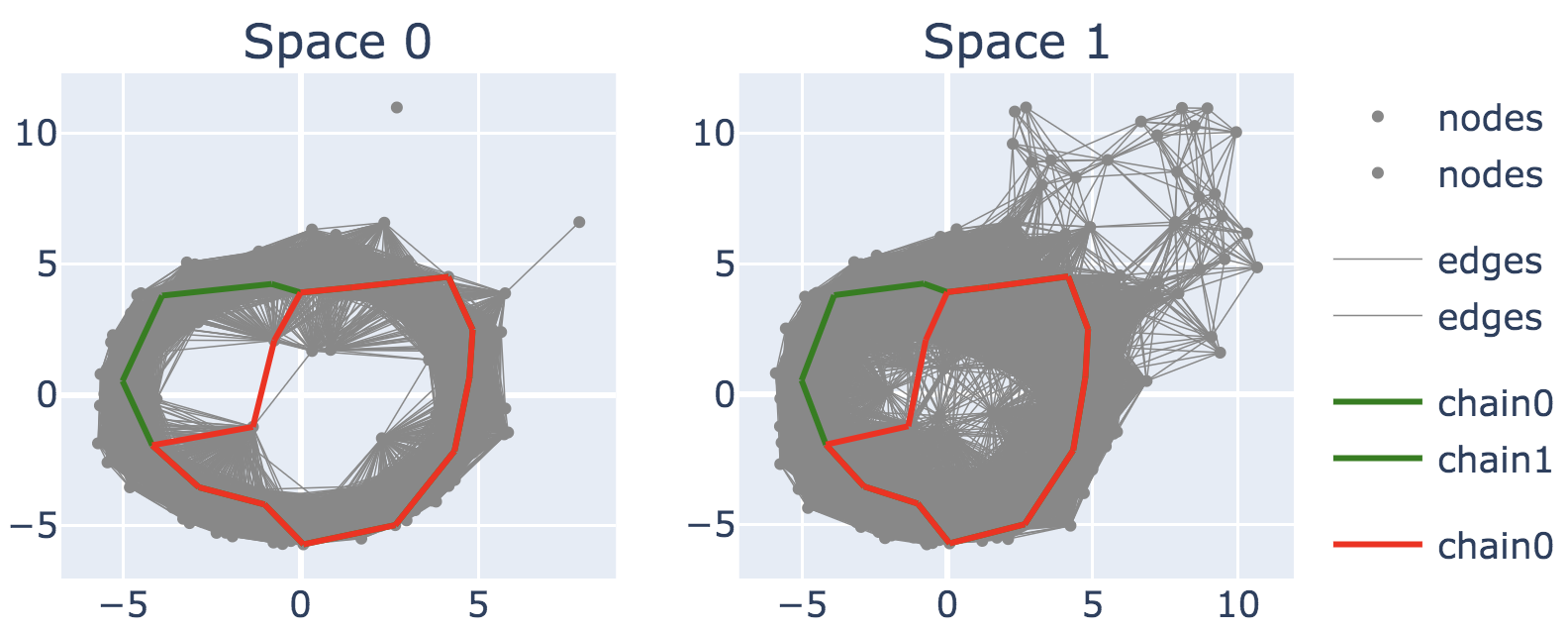}
    \includegraphics[width=0.48\linewidth]{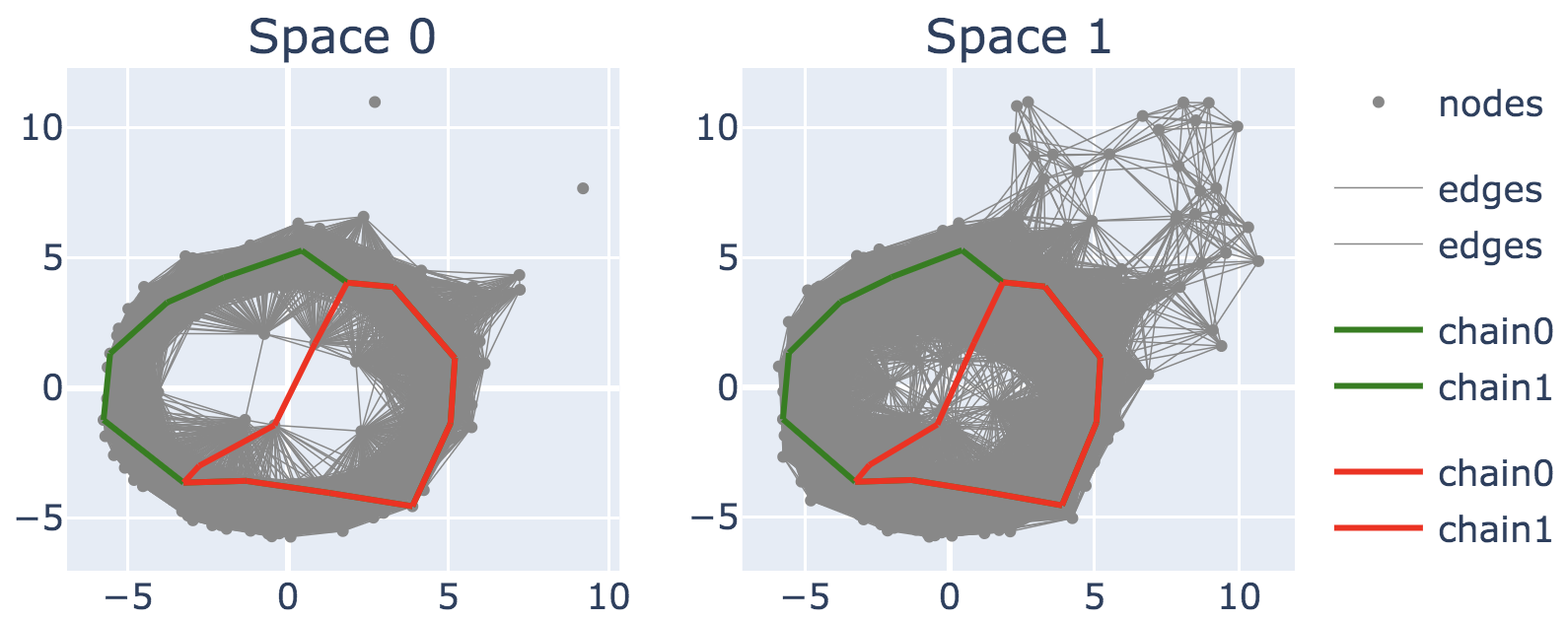}
    \caption{Induced maps of noisy Annulus. Full point cloud size is $1000$. Sample size is $500$.}
    \label{fig:annulus_500_im}
\end{figure}

\begin{figure}
\centering
    \includegraphics[width=0.48\linewidth]{figs/annulus/800hd1.png}
    \includegraphics[width=0.48\linewidth]{figs/annulus/800hd2.jpg}
    \includegraphics[width=0.48\linewidth]{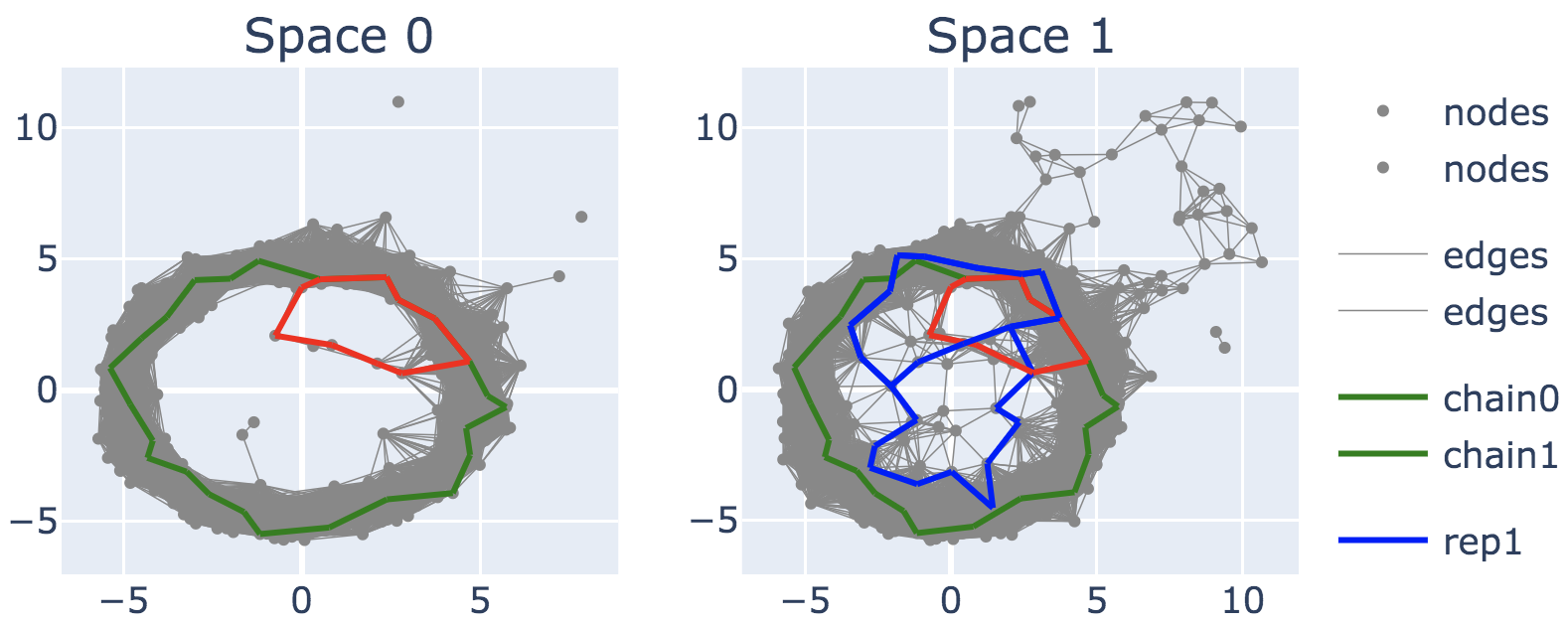}
    \includegraphics[width=0.48\linewidth]{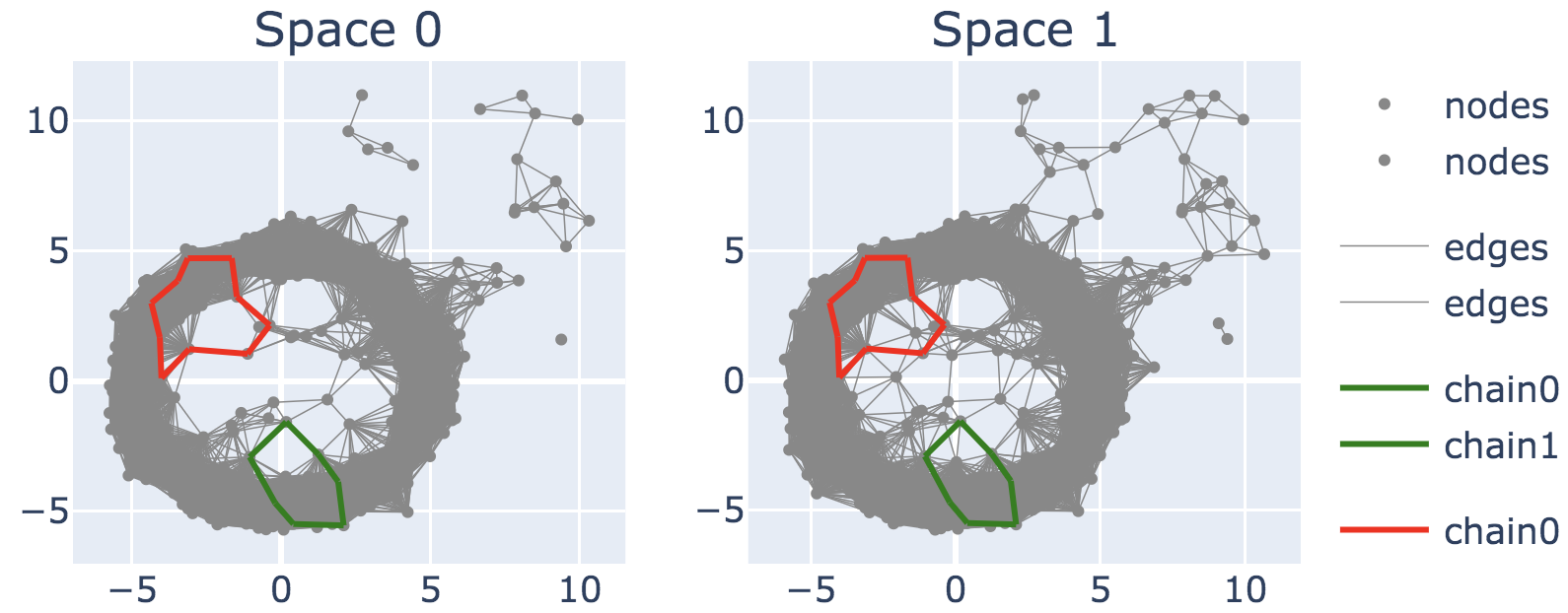}
    \includegraphics[width=0.48\linewidth]{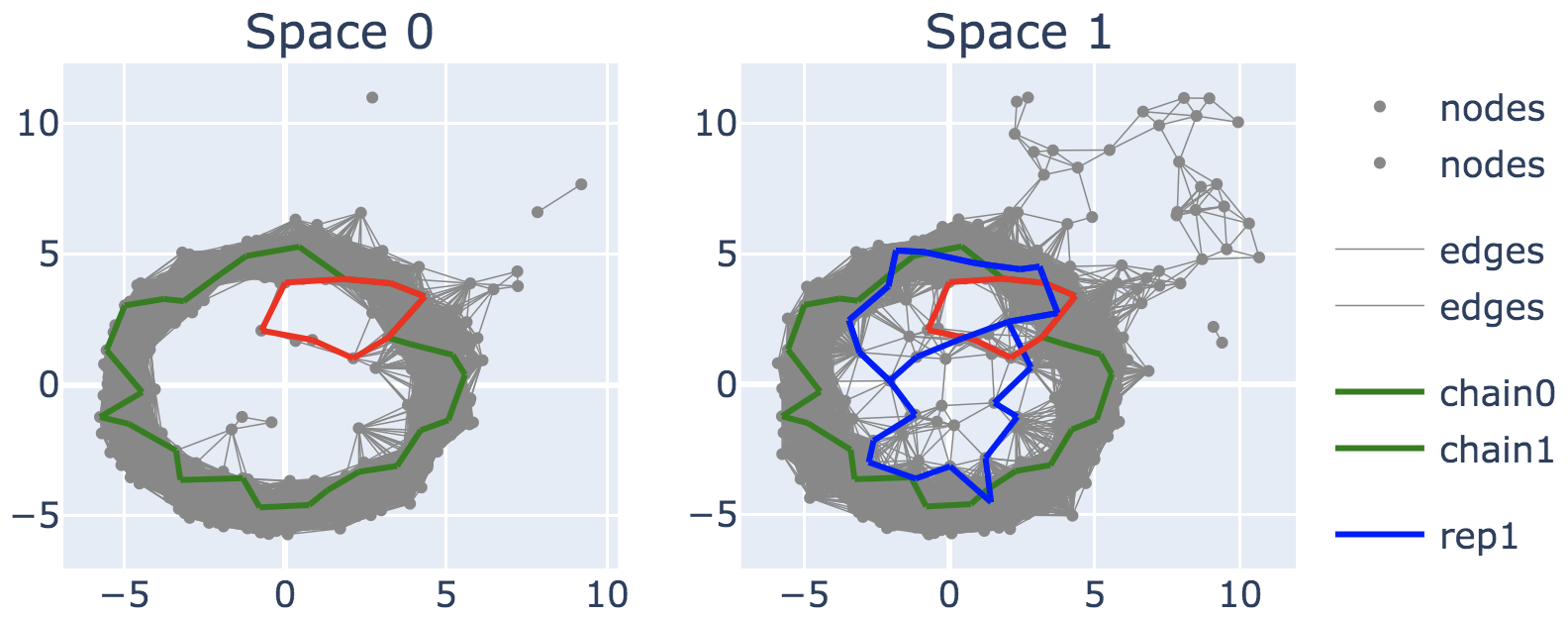}
    \includegraphics[width=0.48\linewidth]{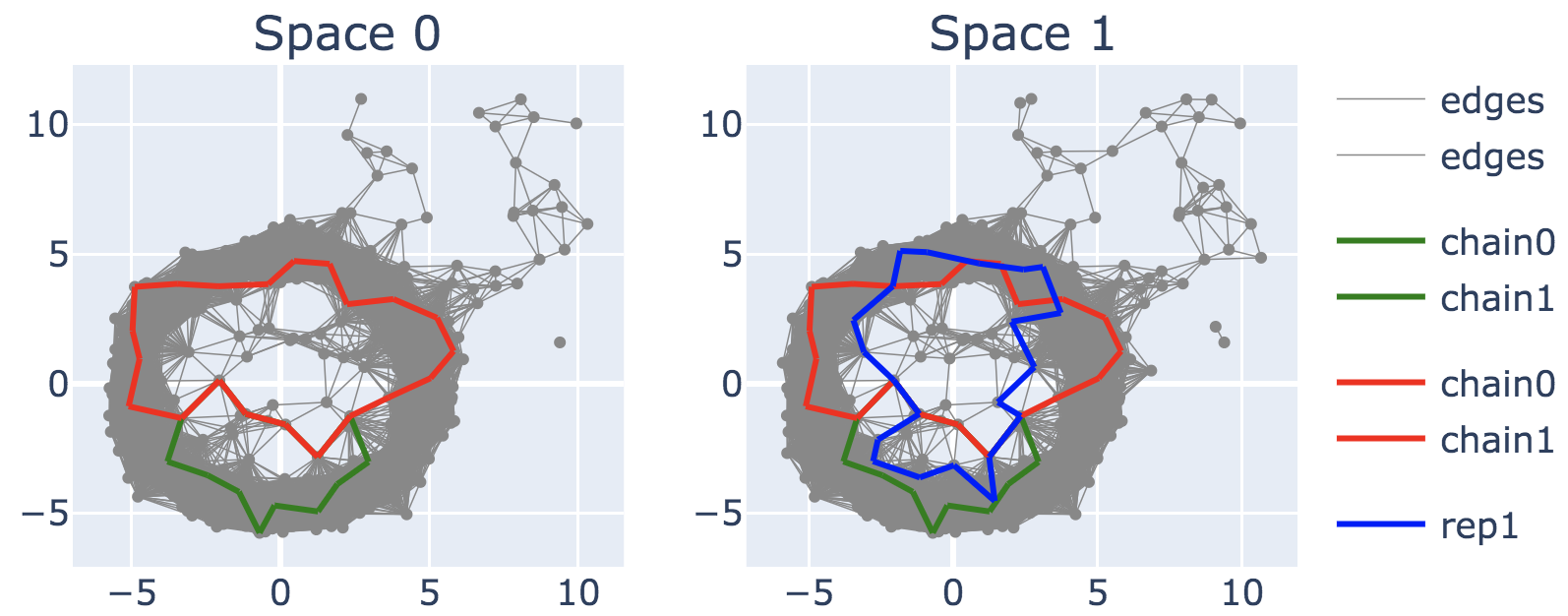}
    \includegraphics[width=0.48\linewidth]{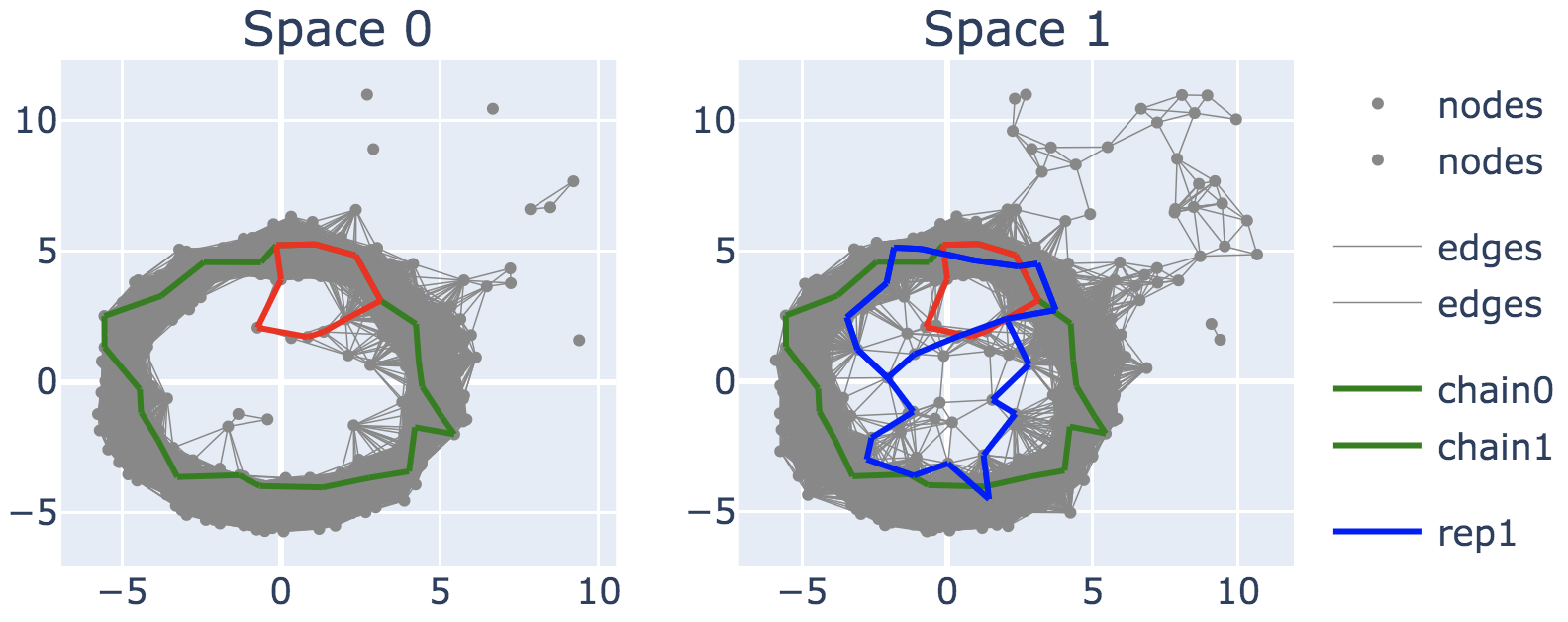}
    \includegraphics[width=0.48\linewidth]{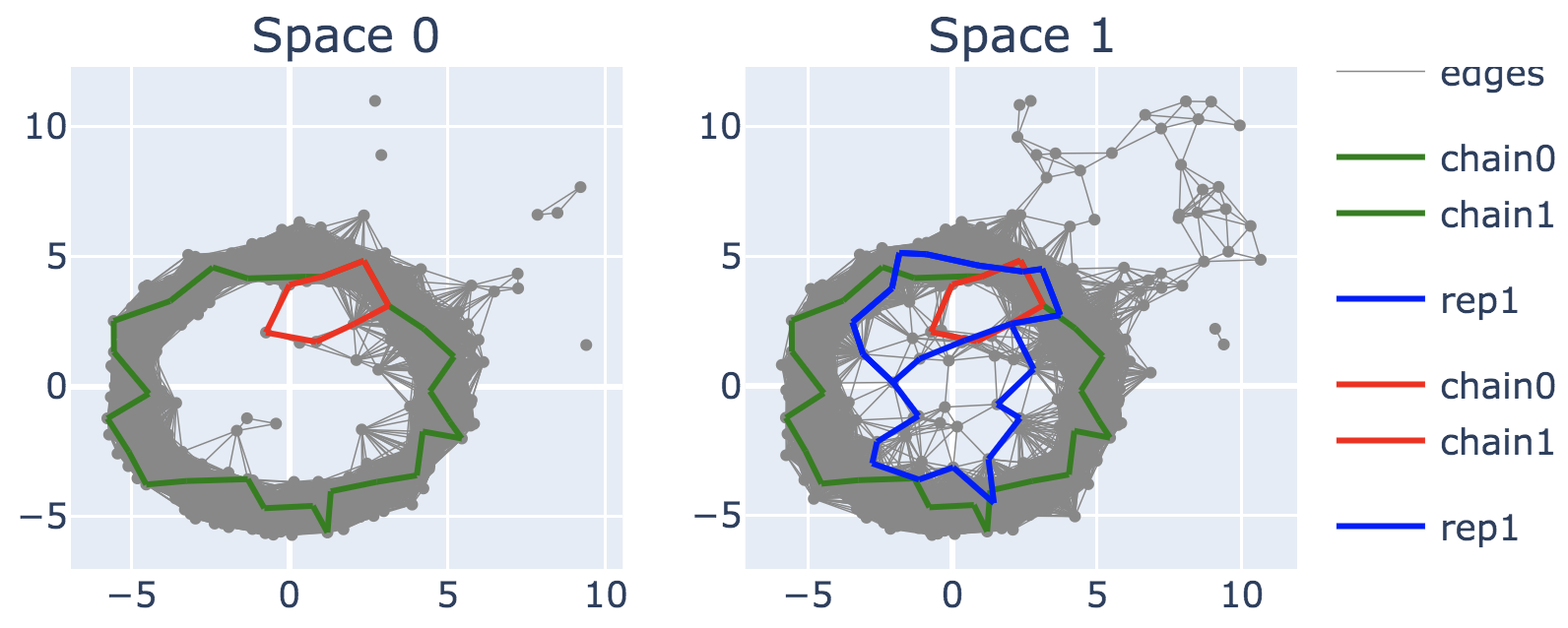}
    \includegraphics[width=0.48\linewidth]{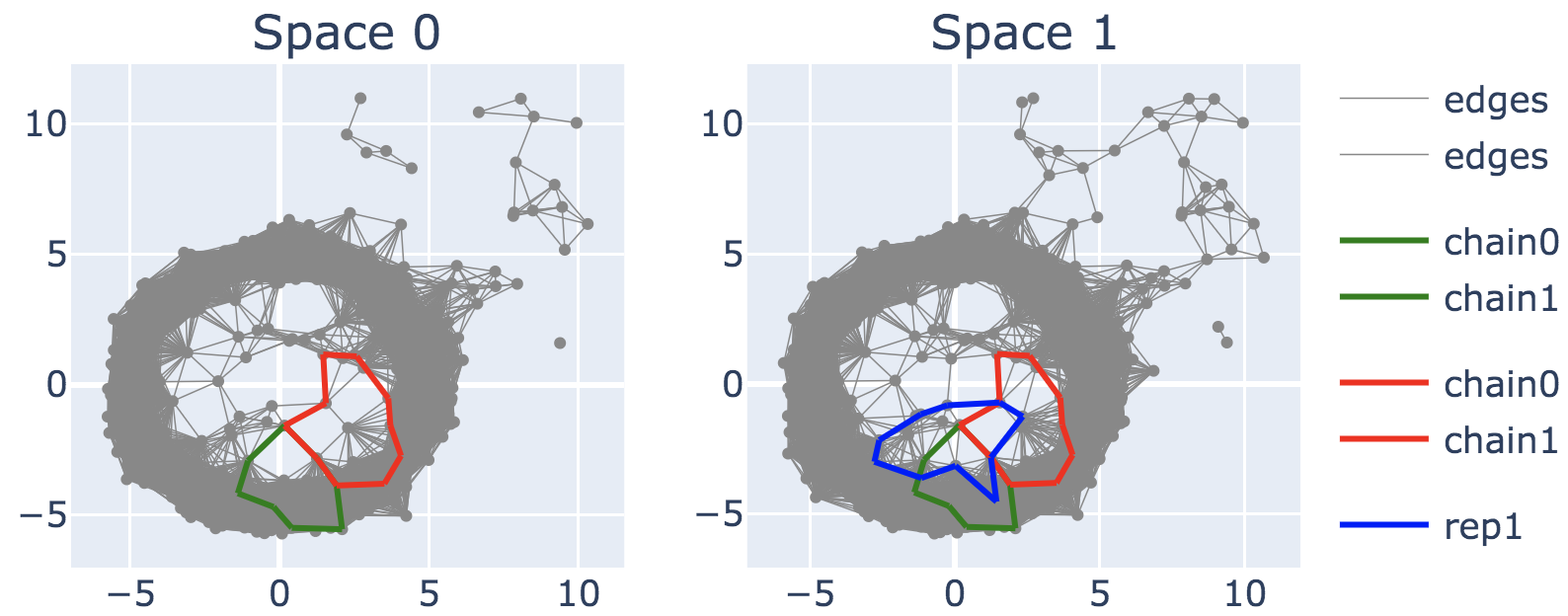}
    \includegraphics[width=0.48\linewidth]{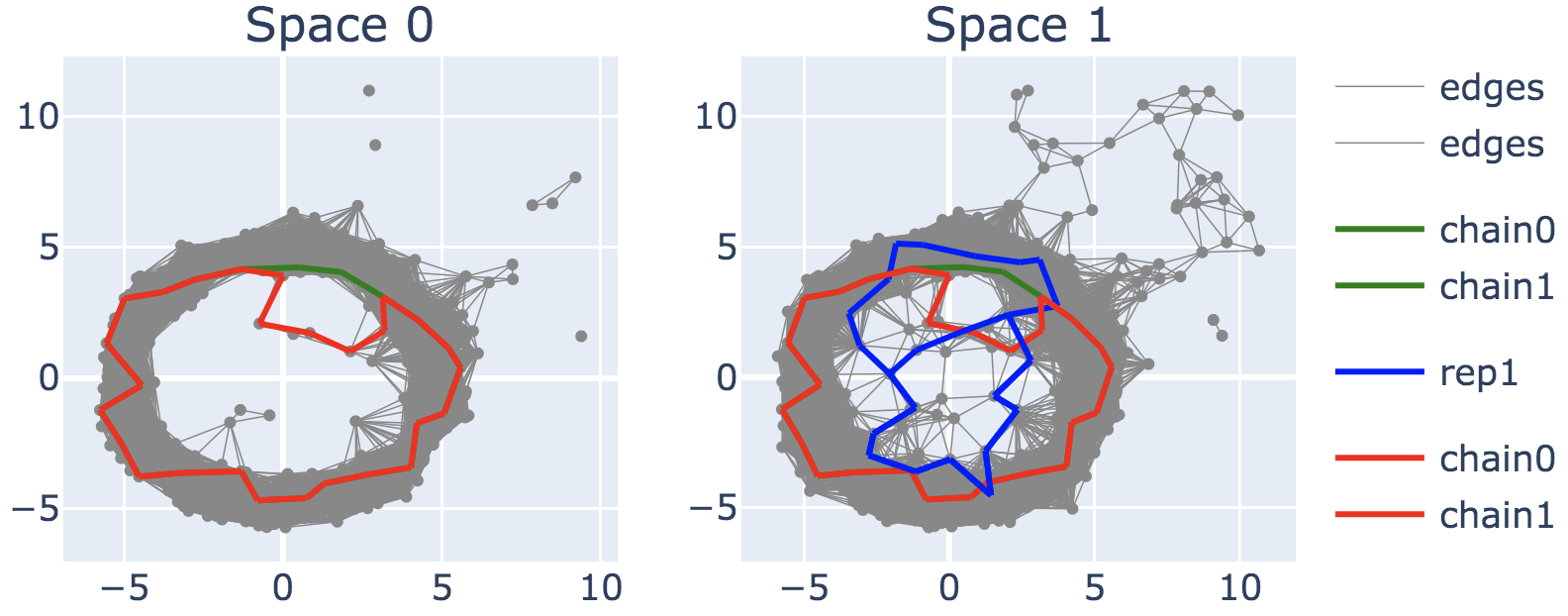}
    \caption{Induced maps of noisy Annulus. Full point cloud size is $1000$. Sample size is $800$.}
    \label{fig:annulus_800_im}
\end{figure}

%% file: 08expectedrank.tex
\begin{figure}
    \centering
    \includegraphics[width=0.32\linewidth]{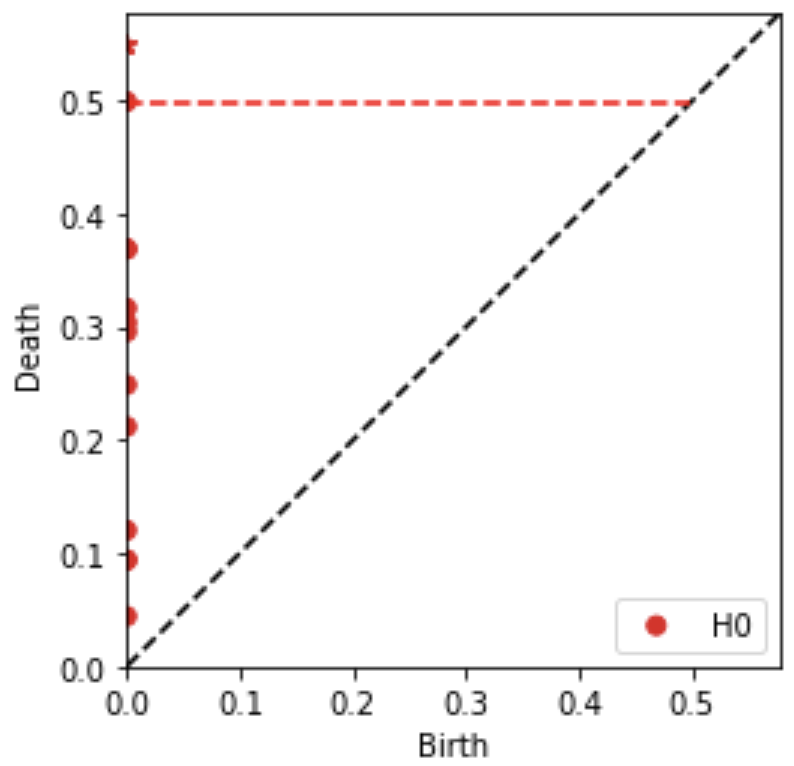}
    \includegraphics[width=0.32\linewidth]{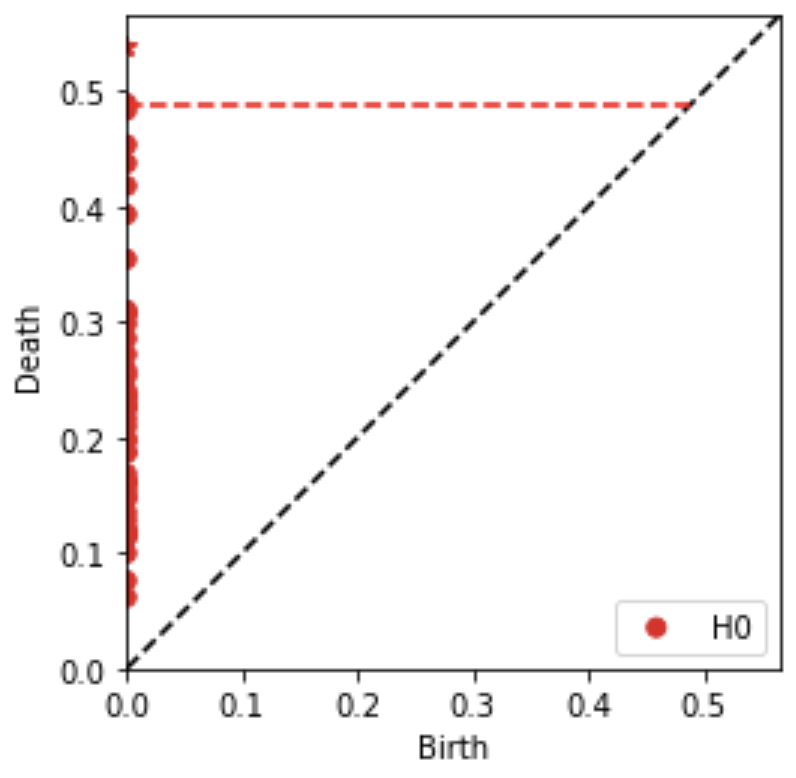}
    \includegraphics[width=0.32\linewidth]{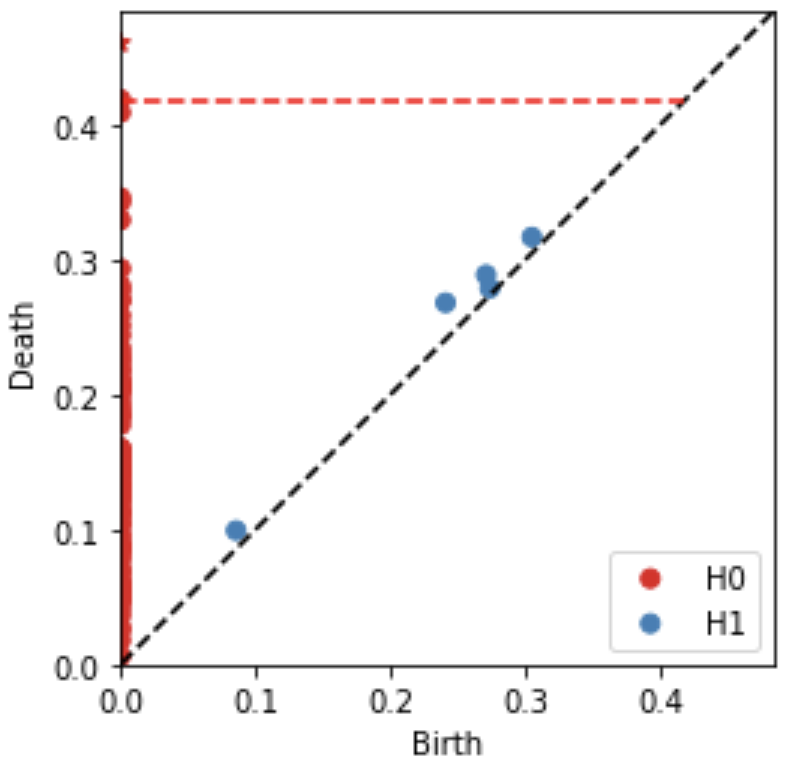}
    \includegraphics[width=0.32\linewidth]{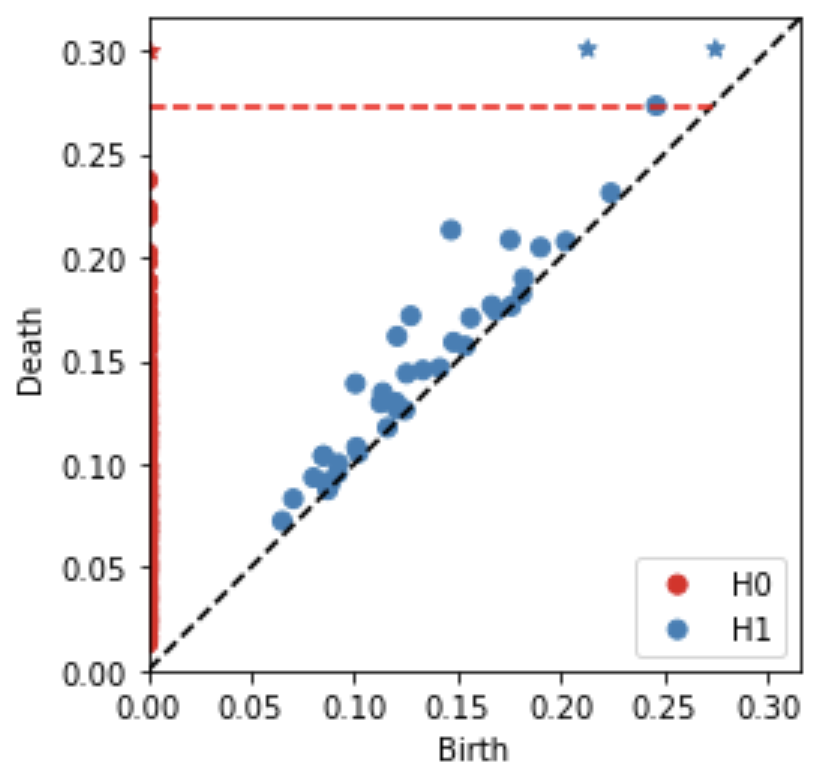}
    \includegraphics[width=0.32\linewidth]{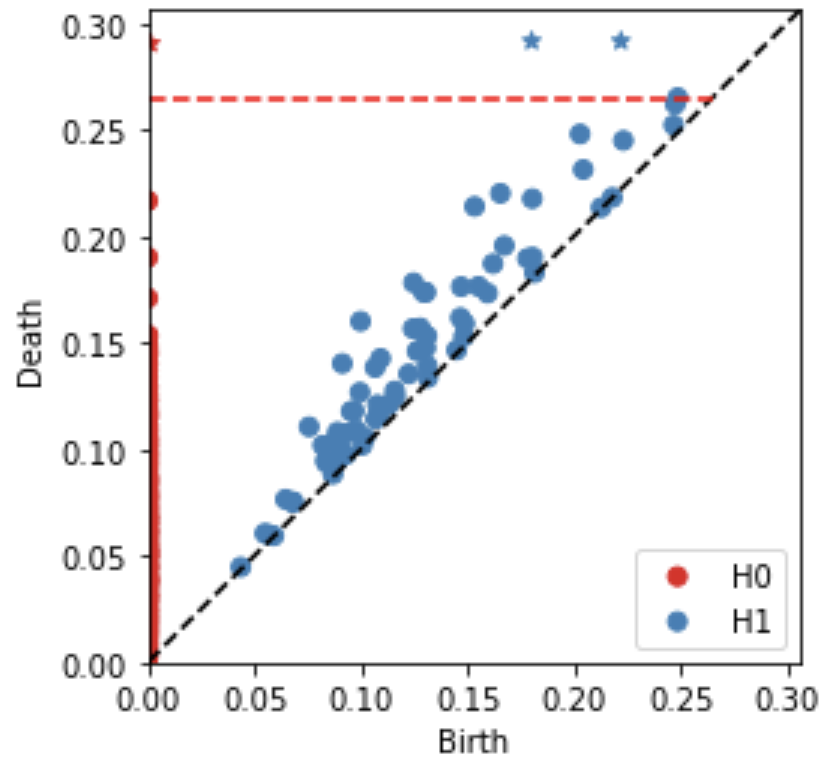}
    \includegraphics[width=0.32\linewidth]{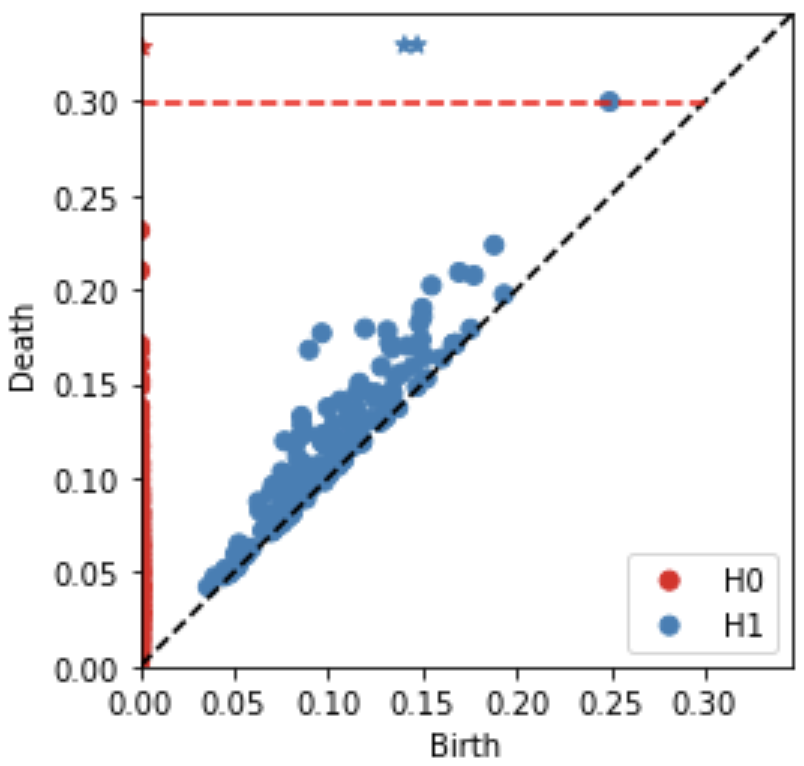}
    \caption{Vietoris-Rips persistence diagrams of Figure-8. Sub-samples' size are $20$, $50$, $100$, $300$, $500$, and $800$ from top to bottom and left to right respectively. We observe that there are two connected components captured in each sample.}
    \label{fig:fig_8_rips}
\end{figure}
\begin{figure}
    \centering
    \includegraphics[width=0.32\linewidth]{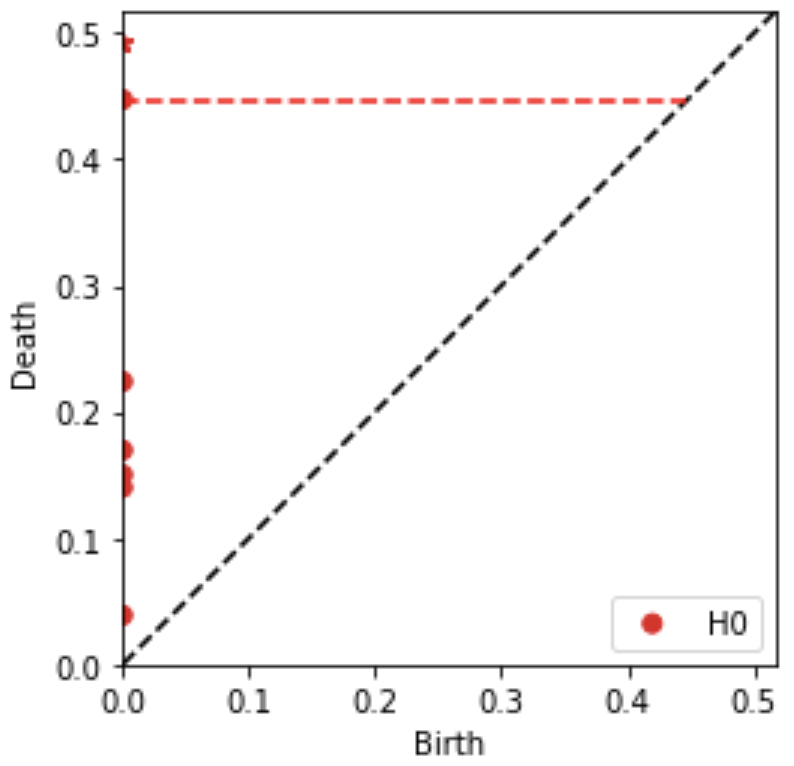}
    \includegraphics[width=0.32\linewidth]{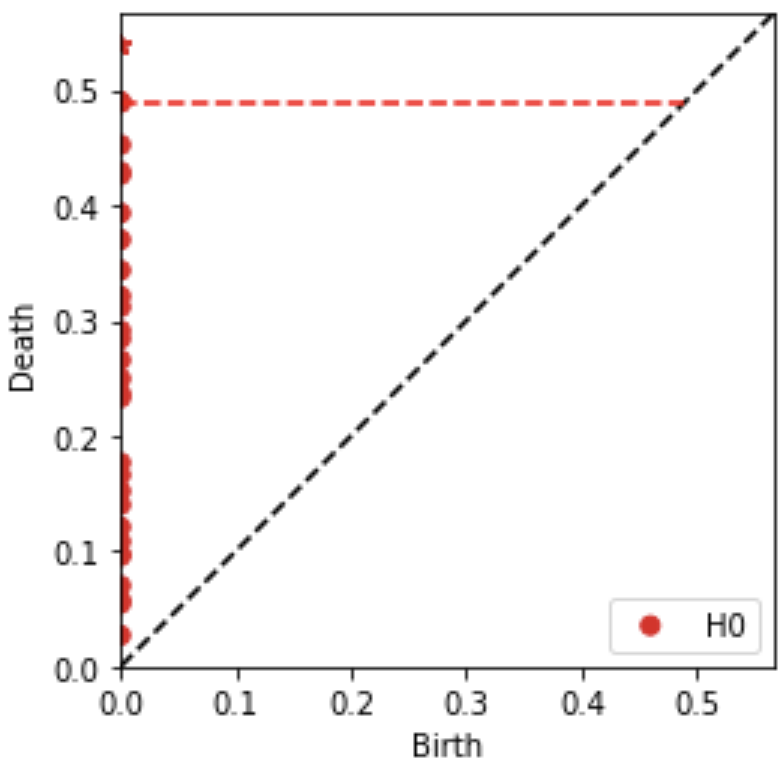}
    \includegraphics[width=0.32\linewidth]{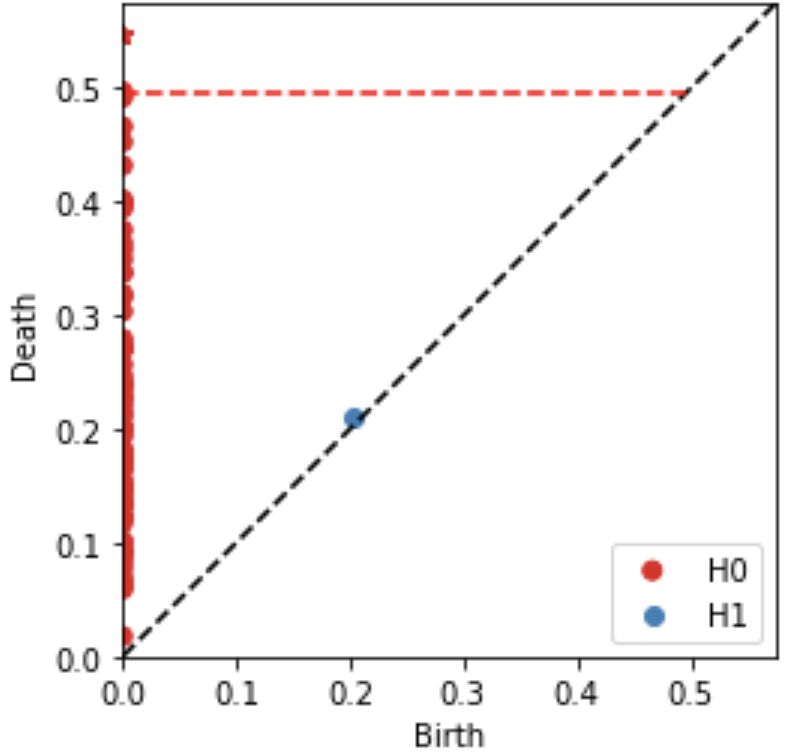}
    \includegraphics[width=0.32\linewidth]{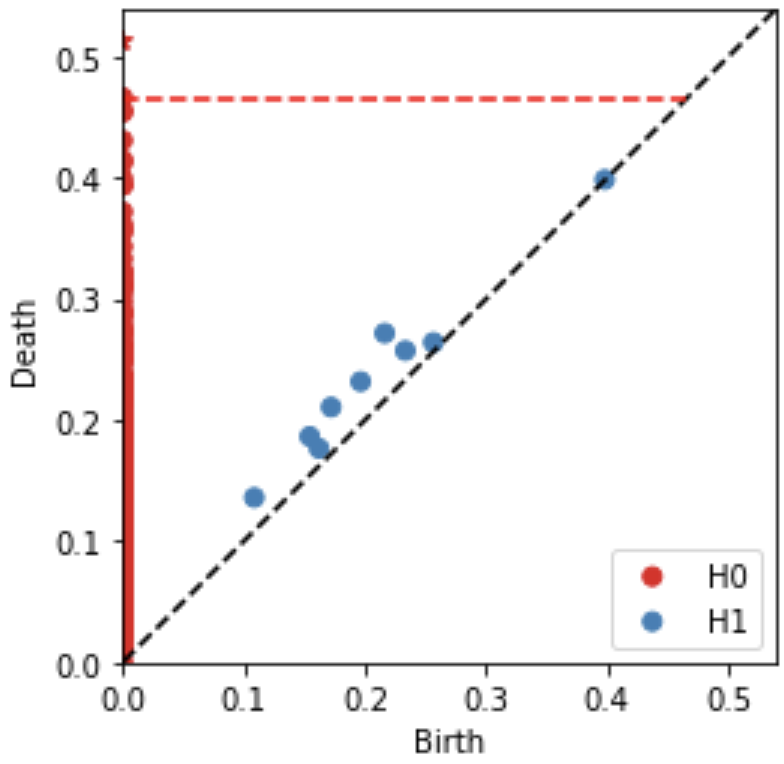}
    \includegraphics[width=0.32\linewidth]{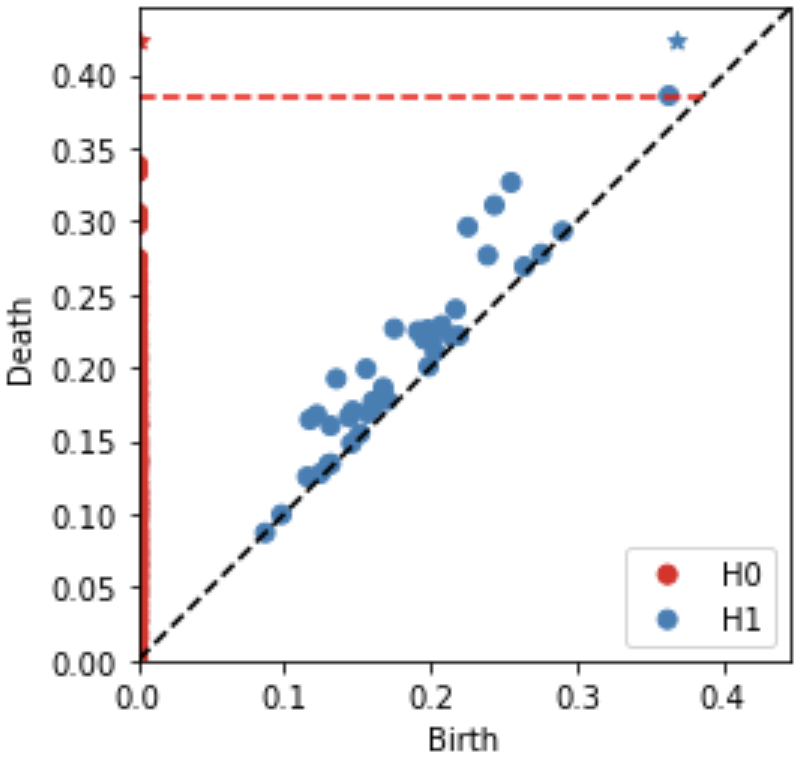}
    \includegraphics[width=0.32\linewidth]{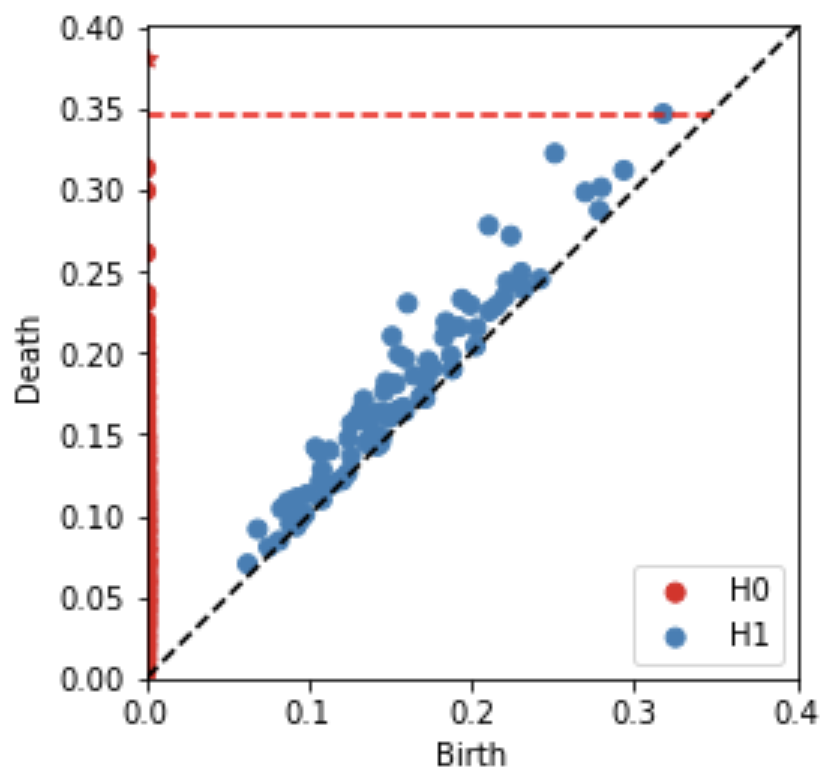}
    \caption{Vietoris-Rips persistence diagrams of Annulus. Sub-samples' size are $20$, $50$, $100$, $300$, $500$, and $800$ from left to right and top to bottom respectively. We observe that there are two connected components captured in each sample.}
    \label{fig:ann_rips}
\end{figure}